\pdfoutput=1
\documentclass[12pt,a4paper,openright]{report}
\usepackage[a4paper]{geometry}
\usepackage[italian,english]{babel}
\usepackage[latin1]{inputenc}
\usepackage[sc]{mathpazo} 
\usepackage{fancyhdr}
\usepackage{indentfirst}
\usepackage{graphicx}
\usepackage{placeins}
\usepackage{url}                                        
\usepackage{newlfont}
\usepackage{amssymb}
\usepackage{amsmath}
\usepackage{latexsym}
\usepackage{amsthm}
\usepackage{amsfonts}
\usepackage{braket}
\usepackage{axodraw4j}
\usepackage{xcolor}
\usepackage[colorlinks,
bookmarksnumbered=true,
bookmarks=true,
hyperfootnotes=true,
backref=page
]
{hyperref}
\hypersetup{
	linktocpage = true,
	linkcolor = blue,
	urlcolor = blue,
	colorlinks = teal,
	linkcolor = blue,
	anchorcolor = blue,
	citecolor = teal,
	urlcolor = teal,
	pdftitle = {PhD_Thesis_Dafne},                
	pdfauthor = {A.D. Bolognino},                 
	pdfsubject = {High-energy phenomenology},               
	pdfcreator = {A.D. Bolognino},               
	pdfproducer = {A.D. Bolognino},
	pdfstartview = {XYZ null null 1.25},
	pdfkeywords = {hep-ph, QCD, BFKL, LHC, hadronic processes, inclusive processes, DIS, rho meson, phi meson, unintegrated gluon distribution, UGD, phenomenology, physics, hep-th ,theoretical physics, theory},
                        }

\newcommand{\beq}{\begin{equation}}
\newcommand{\eeq}{\end{equation}}
\newcommand{\bea}{\begin{eqnarray}}
\newcommand{\eea}{\end{eqnarray}}
\usepackage{verbatim}
\usepackage{slashed}
\usepackage{backrefx}
 \renewcommand{\backrefpagesname}{Cited on page~}
 \renewcommand{\backrefpagesnames}{Cited on pages~}
\usepackage{psfrag}
\usepackage{epsfig,graphics}
\usepackage{siunitx}
\usepackage{cite}
\usepackage{listings}
\newcommand{\bkappa}{\mbox{\boldmath $\kappa$}}
\def\beqa{\begin{eqnarray}}
\def\eqa{\end{eqnarray}}

\newcommand{\tarr}{
	\begin{array}}
	\newcommand{\earr}{\end{array}}
\newcommand*\rfrac[2]{{}^{#1}\!/_{#2}} 
\def\beq{\begin{equation}}
\def\eq{\end{equation}}
\newcommand{\numberset}{\mathbb}
\newcommand{\PP}{\numberset{P}}
\theoremstyle{plain}                    
\theoremstyle{definition}               
\theoremstyle{remark}                   
\renewcommand{\chaptermark}[1]{\markboth{\thechapter.\ #1}{}}

\begin{document}
	
\begin{titlepage}                       
\thispagestyle{plain}
\newgeometry{left=0.5cm,right=0cm,top=0cm,bottom=0cm}
\begin{figure}[h]
	\centering
	\includegraphics[scale=0.99]{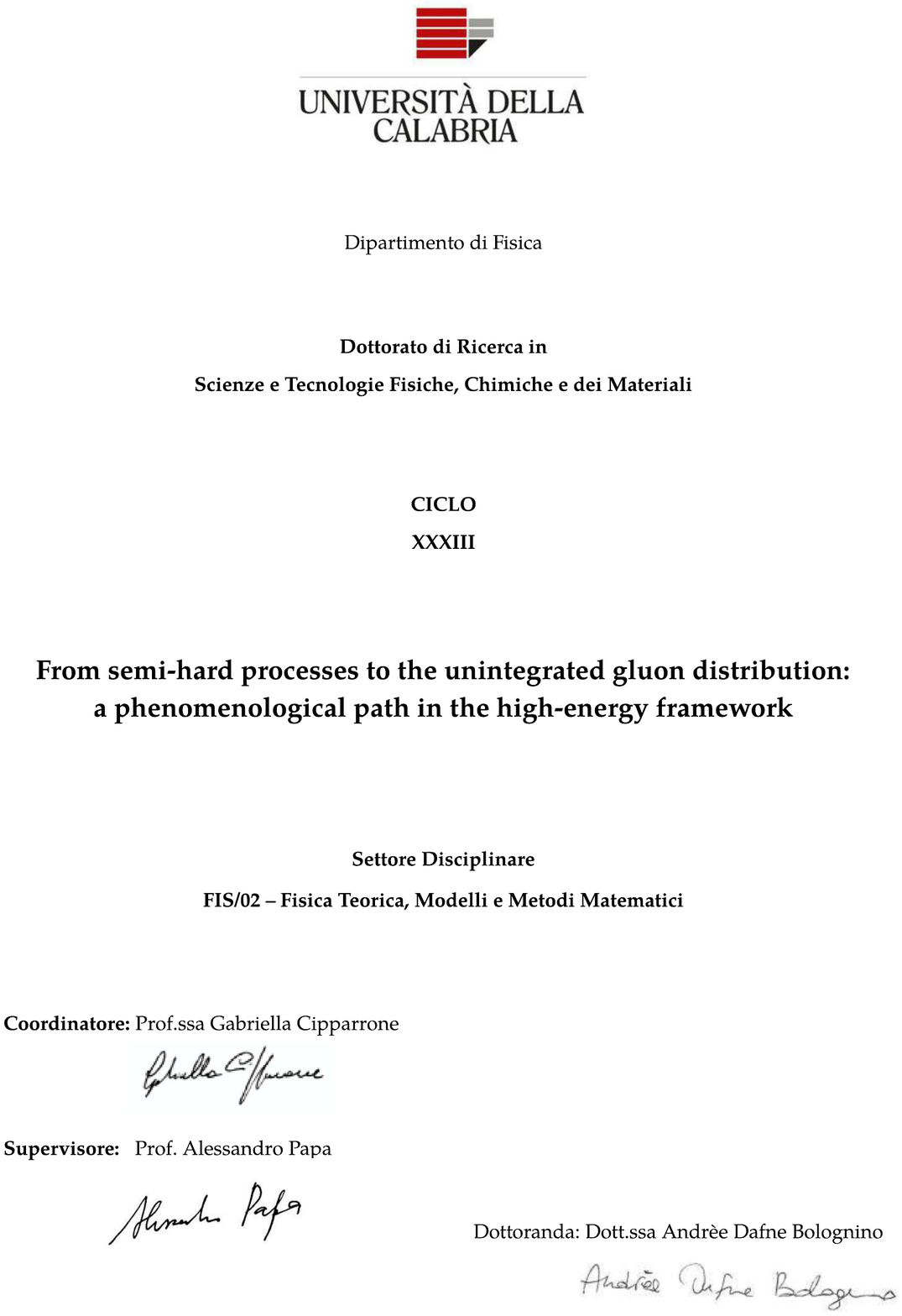}
\end{figure} 
\restoregeometry	
\end{titlepage}	

\input epsf
\pagenumbering{roman}     	

 

\newpage
\begingroup
\pdfbookmark[1]{Contents}{toc}
\rhead[\fancyplain{}{\bfseries\leftmark}]{\fancyplain{}{\bfseries\thepage}}
 \lhead[\fancyplain{}{\bfseries\thepage}]{\fancyplain{}{\bfseries
 		CONTENTS}}
 \newgeometry{left=127.464pt,right=50pt,top=102.464pt,bottom=151.536pt}
\tableofcontents
 \restoregeometry
\newpage
\pdfbookmark[1]{List of Figures}{figures}
\rhead[\fancyplain{}{\bfseries
	LIST OF FIGURES}]{\fancyplain{}{\bfseries\thepage}}
\lhead[\fancyplain{}{\bfseries\thepage}]{\fancyplain{}{\bfseries
		LIST OF FIGURES}}
\listoffigures
\endgroup
\newpage


\pagenumbering{arabic} 
\chapter*{Abstract}
\label{Abstract}
\addcontentsline{toc}{chapter}{Abstract}  
\rhead[\fancyplain{}{\bfseries
	ABSTRACT}]{\fancyplain{}{\bfseries\thepage}}
\lhead[\fancyplain{}{\bfseries\thepage}]{\fancyplain{}{\bfseries
		ABSTRACT}}

The class of semi-hard reactions represents a promising venue where to enhance our knowledge of strong interactions and deepen the aspects related to this theory in kinematical regimes so far unexplored. In particular, the high energies reached in electron-proton and in proton-proton collisions first at HERA and then at the LHC, allow us to study scattering amplitudes of hard and semi-hard processes in perturbative QCD. The structure of this thesis can be considered twofold.

On one hand, the possibility to distinguish those channels where at least two final-state particles are emitted with large separation in rapidity and with an untagged system, permits to test the BFKL dynamics as resummation energy logarithms in the $t$-channel. Indeed, these inclusive reactions occur in the Regge limit, $s \gg |t|$, and fixed-order calculations in perturbative QCD miss the effects of energy logarithms, which are so large to compensate the smallness of the strong coupling constant and must be resummed order by order in perturbation theory. The most powerful theoretical tool to provide the resummation of terms proportional to powers of these logarithms is the Balitsky-Fadin-Kuraev-Lipatov (BFKL) approach. In this framework, the hadroproduction of two forward jets with high transverse momenta separated by a large rapidity at the LHC, known as Mueller--Navelet jets process, has been one of the most investigated reactions. With the idea of deepening our understanding of the BFKL formalism, a new channel belonging to the semi-hard processes category is proposed: the inclusive production of a light charged hadron and a
jet with high transverse momentum widely separated in rapidity, whose calculation is performed at NLO accuracy. The importance of this process relies in the possibility to probe a complementary region to one analyzed for the Mueller--Navelet jets. Hadrons, indeed, can be tagged at much smaller values of the transverse momentum than jets. It is proved how asymmetric cuts for the transverse momentum (naturally occurring because
the final state is featured by objects of different nature) enhance the BFKL effects and how it is possible to discriminate between different parametrizations of fragmentation functions for the
hadron in the final state. Additionally, another reaction is proposed: the hadroproduction of heavy-quark pairs well separated in rapidity.

 On the other hand, also the class of processes featured by the detection of a single forward object in lepton-proton collision can provide us useful ingredients to develop intriguing phenomenological studies. In particular, the exclusive leptoproduction of light vector mesons, $\rho$ and $\phi$, is exhaustively investigated. In this context, the study of helicity-dependent observables allows us to discriminate among several unintegrated gluon distribution models, whose original definition naturally encodes the BFKL-equation evolution dynamics. This kind of parton density allows us to get access to the hadronic structure at small-$x$.
\newpage

\begin{otherlanguage*}{italian}

\chapter*{Sintesi in lingua italiana}
\addcontentsline{toc}{chapter}{Sintesi in lingua italiana}  

\rhead[\fancyplain{}{\bfseries
	SINTESI IN LINGUA ITALIANA}]{\fancyplain{}{\bfseries\thepage}}
\lhead[\fancyplain{}{\bfseries\thepage}]{\fancyplain{}{\bfseries
		SINTESI IN LINGUA ITALIANA}}
La classe di processi semiduri rappresenta un promettente ambito di ricerca per l'estensione della conoscenza delle interazioni forti in regimi cinematici ad oggi inesplorati. In particolare, le alte energie ottenute nelle collisioni leptone-protone e protone-protone, prima a HERA e poi a LHC, consentono lo studio di ampiezze di diffusione di processi duri e semiduri in teoria di Cromodinamica Quantistica (QCD) perturbativa. La struttura di questa tesi è da considerarsi duplice.

Da un lato, la possibilità di distinguere quei processi in cui nello stato finale sono emesse almeno due particelle, caratterizzate da un'ampia separazione in rapidità e da un sistema non rilevato, permette di testare la dinamica BFKL come risommazione dei logaritmi dell'energia nel canale $t$. Tali reazioni inclusive si verificano, infatti, nel \emph{limite di Regge}, $s \gg |t|$, ed i calcoli ad ordine fissato in QCD perturbativa non possono tener conto dell'effetto
non trascurabile dei logaritmi dell'energia, il cui contributo è tale da compensare la piccolezza della costante d'accoppiamento della QCD che, pertanto, devono essere risommati a tutti gli ordini in teoria perturbativa. Lo strumento teorico più potente, in grado di fornire la risommazione dei termini proporzionali alle potenze di questi logaritmi, è l'approccio Balitsky-Fadin-Kuraev-Lipatov (BFKL). In questo contesto, l'adroproduzione inclusiva di due jet ``in avanti" caratterizzati da un elevato valore del momento trasverso e ampiamente separati in rapidità ad LHC, conosciuta come produzione di jet di Mueller--Navelet, è uno dei processi ad oggi maggiormente studiati. Perseguendo lo scopo di approfondire la comprensione del formalismo BFKL, si propone l'analisi di un nuovo canale di produzione appartenente alla categoria dei processi semiduri: la produzione inclusiva di un adrone leggero carico e un jet, entrambi caratterizzati da alto momento trasverso e fortemente separati in rapidità, le cui predizioni teoriche sono fornite all'ordine sottodominante (NLO). L'importanza di tale processo risiede nella possibilità di sondare una regione cinematica complementare rispetto a quella investigata per i jet di Mueller--Navelet. Gli adroni, infatti, possono essere rilevati a valori del momento trasverso inferiori rispetto a quelli dei jet. Si prova come i tagli asimmetrici per il momento trasverso (naturalmente presenti poiché lo stato finale è caratterizzato da oggetti diversi) amplifichino gli effetti di BFKL, e come sia possibile discriminare tra differenti parametrizzazioni delle funzioni di frammentazione per l'adrone prodotto nello stato finale. Si propone, inoltre, un nuovo processo: l'adroproduzione di coppie di quark pesanti fortemente separati in rapidità.

D'altra parte, anche la classe di processi che presentano un singolo oggetto ``in avanti" prodotto nelle collisioni leptone-protone contribuiscono a fornire ingredienti utili per poter sviluppare interessanti studi fenomenologici. In particolare la leptoproduzione esclusiva di mesoni vettori leggeri, $\rho$ e $\phi$, è investigata in modo esaustivo. In tale contesto, lo studio di osservabili dipendenti dall'elicità consente di discriminare tra alcuni modelli di distribuzione non integrata del gluone, la cui definizione originale racchiude la dinamica dell'evoluzione dell'equazione BFKL in modo naturale. Questa tipologia di densità partonica è rilevante per lo studio della struttura adronica nel limite di piccolo~$x$.

\end{otherlanguage*}

\chapter*{Introduction}
\addcontentsline{toc}{chapter}{Introduction}  
\rhead[\fancyplain{}{\bfseries
	INTRODUCTION}]{\fancyplain{}{\bfseries\thepage}}
\lhead[\fancyplain{}{\bfseries\thepage}]{\fancyplain{}{\bfseries
		INTRODUCTION}}

The Standard Model (SM) of elementary particles, based on the existence of fermionic fundamental constituents and their interaction mediated through the exchange of vector bosons, is one of the most relevant and complex theories of our time. One of the core pillars of the SM is the Quantum Chromodynamics (QCD), the universally accepted theory of strong interactions physics, which describes \emph{how} the fermionic \emph{quarks} and the mediator bosonic \emph{gluons}, the elementary constituents of hadrons, interact with each other. The strong interacting particles are endowed with the so-called color charge and the simmetry group which allows to describe the possible states for quarks and gluons is the $SU(3)$ group. QCD is thus based on a non-Abelian group, which provides the fundamental and challenging features of the theory: the \emph{asymptotic freedom}, according to which the partons behave as point-like and non-interacting particles at short distances, and the \emph{confinement}, which states that quarks and gluons cannot be isolated at large distances and are therefore bounded into hadrons. In the short distance (or high-energy) regime, the perturbative approach is a valid description tool.
Conversely, at large distance, in the regime where confinement takes place, the perturbative technique is unapplicable. The duality between perturbative and non-perturbative aspects underlies the complexity of QCD. A way to investigate strong interactions in the unexplored kinematical region of large center-of-mass energy is given by a particular wide class of collision processes, called \emph{diffractive semi-hard} reactions. These latter are processes featured by the scale hierarchy, $s \gg Q^2 \gg \Lambda^2_{QCD}$, where $s$ is the center-of-mass energy squared, $Q$ indicates one of the hard scale characteristic of the process and $\Lambda_{QCD}$ is the QCD scale. In the kinematical regime, known as \emph{Regge limit}, where $s$ is larger than the $t$-channel Mandelstam variable, fixed order calculations in perturbative QCD are unable to take into account the effects due to the large energy logarithms, occurring in the perturbative series order by order with increasing powers, which compensate the smallness of the coupling $\alpha_s$. So far the Balitsky-Fadin-Kuraev-Lipatov (BFKL) approach~\cite{Fadin,elf,elf2,BL} has been the most powerful theoretical tool in the hadronic sector to provide the resummation of terms proportional to powers of these large energy logarithms. This method permits the resummation of all contributions both in the so called leading logarithmic approximation (LLA), \emph{i.e.} those terms proportional to $(\alpha_s \, \ln(s))^n$ and also in the next-to-leading logarithmic approximation (NLA), \emph{i.e.} those terms proportional to $\alpha_s(\alpha_s \, \ln(s))^n$. The BFKL formalism allows us to express the cross section of a hadronic process through the convolution between two impact factors, which depend on the specific reaction and represent the transition from a colliding parton to the respective object in the final state, and the Green's function, which has universal validity and is process independent.
The most popular reaction, studied via the BFKL approach to investigate semi-hard collisions at a hadron collider, is the inclusive production of Mueller--Navelet jets, \emph{i.e.} two jets with transverse momenta much larger than the mass scale $\Lambda^2_{QCD}$ and widely separated in rapidity. The Mueller--Navelet jets process~\cite{Mueller:1986ey} can be considered a challenging reaction due to the coexistence of two main resummations, the fixed-order DGLAP and the BFKL ones, in the framework of perturbative QCD. Therefore, jet-impact factors have been exhaustively used to describe the inclusive production of two
jets tagged with high transverse momenta and large separation in rapidity. Consequently, numerous analyses dedicated to NLA predictions for this process represent the fundamental step to collect essential information. With the aim of deepening our understanding of the dynamical mechanism encoding partonic interactions in the high-energy limit, a wide class of semi-hard reactions has been presented so far, such as: $J/\Psi$-jet~\cite{Boussarie:2017oae} and Higgs-jet processes~\cite{Celiberto:2020tmb}, quarkonium-state ($J/\Psi$ or $\Upsilon$) photoproduction~\cite{Bautista:2016xnp,Garcia:2019tne} in the NLA as well as the forward Drell--Yan dilepton production~\cite{Celiberto:2018muu,Golec-Biernat:2018kem} with a backward jet in the LLA.
In order to enrich this collection, it is interesting to present less inclusive channels than the Mueller--Navelet jets, which can be proposed and probed in the future LHC analyses. For this reason, following the NLA analysis of the inclusive production of two identified hadrons~\cite{Celiberto:2016hae,Celiberto:2016zgb,Celiberto:2017ptm}, the forward-hadron impact factors allows us to provide with predictions also for the inclusive production of a hybrid process: an identified light charged hadron and a jet with high transverse momentum and largely separated in rapidity.
Theoretical predictions will be provided at 7 and 13 TeV, in the NLA, in the typical kinematics of the LHC, particularly taking into account the configurations due to CMS and CASTOR detectors. Phenomenological results will emphasize how asymmetric cuts for the transverse momentum, naturally occurring because the final state is featured by objects of different nature, enhance the BFKL effects. Moreover, this channel serves as a testing ground to discriminate between different parametrizations of parton distribution functions (PDFs) for the initial proton and parton fragmentation functions (FFs) for the hadron in the final state.\\
Another proposal to continue the investigation of BFKL dynamics,
considering a new semi-hard channel, is represented by the inclusive production of forward heavy-quark pair separated in rapidity in the collision of two protons (hadroproduction). In Ref.~\cite{Celiberto:2017nyx} a process with the same final state was considered, but produced via the collision of two (quasi-)real photons (photoproduction) emitted by two interacting electron and positron beams according to the
equivalent-photon approximation (EPA). In a similar way to the photoproduction, predictions for cross section and azimuthal coefficients of the expansion in the cosine of the relative angle in the transverse plane between the directions of the two tagged heavy quarks will be provided. It is worth to note that the results related to the hadroproduction, in the same kinematical conditions of the photoproduction, reveal higher cross section values.\\
The channels presented until now represent all those reactions where at least two final-state particles are always emitted with large mutual separation in rapidity and with an undetected system. However, another challenging semi-hard processes category is provided by those final states characterizing the detection of a single forward object in lepton-proton or proton-proton scatterings.
This kind of configurations gives us the chance to define the \emph{unintegrated gluon distribution} in the proton (UGD), in the context of the high-energy factorization, also known as $\kappa$-factorization. This scheme was originally developed in Ref.~\cite{Catani:1990eg}, which holds in high-energy limit where amplitudes and/or cross sections are factorized into a convolution among off-shell matrix elements and transverse-momentum-dependent PDFs. Hence, it is clear the relation with the BFKL approach: the matrix element corresponds to the forward impact factor describing
the emission of the final-state particle, while the parton content, governed by gluon evolution, is encoded by the
UGD. The standard definition of the UGD is thus given in terms of a convolution between the BFKL gluon Green's function, hence including the BFKL dynamics, and the proton impact factor. Because of the non-perturbative nature of the UGD, this type of parton density has not a universal formulation, thus in literature several parametrizations can be developed.\\In this thesis we will present studies of the single exclusive production of two kinds of light vector mesons, $\rho$ and $\phi$, as interesting testfields to discriminate among several different UGD models. In particular, we will focus on helicity-dependent observables for the forward polarized $\rho$-meson electroproduction~\footnote{Among the main subjects of investigation on the diffractive production of $\rho$ mesons in the high energy factorization, we remind Refs.~\cite{Anikin:2009bf,Anikin:2011sa,Besse:2012ia,Besse:2013muy}.} in order to constrain the $\kappa$-dependence of the UGD, as well as to provide the most suitable choice of UGD model to describe HERA data. The study of polarized cross sections will be extended to the emission of the single forward $\phi$-meson, obtaining improved predictions via the inclusion of the strange-quark mass and calculating the $\gamma^* \rightarrow \phi$ impact factor endowed with the quark mass, in the light-cone wave-function (LCWF) approach.\\
This thesis is focused on a phenomenological overview of semi-hard reactions in the high-energy regime. Its structure is twofold. Actually, two fundamental aspects occur, which allow us to classify semi-hard processes. On one side our attention is devoted to inclusive forward/backward productions, which permit us to investigate the BFKL dynamics via a hybrid factorization obtained through the simultaneous presence of collinear PDFs and high-energy resummation in the $t$-channel. On the other side we propose single forward emissions, which serve as testing ground for the study of parton densities. 
Both channels play a key role: they give us the possibility to conduct stringent tests of the strong interactions in the high-energy limit and of the BFKL resummation. The BFKL approach, in addition to the previous discussions, can be considered a powerful technique to investigate the proton structure at small-$x$. Indeed, it gave access to the definition and to the study of the  UGD, as well as for improving the description of collinear PDFs at NLO and next-to-NLO (NNLO) through the inclusion of NLA resummation effects~\cite{Ball:2017otu,Abdolmaleki:2018jln,Bonvini:2019wxf}, including the possibility to predict the small-$x$ behavior of trasverse momentun dependent (TMD) gluon distributions~\cite{Bacchetta:2020vty}.

 \begin{center}
 	***
 \end{center}
 This thesis is organized as follows: in Chapter~\ref{Chap:BFKL} we provide a brief discussion on the BFKL formalism, while Chapter~\ref{Chap:Incl_emiss} presents a useful overview on Mueller--Navelet jets results and relative theoretical aspects, relevant to motivate the pursuing of phenomenological predictions, obtained at LHC energies, of new inclusive semi-hard processes, which are considered in the next subsections. In particular, we propose NLA accuracy results for the cross section and azimuthal correlations of jet-hadron production process in Sec.~\ref{jethad}, using different PDF and FF parametrizations. The heavy-quark pair production separated in rapidity in the collision of two quasi real gluons coming from two protons is investigated in Sec.~\ref{heavy}. In Chapter~\ref{chap3} we investigate on the UGD, giving theoretical results in comparison with HERA data for helicity-amplitude ratio and polarized cross sections for the $\rho$-meson leptoproduction. With the interest to extend the study of UGD models, predictions tailored of quark-mass effects are presented for observables describing the $\phi$-meson leptoproduction process.

\clearpage{\pagestyle{empty}\cleardoublepage}
\chapter{A glance on the BFKL resummation}  
\label{Chap:BFKL}              
\lhead[\fancyplain{}{\bfseries\thepage}]{\fancyplain{}{\bfseries\rightmark}}
\section{The Regge theory}
In 1959 the Italian physicist T. Regge~\cite{Regge} proved that it is relevant to regard the angular momentum, $l$, as a complex variable, when considering solutions of the Schr\"odinger equation for non-relativistic potential scattering. He demonstrated that, for a wide class of potentials, the singularities of the scattering amplitude in the complex $l$-plane are poles, now known as "Regge poles"~\cite{Collins,Predazzi}. If these poles appear for integer values of $l$, they represent bound states or resonances and are crucial in determining analytic properties of the amplitudes. The presence of the poles is ruled by the following relation
\begin{equation}
\label{Regg}
	l = \alpha(k)\,,
\end{equation}
where $\alpha(k)$ is a function of the energy, known as Regge trajectory or Reggeon.\\
A single trajectory described by \eqref{Regg} is referred to specific class of bound states and resonances. The energies of these states come from~\eqref{Regg} giving physical integer values to the angular momentum $l$. The application of the Regge's method to the high-energy particle physics was mainly due to Chew and Frautschi~\cite{ChewandF} and Gribov~\cite{Gribov0}, but many other physicists developed and deepened this theory and its applications.\\
Exploiting the properties of $S$-matrix, it is possible to analytically continue the relativistic partial wave amplitude $\cal{A}_l(t)$ to complex $l$ values in a unique way. The resulting function, $\cal{A}(l, t)$, has simple poles at
\begin{equation}
  l = \alpha(t)\,.
\end{equation}
The Regge theory predicts that poles contribute to the scattering amplitude with terms which asymptotically behave (\emph{i.e.} for $s\rightarrow\infty$ and $t$ fixed) as
\begin{equation}
\label{simpleRegge}
	\cal{A}(l, t) \sim s^{\alpha(t)}\,,
\end{equation}
where $s$ and $-t$ are the squared of the center-of-mass energy and of the momentum transfer, respectively.
The leading singularity in the $t$-channel, which is the one with the largest real part, leads to the asymptotic behavior of the scattering amplitude in the $s$-channel.
Although the extremely easy form, the Regge theory was a successful approach able to accurately describe a large variety of high-energy processes through simple predictions as \eqref{simpleRegge}.
\subsection{The ``soft" Pomeron}
Regge theory is one of the so-called $t$-channel models, which describe hadro-nic processes in terms of the exchange of a \emph{``virtual" particle} in the $t$-channel. In the Regge theory the role of the virtual particle is played by the Reggeon, which represents a whole family of resonances, instead of a single particle. In the limit of high $s$, a hadronic process is ruled by one or more Reggeons in the $t$-channel.\\
The exchange of Reggeons instead of particles leads to scattering amplitudes of the type given by Eq.~\eqref{simpleRegge}. The latter, through the \emph{optical theorem}~\cite{Newton}, allows us to write the total cross section $\sigma_{tot}$ in the Regge theory as
\begin{equation}
\label{tot_sigma_regge}
\sigma_{tot} \simeq \frac{\text{Im}\,\cal{A}(s, t = 0)}{s} \simeq s^{\,\alpha(0) - 1}\,.
\end{equation}
It is known, from the experimental world, that hadronic total cross sections, as a function of $s$, do not vanish asymptotically, but they are rather flat in the range of $\sqrt{s} \simeq (10\div 20)$ GeV and increase slowly at higher-energy values. Attributing this rise to the exchange of a single Regge pole, it follows that the intercepts $\alpha(0)$ of the exchanged Reggeon is greater than one, giving the power growth with energy of the cross section in Eq.~\eqref{tot_sigma_regge}.
This exchanged Reggeon, which represents the leading trajectory in the elastic and diffractive processes, carries the quantum numbers of the vacuum in the $t$-channel and it is called Pomeron, from I.Ya. Pomeranchuk.\\ Nowadays, the Pomeron trajectory does not correspond to any identified physical particle, but in QCD the possible candidates could be bound states of gluons, named \emph{glueballs}.\\
It is clear that the power growth of the cross section in Eq.~\eqref{tot_sigma_regge} violates the Froissart constraint if $\alpha(0) > 1$, consequently also the unitarity. This means that hadronic total cross sections at the energies considered so far have not yet reached the asymptotic regime. In order to remove the violation, the unitarity has to be restored using dedicated techniques.
\section{The BFKL approach}
The BFKL equation~\cite{Fadin, elf, elf2, BL} is an integral equation that determines the behavior at high energy $\sqrt{s}$ of the perturbative QCD amplitudes in which vacuum quantum numbers are
exchanged in the $t$-channel. 
The approach, developed by Balitsky, Fadin, Kuraev and Lipatov, was derived in the LLA (\emph{i.e.} leading logarithmic approximation), which means taking into account all terms of the type $(\alpha_s\ln s)^n$. It allows us to calculate the total cross section $\sigma_{tot}^{LLA}$, which grows at large center-of-mass energy values as
\begin{equation}
\label{BFKLappr}
\sigma_{tot}^{LLA} \sim \frac{\omega_B^P}{\sqrt{\ln s}}\,,
\end{equation}
where $\omega_B^P = \frac{(g^2\,N\,\ln(2))}{\pi^2}$ is the LLA position of the rightmost singularity in the complex momentum plane of the $t$-channel partial wave with vacuum quantum numbers.
The BFKL equation holds also in the next-to-leading logarithmic approximation (NLA), where  all the terms of the type $\alpha_s(\alpha_s\ln s)^n$ has to be resummed in the perturbative series.\\
The prediction given by Eq.~\eqref{BFKLappr} became popular when the rising trend of $\gamma^*p$ cross section at increasing energy was experimentally observed at HERA.\\ Therefore this equation is
usually associated with the evolution of the unintegrated gluon distribution. The evolution of the parton distribution functions (PDF) with $\tau = \ln(Q^2/\Lambda^2_{QCD})$ is ruled DGLAP equations~\cite{Gribov, Gribov2, Lipatov, Altarelli1977, Dokshitzer}, which allow us to resum to all orders collinear logarithms $\ln Q^2$, taken from the region of small angles between parton momenta.\\
The BFKL approach provides the description of QCD scattering amplitudes in the so called \emph{Regge limit}, which means small $x$, large $s$, and fixed momentum transfer $t$, such that $s \gg t$. In this framework the unintegrated gluon distribution turns out to be a peculiar result for the imaginary part of the forward scattering amplitude. This approach was built for the description of processes characterized by just one hard scale, for instance $\gamma^*\gamma^*$ scattering with the photon virtualities of the same order, for which the DGLAP evolution is unsuitable. In order to derive the BFKL equation, the \emph{gluon Reggeization} is the building block of the BFKL formalism and it can be represented as the appearance of a modified propagator of the form~\cite{BaronePredazzi}
\begin{equation}
\label{ReggPropag}
D_{\mu\nu}(s,\,q^2) = -i\frac{g_{\mu\nu}}{q^2}\left(\frac{s}{s_0}\right)^{\alpha_g(q^2) - 1} \quad \quad \text{(Feynman gauge)}\,,
\end{equation}
where $\alpha_g(q^2) = 1 + \epsilon(q^2)$ is the Regge trajectory of the gluon.
\subsection{Gluon Reggeization}
The idea of Reggeization of an elementary particle endowed with spin $j_0$ and mass $m$ was first proposed by Gell-Mann \emph{et al.}~\cite{GellMann} and Polkinghorne~\cite{Polki}. It means that, in the Regge limit, a factor $s^{j(t) - j_0}$, with $j_0 \equiv j(m^2)$ occurs in Born amplitudes for a process involving the exchange of this particle in the $t$-channel. This phenomenon was originally discovered in the context of the QED, through the backward Compton scattering~\cite{GellMann}. The name Reggeization is due to the fact that the form of amplitudes is given by the Regge poles (moving poles in the complex angular momentum $j$-plane~\cite{Regge}). Although in QED just the electron reggeizes in perturbation theory~\cite{GellMann}, while the photon remains elementary~\cite{Mandel}, in the QCD both the gluon~\cite{Fadin, elf,Grisaru:1973vw,Grisaru:1974cf,Lipatov:1976zz} and the quark indeed reggeize~\cite{Fadin:1976nw,Fadin:1977jr,Bogdan:2002sr,Kotsky:2002aq}. Hence the QCD is the unique theory where all elementary particles reggeize.\\ 
The Reggeization is the basis for the theoretical description of high-energy reactions with fixed momentum transfer. In particular the gluon Reggeization, since cross sections non-decreasing with energy are given by gluon exchanges, determines the form of QCD amplitudes at large energies and limited transverse momenta.
\begin{figure}[t]
	\centering
	\includegraphics[scale=1.00]{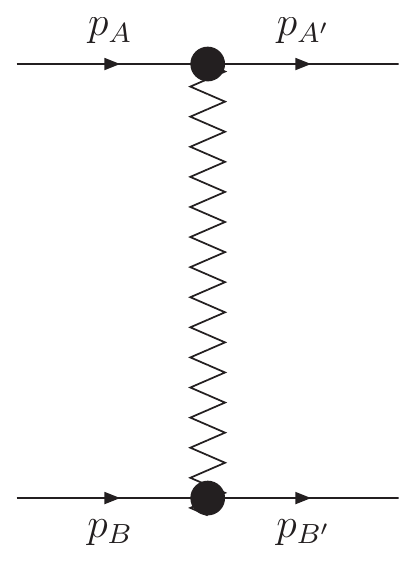}
	\caption
	{\emph{Diagrammatical representation of the process 
		$A + B \to A^\prime + B^\prime$ with
		color octet in the $t$-channel. The waggle line represents the Reggeized gluon exchange.}}
	\label{fig:regge_gluon}
\end{figure}
 Considering the elastic scattering process $A + B \rightarrow A' + B'$ with $s \gg |t|$, where
\begin{equation}
\label{Mandelstam}
s = (p_A + p_B)^2\,, \qquad t = q^2\,, \qquad q = p_A - p_A'\,,
\end{equation}
and in which the amplitudes with the gluon quantum numbers in the $t$-channel have the factorized form,
\begin{equation}
\label{octet_amplitude}
\left(\mathcal{A}_8\right)^{A^\prime B^\prime}_{AB} =
\Gamma^c_{A^\prime  A}\frac{s}{t} 
\left[\left(\left(\frac{s}{-t}\right)^{\omega(t)} + 
\left(\frac{-s}{-t}\right)^{\omega(t)}
\right)\right]
\Gamma^c_{B^\prime B} \; ,
\end{equation}
is possible to realize how the gluon Reggeization occurs.\\
In the Eq.~\eqref{octet_amplitude}, $\omega(t)$ is the gluon trajectory, $c$ is the color index, and $\Gamma^c_{B^\prime B}$ are the
particle-particle-Reggeon (PPR) vertices which are independent of $s$.\\It is easy to observe that the amplitude in Eq.~\eqref{octet_amplitude} is exactly of the form shown in Fig.~\ref{fig:regge_gluon} because there are two
couplings of the particles with the Reggeon and a contribution which behaves as $s^{\alpha(t)}$ . The second term between square brackets comes from the contribution of the $u$-channel (here $u \approx - s$).\\
The PPR vertex can be written as
\begin{equation}
\label{PPRvertex}
 \Gamma^C_{P^\prime P}=g_s\left\langle P^\prime|T^c|P \right\rangle\Gamma_{P^\prime P}\,, 
\end{equation}
where $\left\langle P^\prime|T^c|P \right\rangle$ is the matrix element of the color group generator in the corresponding
representation and $g_s$ is the QCD coupling. In the LLA, the helicity $\lambda_p$ of the scattered particle $P$ is a conserved quantity, so that $\Gamma^{(0)}_{P^\prime P}$ is given by $\delta_{\lambda_{P^\prime}\lambda_P}$ and the Reggeized gluon trajectory, performed with 1-loop accuracy, has the form
\begin{equation}
\label{trajectory_1-loop}
\omega(t) \simeq \omega^{(1)}(t) = 
\frac{g_s^2 t}{(2\pi)^{(D-1)}} \frac{N_c}{2} 
\int \frac{d^{D-2} k_\bot}
{k_\bot^2 (q-k_\bot)^2_\bot}  =
-\frac{g_s^2 N_c \Gamma(1-\epsilon)}
{(4\pi)^\frac{D}{2}}
\frac{\Gamma^2(\epsilon)}
{\Gamma(2\epsilon)}
\left(\vec{q}^2\right)^\epsilon\,,
\end{equation}
where $t = q^2 \approx q_\bot^2$, $D = 4+2$ is the space-time dimension. A non-zero $\epsilon$ has been introduced to regularize infrared divergences. The integration is performed in a $(D-2)$-dimensional space, orthogonal to the initial colliding particles $p_A$ and $p_B$ momentum plane.\\
The gluon Reggeization contributes to the form of inelastic amplitudes in the so called multi-Regge kinematics (MRK), in which all particles are strongly ordered in the rapidity
space with limited transverse momenta and the squared invariant masses $s_{ij} = (k_i + k_j)^2$ of any pair of produced particles $i$ and $j$ are large and increasing with $s$.
This kinematics provides the leading contribution to QCD cross sections.
In the LLA, there are exchanges of vector particles, the gluons, in all channels.
In the NLA, instead, MRK is not the unique kinematics which contributes.
It can happen then just one of the generated particles can have a fixed (not increasing
with $s$) invariant mass, \emph{i.e.} components of this pair can have rapidities of
the same order. This is known as quasi-multi-Regge-kinematics (QMRK)~\cite{Fadin:1989kf}.
\section{Diffusion amplitude in Multi-Regge kinematics}
\begin{figure}[t]
	\centering
	\includegraphics[scale=1.00]{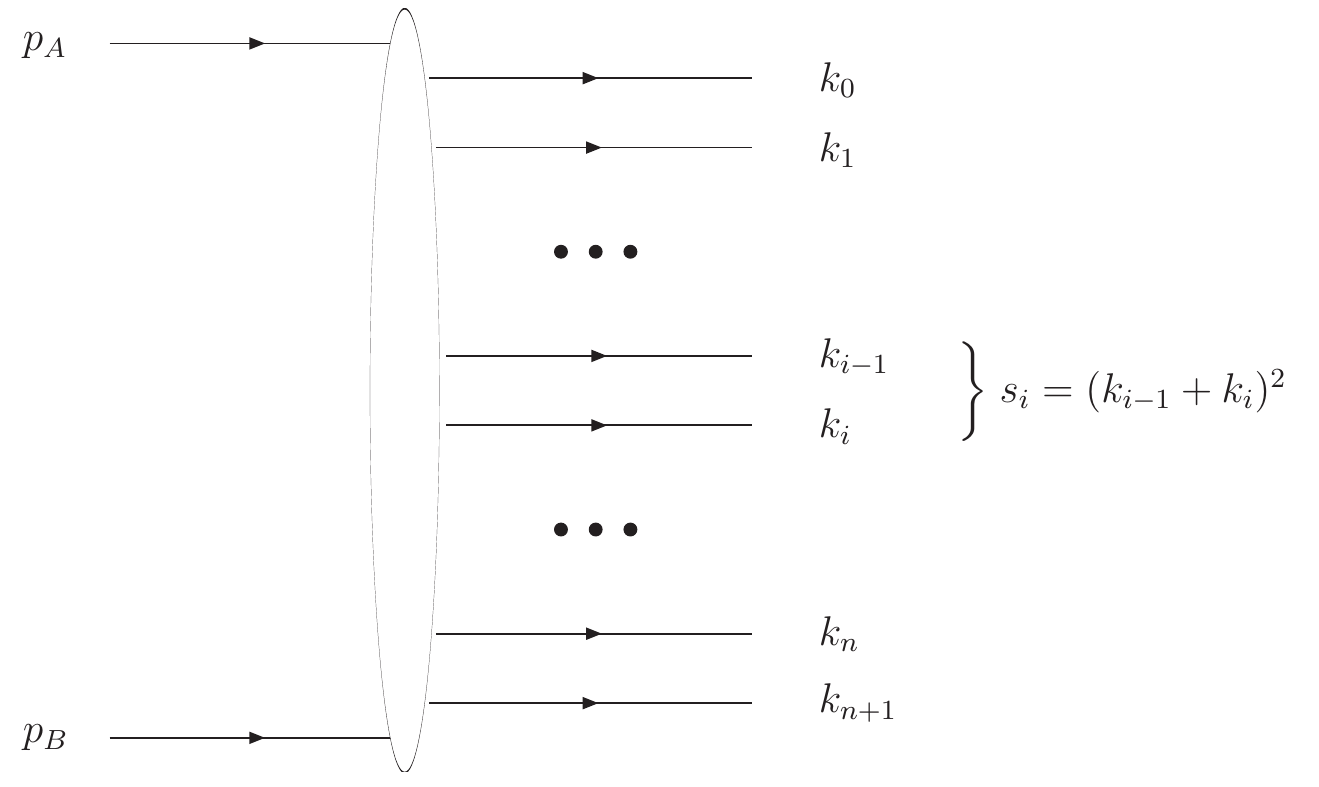}
	\caption{\emph{Production of  $n + 2$ particles in the multi-Regge kinematics.}}
	\label{fig:s_channel_unitarity}
\end{figure}

\begin{figure}[t]
	\centering
	\includegraphics[scale=0.40]{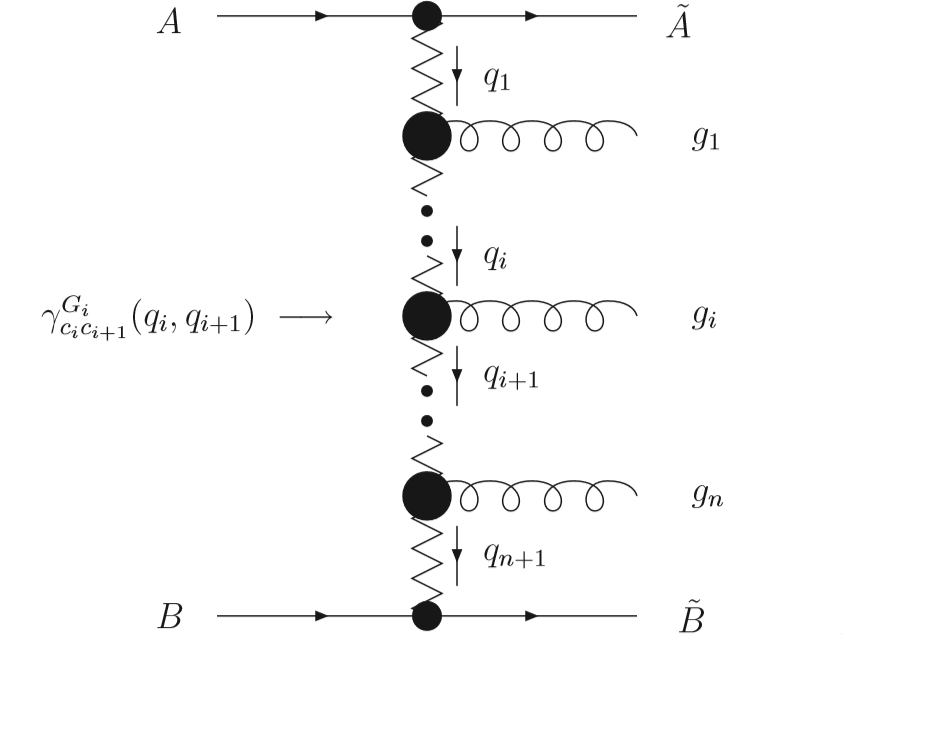}
	\caption{\emph{Schematic representation of the amplitude $\mathcal{A}^{\tilde{A} \tilde{B} + n}_{AB}$.}}
	\label{fig:bfkl_mrk_amplitude}
\end{figure}

In an elastic process $A + B \rightarrow A' + B'$, according to the Cutkosky rules~\cite{Cutkosky:1960sp} and to the unitarity relation in the $s$-channel, the imaginary part of the elastic scattering amplitude $\mathcal{A}^{A^\prime B^\prime}_{AB}$ can be written as
\begin{equation}
\label{s_unitarity_amplitude}
{\cal Im}_s \mathcal{A}^{A^\prime B^\prime}_{AB}
= \frac{1}{2} \sum_{n=0}^{\infty} \sum_{\{f\}} 
\int 
\mathcal{A}^{\tilde{A} \tilde{B} + n}_{AB}\left(\mathcal{A}^{\tilde{A} \tilde{B} + n}_{A^\prime B^\prime}\right)^*
d\Phi_{\tilde{A} \tilde{B} + n} \; ,
\end{equation}
where $\mathcal{A}^{\tilde{A} \tilde{B} + n}_{AB}$ represents the amplitude for the production of $n + 2$ particles (Fig.~\ref{fig:s_channel_unitarity}) with momenta  $k_i$, $i = 0,1,\dots,n,n+1$ in the process $A + B \to \tilde{A} + \tilde{B} + n$, while $d\Phi_{\tilde{A} \tilde{B} n}$ stands for the intermediate particle space of phace and $\sum_{\{f\}}$ means to sum over the discrete quantum numbers $\{f\}$ of the intermediate particles.
Using the light-like vectors $p_1$ and $p_2$, such that $p_1^2 = p_2^2 = 0$ and $(p_1 + p_2)^2 = 2\,p_1 \cdot p_2 = s$, it is possible to decompose the momenta $p_A$ and $p_B$ of the initial particles as follows
\begin{equation}
p_A=p_1+(m^2_A/s)p_2\,, 
\end{equation}
\begin{equation}
p_B=p_2+(m^2_B/s)p_1\,.
\end{equation}
Any $k_i$ momentum, using the Sudakov decomposition, satisfies the relation
\begin{equation}
\label{sudakov_general}
k_i = \beta_ip_1+\alpha_ip_2+k_{i\bot} \; , \qquad 
s\alpha_i \beta_i = 
k_i^2-k_{i\bot}^2 =
k_i^2+\vec{k}_i^2\; ,
\end{equation}
with $\vec{k}_{i\bot}$ transverse component with respect to the plane generated by $p_1$ and $p_2$, and $k_{i\bot}^2=-\vec{k}_i^2$.\\
The expression of the phace-space element can be written as
\begin{equation}
\label{sudakov_ps}
d\Phi_{\tilde{A} \tilde{B} n} = 
\frac{2}{s} (2\pi)^D
\delta\left(1+\frac{m_A^2}{s}-\sum_{i=0}^{n+1}\alpha_i\right)
\delta\left(1+\frac{m_B^2}{s}-\sum_{i=0}^{n+1}\beta_i\right)
\end{equation}
\[ \times \, 
\delta^{D-2}\left(\sum_{i=0}^{n+1}k_{i\bot}\right)
\frac{d\beta_{n+1}}{2\beta_{n+1}}
\frac{d\alpha_0}{2\alpha_0}
\prod_{i=0}^{n} \frac{d\beta_i}{2\beta_i}
\prod_{i=1}^{n+1}\frac{d^{D-2}k_{i\bot}}{(2\pi)^{D-1}} \; , 
\]
where 
$p_{\tilde{A}} = k_0$ ; $p_{\tilde{B}} = k_{n+1}$ and $\alpha_i$ and $\beta_i$ are Sudakov variables. In the Eq.~\eqref{s_unitarity_amplitude} the dominant contribution ($\sim \! s$) in the LLA is provided by the region of limited (not growing with $s$) transverse momenta of produced particles.
Large logarithms come from the integration over longitudinal momenta of the produced particles. In particular, we have a logarithm of $s$ for each particle produced according to the MRK. In this framework transverse momenta of the generated
particles are limited and their Sudakov variables $\alpha_i$ and $\beta_i$ are strongly ordered in the rapidity space, obtaining so
\begin{align}
\label{sudakov_order}
&
\alpha_{n+1} \gg \alpha_n \dots \gg \alpha_0 \; , \\ \nonumber &
\beta_0 \gg \beta_1 \dots \gg \beta_{n+1} \; .
\end{align}
In the MRK the squared invariant masses $s_{ij} = (k_i + k_j)^2$ are large with respect to the squared transverse momenta. In order to obtain the large logarithm after the integration over $\beta_i$ for each produced particle in the space of the phace described by Eq.~\eqref{sudakov_ps}, the amplitude in Eq.~\eqref{s_unitarity_amplitude} must not decrease when the invariant mass increases. This holds just when the exchange objects are vector particles, as previously said, the gluons, in every channel with momentum transfers $q_{i=1,\dots,n+1}$ expressed with
\begin{equation}
q_i = p_a - \sum_{j=0}^{i-1}k_j = - \left(p_B - \sum_{l=i}^{n+1}k_l\right) \simeq
\beta_ip_1-\alpha_{i-1}p_2-\sum_{j=0}^{i-1}k_{j\bot} 
\end{equation}
and
\begin{equation}
q_i^2 \simeq q_{i\bot}^2 = - \vec{q}_i^2\,.
\end{equation}
The amplitudes of this kind of processes, generally, show a complicated analytical form. However, in the BFKL approach at LLA, they assume a easier structure regarding just the real parts of these amplitudes (see Fig.\eqref{fig:bfkl_mrk_amplitude}). Hence the simpler multi-Regge form is
\begin{equation}
\label{bfkl_real_amplitude}
\mathcal{A}^{\tilde{A} \tilde{B}+n}_{AB} = 
2s \Gamma_{\tilde{A}A}^{c_1} 
\left(\prod_{i=1}^n\gamma_{c_ic_{i+1}}^{P_i}(q_i,q_{i+1})\left(\frac{s_i}{s_R}\right)^{\omega(t_i)}\frac{1}{t_i}\right)\frac{1}{t_{n+1}}\left(\frac{s_{n+1}}{s_R}\right)^{\omega(t_{n+1})}\Gamma_{\tilde{B}B}^{c_{n+1}}\,,
\end{equation}
where $s_R$ is an arbitrary energy scale, irrelevant at LLA; $s_i = (k_{i-1}+k_i)^2$, shown in Fig.~\eqref{fig:bfkl_mrk_amplitude}; $t_i = q_i^2 \approx q_{i\bot}^2 = - \vec{q}_i^2$, $\omega(t)$ is the gluon Regge trajectory in LLA (from Eq.~\eqref{trajectory_1-loop}) and $\Gamma_{P^\prime P}^a$ is the PPR vertices given by Eq.~\eqref{PPRvertex}, while $\gamma_{c_ic_{i+1}}^{P_i}$ are the effective vertices for the production of particles $P_i$ with momenta $q_i-q_{i+1}$ in the collisions of Reggeized gluons with momenta $q_i$ and $-q_{i+1}$ and color indices $c_i$ and
$c_{i+1}$.\\ It is worth to remark that only a gluon can be produced by the Reggeon-Reggeon-Particle (RRP) vertex and consequently, the originated particles are massless. The Fig.~\eqref{fig:effective_vertex} illustrates the contributions to the RRP effective vertex for the gluon production. The Reggeon-Reggeon-Gluon (RRG) vertex has the following form~\cite{Fadin, elf, elf2, BL}:
\begin{figure}[t]
	\centering
	\includegraphics[scale=0.50]{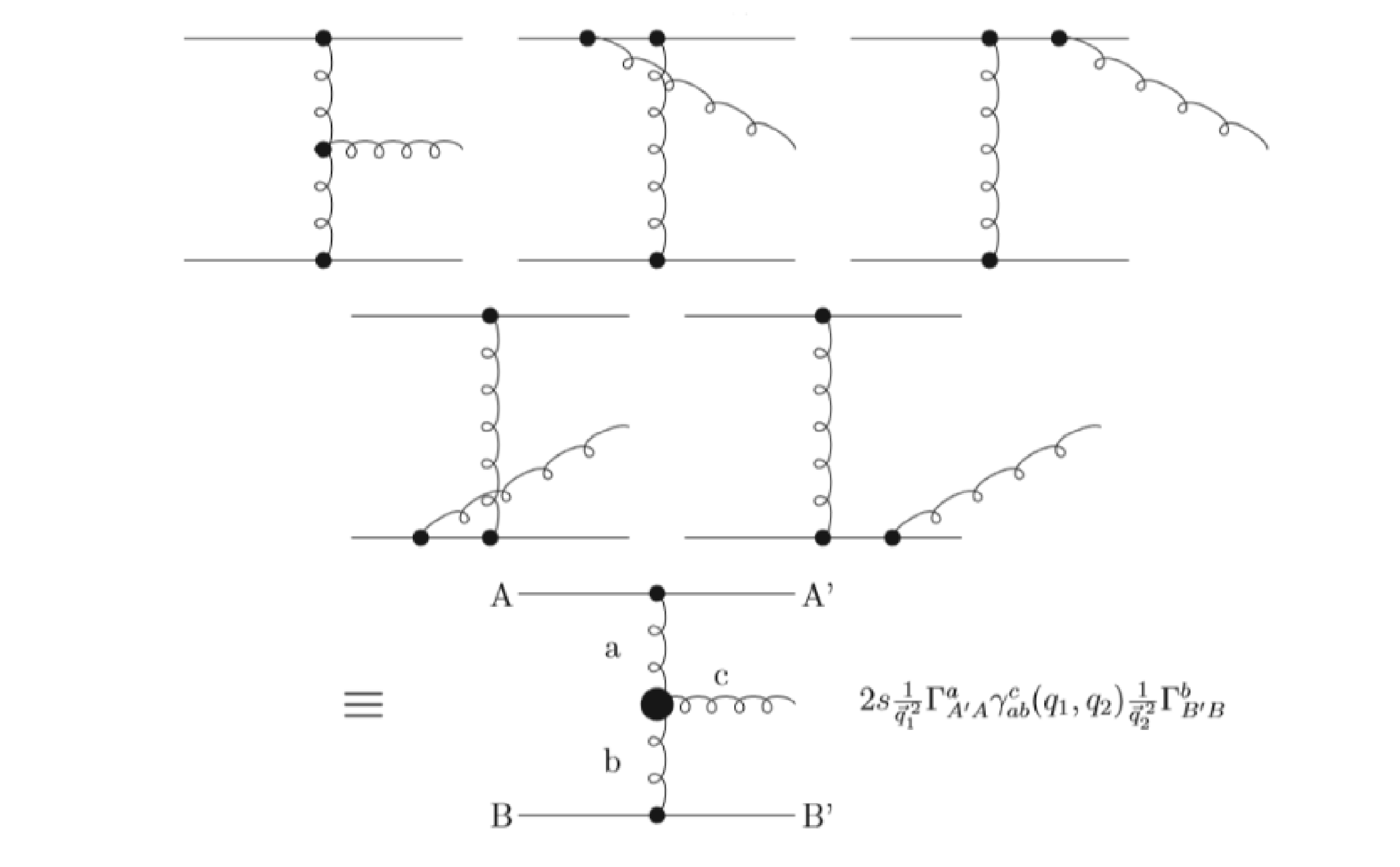}
	\caption{\emph{Contributions to the RRP effective vertex.}}
	\label{fig:effective_vertex}
\end{figure}
\begin{equation}
\label{rrg_vertex}
\gamma_{c_ic_{i+1}}^{G_i}(q_i,q_{i+1}) = 
g_sT_{c_ic_{i+1}}^{d_i}e_{\mu}^{\ast}(k_i)
C^{\mu}(q_{i+1},q_i) \; ,
\end{equation}
with $T_{c_ic_{i+1}}^{d_i}$ the matrix elements of the $SU(N_c)$ group generators in the adjoint representation, while $d_i$ is the color index of the produced gluon, $e_{\mu}^{\ast}(k_i)$ its polarization vector and $k_i=q_i-q_{i+1}$ its momentum.\\ The Lorentz structure of $C^\mu$ given by
\begin{equation}
\label{rrg_current}
C^{\mu}(q_{i+1},q_i) = 
- q_i^{\mu} - q_{i+1}^{\mu} + p_1^{\mu} 
\left(\frac{q_i^2}{k_i \cdot p_1} + 
2\frac{k_i \cdot p_2}{p_1 \cdot p_2}\right)
- p_2 ^{\mu} 
\left(\frac{q_{i+1}^2}{k_i \cdot p_2} + 
2\frac{k_i \cdot p_1}{p_1 \cdot p_2}\right)\,,
\end{equation}
satisfies the current conservation property, $(k_i)_\mu C^{\mu}(q_{i+1},q_i) = 0$, allowing the choice of an arbitrary gauge for each of the produced gluons. It is simple to note that the amplitude $\left(\mathcal{A}^{\tilde{A} \tilde{B} + n}_{A^\prime B^\prime}\right)$ in the Eq.~\eqref{s_unitarity_amplitude} comes from the substitutions
\begin{equation*}
A \rightarrow A^\prime\,, \qquad B \rightarrow B^\prime\,,
\end{equation*}
\begin{equation*}
q_i \rightarrow q_i^\prime \equiv q_i - q\,, \qquad q \equiv p_A - p_{A^\prime} \simeq q_\perp\,.
\end{equation*}
Introducing the decomposition
\begin{equation}
\label{rrg_decomposition}
T_{c_ic_{i+1}}^{d_i}
\left(T_{c_ic_{i+1}}^{d_i}\right)^{\ast} = 
\sum_R c_R \left\langle c_i c_i^\prime \left| 
\hat{\mathcal{P}}_R \right|
c_{i+1} c_{i+1}^\prime \right\rangle\,,         
\end{equation}
where $\hat{\mathcal{P}}_R$ is the projector of the two-gluon color states in the $t$-channel on the irreducible representation $R$ of the color group, it is possible to write
\begin{equation}
\sum_{G_i} \gamma_{c_ic_{i+1}}^{G_i}(q_i,q_{i+1}) 
\left( 
\gamma_{c_ic_{i+1}}^{G_i}(q_i,q_{i+1})
\right)^{\ast} 
\end{equation}
\[ =
2(2\pi)^{D-1} \sum_R 
\left\langle c_i c_i^\prime \left| 
\hat{\mathcal{P}}_R \right|
c_{i+1} c_{i+1}^\prime \right\rangle
\mathcal{K}_r^{(R)}(\vec{q}_i,\vec{q}_{i+1};\vec{q})\,,
\]
where the sum is over color and polarization states of the produced gluon and $\mathcal{K}_r^{(R)}(\vec{q}_i,\vec{q}_{i+1};\vec{q})$ is the so-called \emph{real part} of the kernel.
\subsection{The BFKL equation}
In order to obtain the BFKL equation at LLA, the first step is decomposing the elastic scattering amplitude $\mathcal A_{A\,B}^{A^\prime B^\prime}$ which, using Eq.~\eqref{rrg_decomposition}, assumes the form
\begin{equation}
\mathcal{A}^{A^\prime B^\prime}_{AB} = 
\sum_R\left(\mathcal{A}_R\right)^{A^\prime B^\prime}_{AB} \, ,
\end{equation}
where $\left(\mathcal{A}_R\right)^{A^\prime B^\prime}_{AB}$ is the part of the scattering amplitude corresponding to a definite irreducible representation $R$ of the color group in the $t$-channel.\\ Now it is convenient consider the partial wave function $f_R(\omega, \vec q)^{A'B'}_{AB}$ introducing the Mellin transformation~\cite{Forshaw} of the imaginary part of the amplitude, defined by
\begin{equation}
	f_R(\omega, \vec q)^{A'B'}_{AB} = \int_{s_0}^\infty \frac{ds}{s^2}
	\left(\frac{s}{s_0}\right)^{-\omega}Im_s\left({{\cal A}_R}\right)^{A'B'}_{AB}\,.
	\label{partial_wave}
\end{equation}
Consequently, its inverse given by the partial wave expansion reads
\begin{equation}
\label{im_ampl}
Im_s\left({{\cal A}_R}\right)^{A'B'}_{AB} = \frac{s}{2\pi i} \oint_C d\omega \left(\frac{s}{s_0}\right)^\omega 	f_R(\omega, \vec q)^{A'B'}_{AB}
\end{equation}
\[
 \equiv \frac{s}{2\pi i}\int_{\delta -i\infty }^{\delta +i\infty }\left(\frac{s}{s_0}\right)^\omega 	f_R(\omega, \vec q)^{A'B'}_{AB}\,,
\]
so that, using techniques of dispersion relations it is possible to obtain the total amplitude. Therefore, the total amplitude in the form of a partial wave expansion is
\begin{equation}
\left({{\cal A}_R}\right)^{A'B'}_{AB}=
\frac {s}{2\pi }
\int_{\delta -i\infty }^{\delta +i\infty }\frac{d\omega}{\sin(\pi \omega)}
\left(\left(\frac{-s}{s_0}\right)^\omega -\tau
\left(\frac{s}{s_0}\right)^\omega \right)
f_R(\omega, \vec q)^{A'B'}_{AB}\,,
\label{tot_ampl}
\end{equation}
where $\tau$ is the signature and coincides with the symmetry of
the representation $R$. The function $f_R(\omega, \vec q)^{A'B'}_{AB}$ can be expressed as~\cite{Fadin:1998sh}
\begin{equation}
f_R(\omega, \vec q)^{A'B'}_{AB} = \sum_{n = 0}^{\infty} f_R^{(n)}(\omega,\vec q)^{A'B'}_{AB} = \frac{1}{(2\pi )^{D-2}}
\int \frac{d^{D-2}q _{A\perp}}{\vec q_A^{~2}({\vec q}_{A}-{\vec q})^2 }
\label{a39}
\end{equation}
\[
\hspace{0.65cm}\times\,\frac{d^{D-2}q _{B\perp }}{\vec q_B^{~2}({\vec q}_{B}-{\vec q})^2 }
\sum_{\nu}I_{A'A}^{(R,\nu)}
G^{(R)}_{\omega}({\vec q}_{A}, {\vec q}_{B}; {\vec q})
I_{B'B}^{(R,\nu)} \,.
\]
Here $\vec{q}_A$ and $\vec{q}_B$ are the transverse momenta of the Reggeized gluons, while $s_0$ is an arbitrary energy scale introduced in order to define
the partial wave expansion.
Here the index $\nu$ enumerates the states in the irreducible representation $R$, so that
\begin{equation}
\langle c_ic'_i|\hat{\cal P}_R |c_{i+1}c'_{i+1} \rangle =
\sum_{\nu} \langle c_ic'_i|\hat{\cal P}_R |\nu \rangle
\langle \nu|\hat{\cal P}_R |c_{i+1}c'_{i+1} \rangle\,,
\label{a32}
\end{equation}
and
\[
I_{A'A}^{(R,\nu)} = \sum_{{\tilde A}} \Gamma_{\tilde A A}^{c_1}
\left(\Gamma_{\tilde A A'}^{c'_1}\right)^*
\langle c_1c'_1|\hat{\cal P}_R |\nu \rangle\,,
\]
\begin{equation}
I_{B'B}^{(R,\nu)} = \sum_{{\tilde B}}\Gamma_{\tilde B B}^{c_{n+1}}
\left(\Gamma_{\tilde B B'}^{c'_{n+1}}\right)^*
\langle \nu|\hat{\cal P}_R |c_{n+1}c'_{n+1} \rangle\,.
\label{a33}
\end{equation}
The sum here is taken over the discreet quantum numbers of the states $\tilde A$,
$\tilde B$.\\ It is worth to remind that for the singlet representation the index $\nu$ takes only one value, hence it can be omitted and
\begin{equation}
\langle c c'|\hat{\cal P}_0|0 \rangle =\frac{\delta_{cc'}}{\sqrt {N^2-1}}\,,
\label{a34}
\end{equation}
whereas  for the antisymmetrical octet (gluon) representation the index
$\nu$ coincides with gluon color index and
\begin{equation}
\langle c c'|\hat{\cal P}_8|a \rangle =\frac{f_{acc'}}{\sqrt N}\,,
\label{a35}
\end{equation}
with $f_{abc}$ the structure constants of the color group.\\
The function $G_{\omega}^{(R)}$, named Green function for the scattering of two Reggeized gluons, is defined as
\[
G^{(R)}_{\omega}({\vec q}_{A}, {\vec q}_{B}; {\vec q})=
\sum_{n=0}^{\infty}\int\left(\prod_{i=1}^{n+1}\frac{d^{D-2}q_{i\perp }}
{ \vec q_i^{~2}({\vec q}_{i}-{\vec q})^2
	(\omega -\omega(t_i)-\omega(t'_i)) }\right)
\]
\[
\times\left(\prod_{i=1}^n{\cal K}^{(R)}_r(\vec q_i,
\vec q_{i+1};\vec q)\right){\vec q}_A^{~2}({\vec q}_A-{\vec q})^2
{\vec q}_B^{~2}({\vec q}_B-{\vec q})^2
\]
\begin{equation}
\times \delta^{D-2}(q_{1\perp }-q_{A\perp })
\delta^{D-2}(q_{n+1\perp }-q_{B\perp })\,.
\label{a40}
\end{equation}
\begin{figure}[t]
	\centering
	\includegraphics[scale=1.00]{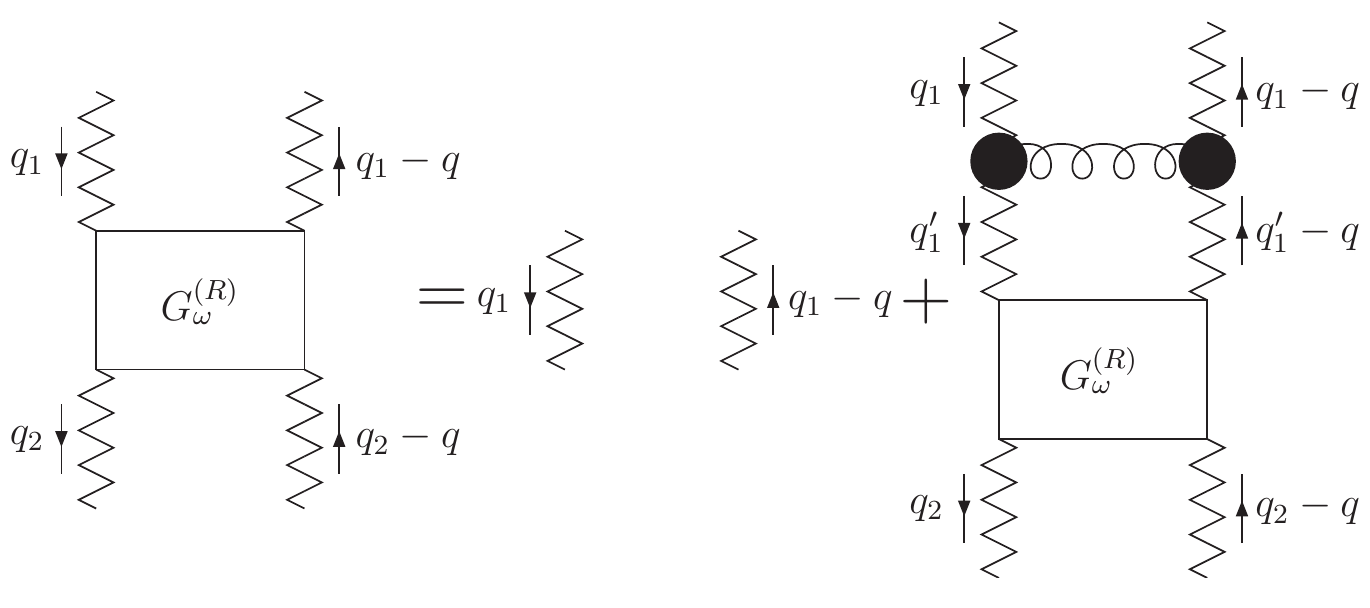}
	\caption
	{\emph{Diagrammatic representation of the generalized BFKL integral equation.}}
	\label{fig:kernel_equation}
\end{figure}
Only the function $G_{\omega}^{(R)}$ depends on $\omega$, so that it determines the $s$-behavior of the scattering amplitude. It obeys  the following integral equation (Fig.~\ref{fig:kernel_equation}), known as \emph{generalized BFKL equation}:
\begin{equation}
\label{bfkl_kernel_equation}
\omega G^{(R)}_{\omega}({\vec q}_1, {\vec q}_2; {\vec q}) =
\vec q_1^{~2}({\vec q}_{1}-{\vec q})^2\delta^{(D-2)}
({\vec q}_1- {\vec q}_2)
\end{equation}
\[
+\int \frac{d^{D-2}q_{1\perp}^{~\prime}}
{\vec q_1^{~\prime 2}({\vec q}_{1}^{~\prime}-{\vec q})^2 }
{\cal K}^{(R)}({\vec q}_1, {\vec q}_1^{~\prime}; {\vec q})
G^{(R)}_{\omega}({\vec q}_1^{~\prime}, {\vec q}_2; {\vec q}) \, ,
\]
where the kernel $\mathcal{K}^{(R)}(\vec{q}_1, \vec{q} \; '_1; \vec{q})$ of the integral function is
\begin{equation}
\label{kernel}
\mathcal{K}^{(R)}(\vec{q}_1, \vec{q}_2; \vec{q}) =  \left[ \omega(q_{1\perp}^2) + \omega((q_1-q)_{\perp}^2) \right] \vec{q}_1^{\;2} \left( \vec{q}_1-\vec{q} \right)^{2} \delta^{(D-2)} (\vec{q}_1-\vec{q}_2)
\end{equation}
\[
+\,\mathcal{K}_r^{(R)}(\vec{q}_1, \vec{q}_2; \vec{q}) \; .
\]
Eq.~\eqref{kernel} consists of two parts: the first one is called \emph{virtual} and it is expressed in terms of the Regge trajectory of the gluon, while the second one, which is related to the production of real particles (see Fig.~\ref{fig:born_real_kernel}) reads: 
\begin{equation}
\label{bfkl_kernel_real}
{\cal K}_r^{(R)}(\vec q_i,\vec q_{i+1};\vec q )=
-\frac{g^2c_R}{2(2\pi )^{D-1}}
C^{\mu }(q_{i+1},q_i) C_{\mu }(q_{i+1}-q,q_i-q)
\end{equation}
\[
=
\frac{g^2c_R}{(2\pi )^{D-1}}
\left(\frac{{\vec q}_i^2{(\vec q_{i+1} -\vec q)}^2
	+{\vec q}_{i+1}^2{(\vec q_{i} -\vec q)}^2}{({\vec q}_i-{\vec q}_{i+1})^2}
-\vec q^{~2} \right) \; .
\]
\begin{figure}[t]
	\centering
	\includegraphics[scale=1.20]{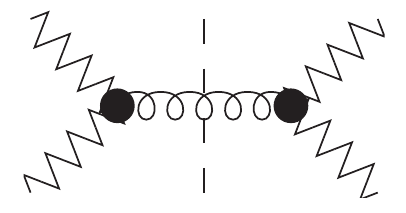}
	\caption
	{\emph{Diagrammatic representation of the real part of the BFKL kernel at the Born level.}}
	\label{fig:born_real_kernel}
\end{figure}
The Eq.~\eqref{bfkl_kernel_equation} is called \textit{BFKL equation} in case of $R=0$ (color-singlet representation) and $t=0$. It is an iterative equation, which means that knowing the kernel in the Born level, it allows us to get all the LLA terms of the Green's function. In the same way, knowing all the NLA corrections to the gluon trajectory and to the real part of the kernel, one can obtain all the NLA terms of the Green's function.
\subsection{The BFKL equation in the NLA}
In the NLA, the expression given in Eq.~(\ref{octet_amplitude}) was checked initially at the first three perturbative orders~\cite{Fadin:1995dd,Fadin:1995km,Kotsky:1996xm,Fadin:1995xg,Fadin:1996tb}, and then assumed to be valid to all orders of perturbation theory. In the this approximation, where all terms of the type $\alpha_s [\alpha_s \ln (s)]^n$ need be collected, the PPR vertex takes the following form
\begin{equation}
\Gamma_{P^\prime P} = 
\delta_{\lambda P, \lambda P^\prime}\Gamma_{PP}^{(+)} + 
\delta_{\lambda P^\prime, - \lambda P}\Gamma_{PP}^{(-)} \; .
\end{equation}
Here a term in which the helicity of the scattering particle $P$ is not conserved occurs. In order to get production amplitudes in the NLA it is needed to take one of
the vertices or the trajectory in Eq.~\eqref{bfkl_real_amplitude} in the next-to-leading order (NLO). In the LLA, the Reggeized gluon trajectory is sufficient at 1-loop accuracy and from the production of one gluon at Born level in the collision of two Reggeons ($K_{RRG}^B$~\cite{Fadin:1998sh}) one obtains the only contribution to the real part of the kernel.
In the NLA the gluon trajectory is taken in the NLO (2-loop accuracy~\cite{Fadin:1995dd,Fadin:1995km,Kotsky:1996xm,Fadin:1995xg,Fadin:1996tb}) and the real part includes the contributions due to: 
one-gluon ($K_{RRG}^1$)~\cite{Fadin:1992rh}, 
two-gluon ($K_{RRGG}^B$)~\cite{Fadin:1996nw,Fadin:1997zv,Kotsky:1998ug},
and quark-antiquark pair ($K_{RRQ\bar{Q}}^B$)~\cite{Catani:1990eg,Fadin:1997hr,Catani:1990xk} production at Born level~\cite{Fadin:1998sh}.

\subsection{The BFKL cross section and the impact factors}
\label{subsec:BFKLcrosssec-IF}
The total cross section is directly related to the imaginary part of the forward scattering amplitude 
($\vec{q}=0$) via the optical theorem and it can be expressed by 
\begin{equation}
	\label{sigma_s}
	\sigma(s) = \frac{{\mathcal Im}_s \mathcal{A}_{AB}^{AB}}{s} \,,
\end{equation}
with ${\mathcal Im}_s \mathcal{A}_{AB}^{AB}$ given in Eq.~\eqref{im_ampl} using the Eq.~\eqref{a39}. Redefining the Green's function as
\begin{equation}
\label{ggf_redefinition}
G_\omega (\vec{q}_1,\vec{q}_2)=
\frac{ G^{(0)}_{\omega}({\vec q}_{1}, {\vec q}_{2}; 0)}
{\vec q_{1}^{~2}\vec q_{2}^{~2}} \; ,
\end{equation}
where $\vec{q}_{1,2}$ are two-dimensional vectors, and $G^{(0)}_{\omega}({\vec q}_{1}, {\vec q}_{2}; 0)$ is the forward Green's function in the singlet-color representation. It is ruled by the BFKL equation given in Eq.~(\ref{bfkl_kernel_equation}) with $R=0$. This leads to a simplification of the expressions of the BFKL equation and of the BFKL kernel~\eqref{kernel}, which now read
\begin{equation}
\omega \,G_\omega (\vec{q}_1,\vec{q}_2)=\delta
^{D-2}(\vec{q}_1-\vec{q}_2)+\int d^{D-2}%
\widetilde{q}\,\,\,{\cal K}(\vec{q}_1,\vec{\widetilde{q}})
\,G_\omega (\vec{\widetilde{q}},\vec {q}_2)\,
\label{bfkl_equation_forward}
\end{equation}
and
\begin{equation}
{\cal K}(\vec{q}_1,\vec{q}_2) =
\frac{{\cal K}^{(0)}({\vec q}_1, {\vec q}_2; 0)}{{\vec q}_1^{~2}{\vec q}_2^{2}}
=
2\,\omega (-\vec q_1^{2})\,\delta^{(D-2)}(\vec{q}_1-\vec{q}_2)
+{\cal K}_r(\vec{q}_1, \vec{q}_2) \, ,
\label{bfkl_kernel_forward}
\end{equation}
respectively.\\
\begin{figure}[t]
	\centering
	\includegraphics[scale=1.00]{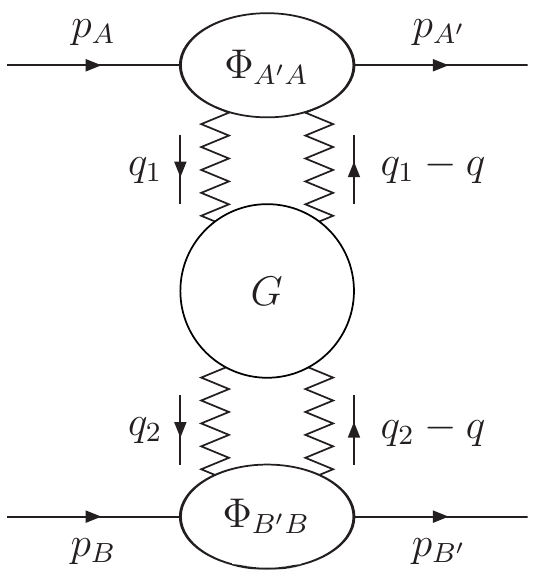}
	\caption{\emph{Diagrammatic representation of the imaginary part of the BFKL amplitude.}}
	\label{fig:bfkl_elastic_amplitude}
\end{figure}
 In Eq.~\eqref{bfkl_kernel_forward} the term $\omega (-\vec {q}^{~2})$ is the gluon Regge trajectory shown in Eq.~\eqref{trajectory_1-loop}. Moreover, as a consequence of the scale invariance of the kernel, it is possible to take its eigenfunctions as powers of one of the two squared momenta $\vec{q}_{1,2}^{~2}$, say $(\vec q_2^{~2})^{\gamma-1}$ with $\gamma$ being a complex number. Therefore, setting the corresponding eigenvalues equal to $\frac{N\alpha_s}{\pi}\chi^B(\gamma)$, one obtains
\begin{equation}
\int d^{D-2}q_2\,\,{\cal K}(\vec{q}_1,\vec {q}_2)
(\vec q_2^{~2})^{\gamma-1}=
\frac{N\alpha_s}{\pi}\chi^B(\gamma)(\vec q_1^{~2})^{\gamma-1} \; ,
\label{eigenfunctions_kernel_averaged}
\end{equation}
with~\cite{Fadin,elf,elf2,BL}
\begin{equation}
\chi^B(\gamma)=2\psi (1)-\psi (\gamma )-\psi (1-\gamma )\,,\,\,\,\psi (\gamma )=\Gamma ^{\prime }(\gamma )/\Gamma (\gamma ) \, .
\label{chi_averaged}
\end{equation}
Here the set of functions $(\vec q_2^{~2})^{\gamma-1}$ with $\gamma =1/2+i\nu,\,\,\, -\infty <\nu<\infty $ is complete and represent the eigenfunctions of the leading order (LO) BFKL kernel averaged on the azimuthal angle between $\vec{q}_1$ and $\vec{q}_2$.
If we take its projection onto them in Eq.~\eqref{im_ampl}, a simple expression for the cross section can be drawn 
\begin{align}
\label{sigma}
&
\sigma(s)=
\frac{1}{(2\pi)^{2}}\int\frac{d^{2}\vec q_A}{\vec
	q_A^{\,\, 2}}\Phi_A(\vec q_A)\int
\frac{d^{2}\vec q_B}{\vec q_B^{\,\,2}} \Phi_B(-\vec q_B)
\\ & \nonumber \times \,
\int\limits^{\delta +i\infty}_{\delta
	-i\infty}\frac{d\omega}{2\pi i}\left(\frac{s}{s_0}\right)^\omega
G_\omega (\vec q_A, \vec q_B)\, ,
\end{align}
where the momenta are defined on the transverse plane.\\ $\Phi_A(\vec q_A) \equiv I_{AA}^{(0)}$ , $\Phi_B(\vec q_B) \equiv I_{BB}^{(0)}$ are the NLO impact factors specific of the process, whose expression are obtained from Eq.~\eqref{a33}.\\ The Green's function in Eq.~\eqref{sigma} satisfies the BFKL equation in Eq.~\eqref{bfkl_kernel_equation} with $\vec{q}=0$, $\vec{q}_1=\vec{q}_A$ and $\vec{q}_2=\vec{q}_B$ and characterizes the universal, energy-dependent part of the amplitude.\\
Now, it is worth to highlight that in the BFKL approach the diffusion amplitude of a process at high energy can be represented as the convolution of the Green function $G$ for two reggeized gluon with the impact factors $\Phi_{A'A}$ and $\Phi_{B'B}$ of the colliding particles (see Fig.~\ref{fig:bfkl_elastic_amplitude}).\\ The Eq.~\eqref{sigma} offers the possibility to understand that if the Green's function $G_\omega$ has a pole at $\omega^\prime$, the cross section at LLA takes the form
\begin{equation}
\sigma^{(LLA)} \sim \frac{s^{\omega_P^B}}{\sqrt{\ln s}} \; , \quad
\omega^\prime \equiv \omega_P^B \; ,
\end{equation}
where $\omega_P^B = 4 N_c (\alpha_s / \pi)\ln 2$ is the intercept of the Regge trajectory that governs the asymptotic behavior in $s$ of the amplitude with exchange of the vacuum quantum numbers in the $t$-channel. This means that the Froissart bound is violated in LLA, giving rise to a power-like behavior of cross section $\sigma$ with energy. This is one of the motivations for moving to the NLA. Another reason is that, at the LLA, it is not possible to fix the scale for $\alpha_s$ in $\omega_P^B$.
\subsubsection{Pomeron NLA}
\label{PomNLA}
In this Section we remind how large are the effects of the NLO corrections, showing an estimation of the Pomeron intercept, following the discussion in Ref.~\cite{Fadin:1998sh}.

The NLA kernel~\footnote{We refer the reader to the Ref.~\cite{Fadin:1998sh} for the detailed calculations of each single contribution to NLA kernel.} has the same form of the LLA one, where the part ${\cal K}_r(\vec q_1,\vec q_2)$ related with the real particle production contains the contributions from the one-gluon, two-gluon and quark-antiquark
production in the Reggeon-Reggeon collisions. This part can be represented as the sum
\begin{equation}
{\cal K}_r(\vec q_1,\vec q_2)= {\cal K}_r^{(B)}(\vec q_1,\vec q_2)
+{\cal K}_r^{(1)}(\vec q_1,\vec q_2); \,\,\,
{\cal K}_r^{(B)}(\vec q_1,\vec q_2) = \frac{4\overline{g}_\mu
	^2\,\mu ^{-2\epsilon }}{\pi ^{1+\epsilon }\,\Gamma (1-\epsilon )}%
\frac 1{(\,{\vec q_1}-{\vec q_2)}^2}
\label{B1}
\end{equation}
of the LLA contribution, expressed in terms of the renormalized coupling constant, and the one-loop correction ${\cal K}_r^{(1)}(\vec q_1,\vec q_2)$, we have for this correction~\cite{Fadin:1998py} 
\[
K_r^{(1)}(\vec q_1,\vec q_2)=\frac{4\overline{g}_\mu
	^4\,\mu ^{-2\epsilon }}{\pi ^{1+\epsilon }\,\Gamma (1-\epsilon )}%
\left\{ \frac 1{(\,{\vec q_1}-{\vec q_2)}^2}\left[ \left( \frac{11}3-\frac{%
	2n_f}{3N}\right) \frac 1\epsilon \left( 1-\left( \frac{(\vec q_1
	-\vec q_2)^2}{\mu ^2}\right) ^\epsilon  \right. \right. \right. \,
\]
\[
\left. \left. \left.(1-\epsilon ^2%
\frac{\pi ^2}6)\right)+\left( \frac{(\vec q_1-\vec q_2)^2}{%
	\mu ^2}\right) ^\epsilon \left( \frac{67}9-\frac{\pi ^2}3-\frac{10}9\frac{%
	n_f}{N}+\epsilon \left( -\frac{404}{27}+14\zeta (3) \right. \right. \right. \right.
\]
\[
\left. \left. \left. \left.+\frac{56}{27}\frac{
	n_f}{N}\right) \right) \right]-\left( 1+\frac{n_f}{N^3}\right) \frac{2{\vec q_1}^{~2}{\vec q_2}%
	^{~2}-3({\vec q_1}{\vec q_2})^2}{16{\vec q_1}^{~2}{\vec q_2}^{~2}}\left(
\frac 2{{\vec q_2}^{~2}}+\frac 2{{\vec q_1}^{~2}}+(\frac 1{{\vec q_2}%
	^{~2}}-\frac 1{{\vec q_1}^{~2}})\ln \frac{{\vec q_1}^{~2}}{{\vec q_2}^{~2}}%
\right) \right.
\]
\[
\left. -\frac 1{(\vec q_1-\vec q_2)^2}\left( \ln \frac{{\vec q_1}^{~2}}{{%
		\vec q_2}^{~2}}\right) ^2\hspace{-0.25cm}+\frac{2(\vec  q_1^{~2}-\vec  q_2^{~2})}{(\vec q_1-\vec q_2)^2(\vec q_1+\vec
	q_2)^2}\left( \frac 12\ln \hspace{-0.1cm}\left( \frac{{\vec q_1}^{~2}}{{\vec q_2}^{~2}}%
\right) \ln \hspace{-0.1cm}\left( \frac{{\vec q_1}^{~2}{\vec q_2}^{~2}(\vec q_1-\vec q_2)^4%
}{({\vec q_1}^{~2}+{\vec q_2}^{~2})^4}\right)  \right. \right.
\]
\[
\left. \left. +L\left( -\frac{{\vec q_1}^{~2}%
}{{\vec q_2}^{~2}}\right) -L\left( -\frac{{\vec q_2}^{~2}}{{\vec q_1}^{~2}}%
\right) \right)-\left( 3+(1+\frac{n_f}{N^3})\left( 1-\frac{({\vec q_1}^{~2}+{\vec
		q_2}^{~2})^2}{8{\vec q_1}^{~2}{\vec q_2}^{~2}} \right. \right. \right.
\]
\[
\left. \left. \left. -\frac{2{\vec q_1}^{~2}{\vec
		q_2}^{~2}-3{\vec q_1}^{~4}-3{\vec q_2}^{~4}}{16{\vec q_1}^{~4}{\vec q_2}^{~4}%
}({\vec q_1}{\vec q_2})^2\right) \right)\hspace{-0.1cm}\int_0^\infty \frac{dx\,\ln \left|
	\frac{1+x}{1-x}\right| }{{\vec q_1}^{~2}+x^2{\vec q_2}^{~2}} -\hspace{-0.07cm}\left(\hspace{-0.06cm}1-\frac{(\vec  q_1^{~2}-\vec  q_2^{~2})^2}{(\vec q_1-\vec q_2)^2(\vec
	q_1+\vec q_2)^2}\right) \right.
\]
\begin{equation}
\left. \left( \int_0^1-\int_1^\infty \right) \frac{dz\,\ln
	\frac{(z{\vec q_1})^2}{({\vec q_2})^2}}{(\vec q_2-z\vec q_1)^2}\right\} \,,
\label{B2}
\end{equation}
where
\begin{equation}
L(z)=\int_0^z\frac{dt}t\ln (1-t)\,,\,\,\zeta (n)=\sum_{k=1}^\infty k^{-n}\,.
\label{B3}
\end{equation}
The Eq.~\eqref{B2} exhibits the cancellation of the
infrared singularities~\cite{F}
at fixed $\vec k_1=\vec q_1-\vec q_2$, where we can expand
$ ({\vec k_1}^2/{\mu ^2})^{\epsilon}$ in powers of $\epsilon $.
This expansion
is not performed in (\ref{B2}) because for the terms having singularity
at ${\vec k_1}^{~2} =0$  the region of
such small $\vec k_1$, that is $\ln(1/\vec k_1^{~2}) \sim 1/{\epsilon}$,
does contribute in the integral over $ \vec k_1$.  The singular contributions
given by this region are canceled, in turn, in the BFKL equation by the
singular terms in the gluon trajectory~\cite{F}.
As well as in LLA, only
the kernel averaged over the angle between the momenta $\vec q_1$ and $%
\vec q_2$
is relevant until spin correlations are not considered.
Hence, it is possible to obtain the averaged one-loop correction adopting the the cut off
$ |\vec  q_1^{~2}-\vec  q_2^{~2}| > \lambda^2$ instead of the dimensional regularization. Then, through Eqs.~\eqref{bfkl_kernel_forward}, \eqref{B1} and the expansion of the gluon trajectory in Eq.~(71) of Ref.~\cite{Fadin:1998sh} it is possible to get a useful representation of the averaged NLA kernel (for which the cancellation of the dependence of $\lambda$ is evident). The expression reads:
\[
\overline{{\cal K}(\vec q_1,\vec q_2)}=
\frac{\alpha _s(\mu ^2)N}{\pi ^2%
}\hspace{-0.1cm}\int \hspace{-0.1cm}d\vec q^{~2}\frac 1{|\vec q_1^{~2}-\vec q^{~2}|}\hspace{-0.1cm}\left(
\delta (\vec q^{~2}-\vec  q_2^{~2})-2\frac{\min
	(\vec q_1^{~2},\vec q^{~2})}{(\vec q_1^{~2}+\vec q^{~2})}
\delta (\vec q_1^{~2}-\vec  q_2^{~2})\hspace{-0.1cm}\right)
\]
\[
\times \left[ 1-\frac{\alpha _s(\mu ^2)N}{4\pi }\left( \left( \frac{11}3-%
\frac{2n_f}{3N}\right) \hspace{-0.1cm}\ln\hspace{-0.1cm} {\left( \frac{|\vec q_1^{~2}-\vec q^{~2}|^2}{\max
		(\vec q_1^{~2},\vec q^{~2})\mu ^2}\right) }-\left( \frac{67}9-\frac{\pi ^2}3-\frac{10}9\frac{%
	n_f}{N}\right) \right) \right]
\]
\[
\hspace{-0.1cm}-\frac{\alpha _s^2(\mu ^2)N^2}{4\pi ^3}\left[ \frac 1{32}\left( 1+\frac{n_f%
}{N^3}\right) \left( \frac 2{{\vec q_2}^{~2}}+\frac 2{{\vec q_1}^{~2}}+
(\frac 1{{\vec q_2}%
	^{~2}}-\frac 1{{\vec q_1}^{~2}})\ln \frac{{\vec q_1}^{~2}}{{\vec q_2}^{~2}}
\right)
+\frac{\left( \ln {(\vec q_1^{~2}/\vec  q_2^{~2})}\right)^2}
{|\vec q_1^{~2}-\vec  q_2^{~2}|} \right.
\]
\[
\left. \hspace{-0.1cm} +\hspace{-0.1cm}\left( 3+(1+\frac{n_f}{N^3})\left( \frac 34-\frac{(\vec q_1^{~2}+\vec  q_2^{~2})^2}{%
	32\vec q_1^{~2}\vec q_2^{~2}}\right) \right) \int_0^\infty \frac{dx}
{\vec q_1^{~2}+x^2\vec  q_2^{~2}}\ln
\left| \frac{1+x}{1-x}\right| \right] -\frac 1{\vec q_2^{~2}+\vec q_1^{~2}}
\]
\[
\left( \frac{\pi ^2}3-4L\left(\min \left(\frac{\vec q_1^{~2}}{\vec  q_2^{~2}},%
\frac{\vec  q_2^{~2}}{\vec q_1^{~2}}\right)\right)\right)
+\frac{\alpha _s^2(\mu ^2)N^2}{4\pi ^3}\left( 6\zeta (3)-\frac{5\pi ^2}{%
	12}\left( \frac{11}3-\frac{2n_f}{3N}\right) \right) 
\]
\begin{equation}
\times\delta (\vec q_1^{~2}-\vec  q_2^{~2})\,.
\label{B6}
\end{equation}
The $\mu $ - dependence in the right hand side of this equality leads to the
violation of the scale invariance and is related with running the QCD
coupling constant.

The form (\ref{B6}) is useful for finding the action of the kernel
on the eigenfunctions $\vec q_2^{~2(\gamma -1)}$ of the Born kernel:
\begin{equation}
\int d^{D-2}q_2\,\,\,{\cal K}(\vec q_1,\vec q_2)\left(
\frac{\vec  q_2^{~2}}{\vec  q_1^{~2}}\right) ^{\gamma -1}=\frac{\alpha _s(\vec  q_1^{~2})\,N\,}\pi
\left( \chi^B (\gamma )+ \frac{\alpha _s(\vec  q_1^{~2})N}{\pi }
\chi^{(1)} (\gamma )%
\right) \,,\, \label{B7}
\end{equation}
where, within our accuracy, we expressed the result in terms of the running coupling constant
\[
\alpha _s(\vec q^{~2})=\frac{\alpha _s(\mu ^2)}{1+\frac{\alpha _s(\mu
		^2)N}{4\pi }%
	\left( \frac{11}3-\frac{2n_f}{3N}\right) \ln {\left( \frac{\vec q^{~2}}{\mu
			^2}%
		\right) }}
\]
\begin{equation}
\hspace{4.5cm}\simeq \alpha _s(\mu ^2)\left( 1-\frac{\alpha _s(\mu^2)N}{4\pi }\left( \frac{11}3-\frac{2n_f}{3N}\right) \ln {\left( \frac{\vec q^{~2}}{\mu^2}\right) }\right)\,,
\label{B8}
\end{equation}
$\chi^B(\gamma ) $ is given by Eq.~\eqref{chi_averaged} and
and  the correction $\chi^{(1)}(\gamma)$ is:
\[
\chi^{(1)} (\gamma )=-\frac{1}{4}\left[ \left( \frac{11}3-\frac{2n_f}{3N}\right) \frac
12\left( \chi ^2(\gamma )-\psi ^{\prime }(\gamma )+\psi ^{\prime }(1-\gamma
)\right) \right.
\]
\[
\left. -6\zeta (3)+\frac{\pi ^2\cos(\pi \gamma )}{\sin^2(\pi \gamma
	)(1-2\gamma )}\left( 3+\left( 1+\frac{n_f}{N^3}\right) \frac{2+3\gamma
	(1-\gamma )}{(3-2\gamma )(1+2\gamma )}\right) \right.
\]
\begin{equation}
\left.-\left( \frac{67}9-\frac{\pi ^2}3-\frac{10}9\frac{n_f}{N}\right)
\chi (\gamma ) -\psi ^{\prime \prime }(\gamma )-\psi ^{\prime \prime }(1-\gamma )-%
\frac{\pi ^3}{\sin (\pi \gamma )}+4\phi (\gamma )\right] \,,  \label{B9}
\end{equation}
with
\[
\phi (\gamma )=-\int_0^1\frac{dx}{1+x}\left( x^{\gamma -1}+x^{-\gamma
}\right) \int_x^1\frac{dt}t\ln (1-t)
\]
\begin{equation}
=\sum_{n=0}^\infty (-1)^n\left[ \frac{\psi (n+1+\gamma )-\psi (1)}{(n+\gamma
	)^2}+\frac{\psi (n+2-\gamma )-\psi (1)}{(n+1-\gamma )^2}\right] \,.
\label{B10}
\end{equation}
For the relative correction $r(\gamma )$  defined by
$\chi^{(1)} (\gamma )=-r(\gamma )\chi^{B} (\gamma )\,$
in the symmetrical point $\gamma =1/2$, corresponding to the
eigenfunction of the LLA kernel with the largest eigenvalue, one has~\cite{Fadin:1998py}
\[
r\left( \frac 12\right) =\left( \frac{11}6-\frac{n_f}{3N}\right) \ln 2-%
\frac{67}{36}+\frac{\pi ^2}{12}+\frac{5}{18}\frac{n_f}{N}+\frac 1{\ln 2}\left[
\int_0^1\arctan (\sqrt{t})\ln (\frac 1{1-t})\frac{dt}t\right.
\]
\begin{equation}
\left. +\frac{11}{8}\zeta (3)+\frac{\pi ^3}{32}\left( \frac{27}{16}+\frac{11}{16}\frac{n_f%
}{N^3}\right) \right] \simeq 6,46 +0.05\frac{n_f}{N}+2.66\frac{n_f}{%
N^3}\,,
 \label{B11}
\end{equation}
which shows that the correction is very large. This is due to the large value of the  LLA Pomeron intercept $\omega_P
^B=4N\ln 2\,\,\alpha _s(q^2)/\pi\,$. If we express the corrected
intercept $\omega_P$ in terms of the Born one, one obtains
\begin{equation}
\omega_P = \omega_P^B(1-\frac{r\left( \frac 12\right)}{4\ln 2}\omega_P^B)
\simeq \omega_P^B(1-2.4\omega_P^B)\,.
\label{B13}
\end{equation}
The coefficient 2.4 does not look very large. Moreover, it corresponds to
the rapidity interval where correlations become important in
the hadron production processes.
However, these numerical estimates indicate,
that in the kinematical region of HERA
probably it is not enough to take into account only the next-to-leading correction. For example, if $\alpha _s(\vec q^{~2})=0.15$, where the Born intercept is
$\omega_P
^B=4N\alpha _s(\vec q^{~2})/\pi \,\ln 2=.39714$, the relative correction for $%
n_f=0 $ is extremely large:
\begin{equation}
\frac {\omega_P}{\omega_P^B}=1-r\left( \frac 12\right) \,
\frac{\alpha _s(\vec q^{~2})}{%
	\pi }N=0.0747\,.
\label{B12}
\end{equation}
The maximal value of $\omega_P \simeq 0.1$ is obtained
for $\alpha _s(\vec q^{~2})\simeq 0.08$.

Nevertheless it is worth to realize that
the above estimates are quite straightforward and do not take into account
neither the influence of the running coupling  on the eigenfunctions nor the
nonperturbative effects~\cite{Lipatov:1985uk, Camici:1996fr}; besides the value of the correction strongly depends on its representation.
For instance, considering the next-to-leading correction by
the corresponding increase of the argument of the running QCD coupling constant, $\omega_P$ at $\alpha _s(q^2)=0.15$ turns out to be only two times smaller than its Born value.
\subsubsection{The $(\nu,n)$-representation}
\label{nun_repr}
In this section we derive a general form for the cross section in the so-called $(\nu,n)$-representation (see Refs.~\cite{Ivanov:2005gn,Ivanov:2006gt}), useful for the further analysis of this thesis. The first step is to work in the transverse momentum representation, defined by
\begin{equation}
\label{transv}
\hat{\vec q}\: |\vec q_i\rangle = \vec q_i|\vec q_i\rangle\;, 
\quad\quad
\langle\vec q_1|\vec q_2\rangle =\delta^{(2)}(\vec q_1 - \vec q_2) \;,
\end{equation}
\[
\langle A|B\rangle =
\langle A|\vec k\rangle\langle\vec k|B\rangle =
\int d^2k A(\vec k)B(\vec k)\;.
\]
In this representation, the total cross section given in Eq.~(\ref{sigma}) assumes the simple form
\begin{equation}
\label{ampl-transv}
\sigma =\frac{1}{(2\pi)^2}
\int_{\delta-i\infty}^{\delta+i\infty}\frac{d\omega}{2\pi i}
\, \left(\frac{s}{s_0}\right)^\omega
\langle\frac{\Phi_1}{\vec q_1^{\,\,2}}|\hat G_\omega|\frac{\Phi_2}{\vec q_2^{\,\,2}}
\rangle \,,
\end{equation}
while the Eq.~\eqref{bfkl_kernel_equation} can be written in the operator form
\begin{equation}
\omega \hat{G}_{\omega} = 1 + \mathcal{K} \hat{G}_{\omega}\,,
\end{equation} 
which implies
\begin{equation}
\label{Goper}
\hat G_\omega=(\omega-\hat K)^{-1}\,.
\end{equation}
The kernel of the operator $\hat K$ becomes
\begin{equation}
\label{kernel-op}
K(\vec q_2, \vec q_1) = \langle\vec q_2| \hat K |\vec q_1\rangle
\end{equation}
and it is given as an expansion in the strong coupling,
\begin{equation}
\label{kern}
\hat K=\bar \alpha_s \hat K^0 + \bar \alpha_s^2 \hat K^1\;,
\end{equation}
where
\begin{equation}
\label{baral}
{\bar \alpha_s} \equiv \frac{N_c}{\pi} \alpha_s 
\end{equation}
with $N_c$ the number of colors. In Eq.~\eqref{kern} $\hat K^0$ represents the BFKL kernel in the LO, while $\hat K^1$ is in the NLO.\\ In order to determine the cross section in the NLA accuracy, it is necessary to give a solution of Eq.~\eqref{Goper}. With the required accuracy this
solution is
\begin{align}\label{exp}
& \hat G_\omega = (\omega-\bar \alpha_s\hat K^0)^{-1} 
\\ \nonumber & 
+
(\omega-\bar \alpha_s\hat K^0)^{-1}\left(\bar \alpha_s^2 \hat K^1\right)
(\omega-\bar \alpha_s \hat
K^0)^{-1}+ {\cal O}\left[\left(\bar \alpha_s^2 \hat K^1\right)^2\right]
\, .
\end{align}
In Eq.~\eqref{eigenfunctions_kernel_averaged} we wrote the expressions for the eigenfunctions of the LO kernel averaged on the azimuthal angle. 
In the general case the basis of eigenfunctions of the LO kernel,
\begin{align}
\label{KLLA}
&
\hat K^0 |n,\nu\rangle = \chi(n,\nu)|n,\nu\rangle \, ,
\\ & \nonumber
\chi (n,\nu)=2\psi(1)-\psi\left(\frac{n}{2}+\frac{1}{2}+i\nu\right)
-\psi\left(\frac{n}{2}+\frac{1}{2}-i\nu\right)\, ,
\end{align}
is given by the set of functions of the type
\begin{equation}
\label{nuLLA}
\langle\vec q\, |n,\nu\rangle =\frac{1}{\pi\sqrt{2}}
\left(\vec q^{\,\, 2}\right)^{i\nu-\frac{1}{2}}e^{in\phi} \;,
\end{equation}
characterized by the integer $n$, known as \emph{conformal spin}.
Here $\phi$ is the azimuthal angle of the vector $\vec q$ counted from
some fixed direction in the transverse space, $\cos\phi \equiv q_x/|\vec q\,|$.
Then, the orthonormality and completeness conditions take the form
\begin{equation}
\label{ort}
\langle n',\nu^\prime | n,\nu\rangle =\int \frac{d^2 \vec q}
{2 \pi^2 }\left(\vec q^{\,\, 2}\right)^{i\nu-i\nu^\prime -1}
e^{i(n-n')\phi}=\delta(\nu-\nu^\prime)\, \delta_{nn'}
\end{equation}
and
\begin{equation}
\label{comp}
\hat 1 =\sum^{\infty}_{n=-\infty}\int\limits^{\infty}_{-\infty}d\nu \, 
| n,\nu\rangle\langle n,\nu |\ \,.
\end{equation}
Now, it is possible to express the action of the full NLO BFKL kernel on these functions as
\begin{align}
\label{Konnu}
& \hat K|n,\nu\rangle =
\bar \alpha_s(\mu_R) \chi(n,\nu)|n,\nu\rangle
\\ \nonumber &
+ \bar \alpha_s^2(\mu_R)\left(\chi^{(1)}(n,\nu)
+\frac{\beta_0}{4N_c}\chi(n,\nu)\ln(\mu^2_R)\right)|n,\nu\rangle
\\ \nonumber &
+ \bar
\alpha_s^2(\mu_R)\frac{\beta_0}{4N_c}\chi(n,\nu)
\left(i\frac{\partial}{\partial \nu}\right)|n,\nu\rangle \;,
\end{align}
where
\begin{equation}
\label{beta00}
\beta_0=\frac{11 N_c}{3}-\frac{2 n_f}{3}\;,
\end{equation}
with $n_f$ the number of active quark flavours.\\ In Eq.~\eqref{Konnu} $\mu_R$ is the renormalization scale of the QCD coupling; the first term stands for the action of LO kernel, while the second and the third ones are the diagonal and the non-diagonal contributions of the NLO kernel.\\
The function $\chi^{(1)}(n,\nu)$, calculated in~Ref.~\cite{Kotikov:2000pm} (see also Ref.~\cite{Kotikov:2002ab}), is conveniently represented in the form
\begin{equation}
\label{ch11}
\chi^{(1)}(n,\nu)=-\frac{\beta_0}{8\, N_c}\left(\chi^2(n,\nu)-\frac{10}{3}
\chi(n,\nu)-i\chi^\prime(n,\nu)\right) + {\bar \chi}(n,\nu)\, ,
\end{equation}
where
\begin{equation}
\label{chibar}
\bar \chi(n,\nu)\,=\,-\frac{1}{4}\left[\frac{\pi^2-4}{3}\chi(n,\nu)
-6\zeta(3)
-\chi^{\prime\prime}(n,\nu) +\,2\,\phi(n,\nu)
\right.
\end{equation}
\[
+\,2\,\phi(n,-\nu)
+\frac{\pi^2\sinh(\pi\nu)}{2\,\nu\, \cosh^2(\pi\nu)}
\left(
\left(3+\left(1+\frac{n_f}{N_c^3}\right)\frac{11+12\nu^2}{16(1+\nu^2)}\right)
\delta_{n0}
\right.
\]
\[
\left.\left.
-\left(1+\frac{n_f}{N_c^3}\right)\frac{1+4\nu^2}{32(1+\nu^2)}\delta_{n2}
\right)\right] \, ,
\]
and
\begin{equation}
\label{phi}
\phi(n,\nu)\,=\,
-\int\limits_0^1dx\,\frac{x^{-1/2+i\nu+n/2}}{1+x}
\left[\frac{1}{2}\left(\psi'\left(\frac{n+1}{2}\right)-\zeta(2)\right) \right.
\end{equation}
\[
+\mbox{Li}_2(x)+\mbox{Li}_2(-x)
+\ln x \left(\psi(n+1)-\psi(1)+\ln(1+x)+\sum_{k=1}^\infty\frac{(-x)^k}
{k+n}\right)
\]
\[
\left. 
+\sum_{k=1}^\infty\frac{x^k}{(k+n)^2}(1-(-1)^k)\right]
\]
\[
=\sum_{k=0}^\infty\frac{(-1)^{k+1}}{k+(n+1)/2+i\nu}\left[\psi'(k+n+1)
-\psi'(k+1)
\right.
\]
\[
+(-1)^{k+1}(\beta'(k+n+1)+\beta'(k+1))
\]
\[
\left.
-\frac{1}{k+(n+1)/2+i\nu}(\psi(k+n+1)-\psi(k+1))\right] \; ,
\]
\begin{equation}
\beta'(z)=\frac{1}{4}\left[\psi'\left(\frac{z+1}{2}\right)
-\psi'\left(\frac{z}{2}\right)\right] \; ,
\end{equation}
\begin{equation}
\mbox{Li}_2(x)=-\int\limits_0^xdt\,\frac{\ln(1-t)}{t} \; .
\end{equation}
Here and below $\chi^\prime(n,\nu) \equiv d\chi(n,\nu)/d\nu$ and
$\chi^{\prime\prime}(n,\nu) \equiv d^2\chi(n,\nu)/d^2\nu$.
The projection of the impact factors onto the eigenfunctions of the LO BFKL kernel is done as follows:
\begin{equation}
\frac{\Phi_1(\vec q_1)}{\vec q_1^{\,\, 2}}=\sum^{+\infty}_{n=-\infty}
\int\limits^{+\infty}_{-\infty}
d\nu \, \Phi_1(\nu,n)\langle n,\nu| \vec q_1\rangle\, \; , 
\end{equation}
\[
\frac{\Phi_2(-\vec{q_2})}{\vec q_2^{\,\, 2}}=\sum^{+\infty}_{n=-\infty}
\int\limits^{+\infty}_{-\infty} d\nu \, \Phi_2(\nu,n)
\langle \vec q_2 |n,\nu \rangle \; ,
\]
\begin{equation}
\label{nu_rep}
\Phi_1(\nu,n)=
\langle\frac{\Phi_1(\vec q_1)}{\vec q_1^{\,\,2}}|n,\nu\rangle \equiv
\int d^2 q_1 \,\frac{\Phi_1(\vec q_1)}{\vec q_1^{\,\, 2}}
\frac{1}{\pi \sqrt{2}} \left(\vec q_1^{\,\, 2}\right)^{i\nu-\frac{1}{2}}
e^{i n \phi_1}\;,
\end{equation}
\[
\Phi_2(\nu,n)=
\langle n,\nu|\frac{\Phi_2(-\vec q_2)}{\vec q_2^{\,\,2}}\rangle 
\equiv
\int d^2 q_2 \,\frac{\Phi_2(-\vec q_2)}{\vec q_2^{\,\, 2}}
\frac{1}{\pi \sqrt{2}} \left(\vec q_2^{\,\, 2}\right)^{-i\nu-\frac{1}{2}}
e^{-i n \phi_2}\;.
\]
It is worth to stress that the impact factor wave functions in the momentum representation of the Eqs.~\eqref{nu_rep} are real.\\
The impact factors can be represented as an expansion in $\alpha_s$,
\begin{equation}
\label{if}
\Phi_{1,2}(\vec q\,)=\alpha_s(\mu_R)\left[ v_{1,2}(\vec q\, ) 
+ \bar \alpha_s(\mu_R)
v_{1,2}^{(1)}(\vec q\, )\right]
\end{equation}
and
\begin{equation}
\label{vertex-exp}
\Phi_{1,2}(n,\nu)=\alpha_s(\mu_R)\left[ c_{1,2}(n,\nu) 
+ \bar \alpha_s(\mu_R)
c_{1,2}^{(1)}(n,\nu) \right]\, .
\end{equation}
Regarding the Eq.~\eqref{ampl-transv} for the cross section and inserting two identity operators according to the completeness relation in Eq.~\eqref{comp}, one obtains
\begin{equation}
\label{sigma-f}
\sigma = \frac{1}{(2\pi)^2}
\sum^{\infty}_{n=-\infty}
\int\limits^{\infty}_{-\infty} d\nu\sum^{\infty}_{n^\prime =-\infty} 
\int\limits^{\infty}_{-\infty} d\nu^\prime 
\int_{\delta-i\infty}^{\delta+i\infty}
\frac{d\omega}{2\pi i}
\left(\frac{1}{s_0}\right)^\omega
\end{equation}
\[
\langle
\frac{\Phi_1}{\vec q_1^{\,\,2}}|n,\nu\rangle\langle n,\nu|\hat 
G_\omega| n^\prime,\nu^\prime\rangle\langle n^\prime,\nu^\prime |
\frac{\Phi_2}{\vec q_2^{\,\,2}}\rangle \ ,
\]
and, after some algebra and integration by parts, finally we have the representation of the NLA BFKL forward amplitude:
\begin{equation}
\label{sigma-ff}
\sigma = \frac{1}{(2\pi)^2}\sum^{\infty}_{n=-\infty}
\int\limits^{\infty}_{-\infty} d\nu \left(\frac{s}{s_0}\right)^{\bar \alpha_s(\mu_R)
	\chi(n,\nu)} 
\end{equation}
\[
\alpha_s^2(\mu_R) c_1(n,\nu)c_2(n,\nu)\left[1+\bar \alpha_s(\mu_R)\left(\frac{c^{(1)}_1(n,\nu)}{c_1(n,\nu)}
+\frac{c^{(1)}_2(n,\nu)}{c_2(n,\nu)}\right) \right.
\]
\[
+\bar \alpha^2_s(\mu_R)\ln\frac{s}{s_0}\left\{\bar \chi(n,\nu)
+\frac{\beta_0}{8 N_c}\chi(n,\nu)
\right.
\]
\[
\left.\left.
\left(
-\chi(n,\nu)+\frac{10}{3}+2\ln \mu_R^2 +i\frac{d}{d\nu}\ln\frac{c_1(n,\nu)}
{c_2(n,\nu)}\right)\right\}\right] \, .
\]
\subsubsection{Representation equivalence}
\label{subsec:repr-equiv}
The expression for the cross section shown in Eq.~(\ref{sigma-ff}) is not the only possible choice. Actually, several equivalent expressions can be adopted. In what follows the so called \textit{exponential} representation will be used. Starting from the Eq.~\eqref{sigma-ff}, exponentiating some terms and introducing two compensating terms of the form $\ln \sqrt{s_1 s'_1}$, it is easy to get
\begin{equation}
\label{sigma_repr_exp}
\sigma^{exp} = \frac{1}{(2\pi)^2}\sum^{\infty}_{n=-\infty}
\int\limits^{\infty}_{-\infty} d\nu \,
\alpha_s^2(\mu_R) c_1(n,\nu)c_2(n,\nu)
\end{equation}
\[\hspace{3.2cm} \times \,
\left(\frac{s}{s_0}\right)^{\bar \alpha_s(\mu_R)\chi 
	+\bar \alpha^2_s(\mu_R)\left\{\bar \chi
	+\frac{\beta_0}{8 N_c}\chi
	\left(
	-\chi+\frac{10}{3}+2\ln \mu_R^2 +i\frac{d}{d\nu}\ln\frac{c_1(n,\nu)}
	{c_2(n,\nu)}\right)\right\}} 
\]
\[\hspace{1cm} \times \,
\left[1+\bar \alpha_s(\mu_R)\left(\frac{c^{(1)}_1(n,\nu)}{c_1(n,\nu)}
+\frac{c^{(1)}_2(n,\nu)}{c_2(n,\nu)}\right) \right] \;.
\]
We will adopt this representation in both the calculations for the charged light hadron-jet production and for the heavy-quark pair hadroproduction.

\subsection{The scale optimization}
\label{BLMscale}
At this point it is worth to remark that a practical application of the BFKL mechanism to physical reactions encounters some difficulties. On one side, this fact is due to the large NLO corrections, from which the BFKL approach is affected, both in the kernel of the gluon Green's function and in the non-universal impact factors. In this regard, just to have an idea of the weight of NLA corrections with respect to the LLA contribution, one can compare $\chi^{(1)}(n, \nu)$ in Eq.~\eqref{ch11} with $\chi(n, \nu)$ in Eq.~\eqref{KLLA}, for the points that define the energy asymptotic behavior in the LLA framework: $n = 0$ and $\nu = 0$. The result given by the ratio $|\chi^{(1)}(0,0)/\chi(0,0)|$ is absolutely not negligible $(\simeq 6.5)$~\cite{Fadin:1998py,Ciafaloni:1998gs}.\\
On the other side, some instabilities in the BFKL expansion rely in the significant renormalization scale setting uncertainties. All these issues have to be handled by the use of some optimization procedure of the QCD perturbative series. They consist in:
\begin{itemize}
	\item the inclusion of some unknown next-to-NLA terms, such as those prescribed by renormalization group, as in collinear improvement~\cite{Caporale:2013uva,Vera:2007kn,Salam:1998tj,Ciafaloni:1998iv,Ciafaloni:1999yw,Ciafaloni:1999au,Ciafaloni:2000cb,Ciafaloni:2002xk,Ciafaloni:2002xf,Ciafaloni:2003ek,Ciafaloni:2003rd,Ciafaloni:2003kd,Vera:2005jt,Caporale:2007vs} or by energy-momentum conservation~\cite{Kwiecinski:1999yx};
	\item the suppression of the gluon emissions which are close by in rapidity in the BFKL context, \emph{i.e.} the rapidity veto approach~\cite{Schmidt:1999mz,Forshaw:1999xm};
	\item a flawless choice of the energy values and renormalization scales that can have a relevant effect through subleading terms.
\end{itemize}
One of the most popular and widely adopted optimization schemes~\footnote{Other common optimization methods are those inspired by the \emph{principle of minimum sensitivity} (PMS)~\cite{PMS,PMS_2} and the \emph{fast apparent convergence} (FAC)~\cite{FAC,FAC_2,FAC_3}.} is the \emph{Brodsky--Lepage--Mackenzie method} (BLM)~\cite{BLM,BLM_2,BLM_3,BLM_4,BLM_5}, which is based on the removal of the renormalizaton scale ambiguity by absorbing the non-conformal 
$\beta_0$-terms into the running coupling.
The application of the BLM scale setting entails a fair improvement of the QCD perturbative convergence due to the elimination of renormalon terms in the perturbative QCD series. Moreover, with the BLM scale setting, the BFKL Pomeron intercept has a weak dependence on the virtuality of the Reggeized gluon~\cite{BLM_4,BLM_5}.\\ Since the results with the BLM optimization proved to be in great agreement with the only kinematics for which experimental data exist, that is symmetric CMS cuts in the Mueller--Navelet reaction, the use of this optimization procedure was extended to other semi-hard process, such as to the hadron-jet production which will be discussed in Section.~\ref{jethad}.\\ In Section~(\ref{subsub:BLM_scale}) we introduce just the main ingredients of the BLM method~\footnote{We refer the reader to Ref.~\cite{Caporale:2015uva} for a complete treatment of this topic.}, providing its implementation in the exact and approximated cases.
\subsubsection{The BLM procedure}
\label{subsub:BLM_scale}
The starting point to apply the BLM procedure~\cite{Caporale:2015uva} is considering the separate contributions specified in Eq.~\eqref{sigma-ff}, by distinct values of $n$ and denoted in the following by ${\cal C}_n$, the azimuthal coefficient, in the $\overline{\text{MS}}$. The expression reads
\begin{equation}
\label{c_n}
{\cal C}_n
=\frac{1}{(2\pi)^2}\int\limits^{\infty}_{-\infty} d\nu 
\left(\frac{s}{s_0}\right)^{\bar \alpha_s(\mu_R)\chi(n,\nu)} \alpha_s^2(\mu_R) 
c_1(n,\nu)c_2(n,\nu)
\end{equation}
\[ \times \,
\left[1+\bar \alpha_s(\mu_R)\left(\frac{c^{(1)}_1(n,\nu)}{c_1(n,\nu)}
+\frac{c^{(1)}_2(n,\nu)}{c_2(n,\nu)}\right)
\right.
\]
\[
+ \, \bar \alpha^2_s(\mu_R)\ln\frac{s}{s_0}\left\{\bar \chi(n,\nu) 
+\frac{\beta_0}{8 N_c}\chi(n,\nu)\right.
\]
\[ \times \,
\left.\left.
\left(
-\chi(n,\nu)+\frac{10}{3}+2\ln \mu_R^2 +i\frac{d}{d\nu}
\ln\frac{c_1(n,\nu)}{c_2(n,\nu)}\right)\right\}\right] \, .
\]
Then, it is necessary to perform a finite renormalization from the $\overline{\text{MS}}$ to the physical MOM scheme, which means:
\begin{equation}
\label{scheme}
\alpha_s^{\overline{\text{MS}}}=\alpha_s^{MOM}\left(1+\frac{\alpha_s^{MOM}}{\pi}T
\right)\;,
\end{equation}
with 
\begin{equation}
\label{T_Tbeta_Tconf}
T=T^{\beta}+T^{conf} \; ,
\end{equation}
\[
T^{\beta}=-\frac{\beta_0}{2}\left( 1+\frac{2}{3}I \right) \; ,
\]
\[
T^{conf}= \frac{C_A}{8}\left[ \frac{17}{2}I +\frac{3}{2}\left(I-1\right)\xi
+\left( 1-\frac{1}{3}I\right)\xi^2-\frac{1}{6}\xi^3 \right] \; ,
\]
where $C_A \equiv N_c$ is the color factor associated with gluon emission from a gluon, $I=-2\int_0^1dx\frac{\ln\left(x\right)}{x^2-x+1}\simeq2.3439$ and $\xi$ 
is a gauge parameter, fixed at zero in the following.
At this point, the condition for the BLM scale setting for a given azimuthal coefficient, ${\cal C}_n$, is determined by solving the following integral equation:
\begin{equation}
\label{c_nnnbeta}
{\cal C}^{\beta}_n
=\frac{1}{(2\pi)^2}\int\limits^{\infty}_{-\infty} d\nu 
\left(\frac{s}{s_0}\right)^{\bar \alpha^{MOM}_s(\mu^{BLM}_R)\chi(n,\nu)} 
\left(\alpha^{MOM}_s (\mu^{BLM}_R)\right)^3
\end{equation}
\[ \times \,
c_1(n,\nu)c_2(n,\nu) \frac{\beta_0}{2 N_c} \left[\frac{5}{3}
+\ln \frac{(\mu^{BLM}_R)^2}{Q_1 Q_2} +f(\nu)-2\left( 1+\frac{2}{3}I \right)
\right.
\]
\[
+ \, \bar \alpha^{MOM}_s(\mu^{BLM}_R)\ln\frac{s}{s_0} \: \frac{\chi(n,\nu)}{2}
\]
\[ \times \,
\left.
\left(-\frac{\chi(n,\nu)}{2}+\frac{5}{3}+\ln \frac{(\mu^{BLM}_R)^2}{Q_1 Q_2} 
+f(\nu)-2\left( 1+\frac{2}{3}I \right)\right)\right]=0 \, .
\]
The solution of the Eq.~\eqref{c_nnnbeta} provides the value of the scale and the expression for our observable, valid in the NLA BFKL accuracy, is
\begin{equation}
\label{c_BLMmain}
{\cal C}^{BLM}_n
=\frac{1}{(2\pi)^2}\int\limits^{\infty}_{-\infty} d\nu \left(\frac{s}{s_0}
\right)^{\bar \alpha^{MOM}_s(\mu^{BLM}_R)\left[\chi(n,\nu)
	+\bar \alpha^{MOM}_s(\mu^{BLM}_R)\left(\bar \chi(n,\nu) +\frac{T^{conf}}{N_c}\chi(n,\nu)\right)\right]}
\end{equation}
\[ \times \,
\left(\alpha^{MOM}_s (\mu^{BLM}_R)\right)^2 c_1(n,\nu)c_2(n,\nu)
\]
\[ \times \,
\left[1+\bar \alpha^{MOM}_s(\mu^{BLM}_R)\left\{\frac{\bar c^{(1)}_1(n,\nu)}
{c_1(n,\nu)}+\frac{\bar c^{(1)}_2(n,\nu)}{c_2(n,\nu)}+\frac{2T^{conf}}{N_c}
\right\} \right] \, ,
\]
Hence, the BLM-optimized renormalization scale, $\mu_R^{BLM}$, is nothing but the value of $\mu_R$ that makes the non-conformal, $\beta_0$-dependent terms entering the expression of the observable of interest vanish.
Regarding the analytic structure of semi-hard observables, it exhibits two groups of non-conformal contribution: the first one originates from the $\beta_0$ part of NLO impact factor, while the second one from the NLA BFKL kernel. This makes $\mu_R^{BLM}$ dependent on the energy of the process.
The Eq.~(\ref{c_nnnbeta}) given above can be solved only numerically. 
For this reason, we give also two analytic approximate approaches to the BLM scale setting. We consider the BLM scale as a function of $\nu$ and chose it in order to make vanish either the first or the second ($\sim \bar \alpha_s^{MOM}\ln s/s_0$)  group of terms in the Eq.~\eqref{c_nnnbeta}. In these two cases one can obtain expressions for the BLM scales which do not depend on the
energy. We thus have: 
\begin{itemize}
	\item case $(a)$
	\begin{equation}
	\label{casea}
	\left(\mu_{R, a}^{BLM}\right)^2=Q_1Q_2\ \exp\left[2\left(1+\frac{2}{3}I\right)
	-f\left(\nu\right)-\frac{5}{3}\right]\ ,
	\end{equation}
	\begin{equation}
	\label{c_BLMa}
	{\cal C}^{BLM, a}_n
	= \frac{1}{(2\pi)^2}\int\limits^{\infty}_{-\infty} d\nu
	\, 
	\left(\alpha^{MOM}_s (\mu^{BLM}_{R, a})\right)^2 c_1(n,\nu)c_2(n,\nu)
	\end{equation}
	\[ \times \,
	\left(\frac{s}{s_0}
	\right)^{\bar \alpha^{MOM}_s(\mu^{BLM}_{R, a})\left[\chi+\bar \alpha^{MOM}_s
		(\mu^{BLM}_{R, a})\left(\bar \chi +\frac{T^{conf}}{N_c}\chi
		-\frac{\beta_0}{8 N_c}\chi^2\right)\right]}
	\]
	\[ \times \,
	\left[1+\bar \alpha^{MOM}_s(\mu^{BLM}_{R, a})
	\left\{\frac{\bar c^{(1)}_1(n,\nu)}{c_1(n,\nu)}+\frac{\bar c^{(1)}_2(n,\nu)}
	{c_2(n,\nu)}+\frac{2T^{conf}}{N_c}
	\right\} \right] \ ,
	\]
	which corresponds to the removal of the $\beta_0$-dependent terms in the impact factors;
	\item case $(b)$
	\begin{equation}
	\label{caseb}
	\left(\mu_{R, b}^{BLM}\right)^2=Q_1Q_2\ \exp\left[2\left(1+\frac{2}{3}I\right)
	-f\left(\nu\right)-\frac{5}{3}+\frac{1}{2}\chi\left(\nu,n\right)\right]\ ,
	\end{equation}
	\begin{equation}
	\label{c_BLMb}
	{\cal C}^{BLM, b}_n
	=\frac{1}{(2\pi)^2}\int\limits^{\infty}_{-\infty} d\nu 
	\,
	\left(\alpha^{MOM}_s (\mu^{BLM}_{R, b})\right)^2 c_1(n,\nu)c_2(n,\nu)
	\end{equation}
	\[ \times \,
	\left(\frac{s}{s_0}
	\right)^{\bar \alpha^{MOM}_s(\mu^{BLM}_{R, b})\left[\chi+\bar \alpha^{MOM}_s
		(\mu^{BLM}_{R, b})\left(\bar \chi +\frac{T^{conf}}{N_c}\chi
		\right)\right]}
	\]
	\[ \times \,
	\left[1+\bar \alpha^{MOM}_s(\mu^{BLM}_{R, b})\left\{\frac{\bar c^{(1)}_1
		(n,\nu)}{c_1(n,\nu)}+\frac{\bar c^{(1)}_2(n,\nu)}{c_2(n,\nu)}
	+\frac{2T^{conf}}{N_c}+\frac{\beta_0}{4 N_c}\chi(n,\nu)
	\right\}\right]\, ,
	\]
	which corresponds to the removal of the $\beta_0$-dependent terms in the BFKL kernel.
\end{itemize}
\chapter{Inclusive forward/backward emissions}
\label{Chap:Incl_emiss}
 The BFKL dynamics, as we discussed in the previous section (see Sec.~\ref{Chap:BFKL}), draws the amplitude of a reaction as a convolution between a Green's function and two impact factors providing a systematic resummation technique of the large energy logarithms in the LLA and NLA. This approach represents the most suitable and successful theoretical tool in order to investigate the typical processes of the hard sector in high-energy regime. For instance, the inclusive hadroproduction of forward jets with high transverse momenta (much larger than $\Lambda_{QCD}$) separated by
 a large rapidity gap at the LHC, known as Mueller--Navelet jets~\cite{Mueller:1986ey}, is one of the most studied reactions. It has been an advantageous channel which allowed to define infrared-safe observables (for details see Refs.~\cite{Ducloue:2013bva,Caporale:2014gpa}), the azimuthal correlation momenta, whose NLA predictions have been revealed an optimal agreement with the experimental data~\cite{Ducloue:2013bva,Caporale:2014gpa,Khachatryan:2016udy,Marquet:2007xx,Colferai:2010wu,Caporale:2012ih,Ducloue:2013wmi,Caporale:2013uva,Ducloue:2014koa,Ducloue:2015jba,Caporale:2015uva,Celiberto:2015yba,Celiberto:2015mpa,Celiberto:2016ygs,Celiberto:2016vva,Caporale:2018qnm,Chachamis:2015crx}. The same vertices enter the calculation of several observables in the inclusive production of three and four jets, separated in rapidity, at the LHC~\cite{Caporale:2015vya, Caporale:2015int, Caporale:2016soq, Caporale:2016zkc,Caporale:2016xku,Celiberto:2016vhn,Chachamis:2016qct,Caporale:2016vxt,Caporale:2016djm,Caporale:2016lnh,Chachamis:2017vfa}. Other phenomenological analyses have been proposed so far: the inclusive detection of two light-charged rapidity-separated
 hadrons~\cite{Celiberto:2016hae,Celiberto:2016zgb,Celiberto:2017ptm},  $J/\Psi$-jet~\cite{Boussarie:2017oae}, forward Drell--Yan dilepton production~\cite{Celiberto:2018muu,Motyka:2014lya,Brzeminski:2016lwh} with a possible backward-jet
 tag~\cite{Golec-Biernat:2018kem,Deak:2018obv}, the Higgs-jet channel~\cite{Celiberto:2020tmb} and the inclusive emission of $\Lambda$ hyperons~\cite{Celiberto:2020rxb}.
 
  In order to deepen further and expand the knowledge of the strong interactions in the Regge limit, the aim to propose more exclusive processes than the Mueller--Navelet ones has been pursued. 
 For this reason, the study of new reactions needs to be performed.\\
 Hence, a first step to achieve this goal, since the process-dependent vertex describing the production of an identified hadron was obtained with NLA~\cite{hadrons}, it is possible to analyze a hybrid collision process in the NLA BFKL approach. Therefore, in Sec.~\ref{jethad} we introduce the hadron-jet reaction~\cite{Bolognino:2018oth}, where a charged light hadron $(\pi^{\pm}, K^{\pm}, p,\bar p)$ and a jet with high transverse momenta, separated by a large interval of rapidity, are produced together with an undetected hadronic system. This channel represents a challenging context where it is possible to constrain not only the parton distribution functions of the initial parton (the PDFs), but also the fragmentation function (the FFs) describing the detected hadron in the final state. In Section~\ref{heavy} we present our second step in order to probe the BFKL mechanism: the inclusive production of
 two heavy quarks~\cite{Bolognino:2019yls}, separated in rapidity, in the collision of two quasi-real gluons that could be studied at the LHC or at some future colliders as Future Circular Collider (FCC), where the high center-of-mass
 energy available in the proton-proton collisions affords to reproduce the needed kinematic
 conditions.
\section{Mueller--Navelet jets: the foundations}
\label{sec:MNjets}
As already mentioned the Mueller--Navelet jets production is one of the most significant reaction in the BFKL scenario. Therefore, in this section we will lay the foundations for a full comprehension of the new developed channels, presenting the Mueller--Navelet sector and introducing the theoretical framework, which is shared with the other processes.
\begin{figure}[t]
	\centering
	\includegraphics[scale=0.7]{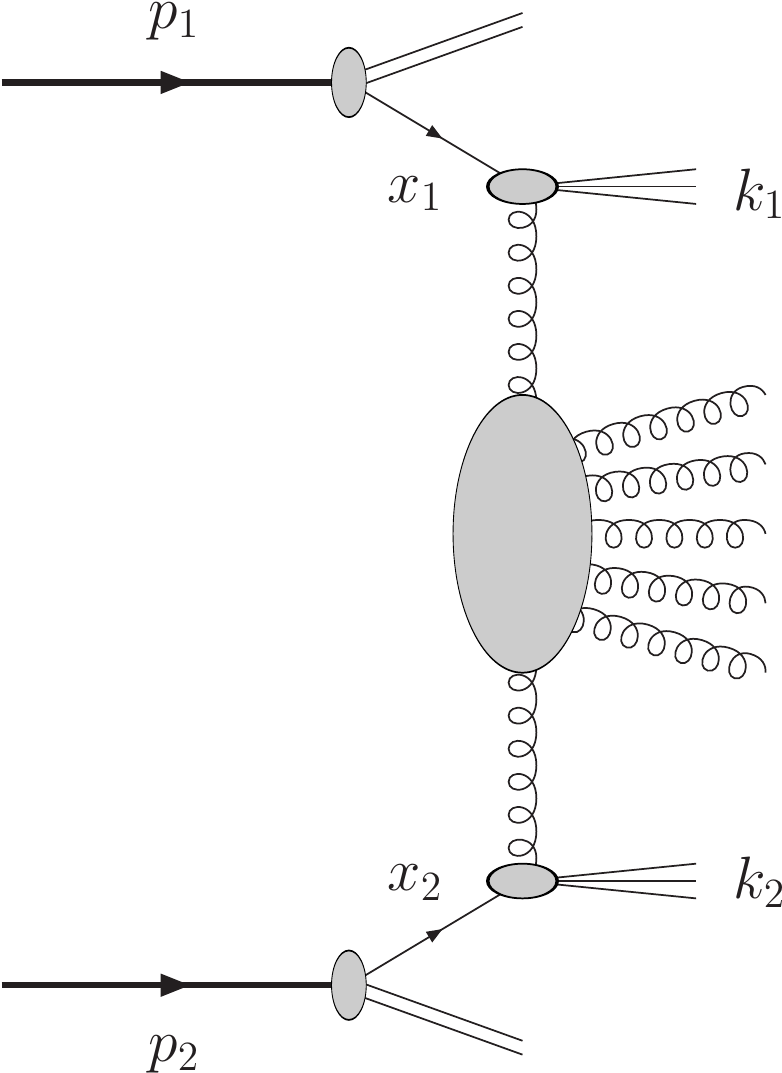}
	\caption{\emph{Mueller--Navelet jet production process in multi-Regge kinematics.}}
	\label{fig:mn-jets}
\end{figure}
\subsection{Theoretical setup}
\label{MNtheory}
The reaction of interest is the inclusive production of two jets (a \emph{dijet} system) in proton-proton collisions
\begin{equation}
\label{process-mn}
\text{p}(p_1) + \text{p}(p_2) \to \text{jet}(k_{J_1}) + \text{jet}(k_{J_2})+ \text{X} \;,
\end{equation}
where the two jets are characterized by high transverse momenta,
$\vec k_{J_1}^2\sim \vec k_{J_2}^2\gg \Lambda_{QCD}^2$ and large separation
in rapidity;  $p_1$ and $p_2$ are taken as Sudakov vectors (see Eq.~(\ref{sudakov_general})) satisfying
$p_1^2=p_2^2=0$ and $2\left( p_1 p_2\right)=s$. At the LHC high energies, the theoretical description of this process predicts the coexistence of two different approaches: collinear factorization and BFKL resummation. According the former, the jet impact factor can be expressed as a convolution between the PDF of the colliding hadron, ruled by DGLAP evolution, with the following hard process: a transition from the parton emitted by the proton to the forward jet tagged in the final state (see Fig.~\ref{fig:mn-jets}). However, also the BFKL approach occurs, allowed by the large center-of-mass energy $\sqrt{s}$ reachable at the LHC, resumming the large energy logarithms to all orders of perturbation.\\ The cross section given in Eq.~\eqref{sigma-ff}, which takes the form a convolution between two, process-dependent impact factors and a process-independent Green's function, holds for a fully inclusive process, that is without
any particle or object identified in the final state.
The inclusiveness condition requested in Eq.~(\ref{sigma-ff}) is relaxed due to the requirement of the the tagging of two forward jets, each of them produced in the fragmentation region of the respective parent proton.
This involves some effects on the form of the process-dependent part of the cross section, \emph{i.e.} the forward jet impact factors.
\begin{figure}[t]
	\centering
	\includegraphics[scale=0.85]{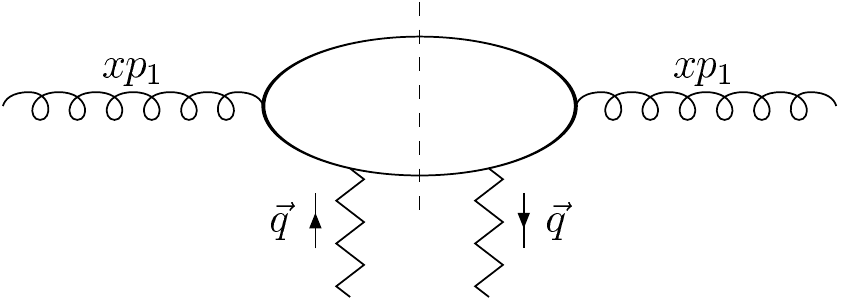}
	\includegraphics[scale=0.85]{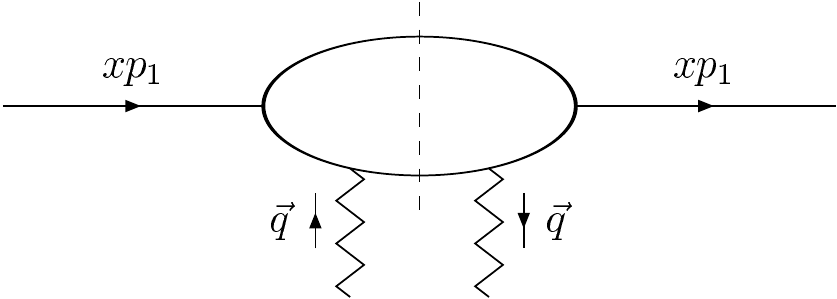}
	\caption{\emph{Representation of the forward gluon (left) and the forward quark (right) impact factor. Here $p_1$ is the proton momentum, $x_1$ is the fraction of proton momentum carried by the parton and $\vec q$ is the transverse momentum of the incoming Reggeized gluon.}}
	\label{fig:if-parton}
\end{figure}
In order to get the impact factor for a tagged jet (see Fig.~\ref{fig:if-jet}), it is necessary to take into account the impact factor for the colliding partons, calculated at the NLO~\cite{Fadin:1999de,Fadin:1999df}. Then the first step is to  `open' one of the integrations over the intermediate-state phase space inserting a proper defined function, which identifies the jet momentum with the momentum of one parton. In this way the jet is generated from the parton. At this point, take the convolution with the parent parton PDFs, following the collinear factorization.\\ The expression in which the jet momentum is identified with the momentum of the parton in the intermediate state $k$ reads~\cite{Ellis:1989vm}
 \begin{equation}
 \label{jetF0}
 S_J^{(2)}(\vec k;x)=\delta(x-x_J)\delta^{(2)}(\vec k-\vec k_J)
 \; ,
 \end{equation}
 where $x$ is the fraction of proton momentum carried by the quark, $x_J$ is the longitudinal fraction of the jet momentum and $\vec k_J$ is the transverse jet momentum.
\begin{figure}[t]
	\centering
	\includegraphics[scale=0.85]{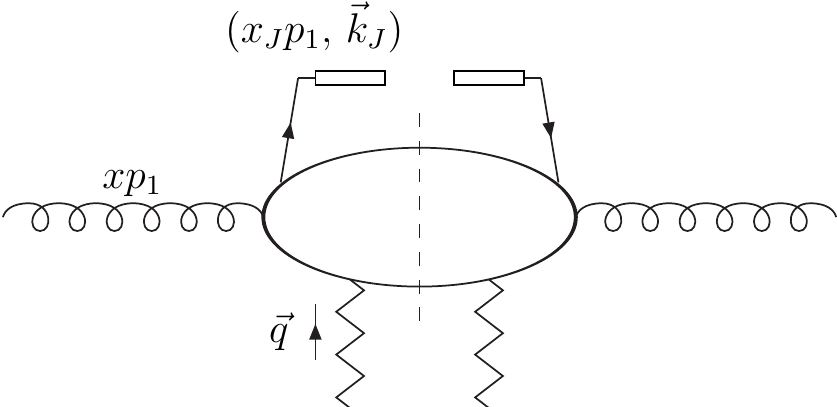}
	\includegraphics[scale=0.85]{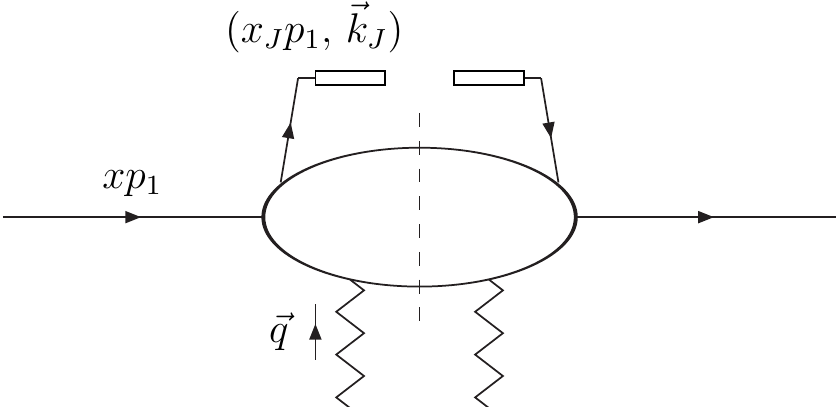}
	\caption{\emph{Diagram of the vertex for the forward jet production in the case of incoming gluon (left) or quark (right).
		Here $p_1$ is the proton momentum, $x$ is the
		fraction of proton momentum carried by the gluon/quark, $x_J p_1$ is the longitudinal jet momentum, $\vec k_J$ is the transverse jet momentum and $\vec q$ is the transverse momentum of the incoming Reggeized gluon.}}
	\label{fig:if-jet}
	 \end{figure}
	 The expression for the jet impact factor, differential with respect to the variables parameterizing the jet phase space, at the LO level can be written as
	 \begin{equation}
	 \label{if-jet-lo}
	 \frac{d\Phi^{(0)}_J(\vec q\,)}{dx_J dk_J}=
	 2\pi\alpha_s\sqrt{\frac{2C_F}{C_A}}
	 \int_0^1 dx
	 \left(\frac{C_A}{C_F}f_g(x)+\sum_{a=q, \bar q} f_{a}(x)\right)S_J^{(2)}
	 (\vec q;x)\;,
	 \end{equation}
	 Here $C_F$ is the color factor related to the gluon emission from a quark and the sum is over the gluon, all possible quark and antiquark PDF contributions $f_g(x)$, $f_a(x)$. The last step is the projection of the Eq.~\eqref{if-jet-lo} onto the eigenfunctions given by Eq.~\eqref{nuLLA} of the LO BFKL kernel in Eq.~\eqref{KLLA}.
\subsection{Cross section and azimuthal correlations}
\label{MNcrossaz}
The cross section of the Mueller--Navelet production, according the QCD collinear factorization, in \eqref{process-mn} is
\begin{equation}
\frac{d\sigma}{dx_{J_1}dx_{J_2}d^2k_{J_1}d^2k_{J_2}}
=
\sum_{r,s=q,{\bar q},g}\int_0^1 dx_1 \int_0^1 dx_2\ f_r\left(x_1,\mu_F\right)
\ f_s\left(x_2,\mu_F\right)
\end{equation}
\[ \hspace{-8.5cm} =
\frac{d{\hat\sigma}_{r,s}\left(x_1x_2s,\mu_F\right)}
{dx_{J_1}dx_{J_2}d^2k_{J_1}d^2k_{J_2}} \; ,
\]
where the $r, s$ indices are referred to the parton types (quarks $q = u, d, s, c, b$; antiquarks $\bar q = \bar u, \bar d, \bar s, \bar c, \bar b$; or gluon $g$), $f_i\left(x, \mu_F \right)$ indicates the initial proton PDFs; $x_{1,2}$ denote the longitudinal fractions of the partons involved in the hard subprocess,while $x_{J_{1,2}}$ are the jet longitudinal fractions; $\mu_F$ is the factorization scale; $d\hat\sigma_{r,s}\left(x_1x_2s, \mu_F \right)$ is
the partonic cross section for the production of jets and
$x_1x_2s \equiv \hat s$ is the squared center-of-mass energy of the
parton-parton collision subprocess (see Fig.~\ref{fig:mn-jets}).\\ According the derivation in the BFKL approach seen in (\ref{subsec:BFKLcrosssec-IF}), the cross section can presented as follows
\begin{equation}
\frac{d\sigma}
{dy_{J_1}dy_{J_2}\, d|\vec k_{J_1}| \, d|\vec k_{J_2}|
	d\phi_{J_1} d\phi_{J_2}}
=\frac{1}{(2\pi)^2}\left[{\cal C}_0+\sum_{n=1}^\infty  2\cos (n\phi )\,
{\cal C}_n\right]\, ,
\end{equation}
where $\phi=\phi_{J_1}-\phi_{J_2}-\pi$, ${\cal C}_0$ gives the total
cross section and the other coefficients ${\cal C}_n$ define the distribution of the azimuthal angle of the two jets. In what follows we will focus on one representation for ${\cal C}_n$ (although many equivalent ones in the NLA are available). We will use exponentiated representation, explained in (\ref{subsec:repr-equiv}), together with the so called BLM optimization method, whose theoretical lines have been presented in Section~(\ref{BLMscale}) and which will be applicated (only in the so called ``exact" case) to the Mueller--Navelet jets reaction in the next section.
\subsection{The BLM scale setting}
\label{BLM-MN}
In this Section we will provide the expression of the azimuthal coefficients, $\cal C_n$, for the process of our interest through the BLM procedure (see Section~\ref{BLMscale} and Ref.~\cite{Caporale:2015uva}). It is useful to define the rapidity distance between two emitted jets as $Y=y_{1}-y_{2} = \ln\frac{x_{1}x_{2}s}{|\vec k_{1}||\vec k_{2}|}$.
As we already discussed, the BLM optimal scale $\mu_R^{BLM}$ is defined as the value of $\mu_R$ that makes all contributions to the considered observables which are proportional to the QCD $\beta_0$ function, vanish. Hence, solving the equation
\begin{equation}
\label{int_Cn_beta}
C_n^{(\beta)}(s,Y) = 
\int_{k_1^{min}}^{k_1^{max}} d k_1
\int_{k_2^{min}}^{k_2^{max}} d k_2
\int_{y_1^{min}}^{y_1^{max}} d y_1
\int_{y_2^{min}}^{y_2^{max}} d y_2
\, \delta \left( Y - (y_1 - y_2) \right)
\, {\cal C}_n^{(\beta)}  = 0 \, ,
\end{equation}
the final expression, called ``exact", for the azimuthal coefficients is
\[
C_n^{NLA} = 
\int_{k_1^{min}}^{k_1^{max}} d k_1
\int_{k_2^{min}}^{k_2^{max}} d k_2
\int_{y_1^{min}}^{y_1^{max}} d y_1
\int_{y_2^{min}}^{y_2^{max}} d y_2
\,\; \delta \left( Y - (y_1 - y_2) \right)
\]
\[
\hspace{2cm}\times \,
\int_{-\infty}^{+\infty} d \nu \,
\frac{e^Y}{s}\,
e^{Y \bar \alpha^{MOM}_s(\mu^{BLM}_R)\left[\chi(n,\nu)
	+\bar \alpha^{MOM}_s(\mu^{BLM}_R)\left(\bar \chi(n,\nu) +\frac{T^{conf}}
	{3}\chi(n,\nu)\right)\right]}
\]
\[
\label{Cn_NLA_int_blm}
\hspace{-0.5cm}\times \, c_1(n,\nu,k_1, x_1)[c_2(n,\nu,k_2,x_2)]^*
\left(\alpha^{MOM}_s (\mu^{BLM}_R)\right)^2
\]
\begin{equation}
\label{C_nNLA}
\times \left\{1 + \alpha^{MOM}_s(\mu^{BLM}_R)\left[\frac{\bar c_1(n,\nu,k_1,x_1)}{c_1(n,\nu,k_1, x_1)}
+\left[\frac{\bar c_2(n,\nu,k_2, x_2)}{c_2(n,\nu,k_2,x_2)}\right]^*
+\frac{2T^{conf}}{3} \right] \right\} \, ,
\end{equation}
in which $\alpha_s^{MOM}$ is the QCD coupling 
in the physical momentum subtraction (MOM) scheme, related to 
$\alpha_s^{\overline{\text{MS}}}$ ruled by the Eq.~\eqref{scheme}.\\ In Eq.~\eqref{C_nNLA} the terms $\bar c_{i}(n,\nu,k_i,x_i)$ are the NLO impact-factor corrections after the subtraction 
of those terms, entering their expressions, which are proportional to $\beta_0$ and can be universally expressed through the LO impact factors:
\[
\label{IF_LO_sub}
\bar c_{i}(n,\nu,k_i,x_i) = \hat c_{i}(n,\nu,k_i,x_i) - \frac{\beta_0}{4 N_c} \left[ i \frac{d}{d \nu} \ln c_{i}(n,\nu,k_i,x_i) \right.
\]
\begin{equation}
\hspace{0.5cm}\left.+ \left( \ln \mu_R^2 + \frac{5}{3} \right) c_{i}(n,\nu,k_i,x_i) \right] \, .
\end{equation}
In order to provide a comparison between BFKL predictions and fixed-order
calculations (see Section~\ref{subsec:MNCMS}), we present also the expressions in the DGLAP approach at the NLO. The  $\cal C_n^{DGLAP}$ azimuthal coefficients are the truncation to the ${\cal O}(\alpha_s^3)$ order of the corresponding NLA BFKL ones. Therefore the analogous BLM-MOM scheme formula for the fixed-order case in high-energy regime reads:
\[
C_n^{DGLAP} = 
\int_{k_1^{min}}^{k_1^{max}} d k_1
\int_{k_2^{min}}^{k_2^{max}} d k_2
\int_{y_1^{min}}^{y_1^{max}} d y_1
\int_{y_2^{min}}^{y_2^{max}} d y_2
\,\; \delta \left( Y - (y_1 - y_2) \right)
\]
\[
\times \,
\int_{-\infty}^{+\infty} d \nu \,
\frac{e^Y}{s}\,
c_1(n,\nu,k_1, x_1)[c_2(n,\nu,k_2,x_2)]^*
\left(\alpha^{MOM}_s (\mu^{BLM}_R)\right)^2
\]
\[
\times \,
\left\{1 + \alpha^{MOM}_s(\mu^{BLM}_R)\left[Y \frac{C_A}{\pi} \chi(n,\nu) + \frac{\bar c_1(n,\nu,k_1,x_1)}{c_1(n,\nu,k_1, x_1)}\right.\right.
\]
\begin{equation}
\label{Cn_DGLAP_int_blm}
\left.\left.+\left[\frac{\bar c_2(n,\nu,k_2, x_2)}{c_2(n,\nu,k_2,x_2)}\right]^*
+\frac{2T^{conf}}{3} \right] \right\} \, .
\end{equation}
Although expressions in Eqs.\eqref{C_nNLA} and \eqref{Cn_DGLAP_int_blm} are performed in terms of $\alpha_s^{MOM}$ in the MOM scheme, it is possible to switch to the $\overline{\text{MS}}$ scheme. The proper way to accomplish this step is considering the general expression, then switching from the $\overline{\text{MS}}$ to the MOM scheme, and after fixing the BLM scales, if needed, moving back to the $\overline{\text{MS}}$ scheme. In Section~\ref{subsec:MNCMS} we will present predictions in both frameworks.
\subsection{Integration over the final-state phase space}
\label{subsec:mn-jets-phase-space}
 In what follows, we will consider the \emph{integrated coefficients} in order to find a correspondence with the kinematical cuts used by the CMS collaboration (see Ref.~\cite{Khachatryan:2016udy} for details),
 \begin{equation}
 	\label{Cm_int-mn-jets}
 	\int_{y_{J_2}^{min}}^{y_{J_2}^{max}}dy_{J_2}\int_{k_{J_1}^{min}}^{\infty}dk_{J_1}\int_{k_{J_2}^{min}}^{\infty}dk_{J_2} \, \delta\left(y_{J_1}-y_{J_2}-Y\right)\times \,
 {\cal C}_n\left(y_{J_1},y_{J_2},k_{J_1},k_{J_2} \right)
  \end{equation}
 and their ratios $R_{nm}\equiv C_n/C_m$. The physical meaning of the ratios of the form $R_{n0}$ is that one of the azimuthal correlations $\langle \cos(n\phi)\rangle$. The jet rapidities were fixed in the range delimited by $y_{J_1}^{min}=y_{J_2}^{min}=-4.7$ and 
 $y_{J_1}^{max}=y_{J_2}^{max}=4.7$ and the dependence of the $R_{nm}$ ratios as function of the jet rapidity separation $Y \equiv y_{J_1}-y_{J_2}$ was investigated. Regarding the the jet transverse momenta $k_{J_{1,2}}$, both the \emph{symmetric} and the \emph{asymmetric} choices for the cuts are adopted. The analyses are presented for two characteristic values for the center-of-mass energy, \emph{i.e.} $\sqrt s=7$ TeV, for which experimental data with \emph{symmetric} configuration for the outgoing jet momenta
 already exist, and $\sqrt s=13$ TeV.
 
\subsection{Mueller--Navelet jets at CMS}
\label{subsec:MNCMS}
In this Section we propose predictions for the different azimuthal correlations $R_{nm} \equiv C_n/C_m$ for the Mueller--Navelet channel in the symmetric CMS configuration:
\begin{equation*}
35\,\mbox{GeV} < k_{J \, 1,2} < k_J^{max} \;\qquad \mbox{where}\qquad k_J^{max} \equiv k_{J,{(CMS)}}^{max}= 60 \,\mbox{GeV}\,,
\end{equation*}
\begin{equation*}
|y_{J \, 1,2}| < 4.7 \; \qquad \mbox{with} \qquad Y < 9.4 \;,
\end{equation*}
and compared with the experimental CMS data at $\sqrt{s} = 7$ TeV~\cite{Khachatryan:2016udy}.
 We refer to the improved version of the analysis provided in Ref.~\cite{Caporale:2014gpa}, shown in the recent project given by Ref.~\cite{Celiberto:2020wpk}. Here the BLM ``exact" optimization scale method is adopted and the final calculations are done in the $\overline{\text{MS}}$ renormalization scheme. We remind that the formulae cited in the previous Section are in the MOM scheme and it is possible to obtain the corresponding ones in the $\overline{\text{MS}}$ scheme, through the substitutions:
\[
T^{conf} \to - T^\beta \;, \qquad \alpha_s^{(MOM)} \to \alpha_s^{(\overline{\text{MS}})} \;,
\]
with $T^{conf}$, $T^\beta$, $\alpha_s^{(\overline{\text{MS}})}$ and $\alpha_s^{(MOM)}$ in Eqs.~\eqref{scheme} and \eqref{T_Tbeta_Tconf}.\\
The analysis is performed using the NLO PDF4LHC15 parametrization~\cite{Butterworth:2015oua}, which represents a statistical combination of the three most popular NLO PDF sets: MMHT 2014~\cite{Harland-Lang:2014zoa}, CT 2014~\cite{Dulat:2015mca} and NNPDF3.0~\cite{Ball:2014uwa}. In particular, a sample of $100$ replicas of the central value of the PDF4LHC15 set is used. This collection of replicas is obtained by altering the central value, in a random way, with a Gaussian background characterizing by the original variance.

\begin{figure}[!h]
	\centering
	\includegraphics[scale=0.38]{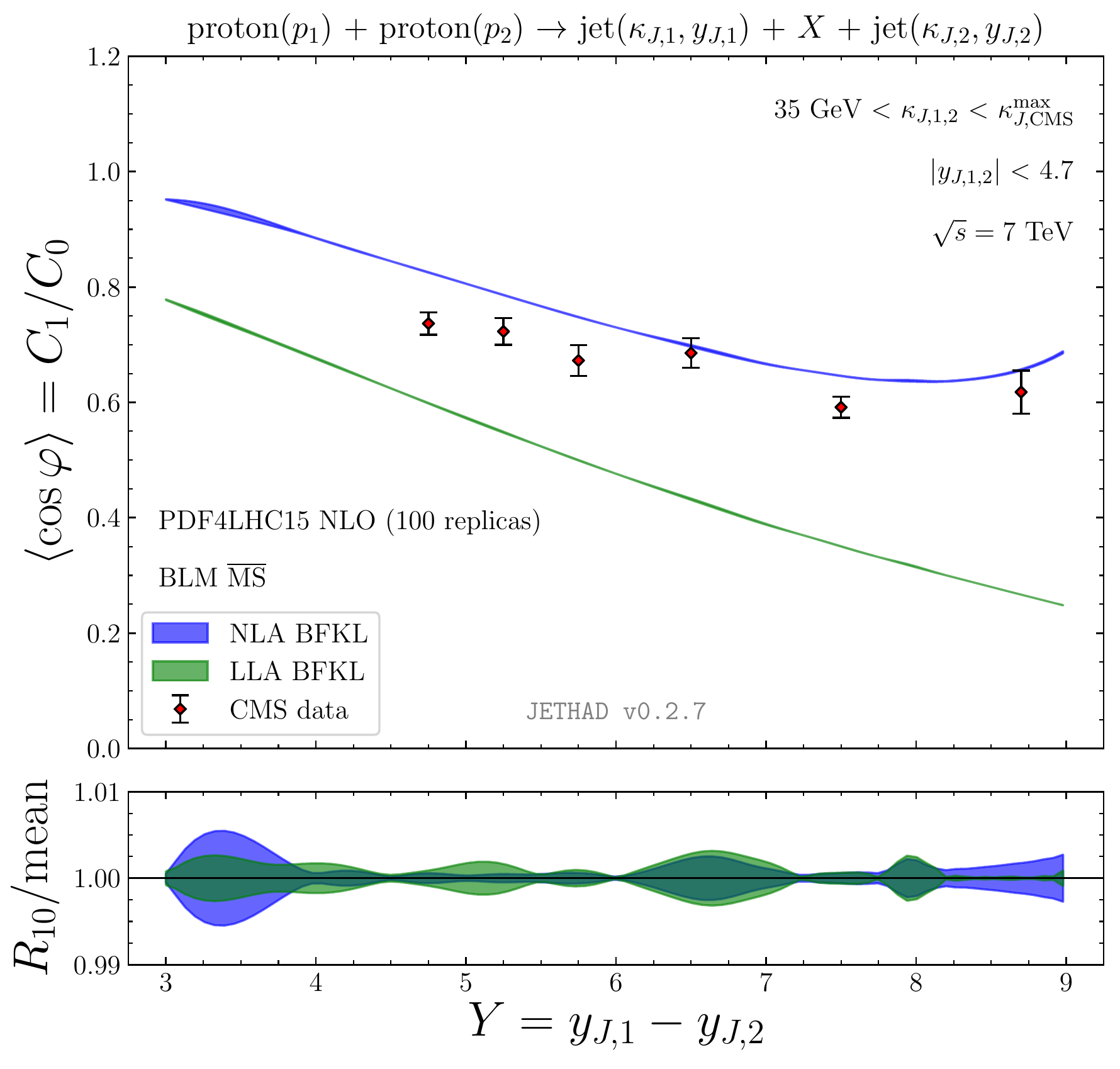}
	\includegraphics[scale=0.38]{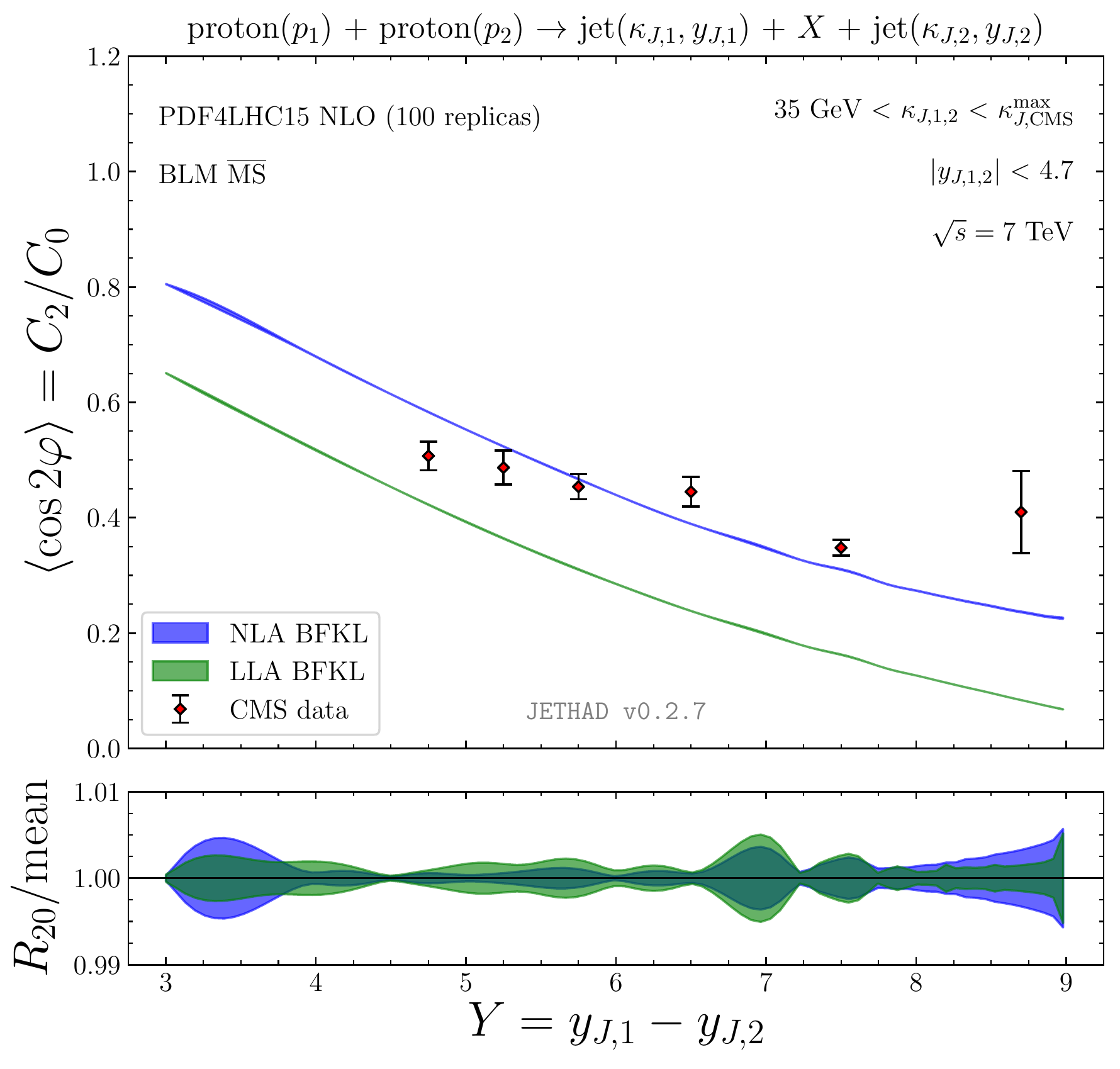}
	
	\includegraphics[scale=0.38]{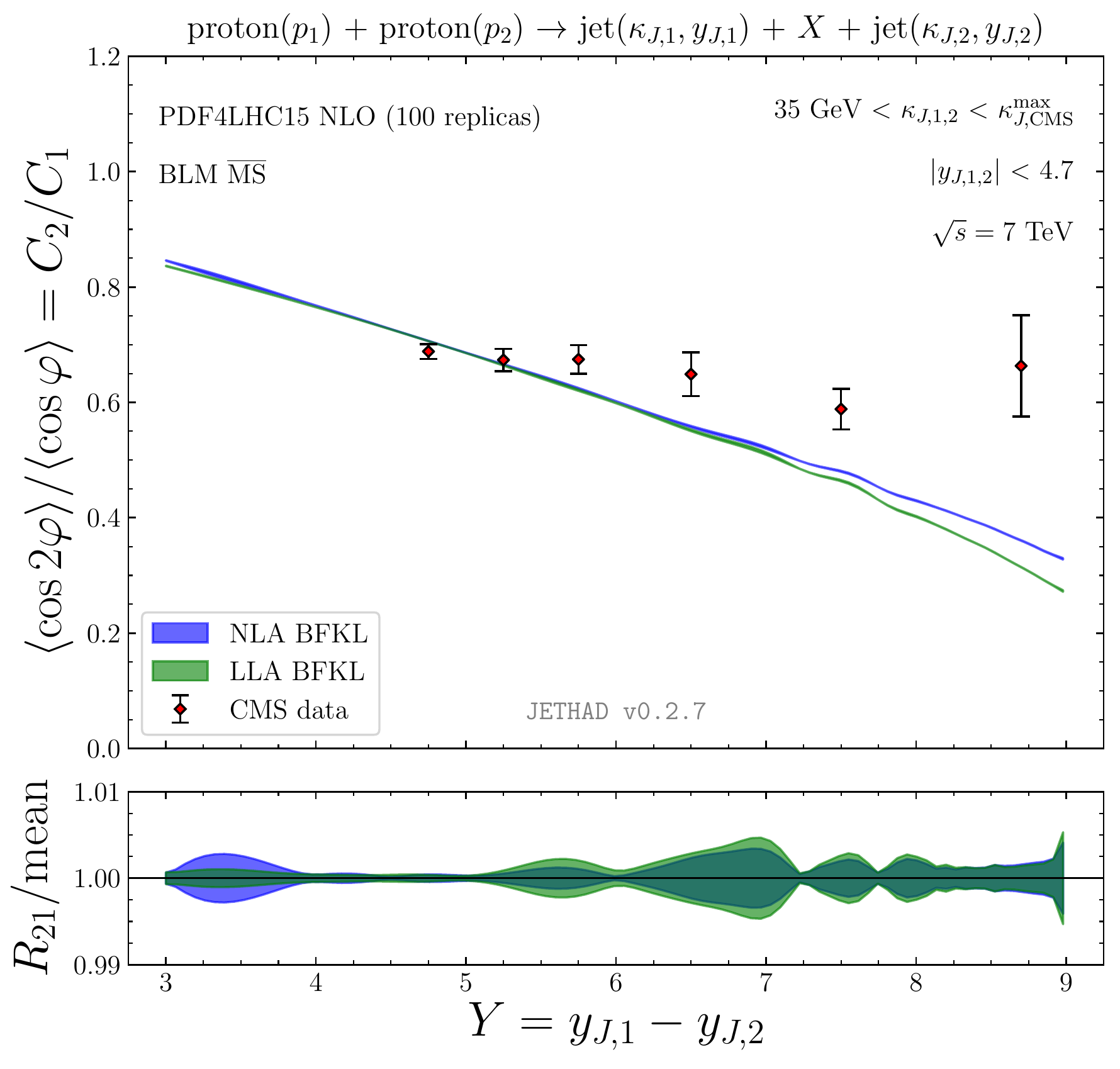}
	\includegraphics[scale=0.38]{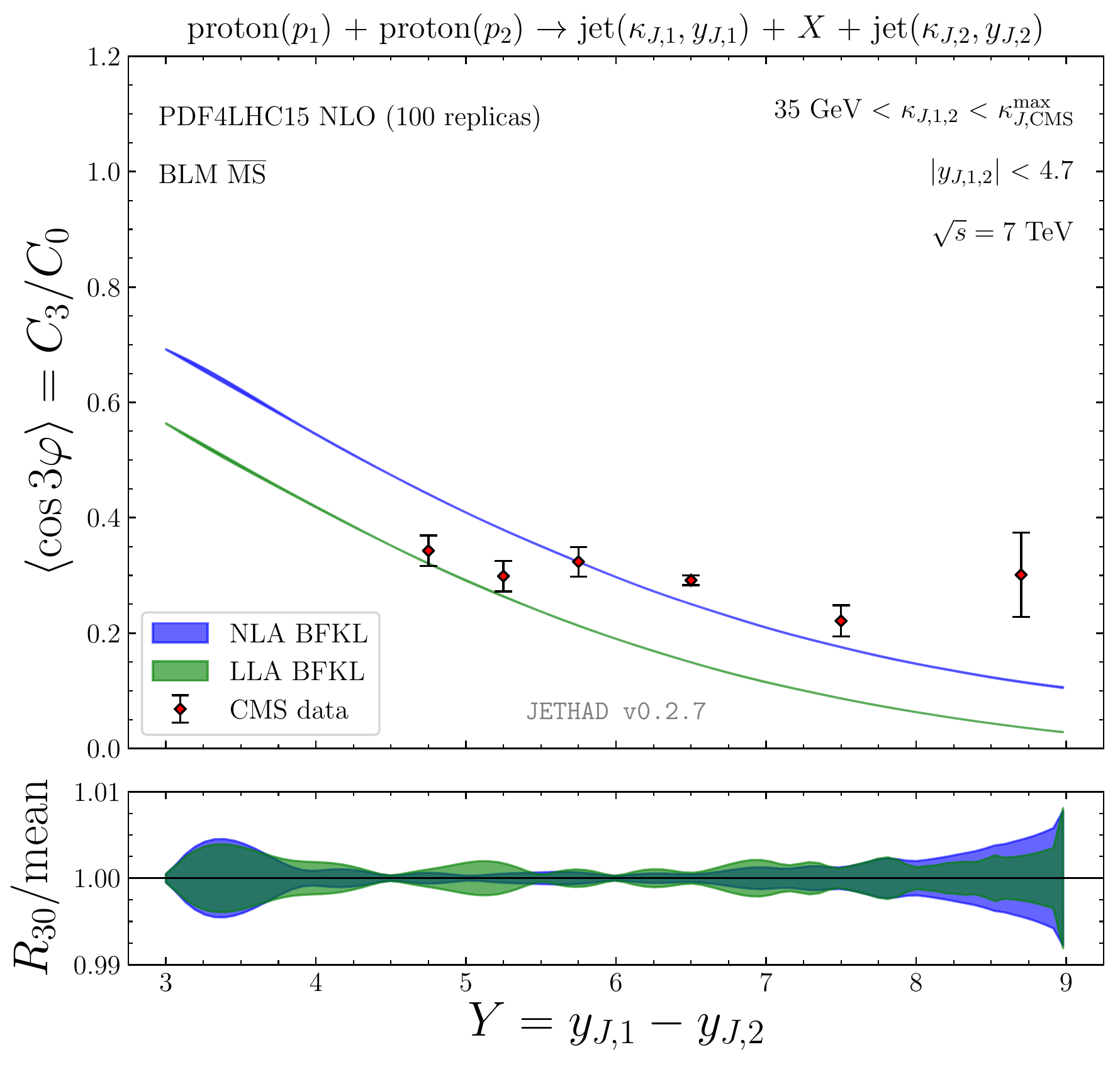}
	
	\includegraphics[scale=0.38]{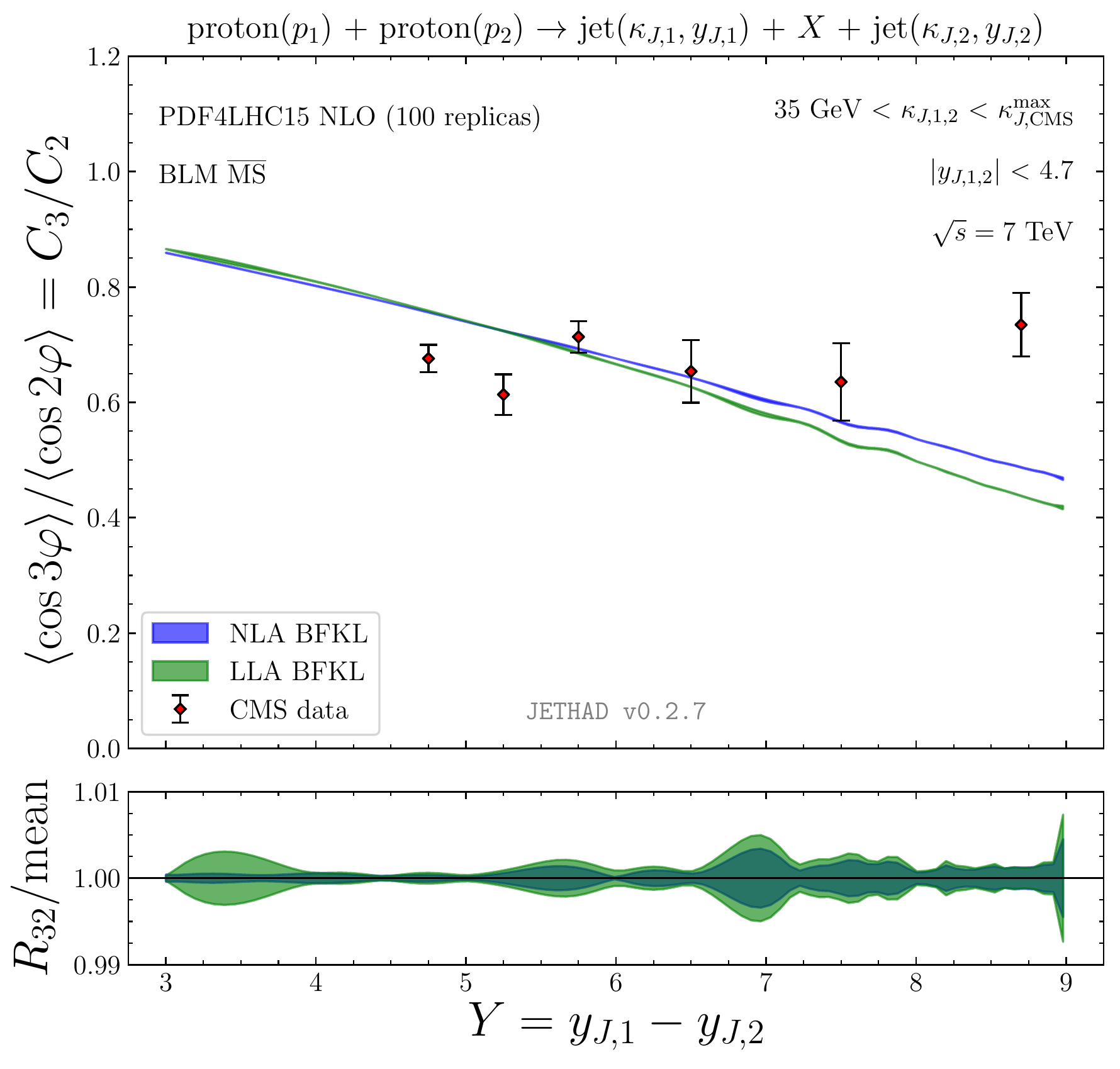}
	\caption{\emph{$Y$-dependence of 
			$R_{10}$, $R_{20}$, $R_{30}$, $R_{21}$, and $R_{32}$ 
			at $\sqrt{s} = 7$ TeV~\cite{Khachatryan:2016udy} 
			in the \emph{symmetric} configuration in $\overline{\text{MS}}$ scheme. Supporting plots below the primary pannels show the $R_{nm}$ replicas' results divided by their mean value. Plots from Ref.~\cite{Celiberto:2020wpk}.}}
	\label{fig:tve_BLM}
\end{figure}

In Fig.~\ref{fig:tve_BLM} we can observe the loss of correlation at increasing of the rapidity interval $Y$.
Moreover, as we can see, LLA calculations overestimate undoubtedly the decorrelation between the two tagged jets. The agreement with experimental data is clearly improved, when predictions take into account also the NLA BFKL corrections, together with the BLM prescription. Supporting plots below the main panels in Fig.~\ref{fig:tve_BLM} illustrate that the relative standard deviation of the replicas' results is extremely small ($\ll$ 1\%).

The analysis on azimuthal-angle decorrelations in the Mueller--Navelet channel, conducted by the CMS collaboration at $\sqrt{s} = 7$ TeV, provided a valid indication that the considered kinematic domain lies in between
the sectors described by the BFKL and the DGLAP evolution, whereas clearer manifestations of high-energy effects are expected to be more definite at increasing collision energies. Then, recent studies~\cite{Celiberto:2015yba,Celiberto:2015mpa} have highlighted how the use of partially asymmetric configurations for the transverse momenta of the two tagged jets allows for a clear separation between BFKL-resummed and fixed-order predictions. In what follows it will be clear that the use of $k$-asymmetric windows has various advantages.

\subsection{BFKL versus fixed order}
\label{subsec:BFKL_DGLAP}

In addition to the previous discussion, it is worth to remark that the use of the symmetric cuts in the values of $k_{J_i}^{min}$
maximizes the contribution of the Born term in $C_0$, which exists and is quite large for  back-to-back jets, hence hiding the effect of BFKL resummation in the cross section. The choice of an asymmetric configuration, instead, permits a reduction of the Born contribution, increasing the effects of the undetected gluon radiation, which means enhancing the BFKL effects compared to the fixed-order DGLAP ones. These reasons are the key of the following analysis. Therefore, this Section is devoted to the comparison between the predictions for the azimuthal correlations and their ratios obtained both in the BFKL NLA framework and in the fixed-order DGLAP approach. We want to remind that the calculation for the $\cal C_n^{DGLAP}$ at NLO is approximated: the expression is nothing but the NLA BFKL one truncated to the ${\cal O}(\alpha_s^3)$. This approximation is justified in the regime of large rapidity distance $Y$, studied in the present analysis. It allows us to take into account the leading power of the exact fixed-order DGLAP calculations, but neglecting those factors which are suppressed by the inverse power of the energy of the partonic subprocess.

\FloatBarrier
\begin{figure}[!h]
	\centering
	\includegraphics[scale=0.49,clip]{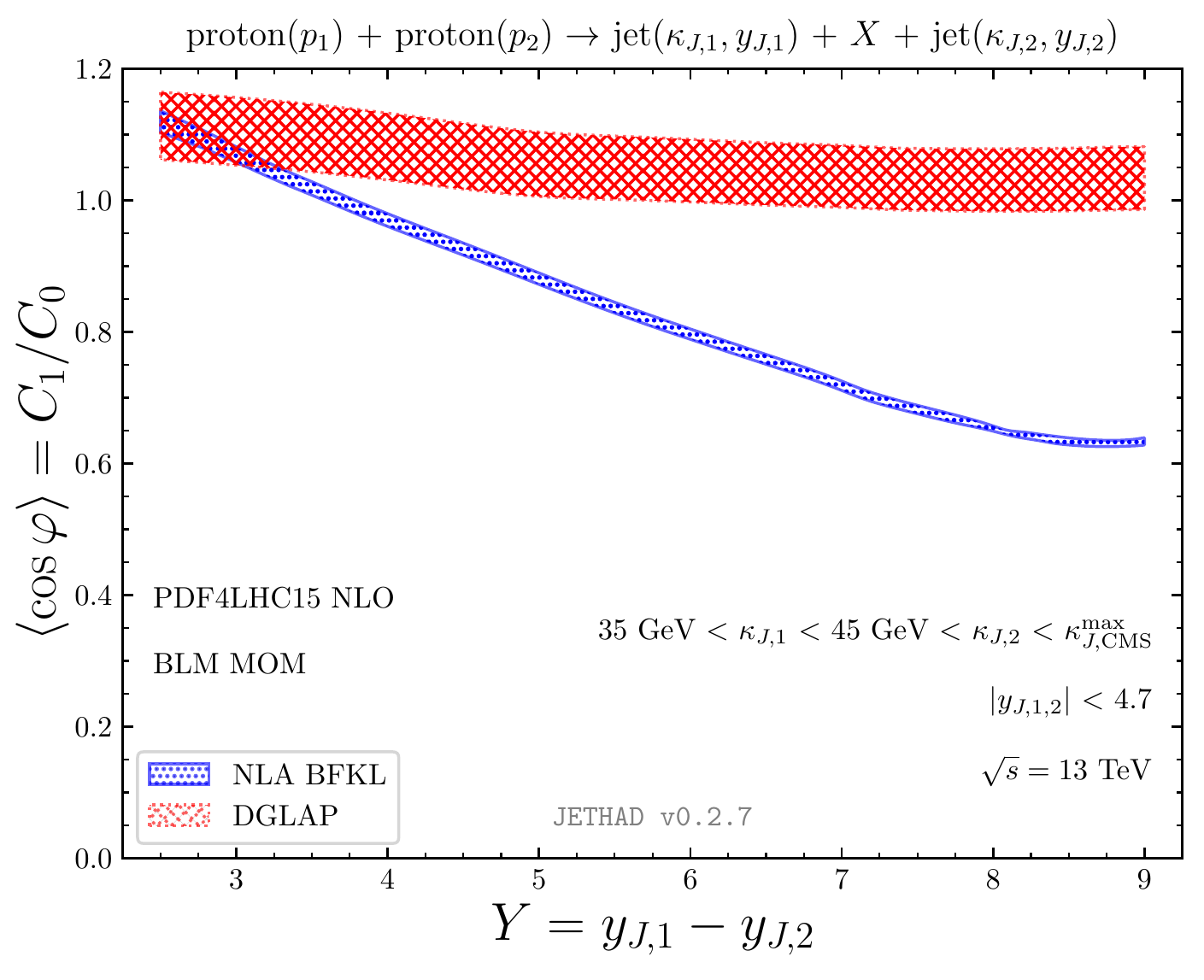}
	\hspace{0.25cm}
	\includegraphics[scale=0.49,clip]{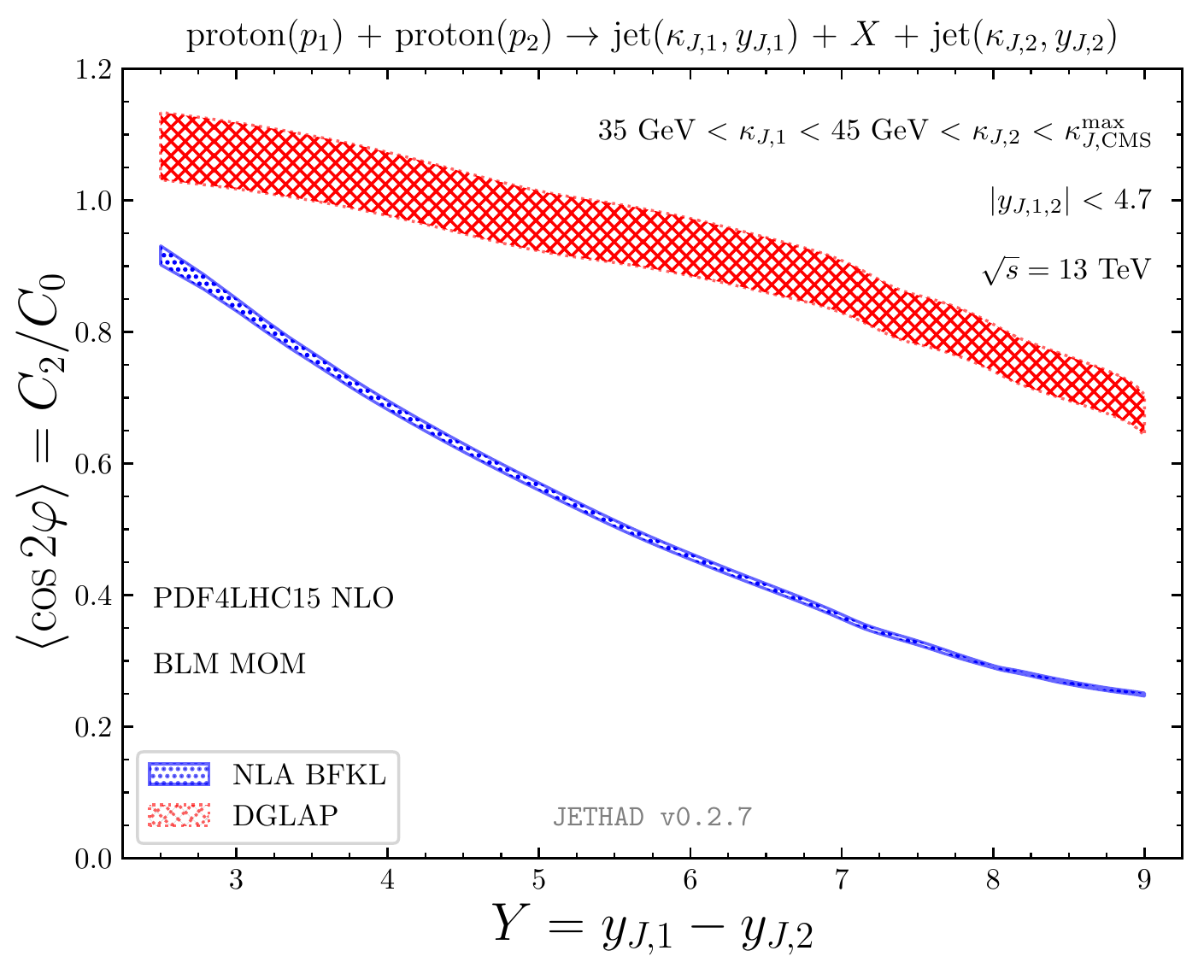}
	
	\includegraphics[scale=0.49,clip]{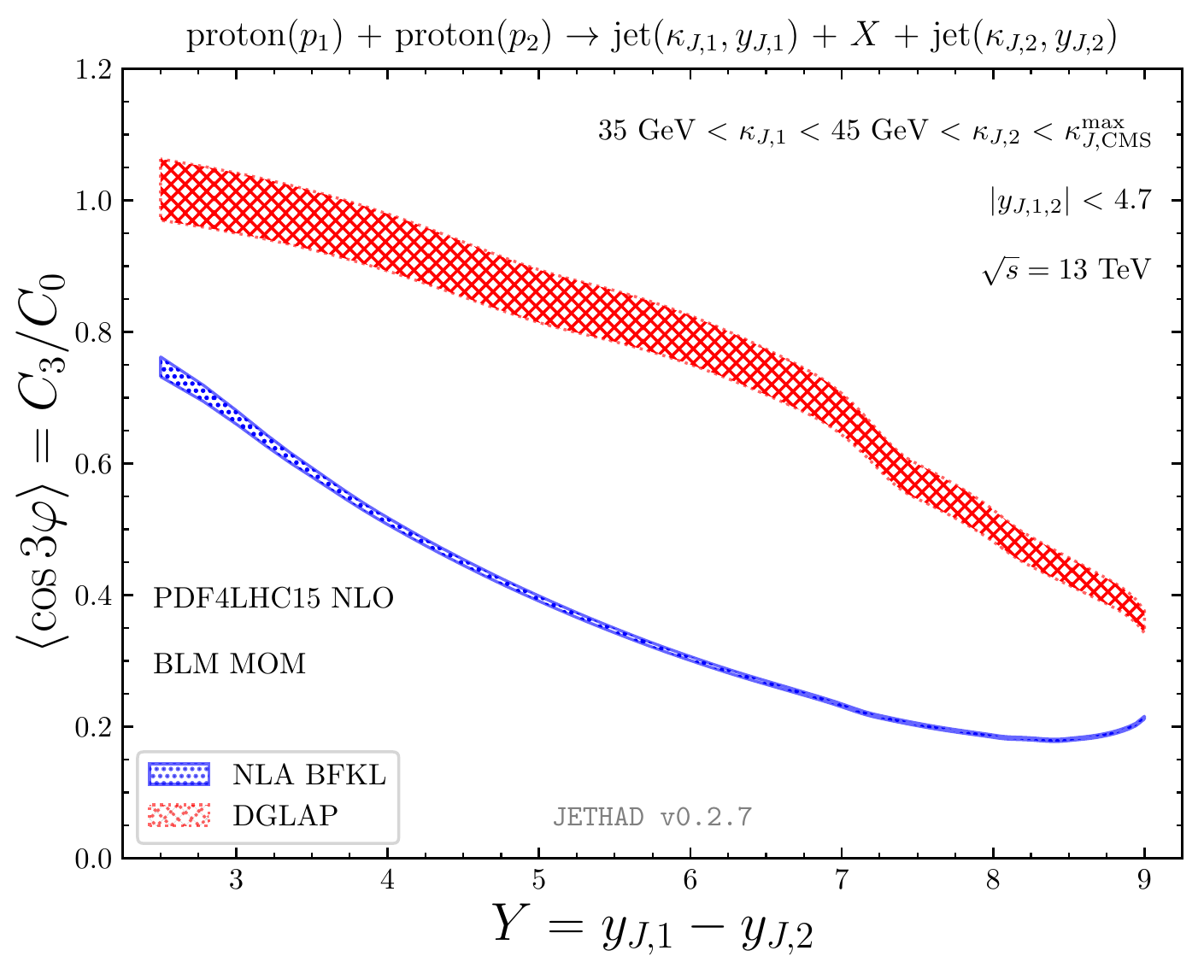}
	\hspace{0.25cm}
	\includegraphics[scale=0.49,clip]{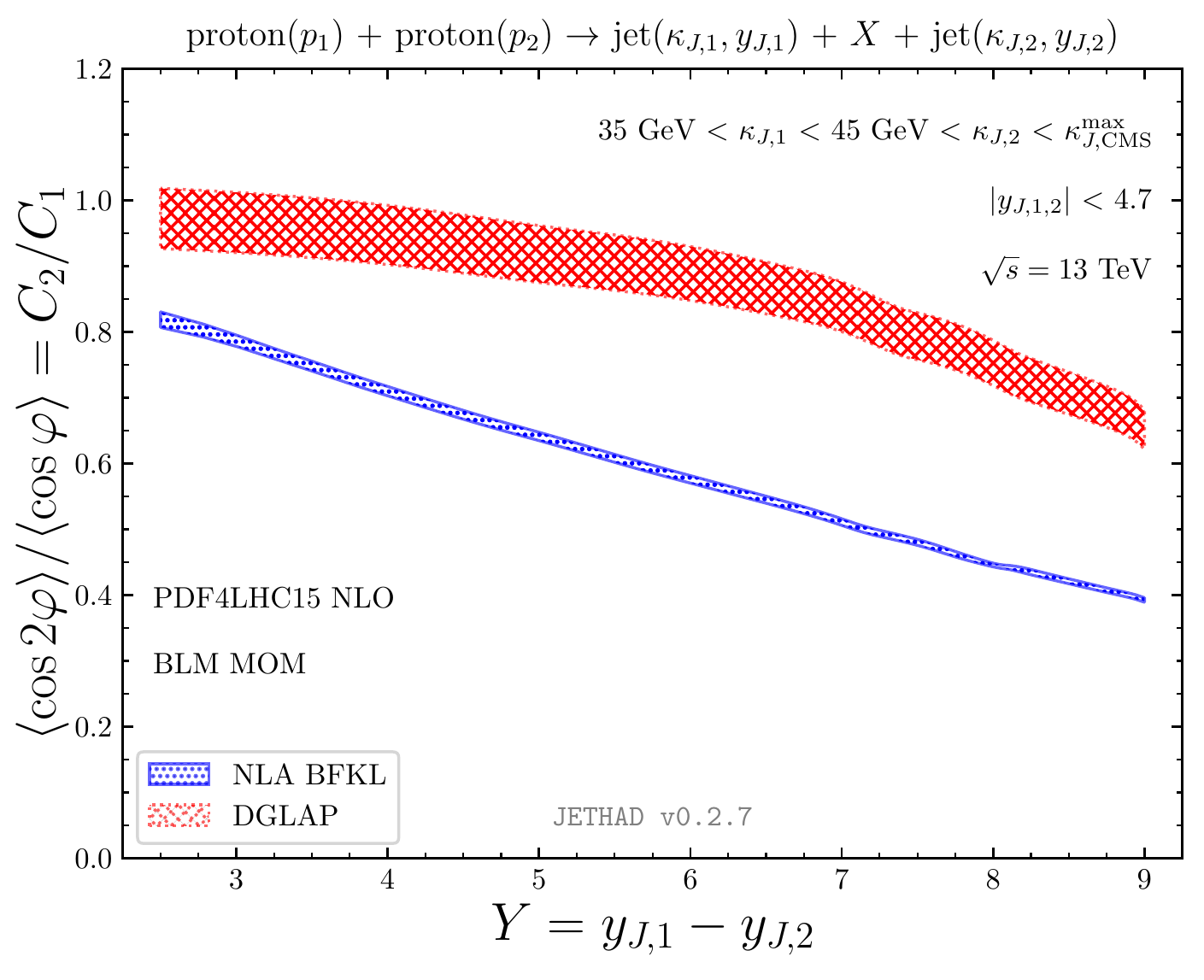}
	
	\includegraphics[scale=0.49,clip]{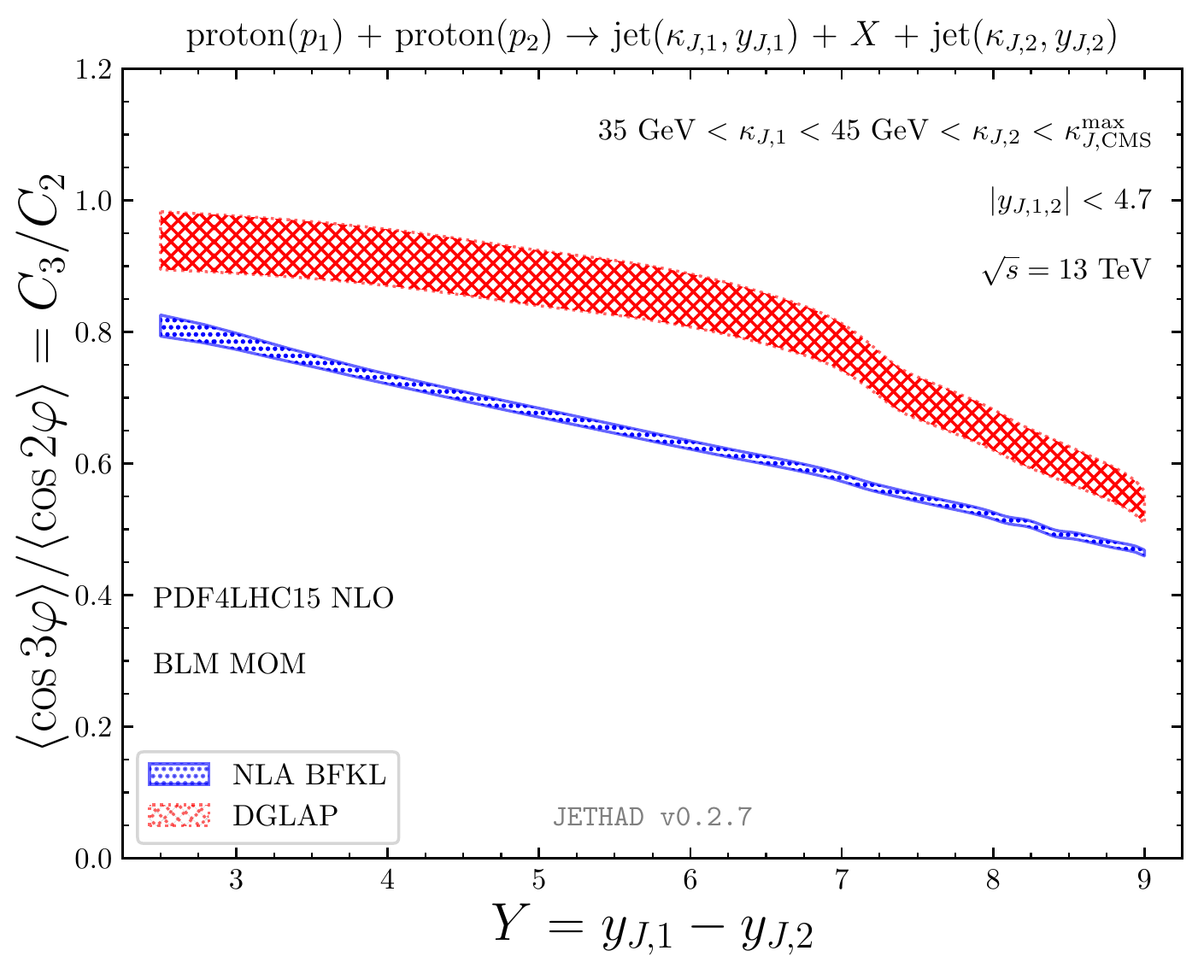}
	
	\caption{\emph{$Y$-dependence of several azimuthal correlations, $R_{nm} \equiv C_n/C_m$, of the Mueller--Navelet jet production for
			$\mu_{F1,2} = \mu_R = \mu_R^{BLM}$ and $\sqrt{s} = 13$ TeV (\textit{asymmetric CMS} configuration). Full NLA BFKL predictions are compared with the corresponding ones in the high-energy DGLAP approach. Plots from Ref.~\cite{Celiberto:2020wpk}.}}
	\label{fig:MN-BvD-CMS}
\end{figure}
\FloatBarrier
 Figure~\ref{fig:MN-BvD-CMS} shows the NLA BFKL predictions for the azimuthal ratios $R_{nm} \equiv C_n/C_m$ in the Mueller--Navelet jets reaction, provided by Eq.~\eqref{C_nNLA}, which is compared with the corresponding DGLAP ones in Eq.~\eqref{Cn_DGLAP_int_blm}.
 Here the exact BLM procedure is applied. The analysis is performed at $\sqrt{s} = 13$ TeV using the asymmetric CMS kinematics configuration, which consists in requiring the emitted jets to be tagged in disjoint intervals for the transverse momenta: $35$ GeV $< k_{J,1} < 45$ GeV $< k_{J, 2} < k_{J}^{max}$. In all results the physical renormalization MOM scheme is adopted. We remind that it is also possible to use the $\overline{\text{MS}}$ scheme.
 The behavior of the $R_{n0}$ series presents a plateau at large rapidiy distance $Y$, which changes into a turn-up for $R_{10}$ (upper left panel of Fig.~\ref{fig:MN-BvD-CMS}). From the kinematic point of view, this is due to the weight of undetected hard-gluon radiation which is suppressed by the large rapidity intervals, giving way to soft-gluon emissions which, because of their nature, do not modify the hard subprocess kinematics. For this reason, the correlation of the two detected jets in the relative azimuthal angle is restored. Instead, from the parton-density point of view, the increase of $Y$ values shifts the parton longitudinal fractions towards the so-called threshold region, where the energy of the di-jet system approaches the value of the center-of-mass energy, $\sqrt{s}$. Therefore, the PDFs are in ranges close to the end-points of their definitions, where they manifest large uncertainties. In these configurations, the formalism does not include the sizeable effect of threshold double logarithms which enter the perturbative series and have to be resummed to all orders. Due to the oscillatory behavior of the $\nu$-integrand in Eq.~(\ref{Cn_DGLAP_int_blm}), not compensated by the exponential factor as in the NLA BFKL case, in Eq.~(\ref{Cn_NLA_int_blm}), the error bands in the DGLAP case are larger with respect to the resummed ones.

 In the next Section we extend the discussion to a more exclusive process in which one of the two jets is replaced by a hadron: the hadron-jet reaction.
\section{Hadron-jet production}
\label{jethad}
\begin{figure}[!h]
	\centering
	\includegraphics[scale=0.5]{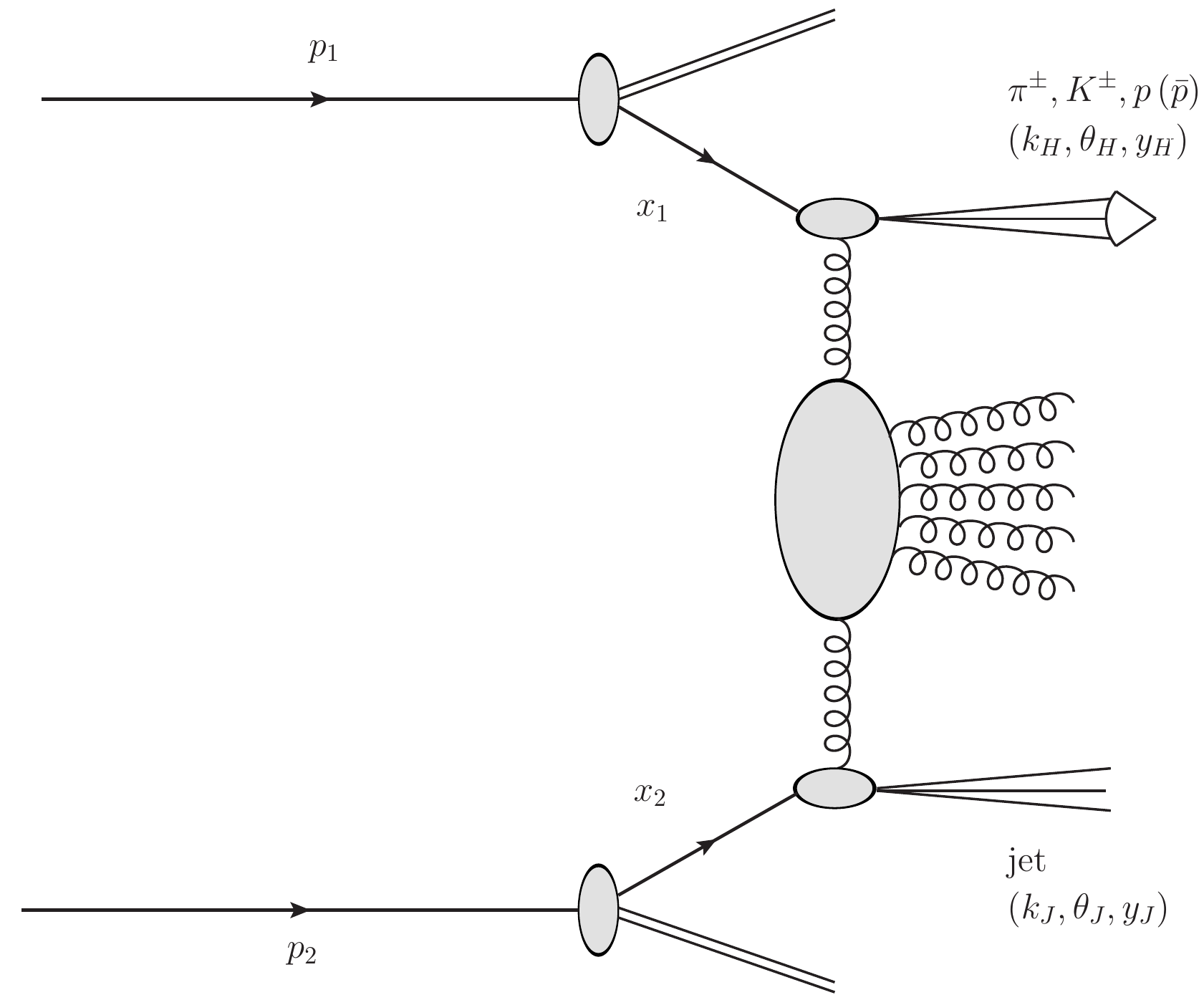}
	\caption{\emph{Inclusive hadroproduction of a charged light hadron and of a jet.}}
	\label{fig:hadron-jet}
\end{figure}

This Section is dedicated to a new process which can be recognized as a testfield for the BFKL dynamics at the LHC: the inclusive hadron-jet production in proton-proton collision~\cite{Bolognino:2018oth,Bolognino:2019yqj,Bolognino:2019cac}
\begin{eqnarray}
\label{processHJ}
{\text{p}}(p_1) + {\text{p}}(p_2) 
\to 
{\text{hadron}}(k_H, y_H) + \text{X} + \text{jet}(k_J, y_J) \;,
\end{eqnarray}
where a charged light hadron, $\pi^{\pm}, K^{\pm}, p \left(\bar p\right)$ and a jet with high transverse momenta, separated by a large interval of rapidity, are produced together with an undetected hadronic system X (see Fig.~\ref{fig:hadron-jet} for a schematic view).

This production channel has many characteristics in common with the inclusive $J/\Psi$-meson plus backward jet production, considered in Ref.~\cite{Boussarie:2017oae}.

 From the experimental point of view, the detection of
the $J/\Psi$-meson seems to be challenging. But from the theory side, there are more uncertainties in this case in comparison to our proposal. The $J/\Psi$-meson production impact factor was considered in LO; moreover, several production mechanisms in the frame of nonrelativistic QCD were discussed. Instead, the light hadron impact factor is well defined in collinear factorization and all ingredients are known with NLO accuracy. Past experience in BFKL calculations for numerous
processes at LHC shows that taking into account NLO corrections to the impact factors leads both to a remarkable change of predictions and to a significant reduction of the theoretical uncertainties.
\FloatBarrier
The process under investigation shares the theoretical framework with the Mueller--Navelet jets production, but with a simple and substantial difference: the replacement of one of the two jet impact factor entering the Mueller--Navelet formulas with the vertex for the proton-to-hadron transition. Although the treatment of this reaction in the BFKL framework could appear an easy goal, there are some relevant aspects which makes this process appealing, such that building numerical predictions for the observables of interest, \emph{i.e.} cross sections and azimuthal angle correlations, and proposing them to attention of both theorists and expermentalists would reveal to be a notable scenario.
In particular, we remember the BFKL resummation implies specific factorization structure for the predicted observables: these ones are calculated as a convolution of the universal BFKL Green's function with the process dependent impact factors, which resembles the factorization in Regge theory. Hence, a first motivation is to test this picture experimentally, considering all possible processes for which the full NLO BFKL description is available. Moreover, for the process under consideration only one hadron in the final state
should be identified, instead of two as in the hadron-hadron inclusive
production (see Refs.~\cite{Celiberto:2016hae},~\cite{Celiberto:2017ptm}), the other identified object being a jet with a typically much larger transverse momentum. This should make simpler the mining of these events out of the minimum-bias ones, which represent the main background in high-luminosity runs at a collider.
As we previously discussed, in the Mueller--Navelet channel, the choice of asymmetric cuts for the transverse momenta of the tagged jets enhances
the effects of the additional undetected hard gluon radiation, suppressing the Born term, present only for back-to-back jets. In this way the impact of the BFKL framework is more evident with respect to the DGLAP contribution. For the inclusive hadron-jet production we are considering this kind of asymmetry would be a natural feature due the two detected objects. In fact, the identified jet
should have transverse momentum not smaller than 20~GeV, whereas the
minimum hadron transverse momentum can be as small as 5~GeV. In this sense, the present reaction allows us to access to a complementary kinematical region of the Mueller--Navelet one.
Last but not least, an advantageous reason to use and test this channel lies in comparing models for FFs or for jet algorithms, exploiting expressions which are linear in the corresponding functions and not quadratic as it would be, respectively, in the hadron-hadron and in the Mueller--Navelet jet case. 
\begin{center}
***
\end{center}
In Section~\ref{JH_setup} we introduce the theoretical framework of the process, providing the main expressions for the cross section and azimuthal correlations (see Section~\ref{JH_cross_section}); a discussion on the BLM procedure (Section~\ref{JH_BLM}) together with the final-state phase space
integration~\ref{JH_int} will follow. Then Section~\ref{JH_num_tools} is devoted to the details of numerical implementation. Finally, we show and discuss our results (see Section~\ref{JH_results}) in the full next-to-leading logarithmic approximation of cross sections and azimuthal correlations, taking advantage of the richness of configurations gained by combining the acceptances of CMS and CASTOR detectors. 
We close with a Summary in Section~\ref{JH_Sum}.

The present analysis given in this Section is based on a work done in Ref.~\cite{Bolognino:2018oth} and presented also in~Refs.~\cite{Bolognino:2019yqj, Bolognino:2019cac}.
\subsection{Theoretical setup}
\label{JH_setup}
The process invoved in this discussion, schematically represented in the Fig.~\ref{fig:hadron-jet}, is featured by a final state configuration in which a charged
light hadron $(k_H, y_H)$ and a jet $(k_J, y_J)$ are detected, with
a large rapidity separation, together with an undetected system of hadrons. Here we will consider the case where the 
hadron rapidity $y_H$ is larger than the jet one  $y_J$, so that $Y\equiv
y_H-y_J$ is always positive. This implies that, for most of the considered
values of $Y$, the hadron is forward and the jet is backward.

The hadron and the jet are also required to possess large transverse momenta,
$\vec k_H^2\sim \vec k_J^2 \gg \Lambda^2_{QCD}$.
The protons' momenta $p_1$ and $p_2$ are taken as Sudakov vectors satisfying $p^2_1= p^2_2=0$ and $2 (p_1p_2) = s$,  so that the momenta of the final-state objects can be decomposed as
\begin{eqnarray}
k_H&=& x_H p_1+ \frac{\vec k_H^2}{x_H s}p_2+k_{H\perp} \ , \quad
k_{H\perp}^2=-\vec k_H^2 \ , \nonumber \\
k_J&=& x_J p_2+ \frac{\vec k_J^2}{x_J s}p_1+k_{J\perp} \ , \quad
k_{J\perp}^2=-\vec k_J^2 \ .
\label{sudakov}
\end{eqnarray}
In the center-of-mass system, the hadron/jet longitudinal momentum fractions 
$x_{H,J}$ are related to the respective rapidities through the relations
$y_H=\frac{1}{2}\ln\frac{x_H^2 s}
{\vec k_H^2}$, and $y_J=\frac{1}{2}\ln\frac{\vec k_J^2}{x_J^2 s}$, 
so that $dy_H=\frac{dx_H}{x_H}$, $dy_J=-\frac{dx_J}{x_J}$,
and $Y=y_H-y_J=\ln\frac{x_Hx_J s}{|\vec k_H||\vec k_J|}$, where the
space part of the four-vector $p_{1\parallel}$ being taken positive.
Studying the Mueller--Navelet jets production, the BFKL mechanism is investigated via a fully inclusive process. Instead, in this context, we consider a less inclusive final-state channel, requiring that a hadron is identified in the final state. Now following the procedure adopted in the Mueller--Navelet case, described in Section~\ref{sec:MNjets}, the first step is considering the NLO forward parton impact factor~\cite{Fadin:1999de, Fadin:1999df} pictorially represented in Fig.~\ref{fig:if-parton}.
 \begin{figure}[t]
 	\centering
 	\includegraphics[scale=0.85]{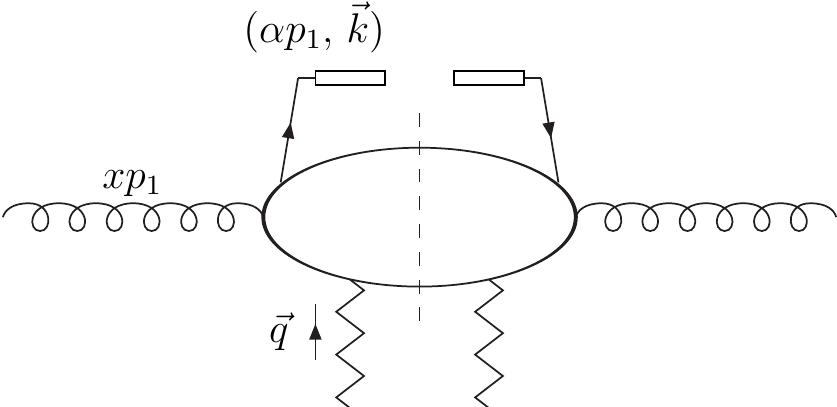}
 	\includegraphics[scale=0.85]{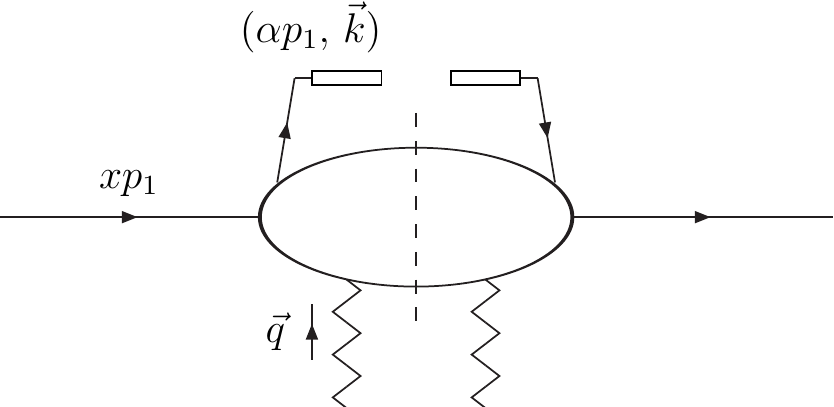}
 	\caption{\emph{Schematic view of the the vertex for the identified hadron production for the case of incoming gluon (left) or quark (right). Here $p_1$ is the proton momentum, $x$ is the fraction of proton momentum carried by the gluon/quark, $\alpha_h p_1$ is the longitudinal momentum of the hadron $h$, $\vec k_h$ is the transverse hadron momentum and $\vec q$ is the transverse momentum of the incoming Reggeized gluon.}}
 	\label{fig:if-hadron}
 \end{figure}
 \FloatBarrier
 To assure the inclusive production of a given hadron, one of these integrations in the definition of parton impact factors is `opened' (see Fig.~\ref{fig:if-parton}). This means that the integration over the momentum of one of the intermediate-state partons is replaced by the convolution with a proper FF. The formula for the hadron impact factor 
 at LO (whose schematic view is shown in Fig.~\ref{fig:if-hadron}) reads
  \begin{equation}
  \label{if-hadron-lo}
  \frac{d\Phi_h^{(0)}}{\vec q^{2}}=
  2\pi\alpha_s\sqrt{\frac{2C_F}{C_A}}
  \frac{d\alpha_h d^2\vec k}
  {\vec k^{\, 2}}\int\limits^1_{\alpha_h} \frac{dx}{x}
  \, \delta^{(2)}\left(\vec k-\vec q\right)
  \end{equation}
  \[ \times \,
  \left(\frac{C_A}{C_F}f_g(x)
  D^h_g\left(\frac{\alpha_h}{x}\right)+\sum_{a=q,\bar q} f_a(x)
  D^h_a\left(\frac{\alpha_h}{x}\right)\right) \,.
  \]
  Here $D^h_a$ is the FF that represents non-perturbative, large-distance part of the transition from the parton $a$ produced with momentum $k$ and longitudinal fraction $x$ to a hadron with momentum fraction $\alpha_h$. Similarly to the jet impact factor, one moves on the $(\nu,n)$-representation (see Section~\ref{nun_repr}): projecting Eq.~\eqref{if-hadron-lo} onto the eigenfunctions Eq.~\eqref{nuLLA} of the LO BFKL kernel~\eqref{KLLA}.
\subsection{Cross section and azimuthal correlations}
\label{JH_cross_section}
In QCD collinear factorization the cross section of the
process~(\ref{processHJ}) reads
\begin{equation}
\frac{d\sigma}{dx_Hdx_Jd^2k_Hd^2k_J}
=\hspace{-0.4cm}\sum_{r,s=q,{\bar q},g}\int_0^1 \hspace{-0.1cm} dx_1 \int_0^1 \hspace{-0.1cm}dx_2\ f_r\left(x_1,\mu_F\right)
\ f_s\left(x_2,\mu_F\right)
\frac{d{\hat\sigma}_{r,s}\left(\hat s,\mu_F\right)}
{dx_Hdx_Jd^2k_Hd^2k_J}\;,
\end{equation}
where the $r, s$ indices specify the parton types 
(quarks $q = u, d, s, c, b$;
antiquarks $\bar q = \bar u, \bar d, \bar s, \bar c, \bar b$; 
or gluon $g$), $f_{r,s}\left(x, \mu_F \right)$ denote the initial proton PDFs; 
$x_{1,2}$ are the longitudinal fractions of the partons involved in the hard
subprocess, while $\mu_F$ is the factorization scale;
$d\hat\sigma_{r,s}\left(\hat s \right)$ is
the partonic cross section and $\hat s \equiv x_1x_2s$ is the squared
center-of-mass energy of the parton-parton collision subprocess.

In the BFKL approach the cross section can be presented 
(see Ref.~\cite{Caporale:2012ih} for the details of the derivation)
as the Fourier sum of the azimuthal coefficients ${\cal C}_n$, 
having so
\begin{equation}
\frac{d\sigma}
{dy_Hdy_J\, d|\vec k_H| \, d|\vec k_J|d\phi_H d\phi_J}
=\frac{1}{(2\pi)^2}\left[{\cal C}_0+\sum_{n=1}^\infty  2\cos (n\phi )\,
{\cal C}_n\right]\, ,
\end{equation}
where $\phi=\phi_H-\phi_J-\pi$, with $\phi_{H,J}$ the hadron/jet 
azimuthal angles, while $y_{H,J}$ and $\vec k_{H,J}$ are their
rapidities and transverse momenta, respectively. 
The $\phi$-averaged cross section ${\cal C}_0$ 
and the other coefficients ${\cal C}_{n\neq 0}$ are given
by
\beq\nonumber
{\cal C}_n \equiv \int_0^{2\pi}d\phi_H\int_0^{2\pi}d\phi_J\,
\cos[n(\phi_H-\phi_J-\pi)] \,
\frac{d\sigma}{dy_Hdy_J\, d|\vec k_H| \, d|\vec k_J|d\phi_H d\phi_J}\;
\eeq
\beq\nonumber
= \frac{e^Y}{s}
\int_{-\infty}^{+\infty} \hspace{-0.3cm} d\nu \, \hspace{-0.17cm}\left(\frac{x_H x_J s}{s_0}
\right)^{\hspace{-0.25cm}\bar \alpha_s(\mu_R)\hspace{-0.06cm}\left\{\chi(n,\nu)+\bar\alpha_s(\mu_R)
	\left[\bar\chi(n,\nu)+\frac{\beta_0}{8 N_c}\chi(n,\nu)\hspace{-0.1cm}\left[-\chi(n,\nu)
	+\frac{10}{3}+2\ln\left(\frac{\mu_R^2}{\sqrt{\vec k_H^2\vec k_J^2}}\right)\right]\right]\right\}}
\eeq
\beq\nonumber
\times \alpha_s^2(\mu_R) c_H(n,\nu,|\vec k_H|, x_H)
[c_J(n,\nu,|\vec k_J|,x_J)]^*\,
\eeq
\beq\label{Cm}
\times \left\{1
+\alpha_s(\mu_R)\left[\frac{c_H^{(1)}(n,\nu,|\vec k_H|,
	x_H)}{c_H(n,\nu,|\vec k_H|, x_H)}
+\left[\frac{c_J^{(1)}(n,\nu,|\vec k_J|, x_J)}{c_J(n,\nu,|\vec k_J|,
	x_J)}\right]^*\right]\right.
\eeq
\beq\nonumber
\left. + \bar\alpha_s^2(\mu_R) \ln\left(\frac{x_H x_J s}{s_0}\right)
\frac{\beta_0}{4 N_c}\chi(n,\nu)f(\nu)\right\}\;.
\eeq
\\
Here $\bar \alpha_s(\mu_R) \equiv \alpha_s(\mu_R) N_c/\pi$, with
$N_c$ the number of colors,
\beq
\beta_0=\frac{11}{3} N_c - \frac{2}{3}n_f
\eeq
is the first coefficient of the QCD $\beta$-function, where $n_f$ is the number
of active flavors,
\beq
\chi\left(n,\nu\right)=2\psi\left(1\right)-\psi\left(\frac{n}{2}
+\frac{1}{2}+i\nu \right)-\psi\left(\frac{n}{2}+\frac{1}{2}-i\nu \right)
\eeq
is the leading-order (LO) BFKL characteristic function,
$c_H(n,\nu)$ is the LO forward hadron impact factor in the
$\nu$-repre\-sen\-ta\-tion,  given as an integral in the parton fraction $x$,
containing the PDFs of the gluon and of the different quark/antiquark flavors
in the proton, and the FFs of the detected hadron,
\bea
c_H(n,\nu,|\vec k_H|,x_H) &=& 2 \sqrt{\frac{C_F}{C_A}}
(\vec k_H^2)^{i\nu-1/2}\,\int_{x_H}^1\frac{dx}{x}
\left( \frac{x}{x_H}\right)
^{2 i\nu-1} 
\nonumber \\
\label{cH}
&\times&\left[\frac{C_A}{C_F}f_g(x)D_g^h\left(\frac{x_H}{x}\right)
+\sum_{r=q,\bar q}f_r(x)D_r^h\left(\frac{x_H}{x}\right)\right] \;,
\eea
$c_J(n,\nu)$ is the LO forward jet vertex in the $\nu$-repre\-sen\-ta\-tion,
\beq
\label{cJ}
c_J(n,\nu,|\vec k_J|,x_J)=2\sqrt{\frac{C_F}{C_A}}
(\vec k_J^{\,2})^{i\nu-1/2}\,\left(\frac{C_A}{C_F}f_g(x_J)
+\sum_{s=q,\bar q}f_s(x_J)\right)
\eeq
and the $f(\nu)$ function is defined by
\beq
\label{fnu}
i\frac{d}{d\nu}\ln\left(\frac{c_H}{[c_J]^*}\right)=2\left[f(\nu)
-\ln\left(\sqrt{\vec k_H^2 \vec k_J^2}\right)\right] \;.
\eeq
The remaining objects are the hadron/jet NLO impact factor corrections in the $\nu$-representation, $c_{H,J}^{(1)}(n,\nu,|\vec k_{H,J}|, x_{H,J})$, their expressions being given in Eqs. (4.58), (4.65) of Ref.~\cite{hadrons}
and in Eq.~(36) of Ref.~\cite{Caporale:2012ih}, respectively.

\subsection{Integration over the final-state phase space}
\label{JH_int}
In order to match the LHC kinematic cuts, we integrate the coefficients
over the phase space for two final-state objects,
\beq
\label{Cn_int}
C_n= 
\int_{y^{\text{min}}_H}^{y^{\text{max}}_H}dy_H
\int_{y^{\text{min}}_J}^{y^{\text{max}}_J}dy_J\int_{k^{\text{min}}_H}^{k^{\text{max}}_H}dk_H
\int_{k^{\text{min}}_J}^{{k^{\text{max}}_J}}dk_J
\, \delta \left( y_H - y_J - Y \right)
\, {\cal C}_n \left(y_H,y_J,k_H,k_J \right)\, .
\eeq
 The rapidity interval, $Y$, between the hadron and the jet is kept fixed. We consider two distinct ranges for the final-state objects:

 \begin{itemize}
 	\item
 	both the hadron and the jet tagged by the CMS detector in their typical
 	kinematic configurations, {\it i.e.}:
 	$k^{\text{min}}_H=5$ GeV, $k^{\text{min}}_J=35$ GeV, 
 	$y^{\text{max}}_H=-y^{\text{min}}_H=2.4$, 
 	$y^{\text{max}}_J=-y^{\text{min}}_J=4.7$~\cite{Khachatryan:2016udy}. For the sake of
 	brevity, we will refer to this choice as the \textit{\textbf{CMS-jet}}
 	configuration; \,
 	\item
 	a hadron always detected inside CMS in the range given above, together with a
 	very backward jet tagged by CASTOR. In this peculiar,
 	\textit{\textbf{CASTOR-jet}} configuration, the jet lies in the typical
 	range of the CASTOR experimental analyses, {\it i.e.}
 	$k^{\text{min}}_J=5$ GeV, 
 	$y^{\text{max}}_J=-5.2$, $y^{\text{min}}_J=-6.6$ ~\cite{CMS:2016ndp}.
 \end{itemize}
 The value of $k_H^{\text{max}}$ is constrained by the lower cutoff of the adopted FF parametri-zations (see below) and is always fixed at 21.5~GeV.
 The value of $k_J^{\text{max}}$ is instead constrained by the requirement that
 $x_J \le 1$ which implies $k_J^{\text{max}} \simeq 60$ GeV for $\sqrt{s} = 7$~TeV
 and $|y_J| < 4.7$ ({\it CMS-jet)} and $k_J^{\text{max}} \simeq 17.68$~GeV for
 $\sqrt{s} = 13$~TeV ({\it CASTOR-jet)}. The rapidity interval, $Y$, is taken to be positive: $0 < Y \leq y^{\text{max}}_H - y^{\text {min}}_J$. We consider two center-of-mass energies, $\sqrt s = 7$ and 13 TeV in the {\it CMS-jet} configuration, while we give predictions for $\sqrt s = 13$ TeV in the {\it CASTOR-jet} case.
 
 In our calculations we use the MMHT 2014 NLO PDF set~\cite{Harland-Lang:2014zoa}
 with two different NLO parametrizations for hadron FFs:
 AKK~2008~\cite{Albino:2008fy} and HKNS 2007~\cite{Hirai:2007cx}. In the results presented below, we sum over the production of forward charged light hadrons: $\pi^{\pm}, K^{\pm}, p,\bar p$.
 \subsection{The BLM scale setting}
 \label{JH_BLM}
 As we know, the renormalization scale $\mu_R$ can be arbitrarily chosen within the NLA, so in order to fix it we use the BLM~\cite{BLM, BLM_2, BLM_3, BLM_4,BLM_5} procedure, useful for semihard processes.
 We remind that it is necessary to perform a finite renormalization from the $\overline{\text{MS}}$ to the physical MOM scheme, applying the substitutions given in Eqs.~\eqref{scheme} and~\eqref{T_Tbeta_Tconf}.
 After that, BLM scale $\mu_R^{\text{BLM}}$ is the value of $\mu_R$
 that makes the $\beta_0$-dependent part in the expression for the observable of interest vanish. Due to the fact that terms proportional 
 to the QCD $\beta_0$-function are present not only in the NLA BFKL kernel, but also in the expressions for the NLA impact factor, the BLM scale is not universal. This implies it is energy-process dependent.
 
 Finally, the condition for the BLM scale setting was found to be 
 \[
 C^{\beta}_n
 \propto \!\!
 \int_{y^{\text{min}}_H}^{y^{\text{max}}_H}dy_H
 \int_{y^{\text{min}}_J}^{y^{\text{max}}_J}dy_J\int_{k^{\text{min}}_H}^{k^{\text{max}}_H}dk_H
 \int_{k^{\text{min}}_J}^{k^{\text{max}}_J}dk_J
 \!\! 
 \int\limits^{\infty}_{-\infty} \!\!d\nu\,e^{Y \bar \alpha^{\text{MOM}}_s(\mu^{\text{BLM}}_R)\chi(n,\nu)}
 \]
 \[
 c_H(n,\nu)[c_J(n,\nu)]^*\left[\frac{5}{3}
 +\ln \frac{(\mu^{\text{BLM}}_R)^2}{|\vec k_H|
 	|\vec k_J|} +f(\nu)-2\left( 1+\frac{2}{3}I \right)
 \right.
 \]
 \[
 \left.
 +\bar \alpha^{\text{MOM}}_s(\mu^{\text{BLM}}_R) Y \: \frac{\chi(n,\nu)}{2}
 \left(-\frac{\chi(n,\nu)}{2}+\frac{5}{3}+\ln \frac{(\mu^{\text{BLM}}_R)^2}{|\vec k_H|
 	|\vec k_J|}\right.\right.
 \]
 \beq{}
 \label{beta0}
 \hspace{4.5cm}\left.\left.+f(\nu)-2\left( 1+\frac{2}{3}I \right)\right)\right]=0 \, .
 \eeq
 The term in the r.h.s. of~Eq.~(\ref{beta0}) proportional to $\alpha^{\text{MOM}}_s$ originates from the NLA part of the kernel, while the remaining ones come from the NLA corrections to the hadron/jet vertices.
 
 In order to find the values of the BLM scales, we introduce the ratios of the 
 BLM to the ``natural'' scale suggested by the kinematic of the process, 
 $\mu_N=\sqrt{|\vec k_H||\vec k_J|}$, so that $m_R=\mu_R^{\text{BLM}}/\mu_N$, 
 and look for the values of $m_R$ which solve Eq.~(\ref{beta0}). 
 
 We finally plug these scales into our expression for the integrated
 coefficients in the BLM scheme
 \beq
 \label{Cn_int_blm}
 C_n =
 \int_{y^{\text{min}}_H}^{y^{\text{max}}_H}dy_H
 \int_{y^{\text{min}}_J}^{y^{\text{max}}_J}dy_J\int_{k^{\text{min}}_H}^{k^{\text{max}}_H}dk_H
 \int_{k^{\text{min}}_J}^{k^{\text{max}}_J}dk_J
 \,
 \int\limits^{\infty}_{-\infty} d\nu 
 \eeq
 \beq \nonumber
 \frac{e^Y}{s}\,
 e^{Y \bar \alpha^{\text{MOM}}_s(\mu^{\text{BLM}}_R)\left[\chi(n,\nu)
 	+\bar \alpha^{\text{MOM}}_s(\mu^{\text{BLM}}_R)\left(\bar \chi(n,\nu) +\frac{T^{\text{conf}}}
 	{3}\chi(n,\nu)\right)\right]}
 \left(\alpha^{\text{MOM}}_s (\mu^{\text{BLM}}_R)\right)^2 
 \eeq
 \[
 \times c_H(n,\nu)[c_J(n,\nu)]^*
 \left\{1+\bar \alpha^{\text{MOM}}_s(\mu^{\text{BLM}}_R)\left[\frac{\bar c^{(1)}_H(n,\nu)}
 {c_H(n,\nu)}+\left[\frac{\bar c^{(1)}_J(n,\nu)}{c_J(n,\nu)}\right]^*
 +\frac{2T^{\text{conf}}}{3} \right] \right\} \, .
 \]
 Here the coefficient $C_0$ gives the $\phi$-averaged cross section, while the ratios $R_{n0} \equiv C_n/C_0 = \langle\cos(n\phi)\rangle$ determine the values of the mean cosines, or azimuthal correlations, of the produced hadron and jet.
 In Eq.~(\ref{Cn_int_blm}), $\bar \chi(n,\nu)$ is the eigenvalue of NLA BFKL kernel~\cite{Kotikov:2000pm} and its expression is given, in Eq.~(23) of Ref.~\cite{Caporale:2012ih}, whereas $\bar c^{(1)}_{H,J}$ are the NLA parts of hadron and jet vertices~\cite{Caporale:2015uva}.
 
 The factorization scale $\mu_F$ is set equal to the renormalization scale
 $\mu_R$, as assumed by the MMHT~2014 PDF. The analytical expressions are calculated in the MOM scheme. We present also results in the $\overline{\text{MS}}$ scheme for the $\phi$-averaged cross section $C_0$.
 In this case, we choose natural values for $\mu_R$, that is
 $\mu_R = \mu_N \equiv \sqrt{|\vec k_H||\vec k_J|}$, and two different values of
 the factorization scale, $(\mu_F)_{1,2} = |\vec k_{H,J}|$, depending on which of the two vertices is considered. We checked that the effect of using natural
 values also for $\mu_F$ fixed to $\mu_F = \mu_N$, is negligible with respect to our two-value choice.
\subsection{Numerical tools and uncertainty estimation}
\label{JH_num_tools}
All numerical calculations presented in Section~\ref{JH_results} were performed in \textsc{Fortran}, choosing a two-loop running coupling setup with $\alpha_s\left(M_Z\right)=0.11707$ and five quark flavors. In particular, we used a recently developed code, called {\tt JETHAD}, implemented for the computation of cross sections and
related observables for semi-hard reactions. In order to perform numerical integrations, {\tt JETHAD} is interfaced with
specific \textsc{CERN} program libraries~\cite{cernlib} and with {\tt Cuba} library integrators~\cite{Cuba:2005,ConcCuba:2015}. 
We use several CERNLIB routines as {\tt Dadmul} and {\tt WGauss}, while the {\tt Cuba} ones were mainly used for crosschecks. 
As we know, the choice of the particular PDF and FF parametrizations can determine a potential source of uncertainty. For this reason, we did
preliminary tests by using three different NLO PDF sets, expressly: 
MMHT~2014~\cite{Harland-Lang:2014zoa},  
CT~2014~\cite{Dulat:2015mca}
and NNPDF3.0~\cite{Ball:2014uwa},
and convolving them with the four following NLO FF routines: 
AKK~2008~\cite{Albino:2008fy}, 
DSS~2007~\cite{DSS}, HKNS~2007~\cite{Hirai:2007cx} and
NNFF1.0~\cite{Bertone:2017tyb}. All PDF parametrizations, as well as the NNFF1.0 set, were performed via the Les Houches Accord PDF Interface (LHAPDF) 6.2.1~\cite{Buckley:2014ana}, while the native routines for the remaining FFs were direclty linked to the corresponding module in our code. Through our tests we proved that no large discrepancy when distinct PDF sets are used in the kinematic range of our interest. Hence in the final calculations we selected the MMHT~2014~PDF set and the FF interfaces mentioned above. We do not show the results with DSS~2007 and NNFF1.0~FF routines, 
since they would be hardly distinguishable from those with 
the HKNS~2007 parametrization. We list below the most important sources of uncertainty:
\begin{itemize}
\item the numerical 4-dimensional integration over the two transverse momenta $|\vec k_{H,J}|$, the hadron rapidity $y_H$, and over $\nu$. Its effect was estimated by {\tt Dadmul} integration routine~\cite{cernlib};
\item the one-dimensional integration over the parton fraction $x$
needed to perform the convolution between PDFs and FFs 
in the LO/NLO hadron impact factors;
\item the one-dimensional integration over the longitudinal momentum fraction
$\zeta$ in the NLO hadron/jet impact factor corrections
\item the upper cutoff in the numerical integrations over $|\vec k_{H,J}|$ and
$\nu$.
\end{itemize}
Error bands of all predictions presented in this work are just those given by the {\tt Dadmul} routine.
\subsection{Results and discussion}
\label{JH_results}
\begin{figure}[!h]
	\centering
	\includegraphics[scale=0.30,clip]{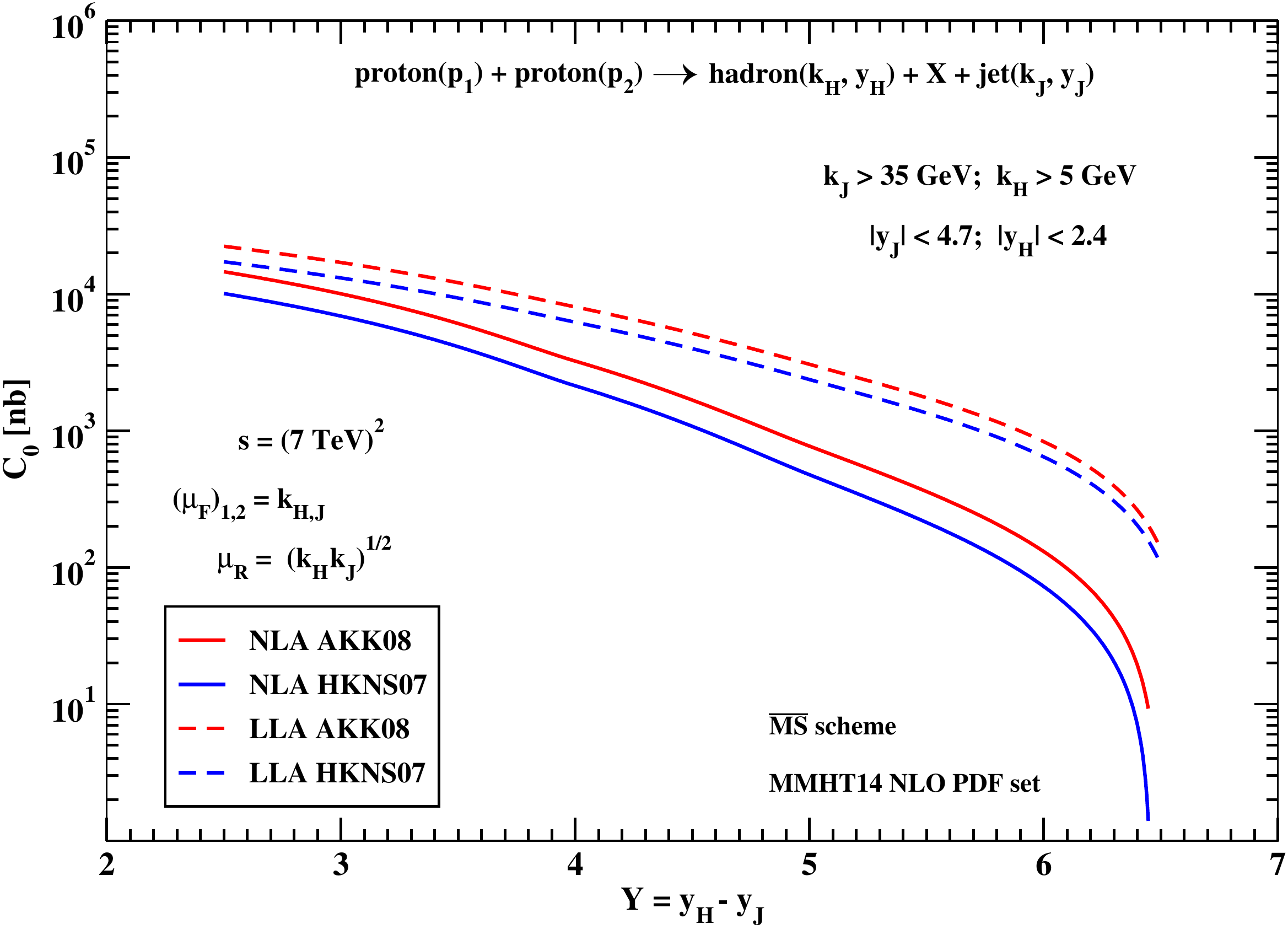}
	\includegraphics[scale=0.30,clip]{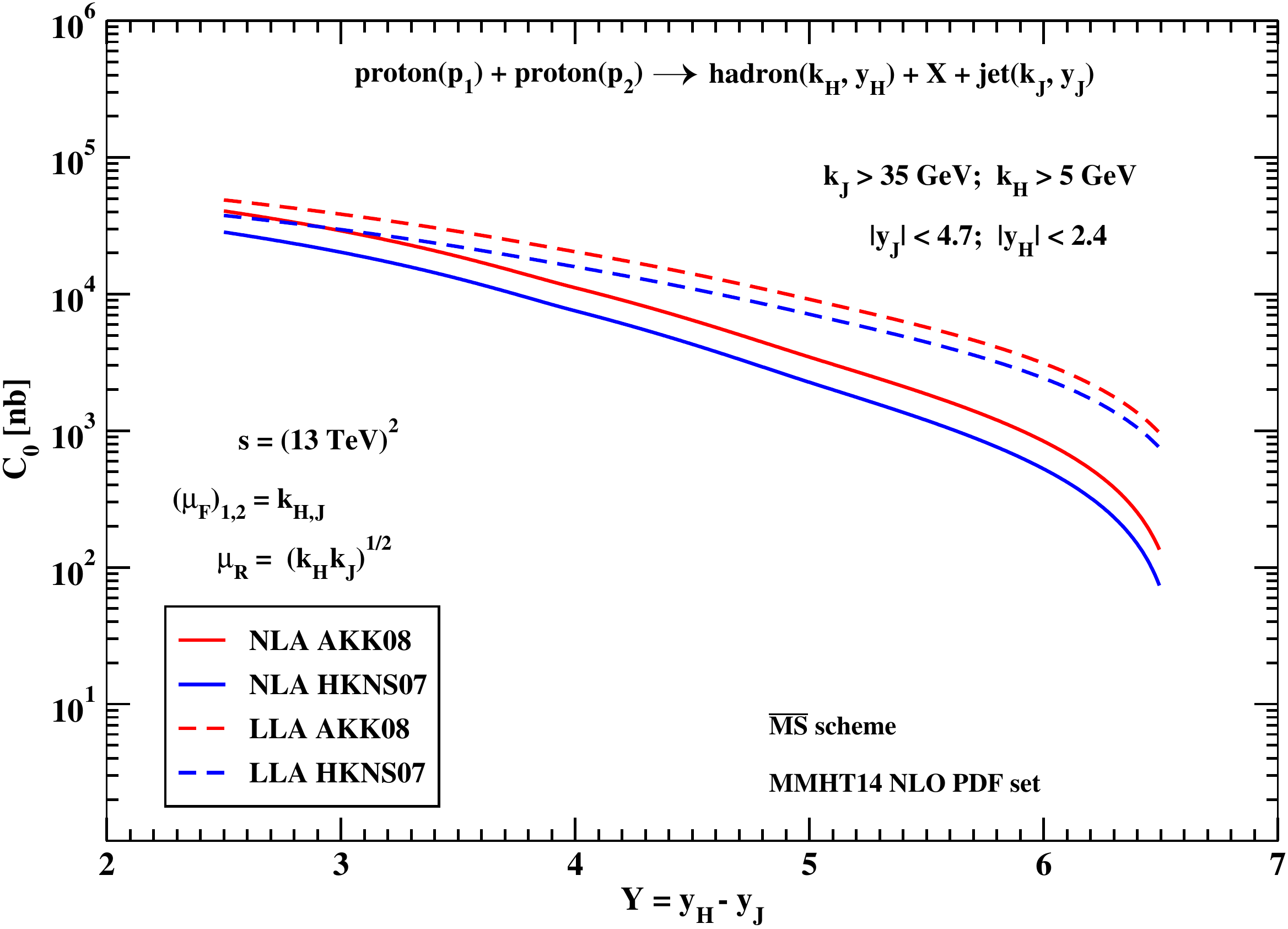}
	\caption{\emph{$Y$-dependence of $C_0$ for $\mu_R = \mu_N =
			\sqrt{|\vec k_H||\vec k_J|}$, $(\mu_F)_{1,2} = |\vec k_{H,J}|$, for $\sqrt{s}= 7$ TeV (left) and $\sqrt{s} = 13$ TeV (right), and
			$Y \leq 7.1$ ({\it CMS-jet} configuration).}}	
	\label{fig:C0_MSb_NS_CMS}	
\end{figure}

\begin{figure}[t]
 	\centering
 	\includegraphics[scale=0.30,clip]{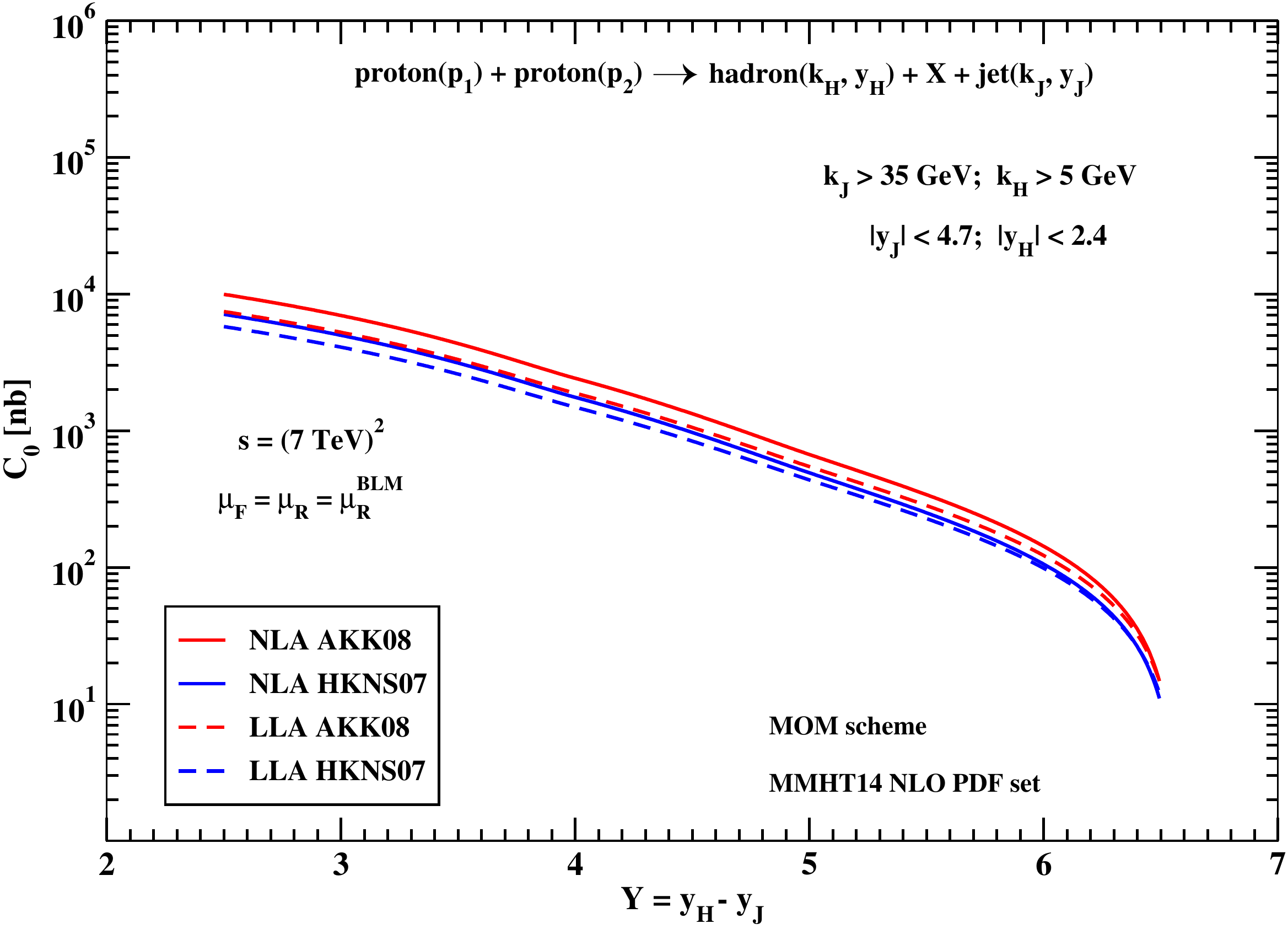}
 	\includegraphics[scale=0.30,clip]{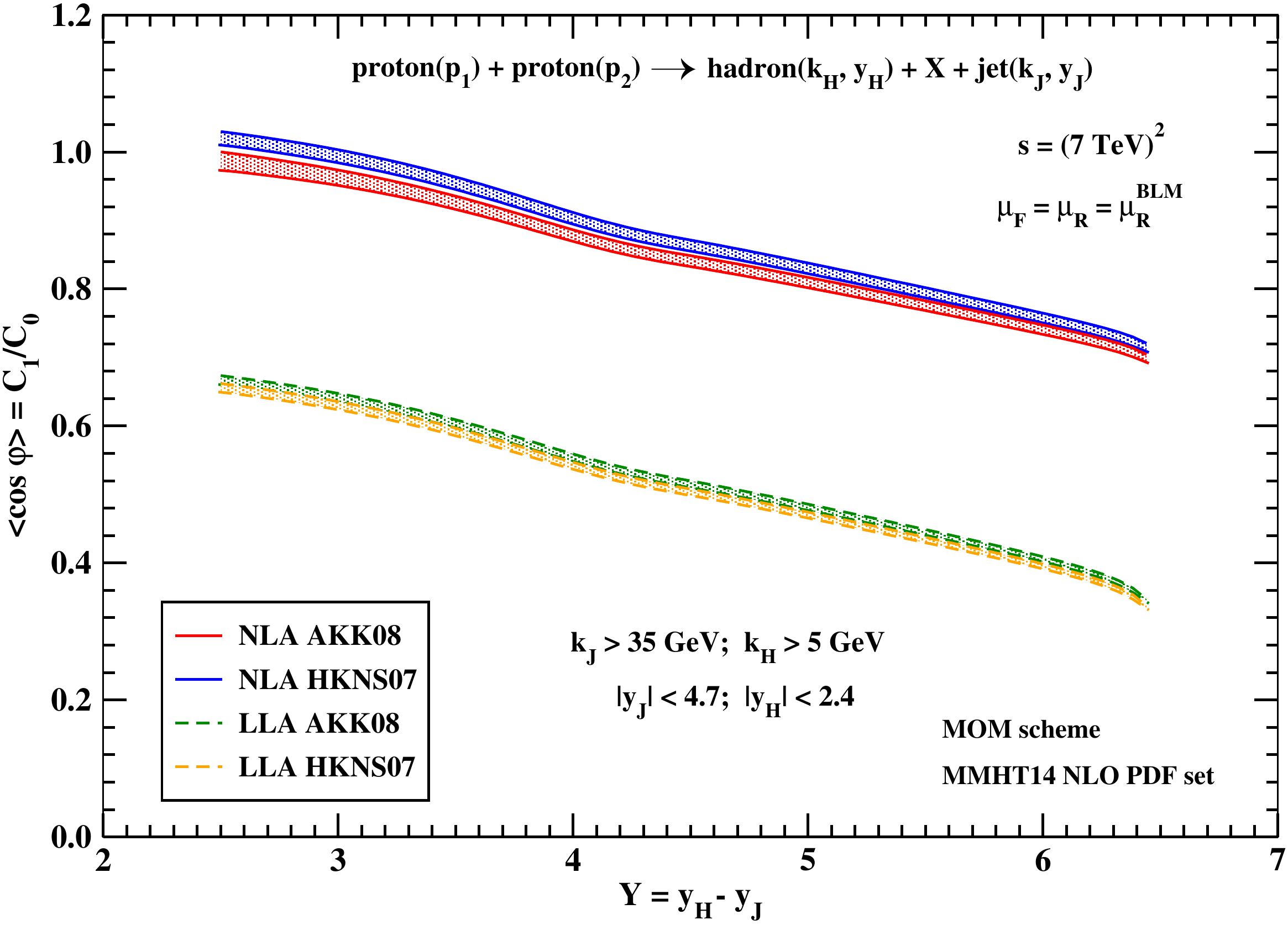}
 	
 	\includegraphics[scale=0.30,clip]{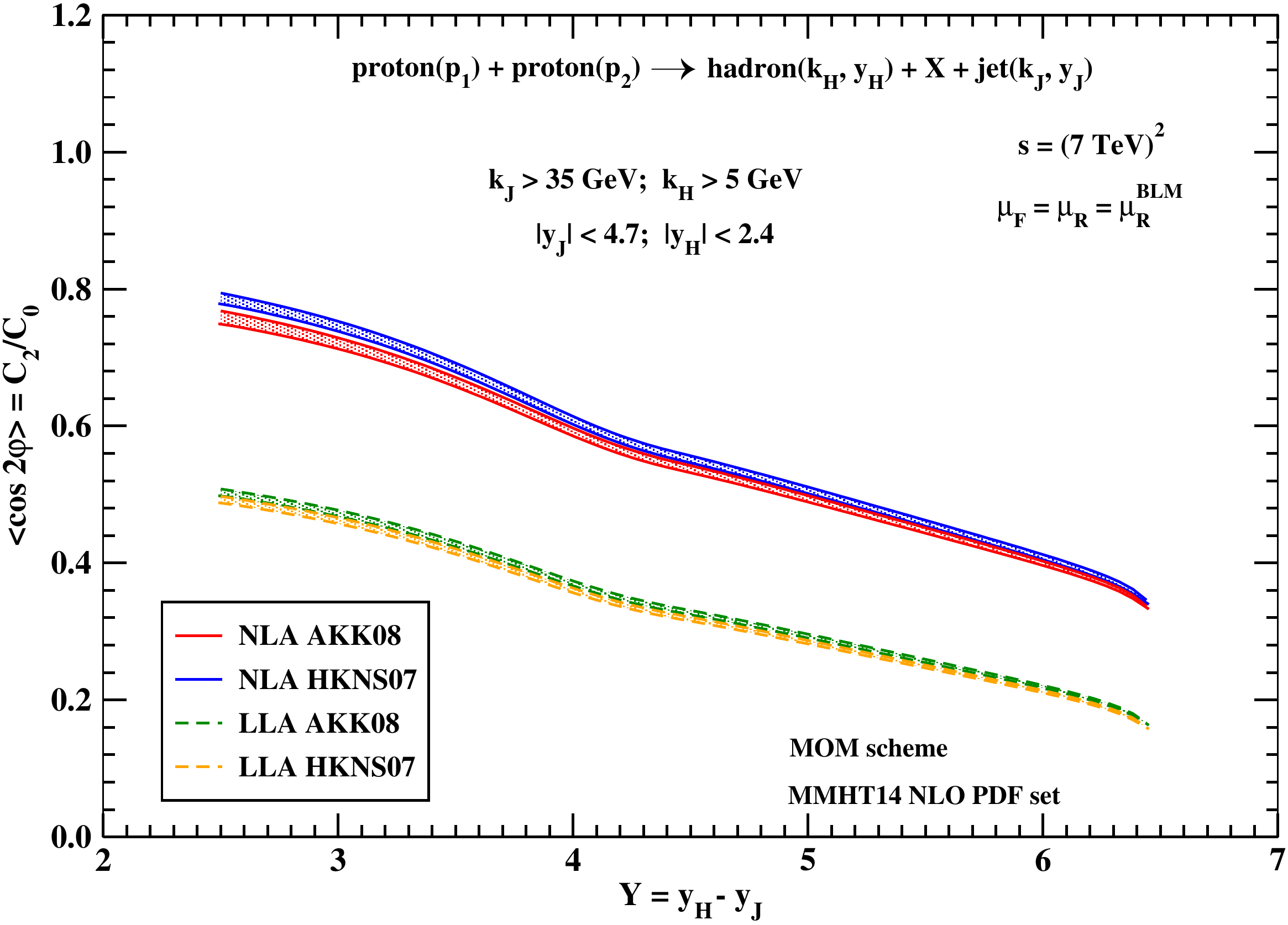}
 	\includegraphics[scale=0.30,clip]{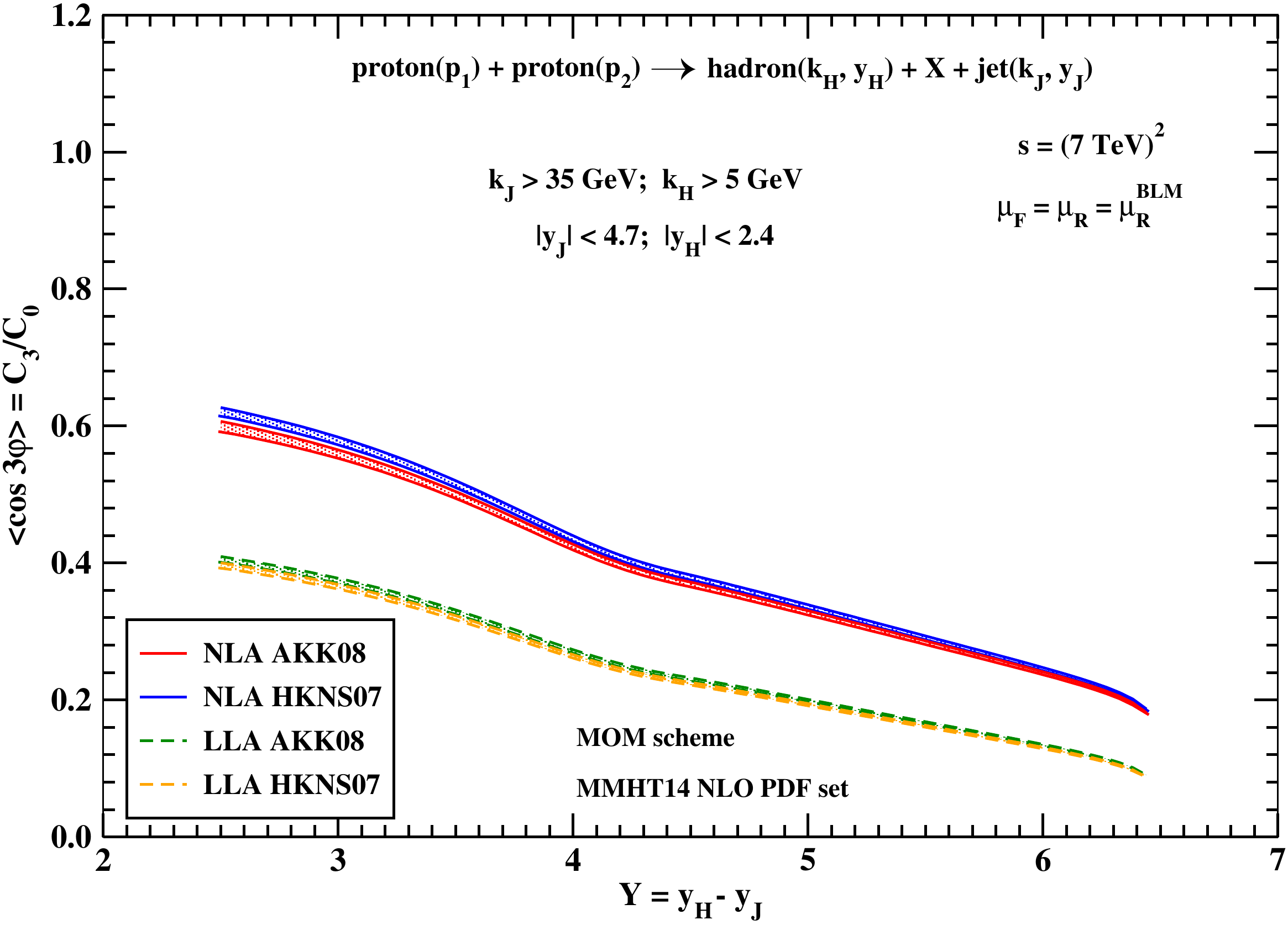}
 	
 	\includegraphics[scale=0.30,clip]{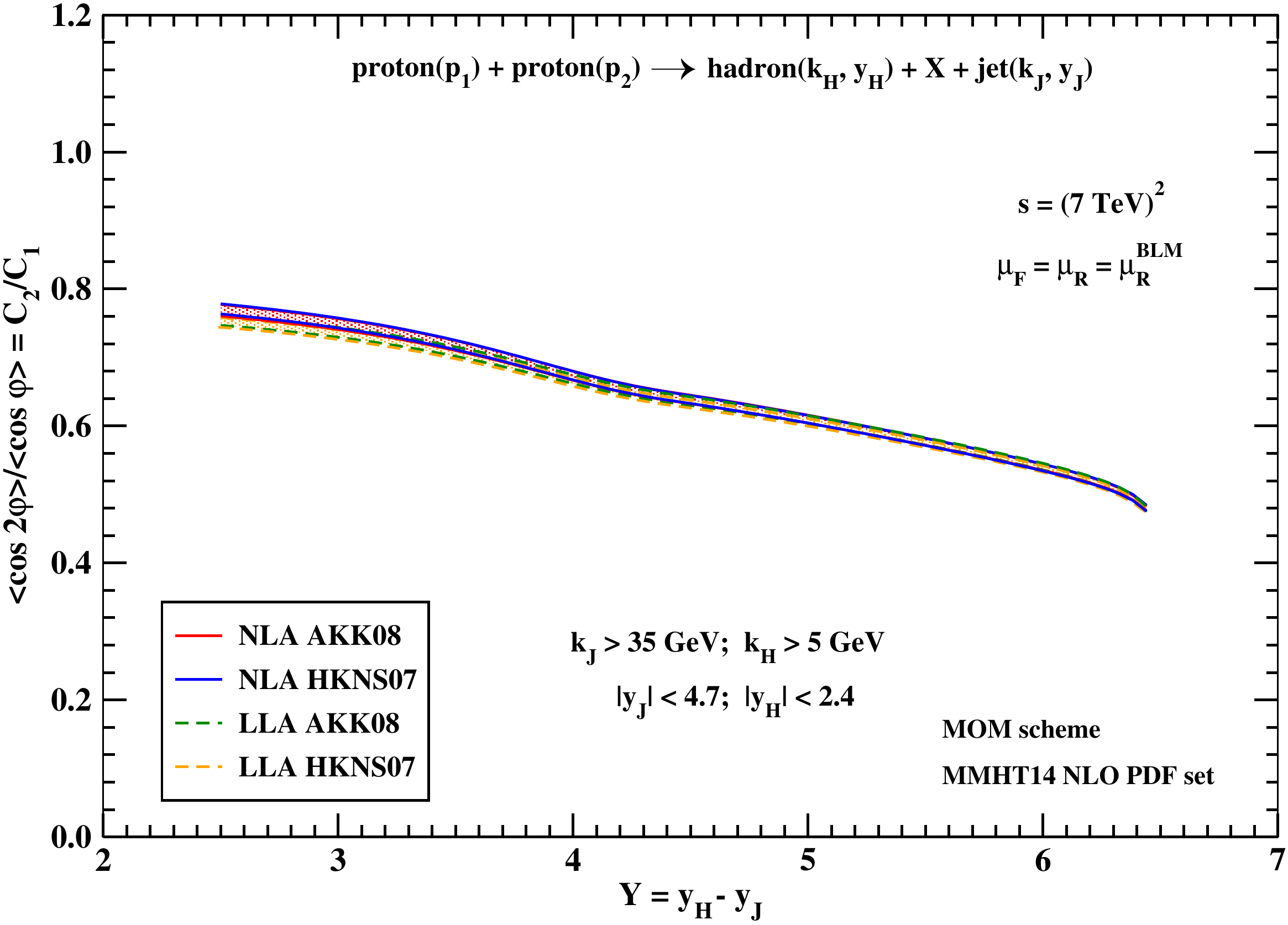}
 	\includegraphics[scale=0.30,clip]{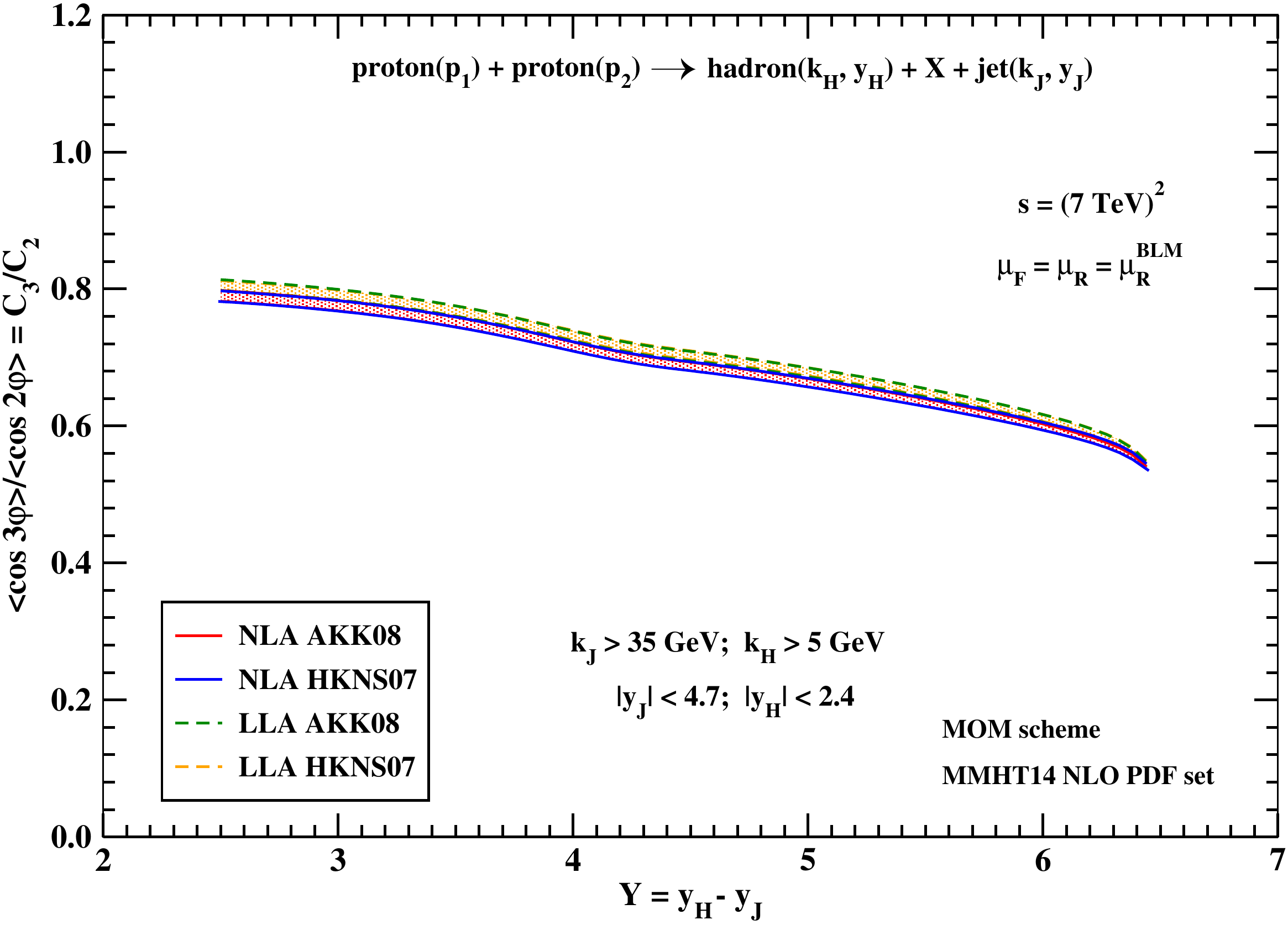}
 	\caption{\emph{$Y$-dependence of $C_0$ and of several ratios $C_m/C_n$ for 
 			$\mu_F = \mu_R^{\text{BLM}}$, $\sqrt{s} = 7$ TeV, and $Y \leq 7.1$
 			({\it CMS-jet} configuration).}}
 	\label{fig:Cn_MOM_BLM_CMS_7}
 \end{figure}

\begin{figure}[!h]
 	\centering
 	\includegraphics[scale=0.30,clip]{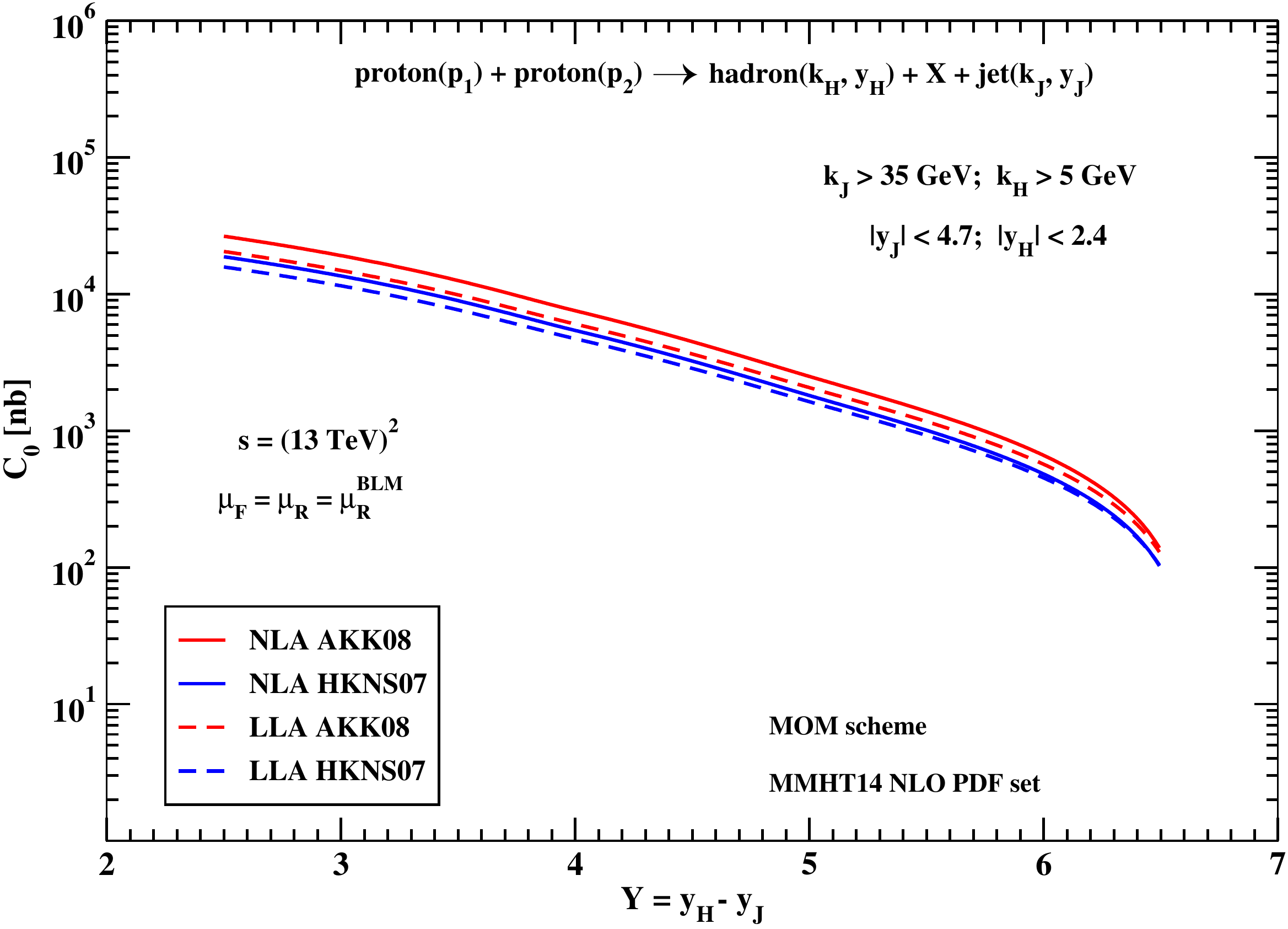}
 	\includegraphics[scale=0.30,clip]{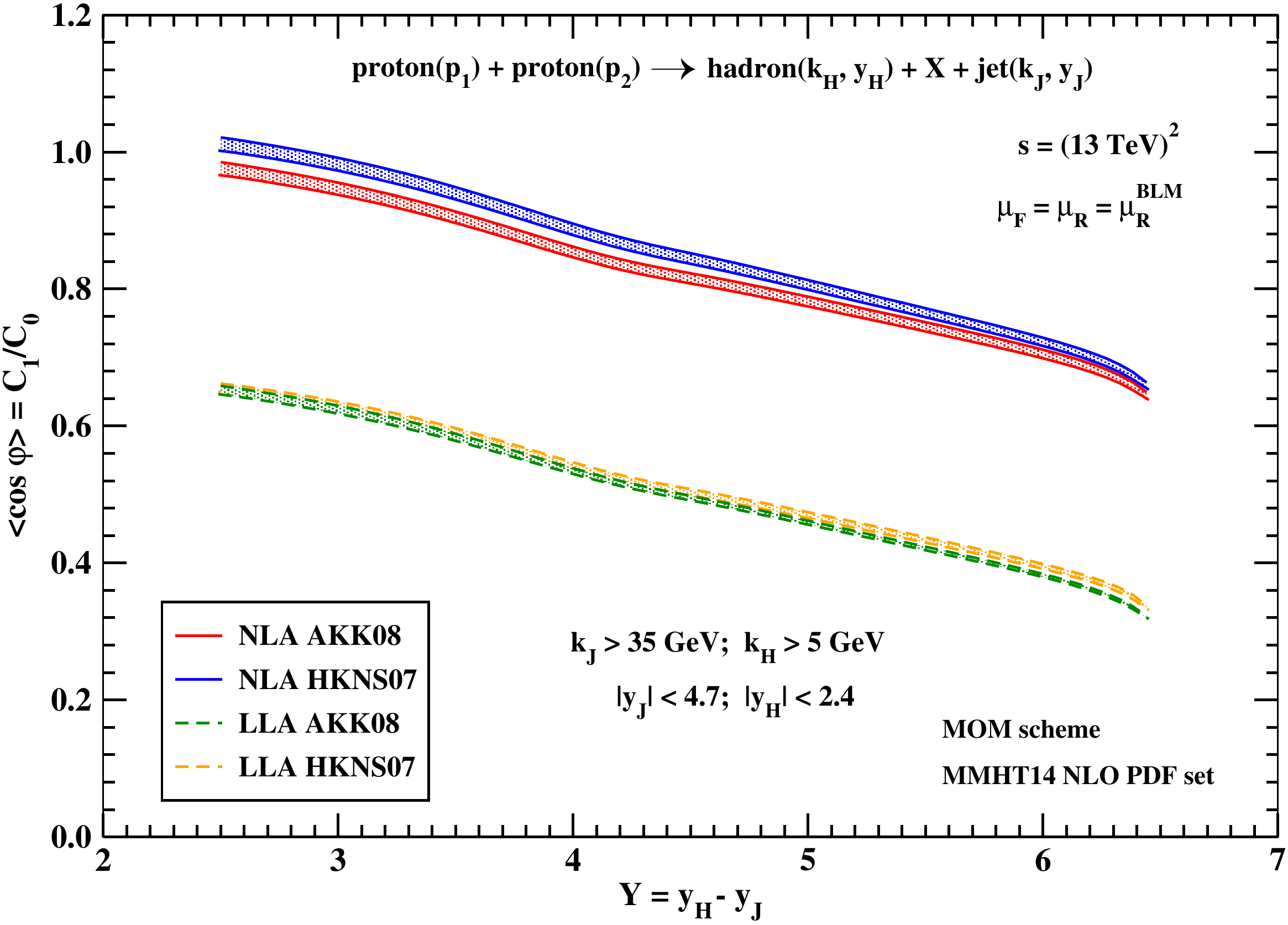}
 	
 	\includegraphics[scale=0.30,clip]{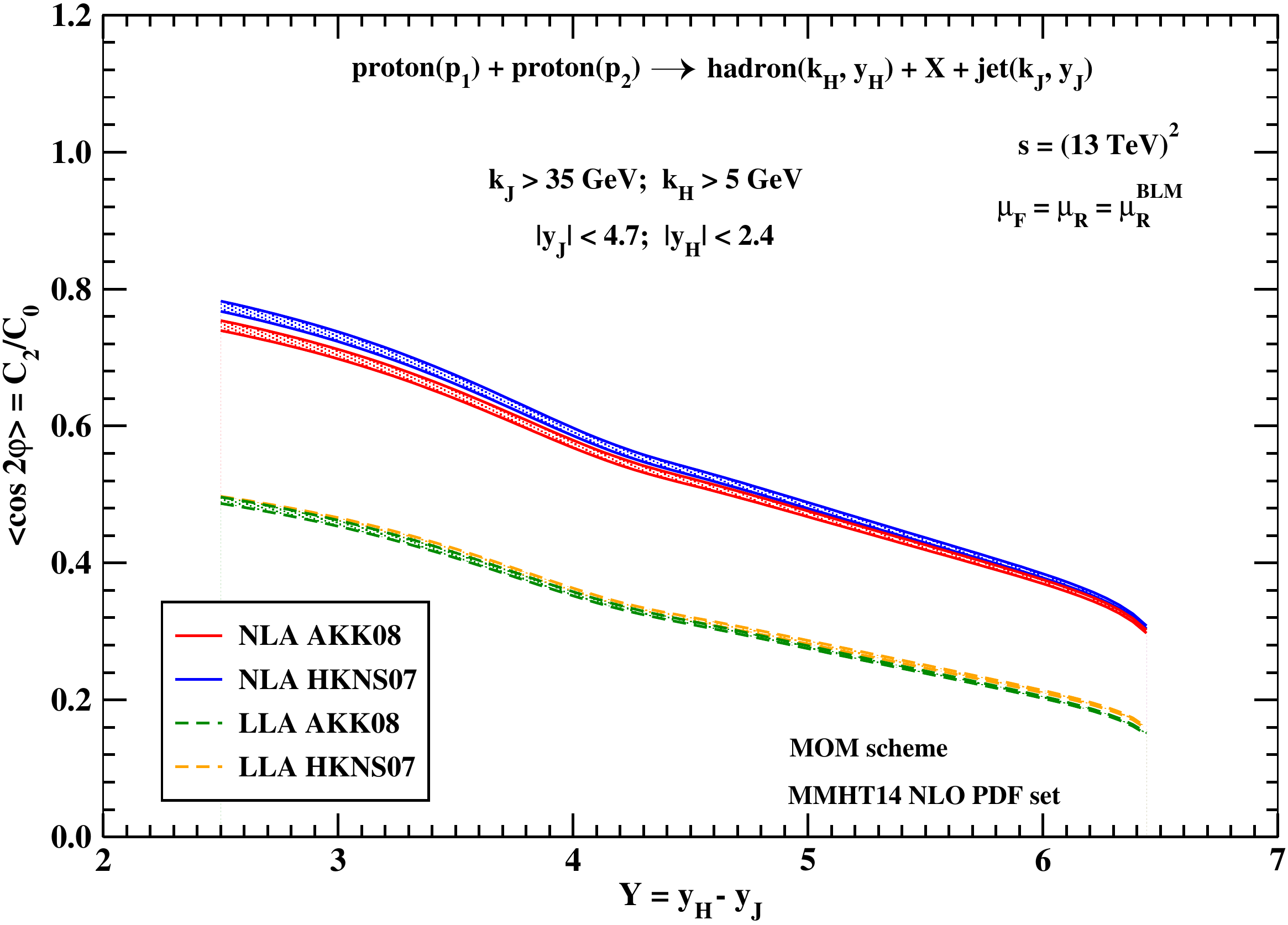}
 	\includegraphics[scale=0.30,clip]{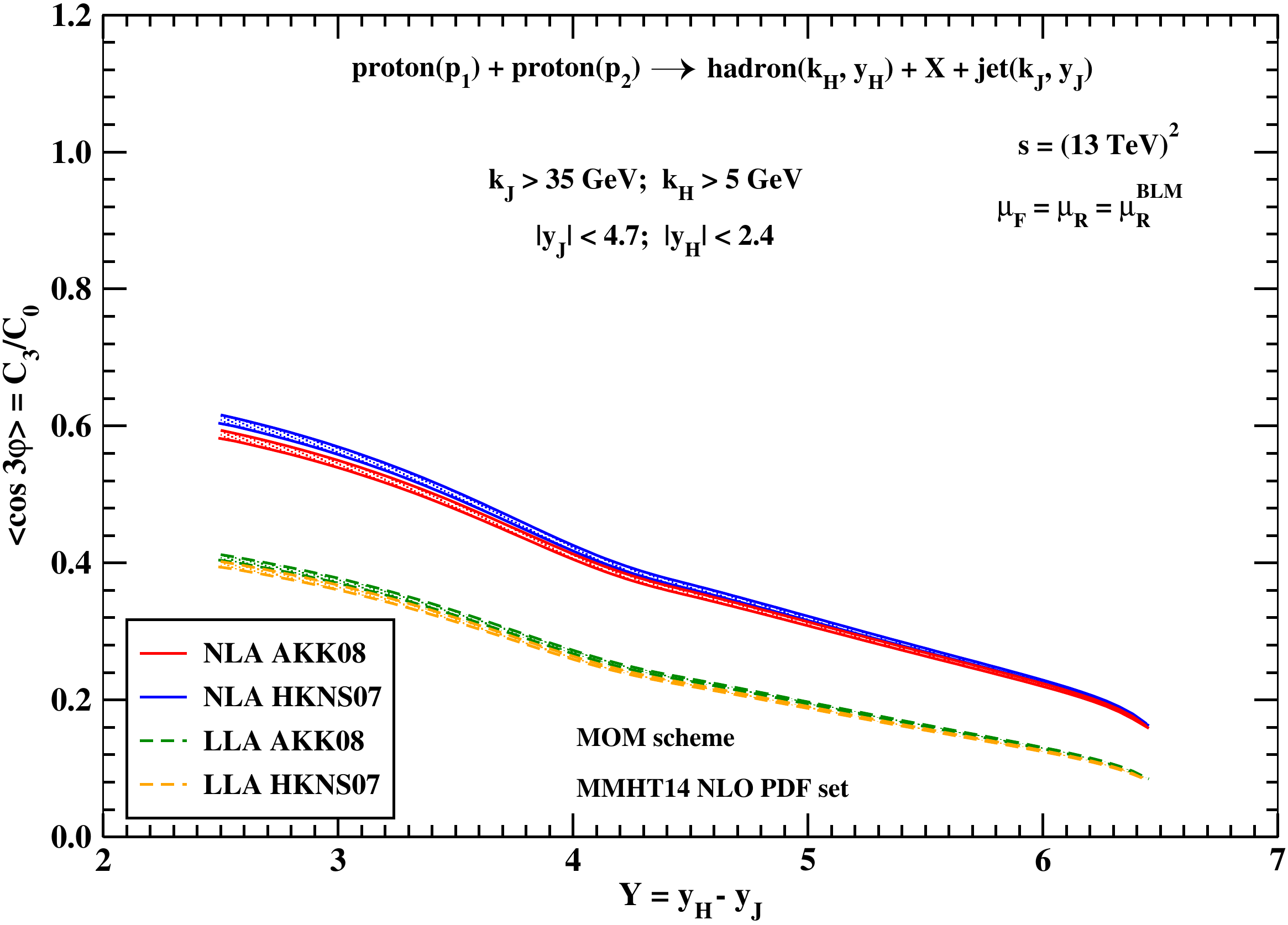}
 	
 	\includegraphics[scale=0.30,clip]{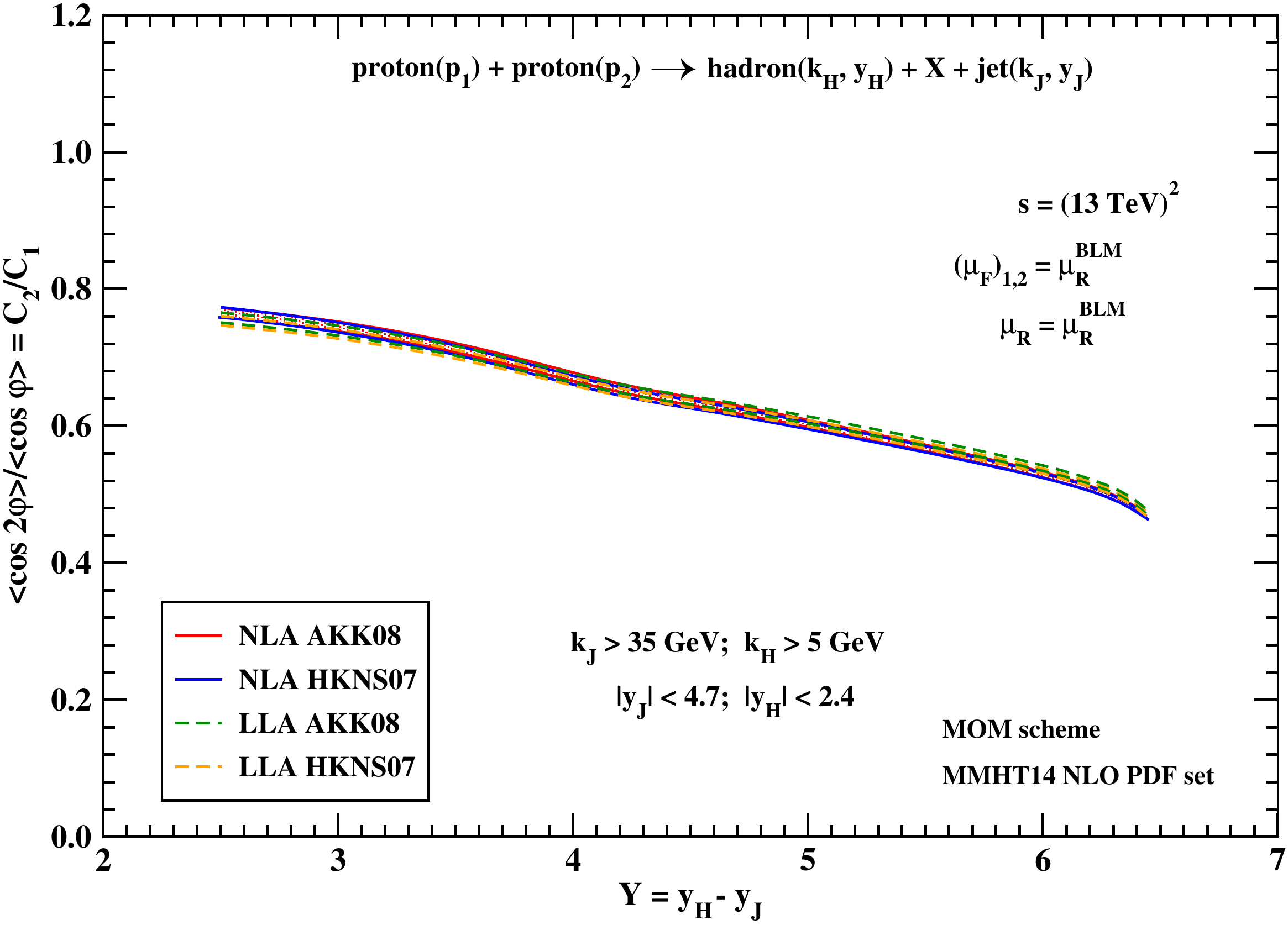}
 	\includegraphics[scale=0.30,clip]{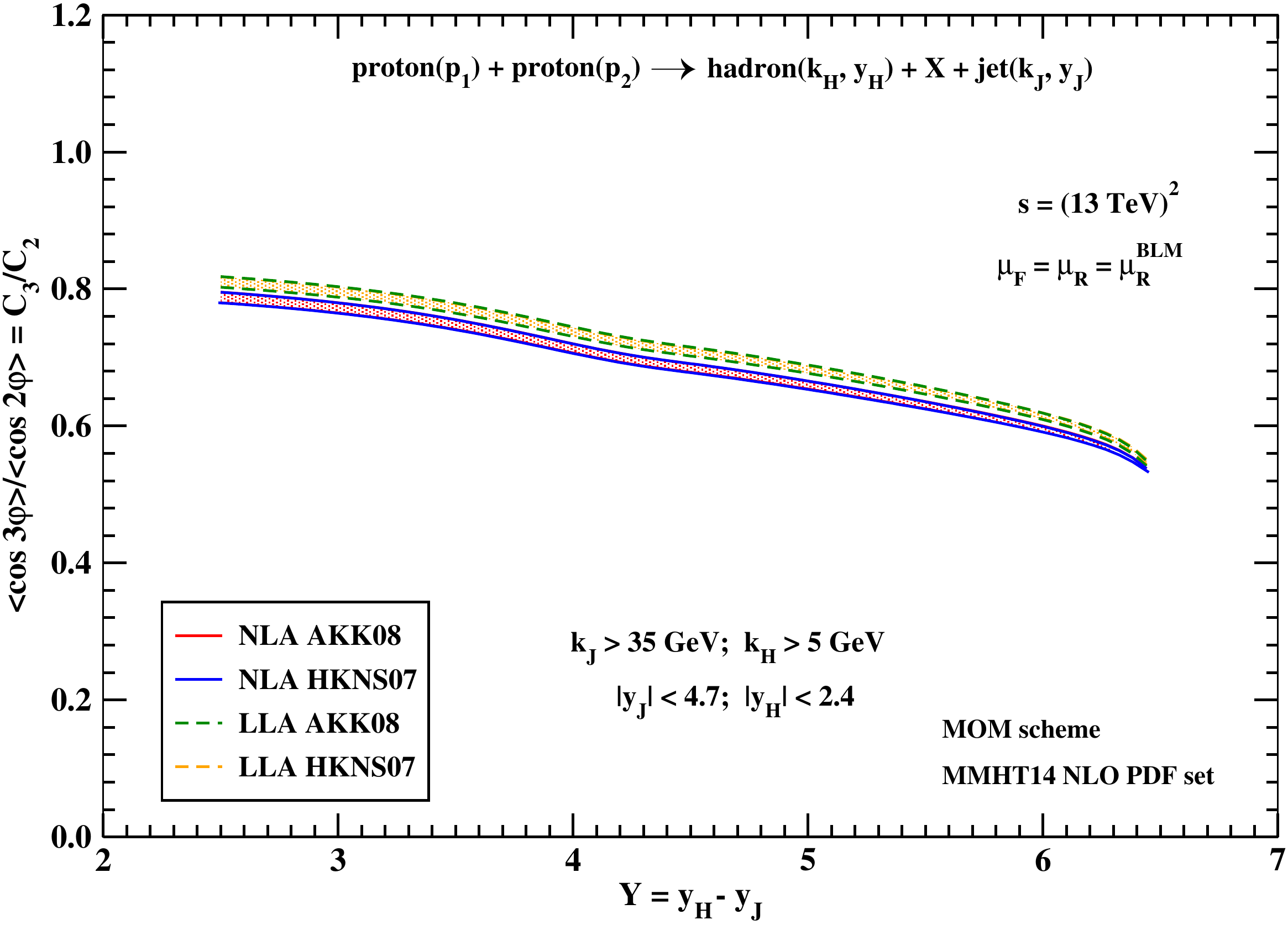}
 	\caption{\emph{$Y$-dependence of $C_0$ and of several ratios $C_m/C_n$ for $\mu_F = \mu_R^{\text{BLM}}$, $\sqrt{s} = 13$ TeV, and $Y \leq 7.1$ ({\it CMS-jet} configuration).}}
 	\label{fig:Cn_MOM_BLM_CMS_13}
 \end{figure}

\begin{figure}[!h]
	\centering
	\includegraphics[scale=0.30,clip]{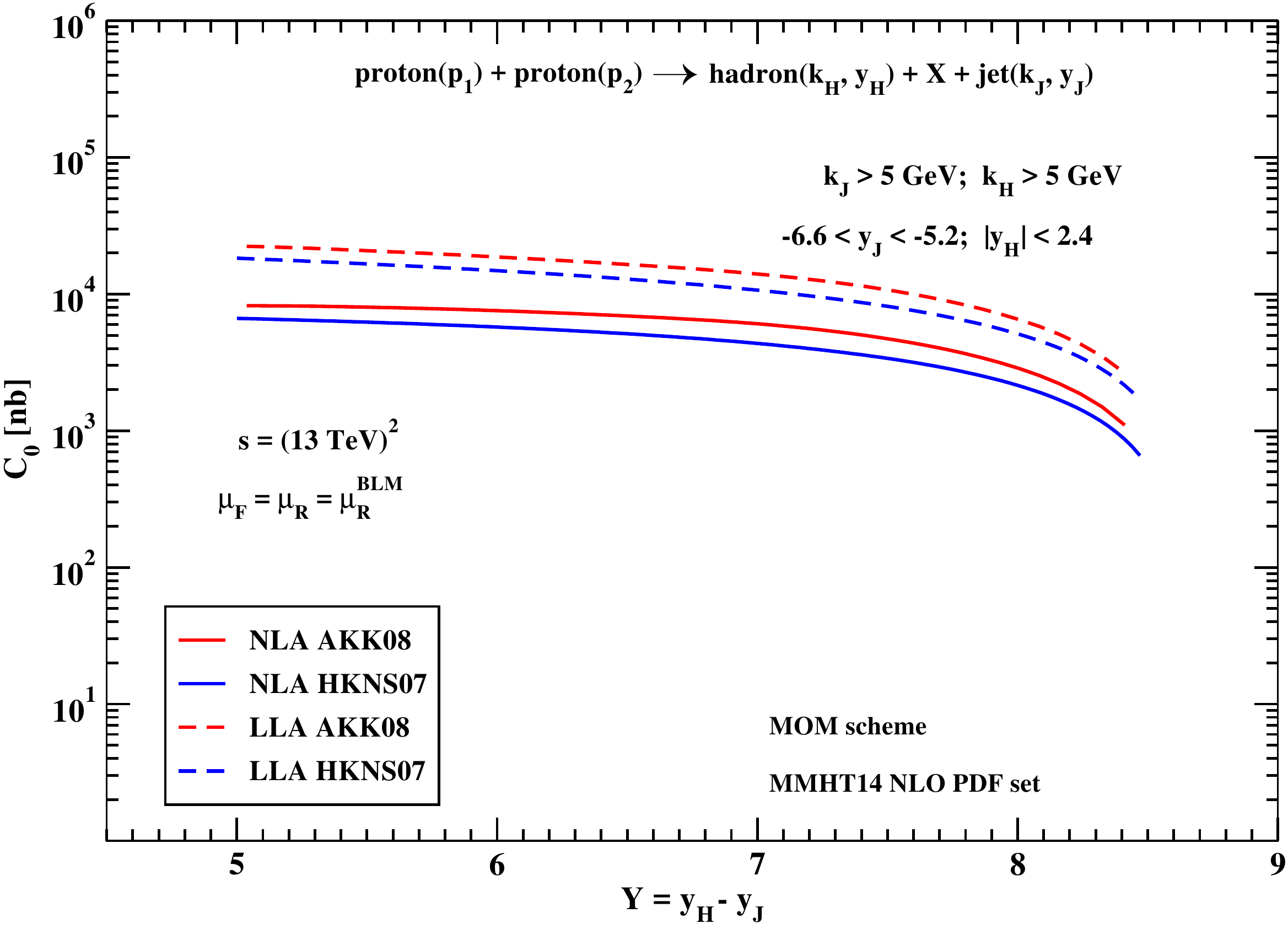}
	\includegraphics[scale=0.30,clip]{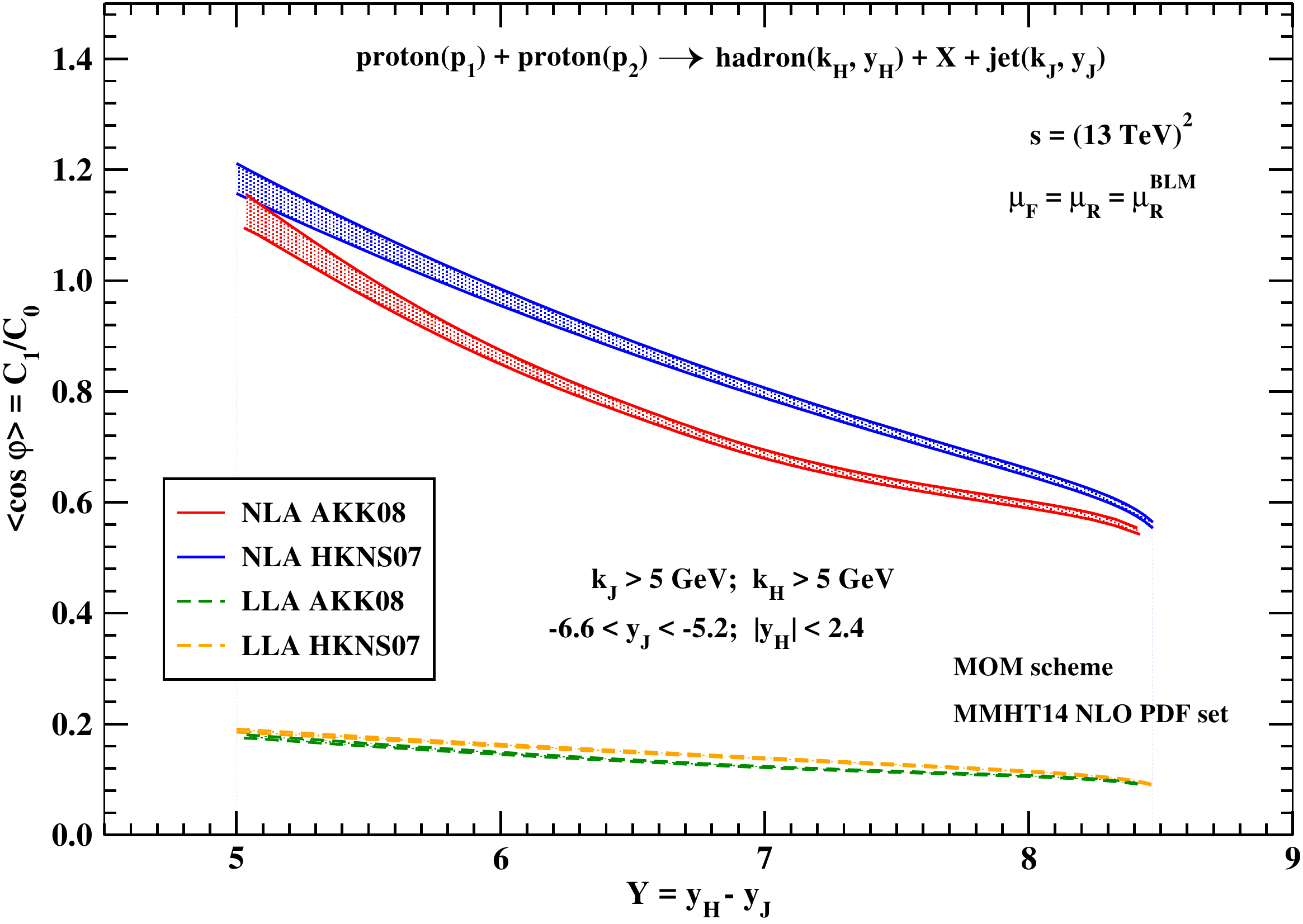}
	
	\includegraphics[scale=0.30,clip]{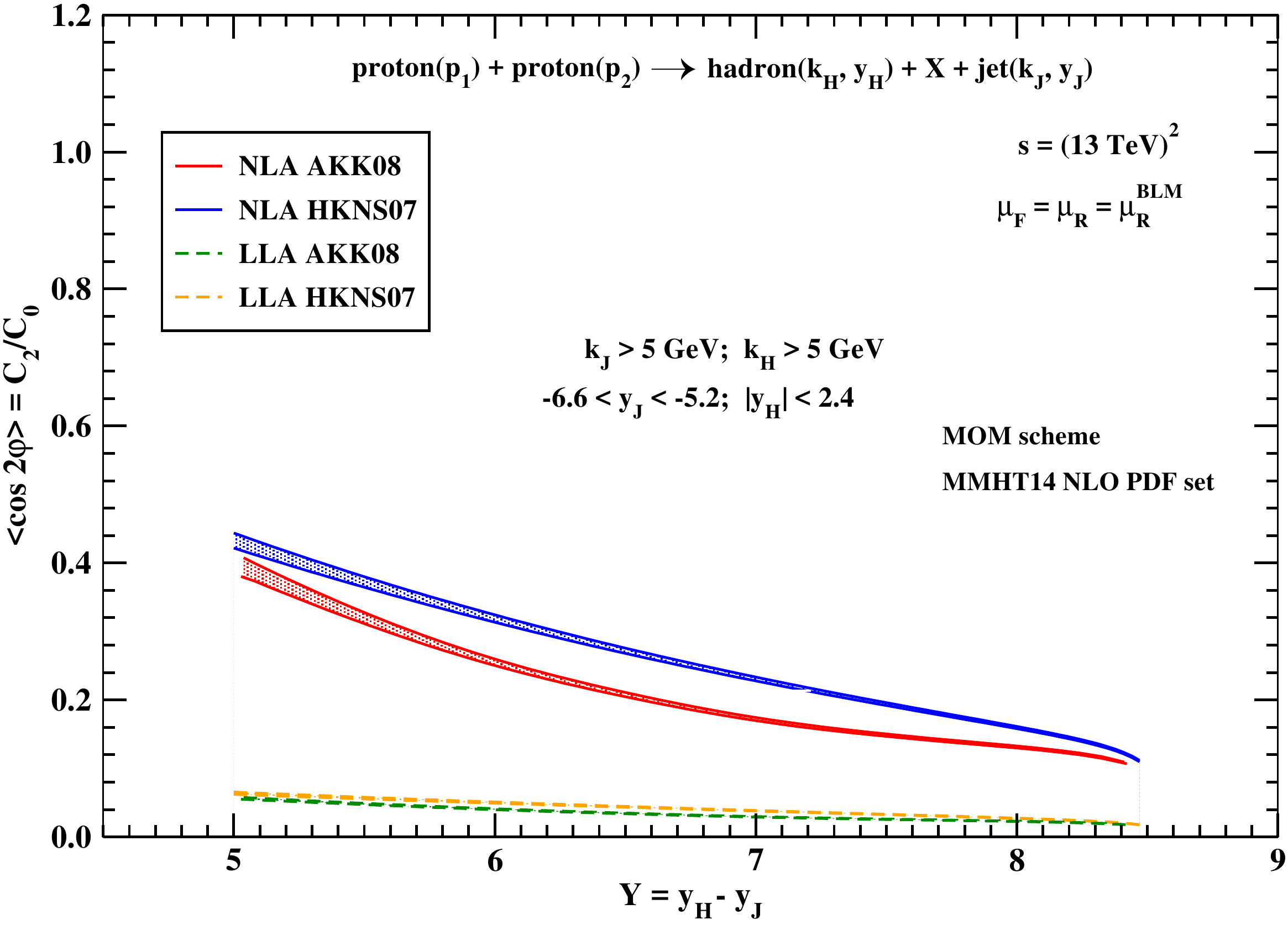}
	\includegraphics[scale=0.30,clip]{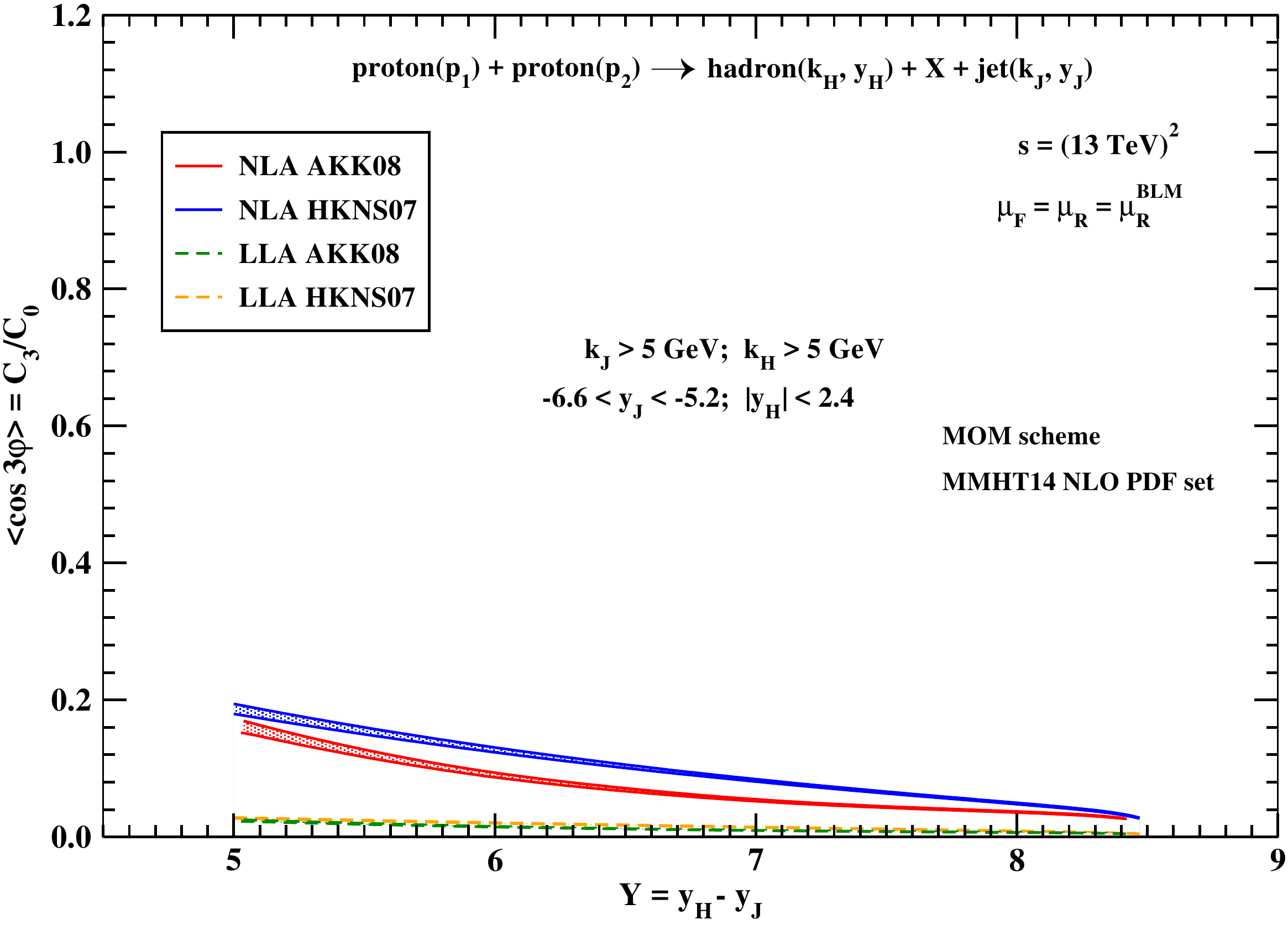}
	
	\includegraphics[scale=0.30,clip]{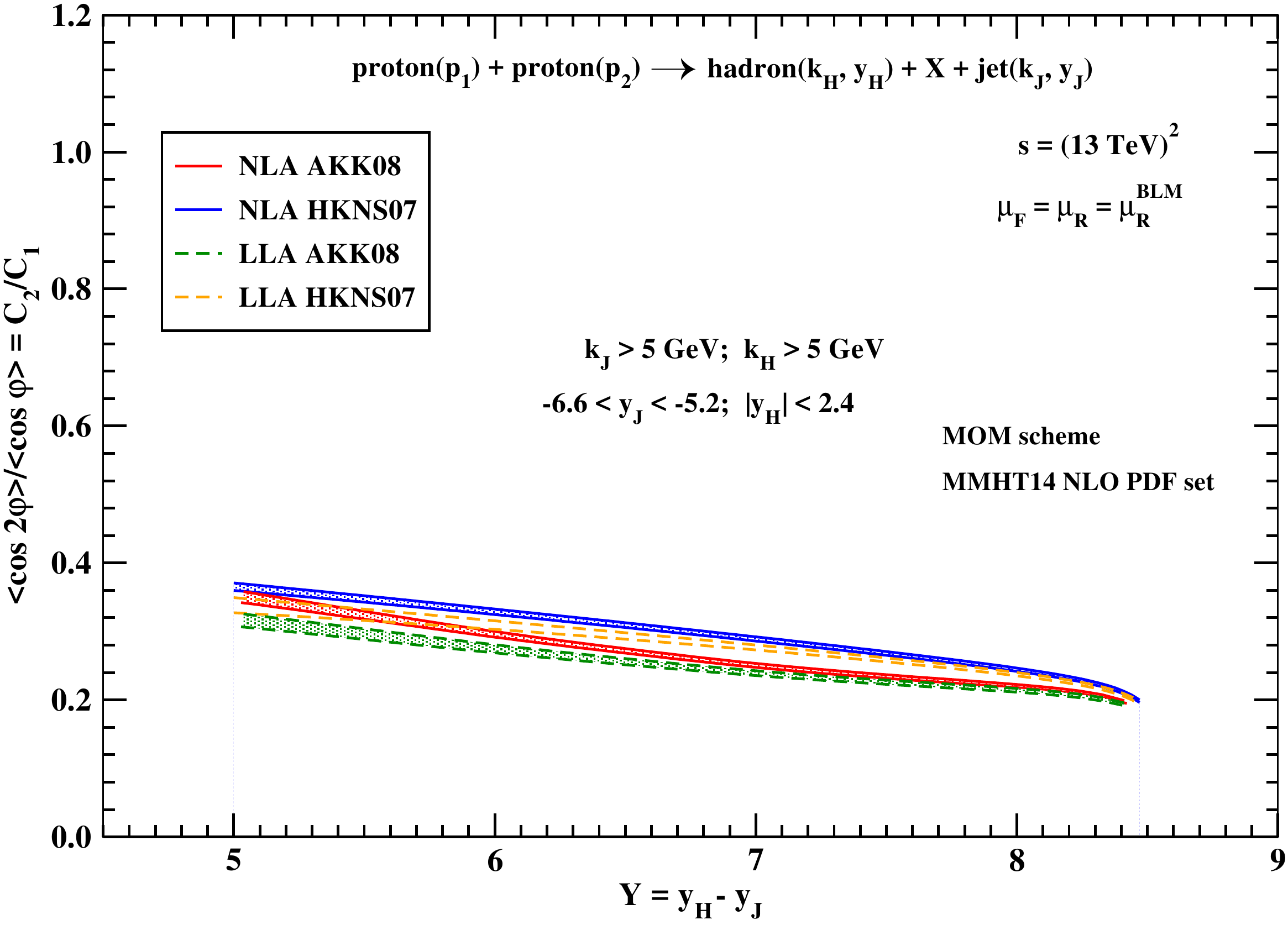}
	\includegraphics[scale=0.30,clip]{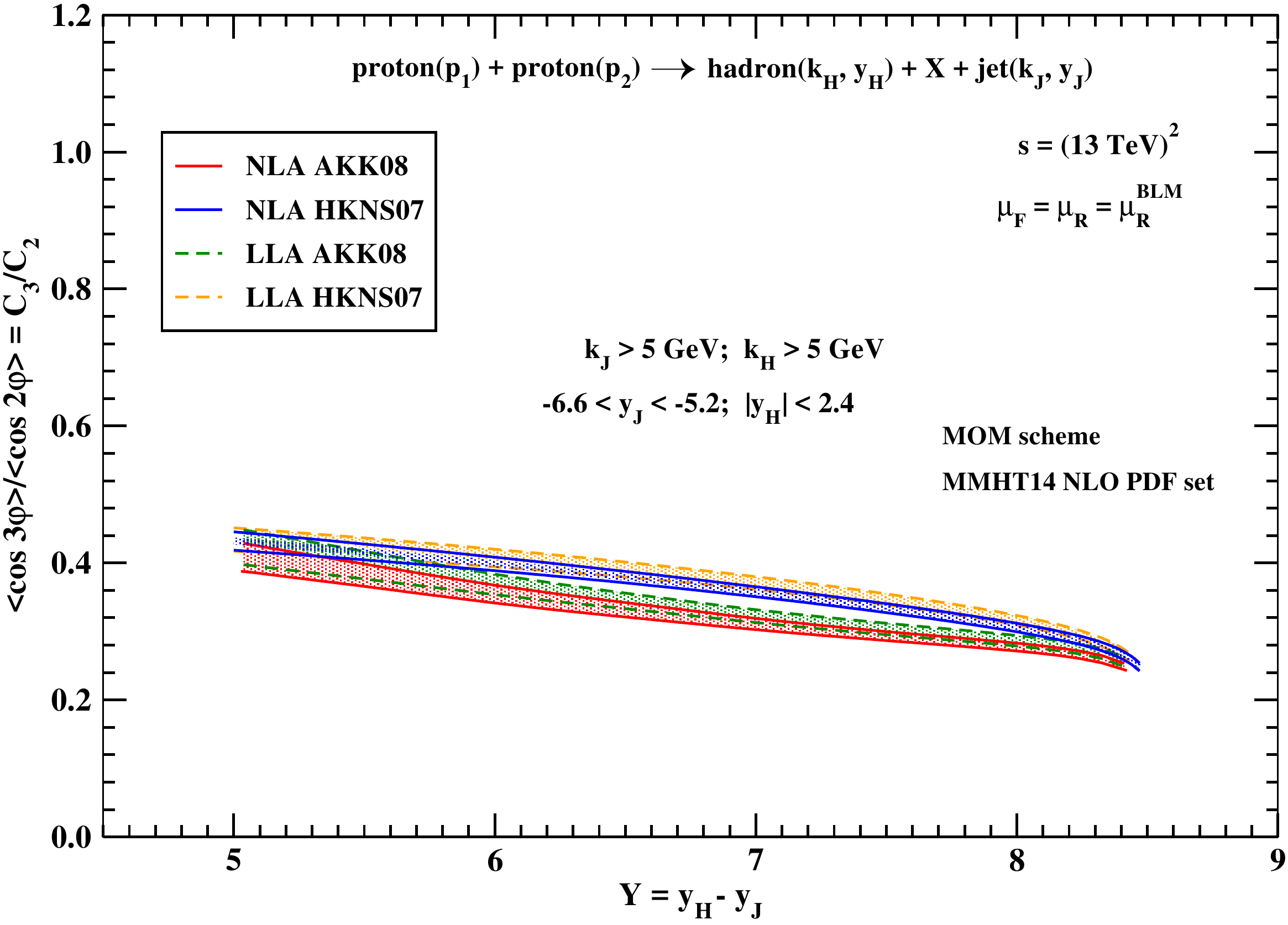}
	\caption{\emph{$Y$-dependence of $C_0$ and of several ratios $C_m/C_n$ for 
			$\mu_F = \mu_R^{\text{BLM}}$, $\sqrt{s} = 13$ TeV, and $Y \leq 9$
			({\it CASTOR-jet} configuration).}}
	\label{fig:Cn_MOM_BLM_CASTOR_13}
\end{figure}

In Fig.~\ref{fig:C0_MSb_NS_CMS} we present our results at natural scales for
the $\phi$-averaged cross section $C_0$ at $\sqrt{s} = 7$ and 13 TeV in the
{\it CMS-jet} kinematic configuration. We can observe that the NLO corrections
increase at large rapidity interval $Y$, an expected phenomenon in the
BFKL approach.
\FloatBarrier
In Figs.~\ref{fig:Cn_MOM_BLM_CMS_7} and~\ref{fig:Cn_MOM_BLM_CMS_13},
predictions with the BLM scale optimization for $C_0$ and
several $R_{nm} \equiv C_n/C_m$ ratios with the jet tagged inside the CMS
detector are shown for $\sqrt{s} = 7$ and 13~TeV, respectively.
Here the benefit of the use of BLM optimization appears, since the LLA and
NLA predictions for $C_0$ are now comparable, a sign of stabilization of the
perturbative series. The trend of ratios of the form $R_{n0}$ is the standard
one and indicates increasing azimuthal decorrelation between the jet and the
hadron as $Y$ goes up, with the NLA predictions systematically above the
LLA ones, as it was also observed in Mueller--Navelet jets and in the
hadron-hadron case. The ratios $R_{21}$ and $R_{32}$ seem to be almost
insensitive to the NLO corrections. Plots in Fig.~\ref{fig:Cn_MOM_BLM_CASTOR_13} show results with BLM scale
optimization for $C_0$ and several $R_{nm}$ ratios in the {\it CASTOR-jet}
configuration at $\sqrt{s} = 13$ TeV.\\They exhibit some new and, to some extent, unexpected features: (i) the two parametrizations for the FFs
lead to clearly distinct predictions, (ii) $\langle \cos \phi\rangle$ exceeds
one at the smaller values for $Y$, a clearly unphysical effect. The
reason for these phenomena could reside in the fact that, the lower values
for $Y$ in the {\it CASTOR-jet} case are obtained for negative values of the
hadron rapidity, {\it i.e.} in final-state configurations where both
jet and hadron are backward.

\begin{figure}[!h]
	\centering
	\includegraphics[scale=0.30,clip]{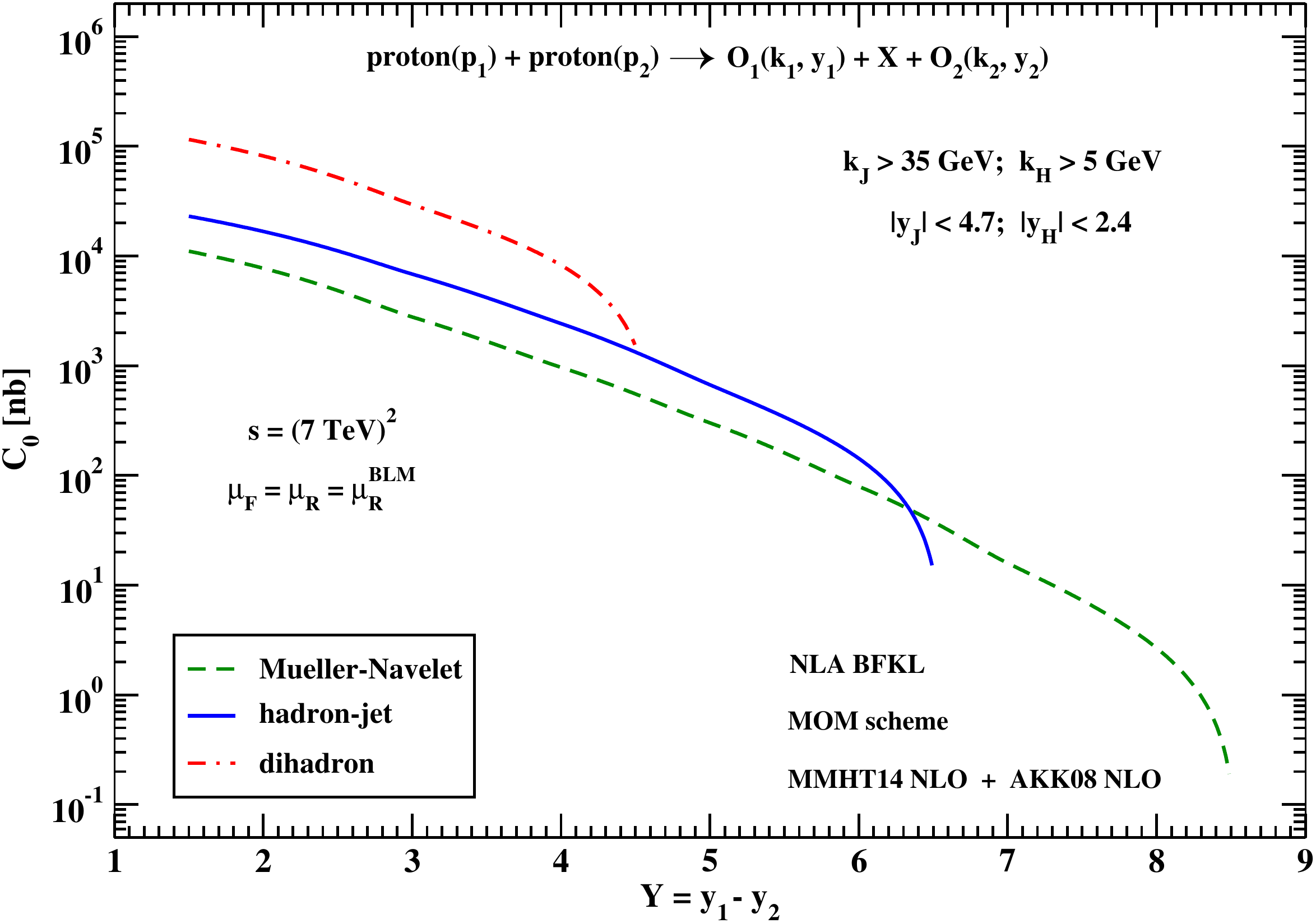}
	\includegraphics[scale=0.30,clip]{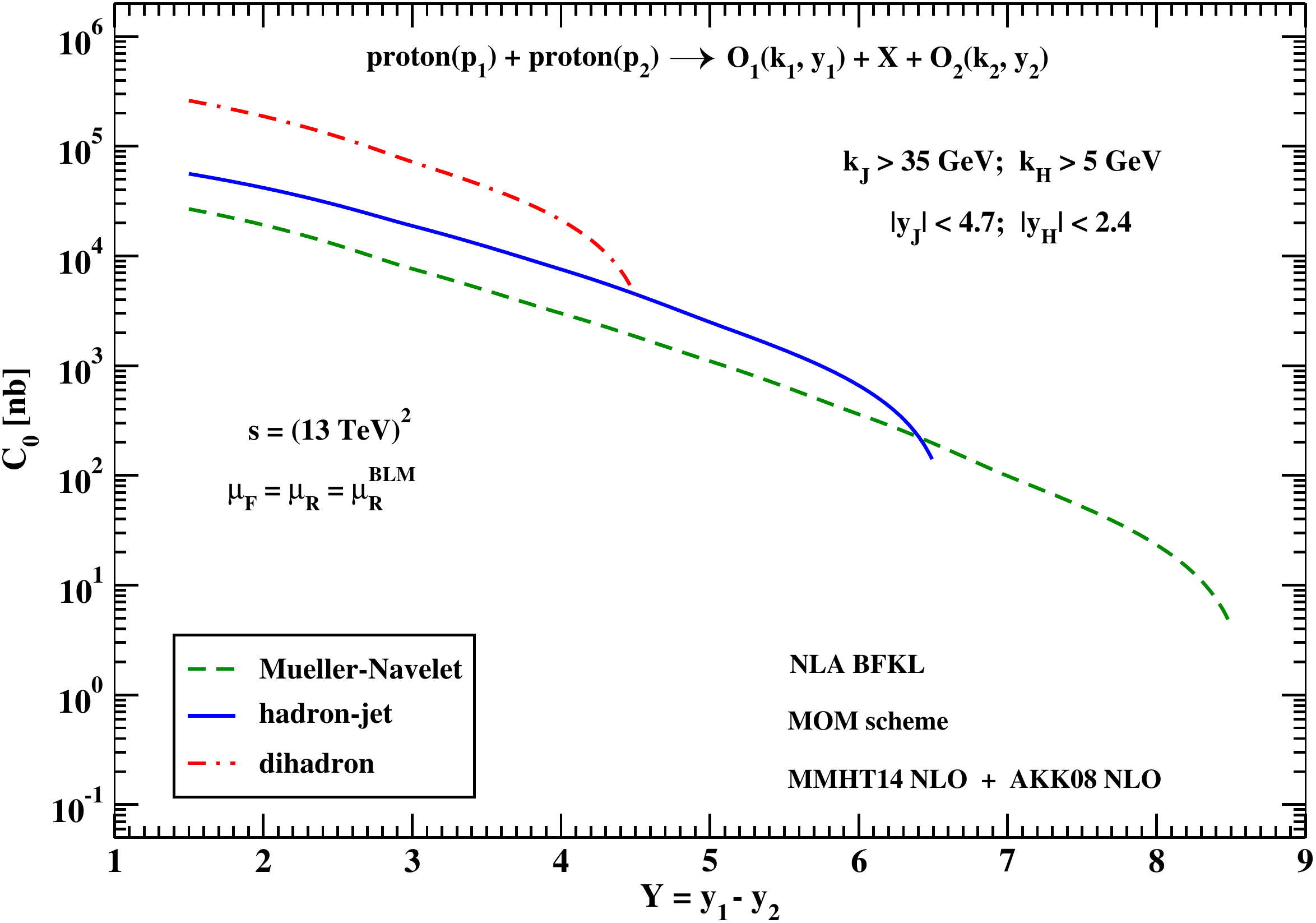}	
	\caption{\emph{Comparison of the $\phi$-averaged cross section $C_0$ in different
			NLA BFKL processes: Mueller--Navelet jet, hadron-jet and dihadron production,
			for $\mu_F = \mu_R^{\text{BLM}}$, $\sqrt{s} = 7$ and 13 TeV, and $Y \leq 7.1$
			({\it CMS-jet} configuration).}}
	\label{fig:C0_comp_NLA_BLM_CMS}
\end{figure}

Finally, in Fig.~\ref{fig:C0_comp_NLA_BLM_CMS} we compare the $\phi$-averaged
cross section $C_0$ in different NLA BFKL processes: Mueller--Navelet jet,
hadron-jet and hadron-hadron production, for $\mu_F = \mu_R^{\text{BLM}}$, at 
$\sqrt{s} = 7$ and 13 TeV, and $Y \leq 7.1$ in the {\it CMS-jet} case.\\The hadron-hadron cross section, with the kinematical cuts adopted,
dominates over the jet-jet one by an order of magnitude, with the hadron-jet
cross section lying, not surprisingly, in-between.
 \FloatBarrier
 
 \FloatBarrier

\FloatBarrier

\subsection{Summary}
\label{JH_Sum}
We have proposed a new candidate probe of BFKL dynamics
at the LHC in the process for the inclusive production of an identified
charged light hadron and a jet, separated by a large rapidity gap.

This process appears as a naive hybridization of two already well studied ones and presents some own characteristics which can make it worthy of consideration in future analyses at the LHC.

For this reason, we have provided some theoretical predictions, with
next-to-leading accuracy, for the cross section averaged over the azimuthal
angle between the identified jet and hadron and for ratios of the azimuthal
coefficients.

The trends observed in the distributions over the rapidity interval between
the jet and the hadron are not different from the cases of Mueller--Navelet
jets and hadron-hadron, when the jet is detected by CMS, whereas some new
features have appeared when the jet is seen by CASTOR, which deserve further
investigation.
\section{Heavy--quark pair hadroproduction}
\label{heavy}
\begin{figure}
	\centering
	\includegraphics[width=0.45\textwidth]{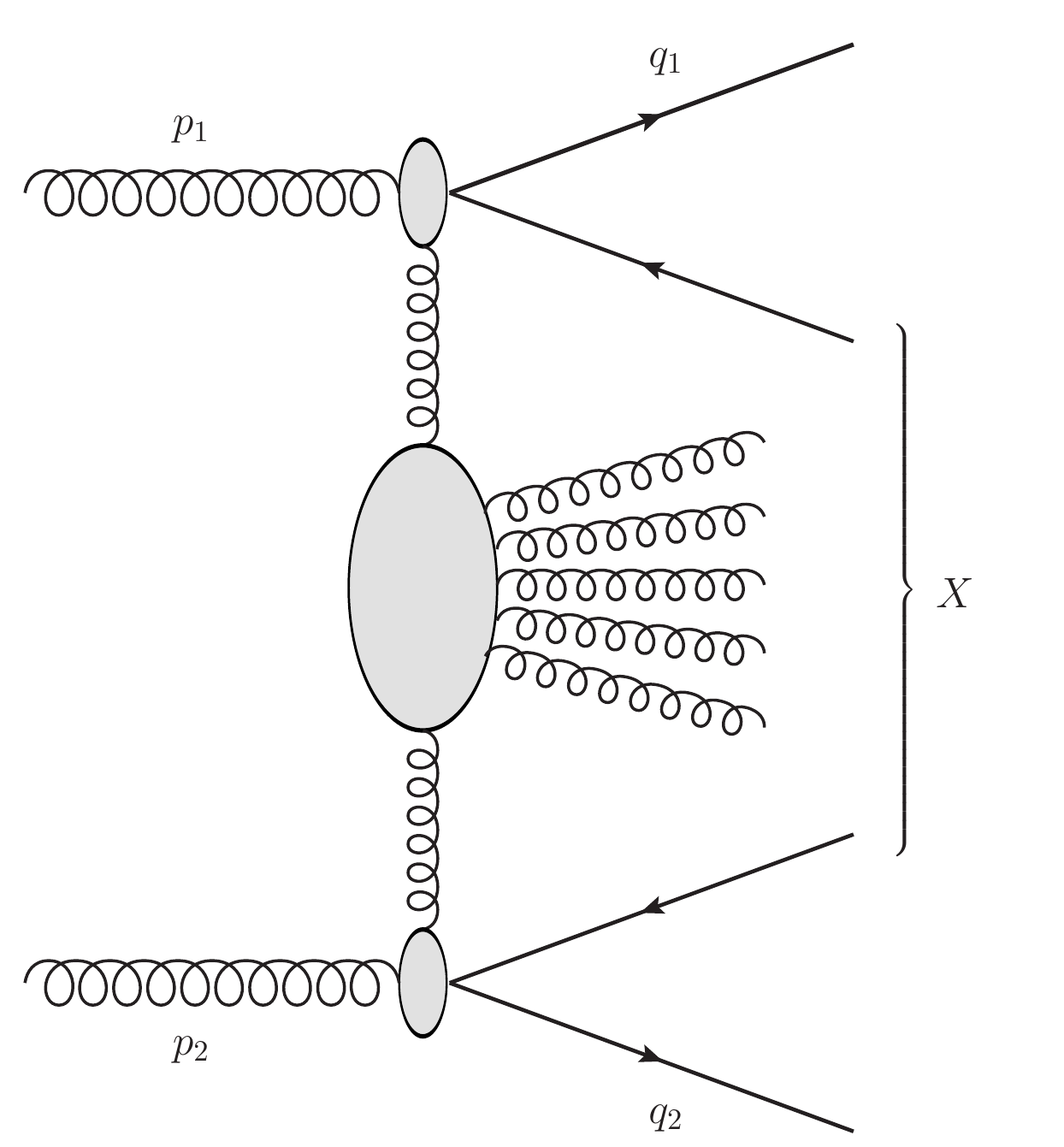}
	\caption{\emph{Schematic representation of the heavy-quark pair hadroproduction in the case when a heavy quark with transverse momentum $q_1(q_2)$ from the upper (lower) vertex is tagged.}}
	\label{fig:hadroproduction}
\end{figure}

This section is dedicated to the study within NLA BFKL accuracy of a novel semi-hard reaction: the hadroproduction, which consists in the inclusive emission of two rapidity-separated heavy quarks in the collision of two quasi-real gluon, coming from two protons~\cite{Bolognino:2019yls,Bolognino:2019ouc}: 
\begin{equation}
\label{processHQ}
g(p_1) \ + \ g(p_2) \ \to \ Q\text{-jet}(q_1) \ + \ X \ + \ Q\text{-jet}(q_2)
\;,
\end{equation}
where $Q$ stands for a charm/bottom quark or the respective antiquark.
A pictorial description of this process is shown in Fig.~\ref{fig:hadroproduction}, in the case when the tagged object are a heavy quark with momentum $q_1$ , detected in the fragmentation region of the gluon with momentum $p_1$ and a heavy quark with momentum $q_2$, detected in the fragmentation region of the gluon with momentum $p_2$.
A similar process with the same final state was considered~\cite{Celiberto:2017nyx,Bolognino:2019ouc}, but produced through the collision of two (quasi-)real photons ({\em photoproduction}) emitted by two interacting electron and positron beams.

The purpose of this work is to provide theoretical predictions for cross section and azimuthal coefficients of the expansion in the (cosine of the) relative angle in the transverse plane between the flight directions of the two tagged heavy quarks. Then, compare the results with experimental analyses at the LHC. It will be interesting to observe that in the same kinematical regime in which also the photoproduction in Refs.~\cite{Celiberto:2017nyx,Bolognino:2019ouc} has been tested, the hadroproduction channel allows us to obtain much higher cross section values. Besides, we will treat the specific case of the inclusive hadroproduction of two bottom quarks and present a phenomenological analysis tailored on the kinematics and the energies of the LHC, proposing it as a new channel for the investigation of the BFKL mechanism at hadron colliders.

\subsection{Theoretical setup}
\label{HQ:setup}
We are interested in the process represented in Fig.~\ref{fig:hadroproduction} in order to build the cross section, differential in some of the kinematic variables of the tagged heavy quark or antiquark, and give some azimuthal correlations between the tagged fermions. As previously discussed in Section~\ref{Chap:BFKL}, the cross section written in the BFKL framework has a factorized form given, in this case, by the convolution of the impact
factors for the transition from a real gluon to a heavy quark-antiquark pair
with the BFKL Green's function $G$.
\subsubsection{Impact factor}
\label{HQ:IF}
In order to give the expression of the impact factor for the hadroproduction of a heavy-quark pair, we consider the leading-order impact factor,
together with the functional form of the amplitude for the
$g + R \to q \bar{q}$ subprocess, where $R$ here means ``Reggeized gluon''. The leading-order impact factor is defined as~\cite{Fadin:1998fv}
\begin{equation}
\begin{split}
d\Phi_{gg}^{\lbrace{Q\bar{Q}\rbrace}}(\vec{q},\vec{k},z) = & \frac{\braket{cc'|\mathcal{\widehat{P}}|0}}{2\left(N^2-1\right)}
\\ &  \hspace{-1.3cm}\times  \sum_{\lambda_Q\lambda_{\bar{Q}}\lambda_G}\sum_{Q\bar{Q}a} \int\frac{ds_{gR}}{2\pi}d\rho_{\lbrace{Q\bar{Q}\rbrace}}\Gamma_{g \rightarrow \lbrace{Q\bar{Q}\rbrace}}^{ca}\left(q,k,z\right)\left(\Gamma_{g \rightarrow \lbrace{Q\bar{Q}\rbrace}}^{ac'}\left(q,k,z\right)\right)^{\ast} \;,
\end{split}
\label{eq:imp.fac}
\end{equation}
where 
\begin{equation}
\braket{cc'|\mathcal{\widehat{P}}|0} = \frac{\delta^{cc'}}{\sqrt{N^2-1}}
\end{equation}
is the projector on the singlet state. We take the sum over helicities,
$\{\lambda_Q,\lambda_{\bar{Q}}\}$, and over color indices, $\{Q,\bar{Q}\}$, of
the two produced particles (quark and antiquark) and average over polarization
and color states of the incoming gluon. In Eq.~(\ref{eq:imp.fac}), $s_{gR}$
denotes the invariant squared mass of the gluon-Reggeon system, while $
d\rho_{\lbrace{Q\bar{Q}\rbrace}}$ is the differential phase space of the outgoing
particles. The amplitude $\Gamma_{g \rightarrow \lbrace{Q\bar{Q}\rbrace}}^{ca}$ describes
the production of quark-antiquark pair in a collision between a gluon and a
Reggeon. The latter can be treated as an ordinary gluon in the so called
``nonsense'' polarization state $\epsilon_R^\mu= p_2^\mu/s$ (for instance see
Ref.~\cite{Fadin:2001dc}).
Having two particles produced in the intermediate state, one can write
\begin{equation}
\label{phasespace}
\frac{ds_{gR}}{2 \pi} d\rho_{\lbrace{Q\bar{Q}\rbrace}} = \frac{1}{2\left(2 \pi \right)^3} \delta \left(1-z-\bar{z}\right) \delta^{(2)} \left( \vec{k}-\vec{q}-\vec{\bar{q}} \right) \frac{dz d\bar{z}}{z \bar{z}} d^2\vec{q} \; d^2\vec{\bar{q}} \; , 
\end{equation}
where $\bar{q}$ the antiquark momentum, and $\bar{z}$ its longitudinal
momentum fraction with respect to the incoming gluon.
\begin{figure}[!h]
	\centering
	\includegraphics[width=1.0\textwidth]{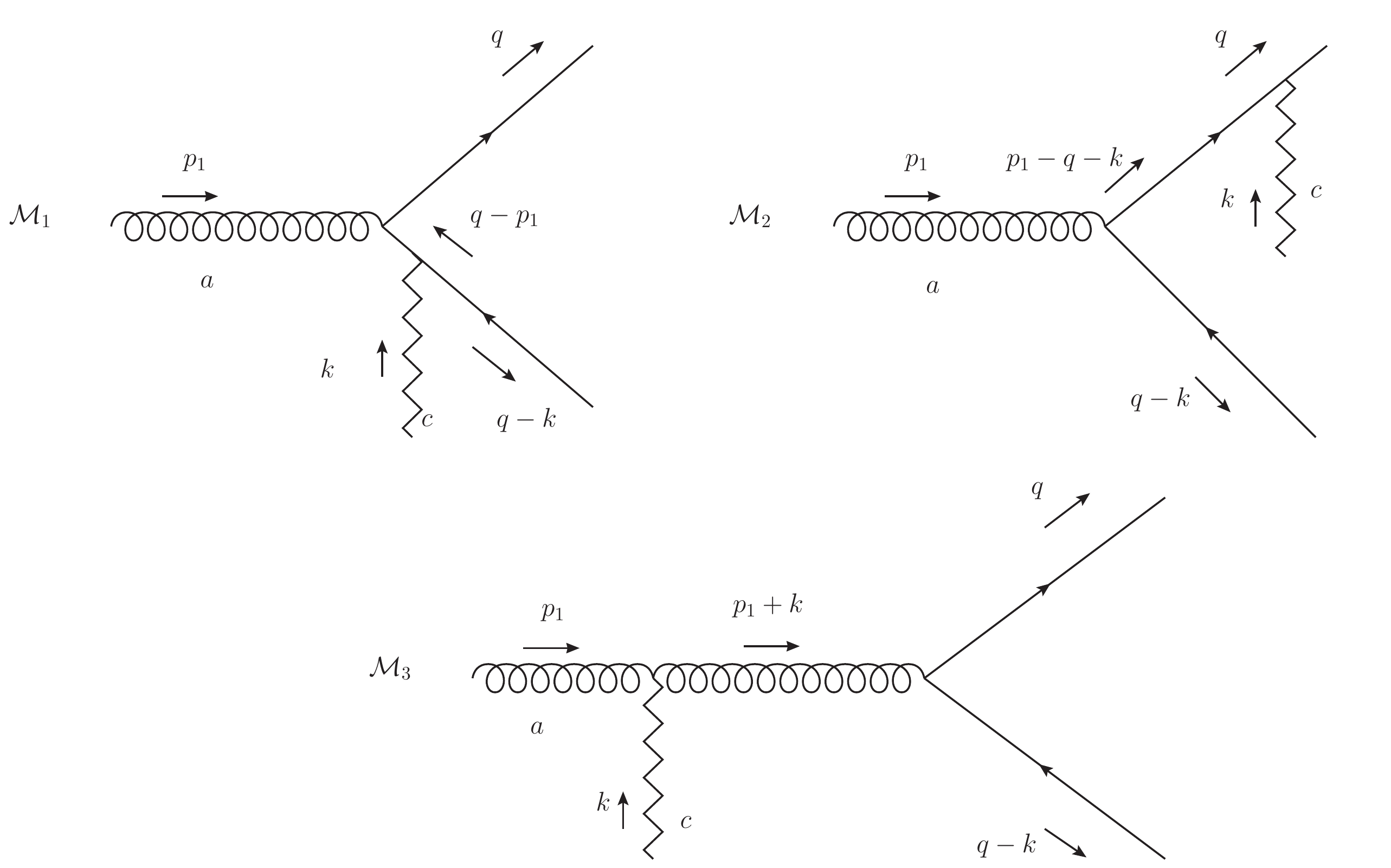}
	\caption{\emph{Feynman diagrams relevant for the calculation of the impact factor for the heavy-quark pair hadroproduction. The zigzag line denotes a Reggeized gluon.}}
	\label{fig:hadroproduction_IF}
\end{figure}
\FloatBarrier
Now we sum over he three contributions, $\{{\cal M}_{1,2,3}\}$, of Fig.~\ref{fig:hadroproduction_IF} and we obtain
\begin{equation}
\label{amp}
\begin{split}
\Gamma_{g \rightarrow \lbrace{Q\bar{Q}\rbrace}}^{ca} & = ig^2\left(\tau^{a}\tau^{c}\right)\bar{u}\left(q\right)\left(mR\slashed{\epsilon} - 2 z \vec{P} \cdot \vec{\epsilon} - \vec{\slashed{P}} \slashed{\epsilon}\right)\frac{\slashed{p}_2}{\hat{s}}v\left(\bar{q}\right) \\ & + ig^2\left(\tau^{c}\tau^{a}\right)\bar{u}\left(q\right)\left(m\bar{R}\slashed{\epsilon} - 2 z \vec{\bar{P}} \cdot \vec{\epsilon}- \vec{\slashed{\bar{P}}} \; \; \slashed{\epsilon} \right)\frac{\slashed{p}_2}{\hat{s}}v\left(\bar{q} \right) \; ,
\end{split}
\end{equation}
where $\hat{s} = W^2$, $\epsilon^\mu$ identifies the gluon polarization vector, $\{\tau\}$ are the $SU(3)$ color matrices and $R, \bar{R}, \vec{P}, \vec{\bar{P}}$ are defined as
\begin{equation}
\label{laR}
R = \frac{1}{m^2+\vec{q}^{\;2}}-\frac{1}{m^2+(\vec{q}-z\vec{k})^{2}} \;,  
\end{equation}
\begin{equation}
\label{LaRbar}
\bar{R} = \frac{1}{m^2+(\vec{q}-z\vec{k})^{2}}-\frac{1}{m^2+(\vec{q}-\vec{k})^{2}} \;, 
\end{equation}
\begin{equation}
\label{laP}
\vec{P} = \frac{\vec{q}}{m^2+\vec{q}^{\;2}}-\frac{\vec{q}-z\vec{k}}{m^2+(\vec{q}
	-z\vec{k})^{2}} \;,  
\end{equation}
\begin{equation}
\label{LaPbar}
\vec{\bar{P}} = \frac{\vec{q}-z\vec{k}}{m^2+(\vec{q}-z\vec{k})^{2}}
-\frac{\vec{q}-\vec{k}}{m^2+(\vec{q}-\vec{k})^{2}} \;.  
\end{equation}
Here $\alpha_s$ denotes the QCD coupling, $N_c$ gives the number of colors, $m$
stands for the heavy-quark mass, $z$ and $\bar{z} \equiv 1-z$ are the
longitudinal momentum fractions of the quark and antiquark produced in the
same vertex and $k$, $q$, $k-q$ represent the transverse momenta with respect
to the gluons collision axis of the Reggeized gluon, the produced quark and
antiquark, respectively.

Using Eq.~(\ref{amp}) together with Eq.~(\ref{eq:imp.fac}) and performing
sums and integrations (the latter ones only on the antiquark variables), we find the expression for the impact factor for the process under consideration
\begin{equation}
\begin{split}
\label{eq:imp.fac2}
d\Phi^{\lbrace{Q\bar{Q}\rbrace}}_{gg}(\vec{k},\vec{q},z)& =\frac{\alpha_s^2
	\sqrt{N_c^2-1}}{2\pi N_c}\left[\left(m^2\left(R+\bar{R}\right)^2
+\left(z^2+\bar{z}^2 \right)\left(\vec{P}+\vec{\bar{P}}\right)^2\right)
\right. \\ & \left. -\frac{N_c^2}{N_c^2-1}\left(2m^2R\bar{R}
+\left(z^2+\bar{z}^2\right)2\vec{P} \cdot \vec{\bar{P}}\right)\right]
\; d^2\vec{q} \; dz\;.
\end{split}
\end{equation}
In order to proceed we need the projection of the impact factors onto the
eigenfunctions of the leading-order BFKL kernel, to get their so called
$(n, \nu)$-representation. Hence, we have
\[
\frac{d\Phi_{gg}^{\lbrace{Q\bar{Q}\rbrace}}\left(n,\nu,\vec{q},z\right)}{d^2\vec{q}
	\; dz} \equiv \int\frac{d^2\vec{k}}{\pi\sqrt{2}}(\vec{k}^{\;2})^{i\nu-\frac{3}{2}}
e^{in\theta}\frac{d\Phi^{\lbrace{Q\bar{Q}\rbrace}}_{gg}(\vec{k},\vec{q},z)}{d^2\vec{q}
	\; dz} 
\]
\[
= \frac{\alpha_s^2 \sqrt{N_c^2-1}}{2\pi N_c}\left\{ m^2 \left( I_3
- 2\frac{I_2(0)}{m^2+\vec{q}^{\;2}} \right) + (z^2 + \bar{z}^2)
\left( -m^2 \left(I_3 - 2\frac{I_2(0)}{m^2+\vec{q}^{\;2}} \right)
 \right.\right.
\]
\[
\left. \left. + \frac{I_2(1)}{m^2 + \vec{q}^{\; 2}} \right) - \frac{N_c^2}{N_c^2-1} \Bigg[ 2 m^2 \left[\left( z^2 + \bar{z}^2
- 1 \right) \left( 1 -  \left(z^2\right)^{\frac{1}{2}-i\nu} \right) \right]
\frac{I_2(0)}{m^2+ \vec{q}^{\; 2}}
\right.
\]
\[
\left. + \left[ 2m^2(z^2 + \bar{z}^2 -1)
\left(z^2\right)^{\frac{1}{2}-i\nu} \right] 
\left( I_3 - \frac{I_4(0)}{\left(z^2\right)^{\frac{1}{2}-i\nu}} \right)
- (z^2 + \bar{z}^2) \right.
\]
\[
\left.\times \bigg[ (1-z)^2 I_4(1) - \frac{\left( 1
	- \left(z^2\right)^{\frac{1}{2}-i\nu} \right)}{m^2 + \vec{q}^{\;2}} I_2(1) \bigg]
\Bigg] \right\}
\]
\begin{equation}
\equiv \alpha_s^2 \; e^{in \varphi} c(n, \nu, \vec{q},z) \;,
\label{eq:imp.fac projected1}
\end{equation}
where the expression of $I_1$, $I_2(\lambda)$, $I_3$ and $I_4(\lambda)$, essential to make the projections of the various terms of the impact factor, will be calculated below.
\begin{flushleft}
\bf{Useful Integrals}
\end{flushleft}
The first integral
\begin{equation}
I_1\equiv\int\frac{d^2\vec{k}}{\pi\sqrt{2}}\left(\vec{k}^{\;2}\right)^{i\nu-\frac{3}{2}}e^{in\theta} \; , \quad {\text{for}} \; n \neq 0 \;,    
\end{equation}
vanishes because of the periodicity condition on the angle $\theta$. The second integral reads
\begin{equation}
\begin{split}
I_2\left(\lambda\right) \equiv \int\frac{d^2\vec{k}}{\pi\sqrt{2}}(\vec{k}^{\;2})^{i\nu-\frac{3}{2}}e^{in\theta}\frac{(\vec{k}^{\;2})^{\lambda}}{m^2+(\vec{q}-\vec{k})^2}
\end{split}
\end{equation}
\begin{equation}
\begin{split}
& = \frac{\left(\vec{q}^{\; 2}\right)^{\frac{n}{2}}e^{in\varphi}}{\sqrt{2}}\frac{1}{\left(m^2+\vec{q}^{\;2}\right)^{\frac{3}{2}+\frac{n}{2}-i\nu-\lambda}} \frac{\Gamma\left(\frac{1}{2}+\frac{n}{2}+i\nu+\lambda\right)\Gamma\left(\frac{1}{2}+\frac{n}{2}-i\nu-\lambda\right)}{\Gamma\left(1+n\right)} \\ & \times \frac{\left(\frac{1}{2}+\frac{n}{2}-i\nu-\lambda\right)}{\left(-\frac{1}{2}+\frac{n}{2}+i\nu+\lambda\right)}\; _2F_1\left(-\frac{1}{2}+\frac{n}{2}+i\nu+\lambda,\frac{3}{2}+\frac{n}{2}-i\nu-\lambda,1+n,\zeta\right) \;.
\end{split}
\end{equation}

The third integral can be presented as
\begin{equation}
\label{I3_start}
I_3
\equiv \int\frac{d^2\vec{k}}{\pi\sqrt2}(\vec{k}^{\,\,2})^{i\nu-\frac{3}{2}}e^{in\theta}
\frac{1}{\left(m^2+(\vec{q}-\vec{k})^2\right)^2} 
\end{equation}
\begin{equation}
\begin{split}
= & \frac{\left(\vec{q}^{\; 2}\right)^{\frac{n}{2}}e^{in\varphi}}{\sqrt{2}} \frac{1}{\left(m^2+\vec{q}^{\;2}\right)^{\frac{5}{2}+\frac{n}{2}-i\nu}} \frac{\Gamma\left(\frac{1}{2}+\frac{n}{2}+i\nu\right)\Gamma\left(\frac{1}{2}+\frac{n}{2}-i\nu\right)}{\Gamma\left(1+n\right)}\frac{\left(\frac{1}{2}+\frac{n}{2}-i\nu\right)}{\left(-\frac{1}{2}+\frac{n}{2}+i\nu\right)}\; \\ & \times \left(\frac{3}{2}+\frac{n}{2}-i\nu\right)\; _2F_1\left(-\frac{1}{2}+\frac{n}{2}+i\nu,\frac{5}{2}+\frac{n}{2}-i\nu,1+n,\zeta \right) \;.
\end{split} 
\end{equation}
For the sake of completeness, we show the entire derivation of the fourth integral, defined as
\begin{equation}
I_4\left(\lambda\right) \equiv \int\frac{d^2\vec{k}}{\pi\sqrt{2}} (\vec{k}^{\;2})^{i\nu-\frac{3}{2}}e^{in\theta}\frac{(\vec{k}^{\;2})^{\lambda}}{(m^2+(\vec{q}-\vec{k})^2)(m^2+(\vec{q}-z \vec{k})^2)} \;,
\end{equation}
which is the most cumbersome one. The strategy to calculate $I_2(\lambda)$ and $I_3$ is the same. To lighten the notation, it is useful to define $\alpha=i\nu+\lambda$, then the integral can be easily put in the following form:
\begin{equation}
I_4\left(\lambda\right)=\frac{1}{z^2} \int\frac{d^2\vec{k}}{\pi\sqrt{2}}\frac{(\vec{k}\cdot\vec{l}\;)^n}{(\vec{k}^{\;2})^{\frac{3}{2}+\frac{n}{2}-\alpha}\left(m^2+\left(\vec{q}-\vec{k}\right)^2\right)\left(\frac{m^2}{z^2}+\left(\frac{\vec{q}}{z}-\vec{k}\right)^2\right)} \;.
\end{equation}
where $\vec{l} \equiv (i,1)$ and the following formula,
\begin{equation}
e^{in\theta}=\left(\cos{\theta}+i\sin{\theta}\right)^n=\left(\frac{k_x+ik_y}{|\vec{k}\;|}\right)^n=\frac{(\vec{k}\cdot\vec{l}\;)^n}{(\vec{k}^{\;2})^{\frac{n}{2}}} \;,
\end{equation}
has been used. Using the Feynman parametrization,
\begin{equation}
\label{feynman param.2}
\begin{split}
\frac{1}{s^Mt^Nw^L}&=\frac{\Gamma\left(M+N+L\right)}{\Gamma\left(M\right)\Gamma\left(N\right)\Gamma\left(L\right)}\int_0^1 dx \int_0^1 dy \int_0^1 dz \frac{x^{M-1}y^{N-1}z^{L-1}\delta\left(1-x-y-z\right)}{\left(xs+yt+zw\right)^{M+N+L}} \\ & =\frac{\Gamma\left(M+N+L\right)}{\Gamma\left(M\right)\Gamma\left(N\right)\Gamma\left(L\right)}\int_0^1 dx \int_0^{1-x} dy \frac{x^{M-1}y^{N-1}\left(1-x-y\right)^{L-1}}{\left(xs+yt+\left(1-x-y\right)w\right)^{M+N+L}} \;.
\end{split}
\end{equation}
one finds that
\begin{equation}
\begin{split}
& I_4\left(\lambda\right)= \frac{1}{z^2}\frac{\Gamma\left(\frac{7}{2}-\alpha+\frac{n}{2}\right)}{\Gamma\left(\frac{3}{2}-\alpha+\frac{n}{2}\right)} \int \frac{d^2\vec{k}}{\pi\sqrt{2}}  \int_0^1 dx\\& \times \int_0^{1-x} dy \frac{\left(1-x-y\right)^{\frac{1}{2}-\alpha+\frac{n}{2}}\left(\vec{k} \cdot \vec{l} \; \right)^n}{\left[x\left(\frac{m^2}{z^2}+\left(\frac{\vec{q}}{z}-\vec{k}\right)^2\right)+y\left(m^2+\left(\vec{q}-\vec{k}\right)^2\right)+\left(1-x-y\right)\vec{k}^{\; 2}\right]^{\frac{7}{2}-\alpha+\frac{n}{2}}} \;.
\end{split}
\end{equation}
Making the following substitution, $\vec{k} \to \vec{k}+\left(\frac{x}{z}+y\right)\vec{q}$, and observing that 
\begin{equation}
\left(\vec{k}\cdot \vec{l}+\left(\frac{x}{z}+y\right) \vec{q}\cdot \vec{l}\;\right)^n = \sum_j{n\choose j}\left(\vec{k} \cdot \vec{l}\;\right)^j \left(\frac{x}{z}+y\right)^{n-j}\left(\vec{q} \cdot \vec{l}\;\right)^{n-j} \;,
\end{equation}
after some trivial calculations, one obtains
\begin{equation}
\label{eq:binomial2}
\begin{split}
I_4\left(\lambda\right)=& \frac{1}{z^2}\left(\frac{5}{2}-\alpha+\frac{n}{2}\right)\left(\frac{3}{2}-\alpha+\frac{n}{2}\right) \int_0^1 dx \int_0^{1-x} dy \left(1-x-y\right)^{\frac{1}{2}-\alpha+\frac{n}{2}} \\ & \times \int \frac{d^2\vec{k}}{\sqrt{2}}\frac{\sum_j{n\choose j}\left(\vec{k} \cdot \vec{l}\;\right)^j \left(\frac{x}{z}+y\right)^{n-j}\left(\vec{q} \cdot \vec{l}\;\right)^{n-j}}{\left[\vec{k}^{\; 2}+L^2 \right]^{\frac{7}{2}-\alpha+\frac{n}{2}}} \;,
\end{split}
\end{equation}
where 
\begin{equation}
L^2=\left(\frac{x}{z^2}+y\right)\left(m^2+\vec{q}^{\; 2}\right)-\left(\frac{x}{z}+y\right)^2\vec{q}^{\; 2} \;.
\end{equation}
The only term which gives non-zero contribute in the binomial in
Eq.~(\ref{eq:binomial2}) is the $0^{\text{th}}$ coefficient, namely
\begin{equation}
\left(\frac{x}{z}+y\right)^n\left(\vec{q}\cdot\vec{l}\;\right)^n=\left(\frac{x}{z}+y\right)^n\left(\vec{q}^{\;2}\right)^{\frac{n}{2}}e^{in\varphi} \;,
\end{equation}
where $\varphi$ is the azimuthal angle of the vector $\vec{q}$.

Hence,
\begin{equation}
\begin{split}
I_4 & \left(\lambda\right)= \frac{\left(\vec{q}^{\; 2}\right)^{\frac{n}{2}}e^{in\varphi}}{z^2}\left(\frac{5}{2}-\alpha+\frac{n}{2}\right)\left(\frac{3}{2}-\alpha+\frac{n}{2}\right) \\ & \times \int_0^1 dx \int_0^{1-x} dy \left(1-x-y\right)^{\frac{1}{2}-\alpha+\frac{n}{2}}\left(\frac{x}{z}+y\right)^n \int \frac{d^2\vec{k}}{\sqrt{2}}\frac{1}{\left[\vec{k}^{\; 2}+L^2 \right]^{\frac{7}{2}-\alpha+\frac{n}{2}}} \;.
\end{split}
\end{equation}
For the integration in $d^2\vec{k}$, we use the formula 
\begin{equation}
\label{integral_k}
\int \frac{d^{2}k}{(2\pi)^{3}}\frac{1}{(\vec{k}^{\,\,2} + L^2)^\rho}=\frac{2}{(4\pi)^{2}} \frac{\Gamma(\rho-1)}{\Gamma(\rho)}(L^2)^{-\rho+1} \; ,
\end{equation}
setting $\rho = \frac{7}{2}-\alpha+\frac{n}{2}$. We then obtain
\begin{equation}
\begin{split}
I_4\left(\lambda\right)=& \frac{\left(\vec{q}^{\; 2}\right)^{\frac{n}{2}}e^{in\varphi}}{z^2\sqrt{2}}\frac{\left(\frac{3}{2}-\alpha+\frac{n}{2}\right)}{\left(m^2+\vec{q}^{\; 2}\right)^{\frac{5}{2}-\alpha+\frac{n}{2}}} \int_0^1 dx \int_0^{1-x} dy \\ & \times \left(1-x-y\right)^{\frac{1}{2}-\alpha+\frac{n}{2}}\left(\frac{x}{z}+y\right)^n \left[\left(\frac{x}{z^2}+y\right)-\zeta\left(\frac{x}{z}+y\right)^2\right]^{-\frac{5}{2}+\alpha-\frac{n}{2}} \;,
\end{split}
\end{equation}
where
\begin{equation}
\label{zeta}
\zeta=\frac{\vec{q}^{\;2}}{m^2+\vec{q}^{\;2}} \;.
\end{equation}
To integrate this expression over one of the two Feynman parameters, we perform
the following change of variables:
\begin{align}
\label{transf.}
& x \,=\, \tau \Delta \; ,
\\ \nonumber 
& y \,=\, \tau \left(1-\Delta \right) \;,
\end{align}  
The Jacobian determinant is simply 
\begin{equation}
||J||=\tau \; ,
\end{equation}
and the integral $I_4\left(\lambda\right)$ becomes
\begin{equation}
\begin{split}
I_4\left(\lambda\right)= \frac{\left(\vec{q}^{\; 2}\right)^{\frac{n}{2}}e^{in\varphi}}{z^2\sqrt{2}} & \frac{\left(\frac{3}{2}-\alpha+\frac{n}{2}\right)}{\left(m^2+\vec{q}^{\; 2}\right)^{\frac{5}{2}-\alpha+\frac{n}{2}}} \int_0^1 d\Delta \left(1+\frac{\Delta}{z}-\Delta\right)^n \left(1+\frac{\Delta}{z^2}-\Delta\right)^{-\frac{5}{2}+\alpha-\frac{n}{2}} \\ & \times \int_0^1 d\tau \left(1-\tau\right)^{\frac{1}{2}-\alpha+\frac{n}{2}} \tau^{-\frac{3}{2}+\alpha+\frac{n}{2}} \left(1-A\tau\right)^{-\frac{5}{2}+\alpha-\frac{n}{2}} \;,
\end{split}
\end{equation}
where
\begin{equation}
A=\zeta \frac{\left(1+\frac{\Delta}{z}-\Delta\right)^2}{\left(1+\frac{\Delta}{z^2}-\Delta\right)} \; .
\end{equation}
Let us now consider the integral representation of the hypergeometric function, 
\begin{equation}
\label{hypergeometric}
\mathcal{B}(b,c-b) \, {_2}F_1(a,b,c,z) = \int_0^1 dx \, x^{b-1} (1-x)^{c-b-1} (1-zx)^{-a} \;,
\end{equation}
with the Euler function 
\begin{equation}
\label{beta_function}
\mathcal{B}(u,\omega) = 
\frac{\Gamma(u)\Gamma(\omega)}{\Gamma(u+\omega)}
\; . 
\end{equation}
For the $\tau$ integration it is enough to consider Eq.~(\ref{hypergeometric})
and the Eq.~(\ref{beta_function}), setting:
\begin{equation*}
a \,=\, \frac{5}{2}-\alpha+\frac{n}{2} \; , \quad \quad b \,=\, -\frac{1}{2}+\alpha+\frac{n}{2} \; , \quad \quad c \,=\, 1+n \; , \quad \quad  z \,=\, A \; .
\end{equation*}
Finally, expressing $A$ and $\alpha$ in their explicit form and making use of
the property
\[
_2F_1(a,b,c,z)= \; _2F_1(b,a,c,z) \;,
\]
one finds
\begin{equation}
\label{I4}
\begin{split}
I_4\left(\lambda\right) = & \frac{\left(\vec{q}^{\; 2}\right)^{\frac{n}{2}}e^{in\varphi}}{z^2\sqrt{2}}  \frac{\left(\frac{3}{2}-i\nu-\lambda+\frac{n}{2}\right)}{\left(m^2+\vec{q}^{\; 2}\right)^{\frac{5}{2}-i\nu-\lambda+\frac{n}{2}}} \frac{\Gamma\left(\frac{1}{2}+\frac{n}{2}+i\nu+\lambda \right)\Gamma\left(\frac{1}{2}+\frac{n}{2}-i\nu-\lambda \right)}{\Gamma\left(1+n\right)} \\ & \times \frac{\left(\frac{1}{2}+\frac{n}{2}-i\nu-\lambda \right)}{\left(-\frac{1}{2}+\frac{n}{2}+i\nu+\lambda \right)} \int_0^1 d\Delta \left(1+\frac{\Delta}{z}-\Delta\right)^n \left(1+\frac{\Delta}{z^2}-\Delta\right)^{-\frac{5}{2}+i\nu+\lambda-\frac{n}{2}} \; \\ & \times \; _2F_1\left(-\frac{1}{2}+i\nu+\lambda+\frac{n}{2},\frac{5}{2}-i\nu-\lambda+\frac{n}{2},1+n,\zeta \frac{\left(1+\frac{\Delta}{z}-\Delta\right)^2}{\left(1+\frac{\Delta}{z^2}-\Delta\right)}\right) \;.
\end{split}
\end{equation}

\subsection{Kinematics of the process}
\label{HQ:kinem}

In order to express the final form for the cross section of the heavy-quark pair hadroproduction, we introduce the standard Sudakov decomposition for the tagged quark momenta using as light-cone basis the momenta $p_1$ and $p_2$ of the colliding gluons,
\begin{equation}
q = zp_1 + \frac{m^2+ \vec{q}^{\; 2}}{zW^2}p_2+q_{\perp} \;,
\end{equation}
with $W^2=(p_1+p_2)^2=2p_1 \cdot p_2=4E_{g_1}E_{g_2}$ ; $p_1=E_{g_1}(1,\vec{0},1)$
and $p_2=E_{g_2}(1,\vec{0},-1)$, so that
\begin{equation}
\label{2q1p2}
2q \cdot p_2=2 z p_1 \cdot p_2= z W^2 = 2E_{g_2} \left(E + q_{\|}\right) \;,
\end{equation}
\begin{equation}
\label{2q1p1}
2q \cdot p_1 = \frac{m^2+\vec{q}^{\; 2}}{z} = 2E_{g_1} \left(E - q_{\|}\right) \;;
\end{equation}
here $q=(E,\vec{q},q_{\|})$ and the rapidity can be expressed as
\begin{equation}
y=\frac{1}{2}\ln\frac{\left(E+q_{\|}\right)}{\left(E-q_{ \|}\right)}
= \ln\left[\frac{2z E_{g_1}} {\sqrt{m^2+\vec{q}^{\; 2}}}\right] \;.
\end{equation}
Consequently, the rapidities of the two tagged quarks in our process are 
\begin{equation}
\label{rapidity1}
y_1=\ln\left[\frac{2z_1 E_{g_1}} {\sqrt{m^2+\vec{q}_1^{\;2}}}\right]
\hspace{0.2 cm} {\text{and}} \hspace{0.2 cm} y_2= -\ln\left[\frac{2z_2 E_{g_2}}
{\sqrt{m^2+\vec{q}_2^{\;2}}}\right]\;,
\end{equation}
and their rapidity difference is 
\begin{equation}
\Delta Y \equiv y_1-y_2= \ln \frac{W^2 z_1 z_2}{\sqrt{\left(m^2+\vec{q}_1^{\; 2}
		\right)\left(m^2+\vec{q}_2^{\; 2}\right)}} \;.
\end{equation}
This process is a semi-hard reaction, so implies that one has to require the semi-hard kinematics: $s \gg s_0 \gg \Lambda_{QCD}^2$, where $s$ is the center of mass energy, $s_0$ is the typical scale of the process and $\Lambda_{QCD}$ is the QCD scale. In this case $s_0$ will be set as $s_0=\sqrt{(m^2+\vec{q}_1^{\; 2})(m^2+\vec{q}_1^{\; '2})}$.
Therefore for the semi-hard kinematics we have the requirement 
\begin{equation}
\frac{W^2}{\sqrt{\left(m^2+\vec{q}_1^{\; 2}\right)\left(m^2+\vec{q}_2^{\; 2}\right)}}=\frac{e^{\Delta Y}}{z_1 z_2} \gg 1 \;,
\end{equation}
so that we will consider the kinematics when $\Delta Y \geq \Delta_0 \sim 1
\div 2$.
Since we will need a cross section differential in the rapidities of the tagged quarks, the further step consists in the following change of variables:
\begin{equation*}
z_1 \to y_1=\ln\left[\frac{2z_1 E_{g_1}} {\sqrt{m^2+\vec{q}_1^{\; 2}}}\right]\;, \quad  dy_1=\frac{dz_1}{z_1}\;,
\end{equation*}
\begin{equation*}
z_2 \to y_2=-\ln\left[\frac{2z_2 E_{g_2}} {\sqrt{m^2+\vec{q}_2^{\; 2}}}\right]\;, \quad  dy_2=-\frac{dz_2}{z_2}\;,
\end{equation*}
which implies
\begin{equation*}
dz_1 dz_2=\frac{e^{\Delta Y}\sqrt{m^2+\vec{q}_1^{\; 2}}\sqrt{m^2+\vec{q}_2^{\; 2}}}{W^2} dy_1dy_2 \;.
\end{equation*}
\subsection{Cross section and azimuthal correlations}
\label{HQ:cross_sec}
The differential cross section for the inclusive production of a pair of heavy
quarks can be cast in the form:
\begin{equation}
\label{crosfin}
\frac{d\sigma_{gg}}{dy_1dy_2d|\vec{q}_1|d|\vec{q}_2|d\varphi_1 d\varphi_2}=\frac{1}{(2\pi)^2} \left[\mathcal{C}_0+2 \sum_{n=1}^{\infty} \cos(n\varphi) \mathcal{C}_n \right]\,,
\end{equation}
where $\varphi=\varphi_1-\varphi_2-\pi$, while $\mathcal{C}_0$ gives the
$\varphi$-averaged cross section summed over the azimuthal angles,
$\varphi_{1,2}$, of the produced quarks. Instead, the other coefficients,
$\mathcal{C}_n$, represent the distribution of the relative azimuthal angle
between the two quarks.

The expression for the $\mathcal{C}_n$ coefficient is~\cite{Caporale:2015uva}:
\begin{equation}
\begin{split}
\label{Cn}
\mathcal{C}_n =& \frac{|\vec{q}_1||\vec{q}_2|\sqrt{m_1^2+\vec{q}_1^{\; 2}}\sqrt{m_2^{2}+\vec{q}_2^{\; 2}}}{W^2} e^{\Delta Y} \\ & \times \int_{- \infty}^{+ \infty} d\nu \left(\frac{W^2}{s_0}\right)^{\bar{\alpha}_s\left(\mu_R\right)\chi\left(n,\nu\right)+\bar{\alpha}_s^2\left(\mu_R\right)\left(\bar{\chi}\left(n,\nu \right)+\frac{\beta_0}{8N_c}\chi\left(n,\nu\right)\left(-\chi \left(n,\nu\right)+\frac{10}{3}+2\ln\frac{\mu_R^2}{\sqrt{s_1s_2}} \right)\right)} \\ & \times \alpha_s^4 \left(\mu_R\right) c_1\left(n,\nu,\vec{q}_1^{\; 2},z_1 \right)c_2\left(n,\nu,\vec{q}_2^{\; 2},z_2 \right) \left \{ 1+\bar{\alpha}_s \left(\mu_R\right) \left(\frac{\bar{c}_1^{(1)}}{c_1}+\frac{\bar{c}_2^{(1)}}{c_2}\right) \right.\\ & \left. + \bar{\alpha}_s \left(\mu_R\right) \frac{\beta_0}{2N_c} \left(\frac{5}{3}+\ln \frac{\mu_R^2}{s_1s_2}+f\left(\nu\right) \right) + \bar{\alpha}_s^2 \left(\mu_R\right) \ln \left(\frac{W^2}{s_0}\right) \frac{\beta_0}{4N_c} \chi \left(n, \nu \right) f\left(\nu\right) \right \} \;,
\end{split}
\end{equation} 
where $\chi \left(n,\nu\right)$ are the eigenvalues of the leading-order BFKL kernel defined in~\eqref{KLLA}, with $\psi = \Gamma'(x)/\Gamma(x)$, and 
\begin{equation}
\beta_0 = \frac{11}{3} N_c - \frac{2}{3} n_f
\end{equation} 
is the first coefficient of the QCD $\beta$-function, responsible for running-coupling effects. The expression of the function $f\left(\nu \right)$ reads
\begin{equation}
i\frac{d}{d\nu}\ln\frac{c_1}{c_2} = 2 \left[f(\nu)-\ln(\sqrt{s_1s_1'}) \right] \; ,
\end{equation}
with $s_i$, $i=1,2$ the hard scales in our two-tagged-quark process, which are chosen to be equal to $m_i^2+\vec{q}_i^{\; 2}$. The functions $c_1$ and $c_2$ result to be
\begin{equation}
\begin{split}
c_1 \left(n,\nu, \vec{q}_1^{\; 2},z_1 \right) = \frac{1}{e^{in\varphi_1} \alpha_s^2} \frac{d\Phi_{gg}^{\lbrace{Q\bar{Q}\rbrace}}\left(n,\nu,\vec{q}_1,z_1\right)}{d^2\vec{q}_1 \; dz_1} \,,
\end{split}
\end{equation}
\begin{equation}
\begin{split}
c_2 \left(n,\nu, \vec{q}_2^{\; 2},z_2 \right) = \frac{1}{e^{-in(\varphi_2+\pi)} \alpha_s^2} \left[\frac{d\Phi_{gg}^{\lbrace{Q\bar{Q}\rbrace}}\left(n,\nu,\vec{q}_2,z_2\right)}{d^2\vec{q}_2 \; dz_2}\right]^{*}\,.
\end{split}
\end{equation}
Moreover we have 
\begin{equation}
\frac{\bar{c}_1^{(1)}}{c_1}+\frac{\bar{c}_2^{(1)}}{c_2}= \chi \left( n,\nu \right) \ln \frac{s_0}{\sqrt{\left(m_1^2+\vec{q}_1^{\; 2}\right) \left(m_2^{2}+\vec{q}_2^{\; 2}\right)}} \;.
\end{equation}
The scale $s_0$ is arbitrary within NLA accuracy and in this calculation, we fixed it as $s_0 = \sqrt{s_1s'_1}$. It is relevant to observe that the Eq.~(\ref{Cn}) is referred to the general case when two heavy quarks of different masses $m_1,m_2$ are detected.
\subsection{From gluon to proton initiated process}
\label{HQ:pp_cross}
In the framework of the perturbative QCD the parton scattering process, characterizing by two colliding gluons can be investigated. Additionally, due to the color confinement property of QCD, it is clear that the proton collision has to be studied. In order to pursue it, the parton distribution functions (PDFs), describing the gluon distribution in the proton, are needed. These objects are non-perturbative and can be extracted by fitting observables to experimental data. 
It is possible to express the differential cross section for the proton-proton process as:
\begin{equation}
\label{dsigma_pp_conv}
d\sigma_{pp} = f_{g_1}(x_1,\mu_{F_1})f_{g_2}(x_2,\mu_{F_2}) d\sigma_{gg} dx_1 dx_2 \;,
\end{equation}
where $f_{g_i}$, $i=1,2$ represent just the gluon collinear parton distribution functions, while the variables $x_1$ and $x_2$ are the fractions of 4-momenta of the protons carried by the two gluons. Here $d\sigma_{gg}$ indicates the cross section for the gluons scattering given by Eq.~\eqref{crosfin}.
Therefore the final expression for our observable is
\begin{equation}
\label{crosfin}
\frac{d\sigma_{pp}}{d(\Delta Y) d\varphi_1 d\varphi_2}=\frac{1}{(2\pi)^2} \left[C_0+2 \sum_{n=1}^{\infty} \cos(n \varphi) C_n \right] \;,
\end{equation}
where
\begin{equation}
\begin{split}
C_n = & 
\int_{q_{1,{\text{min}}}}^{q_{1,{\text{max}}}} d|\vec{q}_1| 
\int_{q_{2,{\text{min}}}}^{q_{2,{\text{max}}}} d|\vec{q}_2| 
\int_{y_{1,{\text{min}}}}^{y_{1,{\text{max}}}} dy_1
\int_{y_{2,{\text{min}}}}^{y_{2,{\text{max}}}} dy_2 \;
\delta(y_1 - y_2 - \Delta Y) \\ & \hspace{-0.4cm}\times\int_{e^{-(y_{1,{\text{max}}}-y_1)}}^{1} dx_1 f_{g_1}(x_1,\mu_{F_1}) \int_{e^{-(y_{2,{\text{max}}}+y_2)}}^{1} dx_2 f_{g_2}(x_2,\mu_{F_2}) \; \mathcal{C}_n
\end{split}
\end{equation}
is the $n^{\text{th}}$ azimuthal coefficient integrated over the
$(\vec{q}_{1,2},y_{1,2})$ phase space and the rapidity separation between the
two tagged quarks is kept fixed to $\Delta Y$~\footnote{We note that, from here on, the rapidity interval $Y$ is now denoted by $\Delta Y$.}.
\subsection{The ``box'' $Q \bar Q$ cross section}
\label{HQ:QQbar}
We consider the lowest-order QCD cross section for the production of a heavy quark-antiquark pair in
proton-proton collisions. This process does not represent a background for the inclusive reaction of interest in this work when the two detected heavy quarks are of different flavors or, being of the same flavors, are both quark or both
antiquarks.
\begin{figure}[!h]
	\centering
	\includegraphics[width=0.7\textwidth]{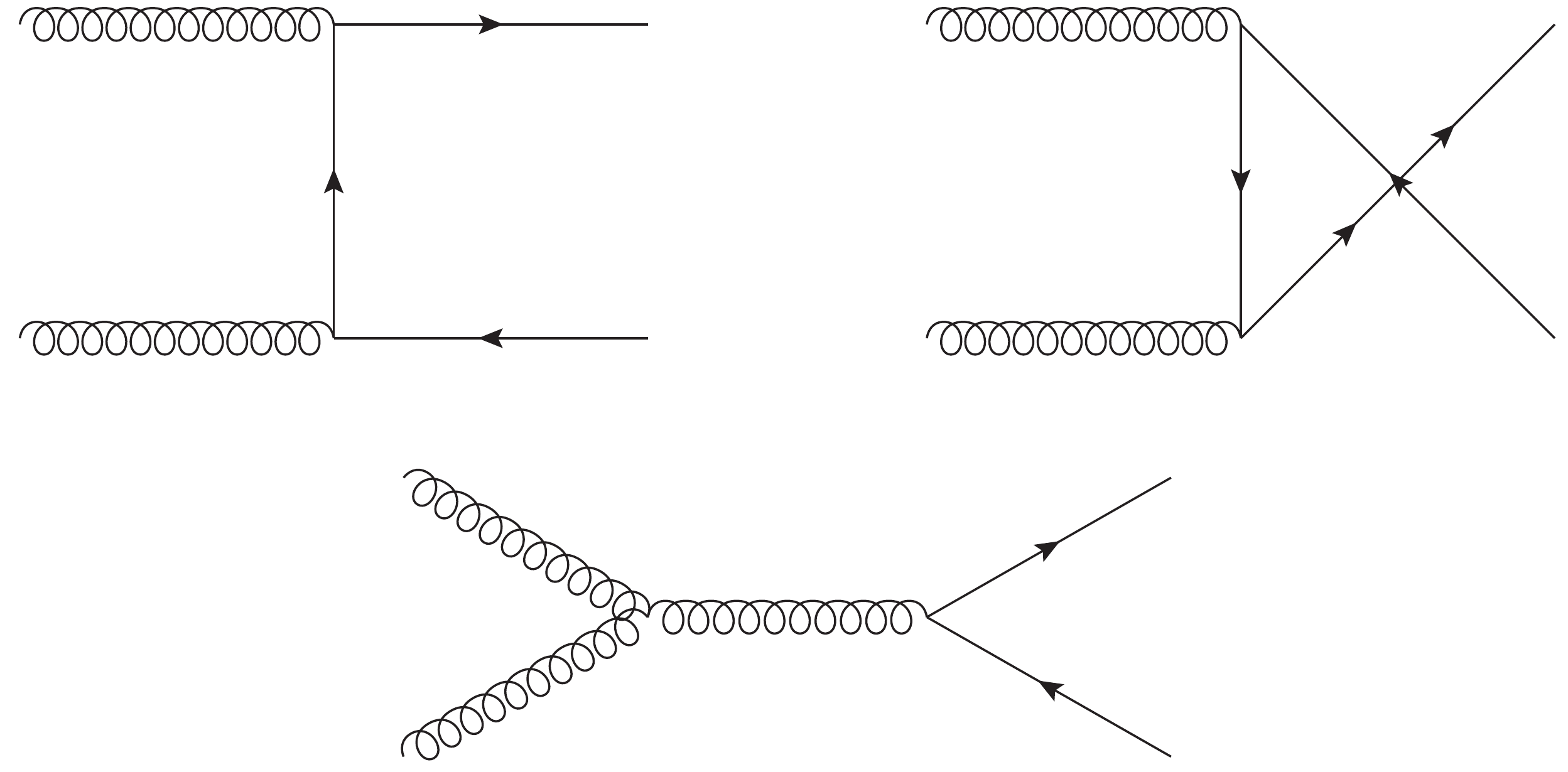}
	\caption{\emph{Feynman diagrams contributing at the lowest order to the $Q \bar{Q}$ hadroproduction.}}
	\label{Box}
\end{figure}
\FloatBarrier
The Feynman diagrams contributing to this process at the leading order
are shown in Fig.~\ref{Box}. The differential cross section is presented,
{\it e.g.}, in Ref.~\cite{Ahrens:2010zv} and in our notation takes the form
\begin{equation}
\begin{split}
\frac{d \sigma_{pp}}{d(\Delta Y)} = \; & \frac{ \pi \alpha_s^2}{s(N_c^2-1)}
\int_{{\text{min}}(q^2_{\text{min}},\frac{s}{4\cosh^2(\Delta Y/2)}-m^2)}^{{\text{max}}(q^2_{\text{max}},
	\frac{s}{4\cosh^2(\Delta Y/2)}-m^2)} \frac{d\vec q^{\: 2}}{M^2}
\int_{\frac{M^2}{s}}^{1}
\frac{dx}{x} f_{g_1}(x, \mu_{F_1}) f_{g_2}(M^2/xs, \mu_{F_2}) \\
& \times \left( C_F \frac{M^4}{t_1 u_1} - C_A \right)
\left[ \frac{t_1^2+u_1^2}{M^4} + 4 \frac{m^2}{M^2}
- 4 \frac{m^4}{t_1 u_1} \right] \; ,
\end{split}
\end{equation}
where
\begin{equation}
M^2 = 4 (m^2 + \vec q^{\: 2}) \cosh^2 (\Delta Y/2) \; ,
\end{equation}

\begin{equation}
t_1 = - \frac{M^2}{2} \left[ 1 - \tanh (\Delta Y/2) \right] \; , 
\end{equation}

\begin{equation}
u_1 = - \frac{M^2}{2} \left[ 1 + \tanh (\Delta Y/2) \right] \; .
\end{equation}
Here, $ C_F = (N_c^2-1)/2 N_c $, $C_A = N_c$, $s$ is the squared center-of-mass
energy of the proton-proton system and $m$ is the heavy-quark mass; $\mu_{F_1}$, $\mu_{F_2}$ are both set to $\sqrt{m_Q^2+\vec q^{\: 2}}$ and $\alpha_s$ is also calculated at this scale. The upper and lower limits in the integration over $\vec q^{\: 2}$ come from the constraint $M^2 \leq s$; in their expression, $q^2_{\text{min}}$ and $q^2_{\text{max}}$ represent the kinematic cuts on the heavy quark/antiquark transverse momentum. There is also a constraint on $\Delta Y$, coming from the requirement that $M^2|_{\vec q^{\: 2}=0} \leq s$, which is however always fulfilled for the values of $\Delta Y$ and $s$ considered in the numerical analysis presented below.

\subsection{Numerical tools and uncertainty estimation}
\label{HQ:num_est}
We propose results for the dependence on the rapidity
interval between the two tagged bottom quarks, $\Delta Y \equiv y_1 - y_2$, of
the $\varphi$-averaged cross section $C_0$ and of the azimuthal ratios $R_{nm}
\equiv C_{n}/C_{m}$ ratios. We fix the $m_{1,2}$ masses at the value $m_b = 4.18$ GeV/$c^2$~\cite{Tanabashi:2018oca}. 

With the idea of matching realistic kinematic configurations, typical of the
current and possible future LHC analyses, we integrate the quark transverse
momenta in the symmetric range 20 GeV $< q_{1,2} <$ 100 GeV, fixing the
center-of-mass energy to $\sqrt{s} = 14$ TeV and studying the behavior of our
observables in the rapidity range $1.5 < \Delta Y < 9$. Ranges of the transverse
momenta of the bottom-jets ($b$-jets) are typical of CMS
analyses~\cite{Chatrchyan:2012dk,Chatrchyan:2012jua}.

\begin{figure}[p]
	\centering
	\includegraphics[scale=0.5,clip]{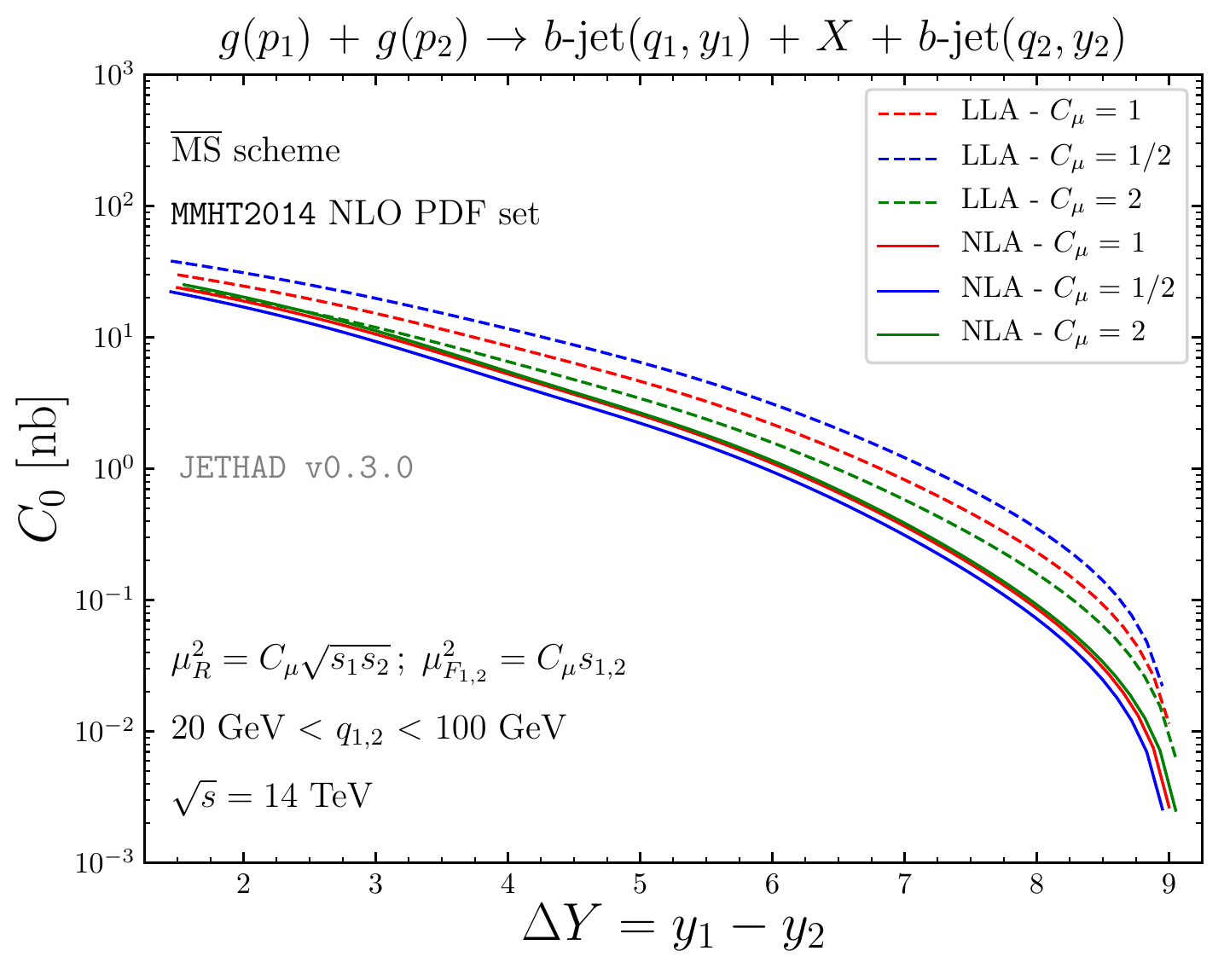}
	\includegraphics[scale=0.5,clip]{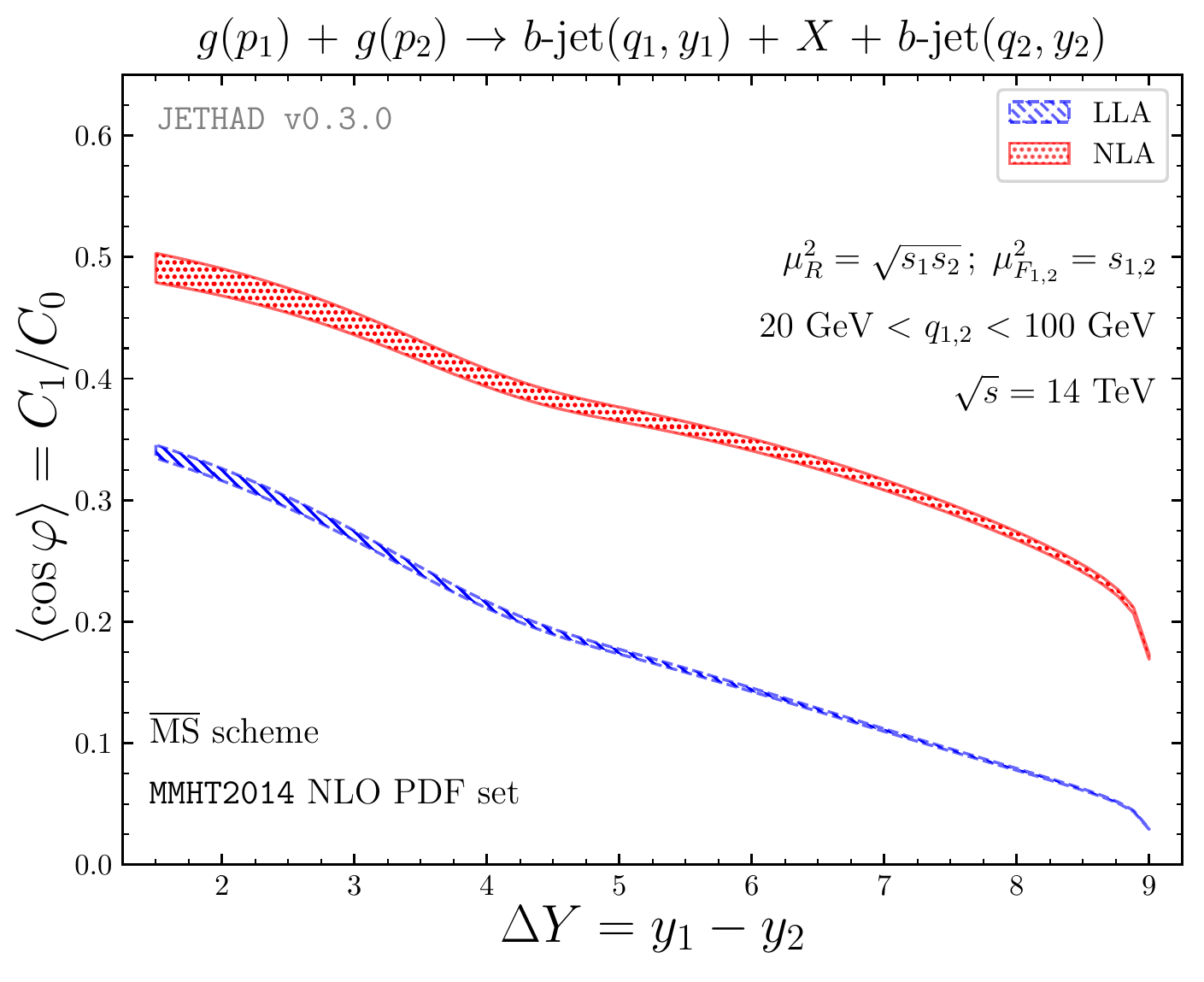}
	
	\includegraphics[scale=0.5,clip]{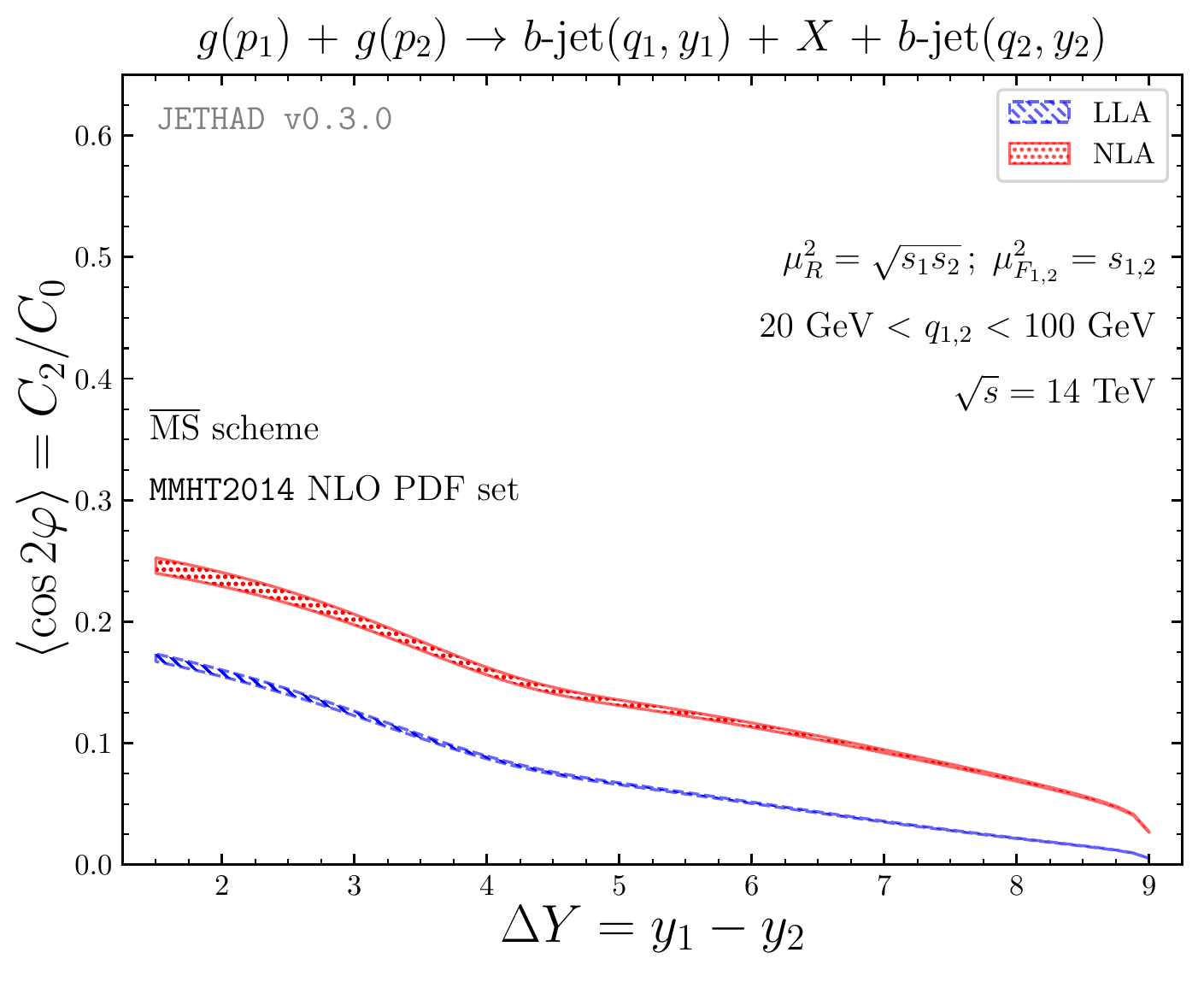}
	\includegraphics[scale=0.5,clip]{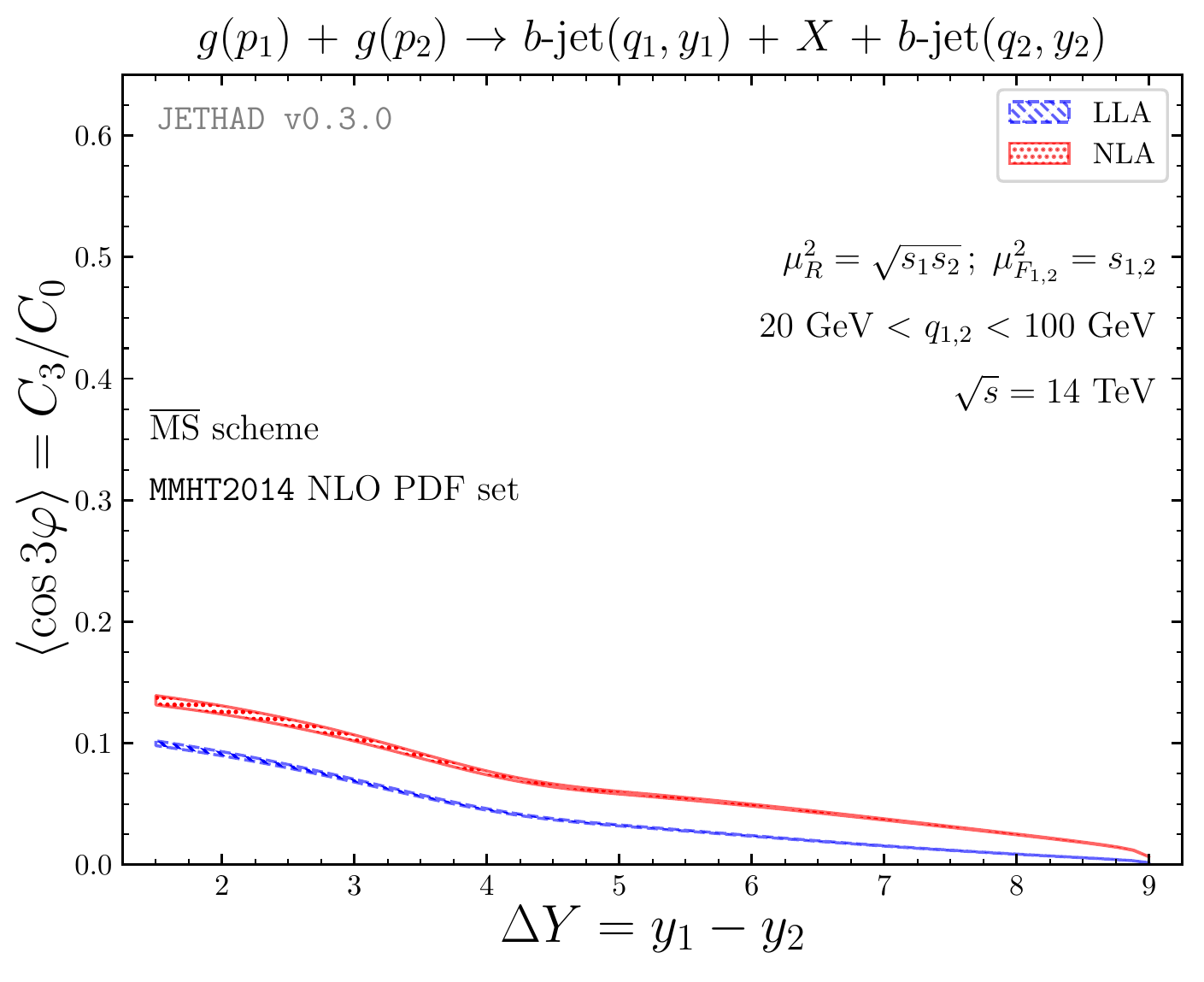}
	
	\includegraphics[scale=0.5,clip]{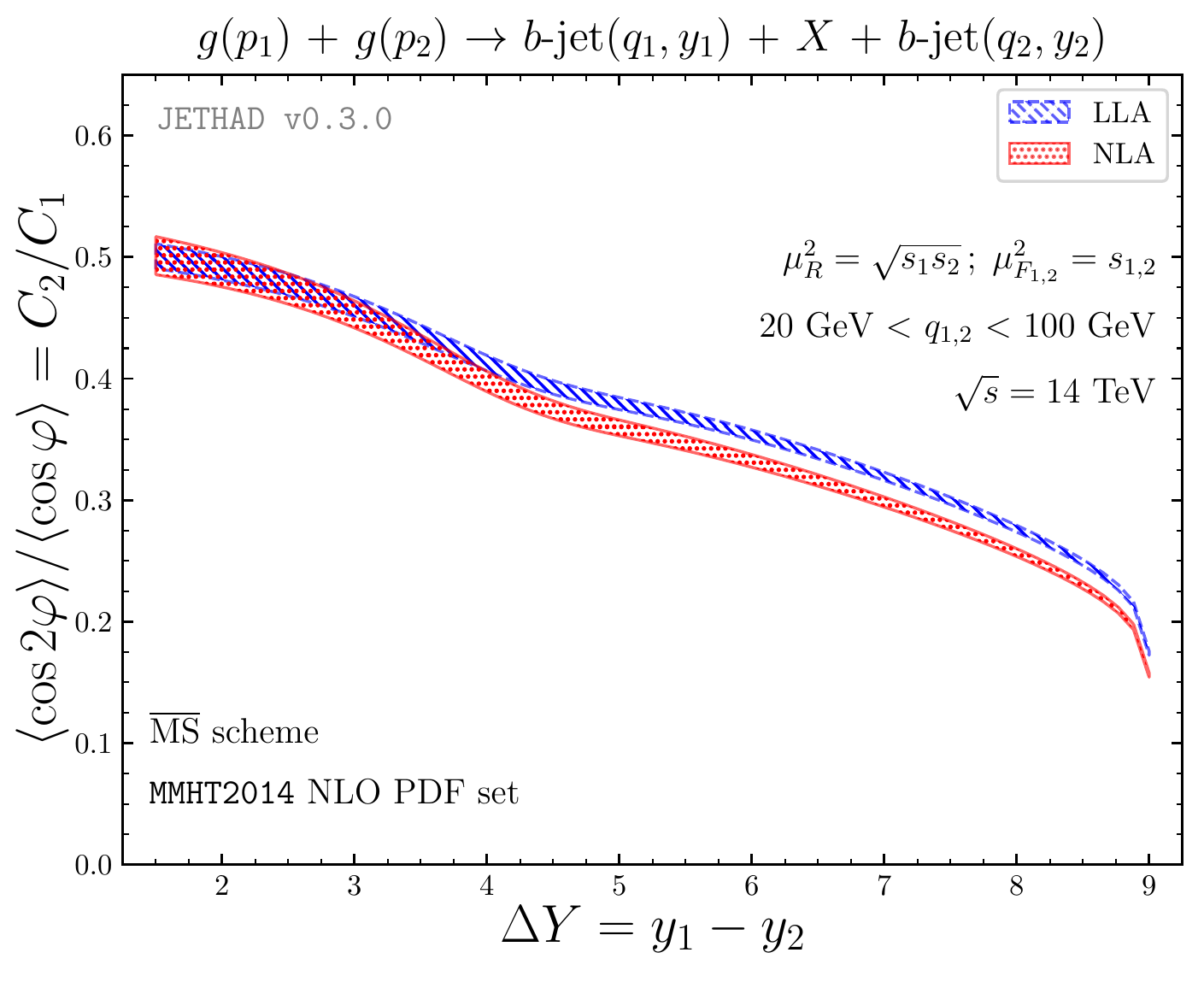}
	\includegraphics[scale=0.5,clip]{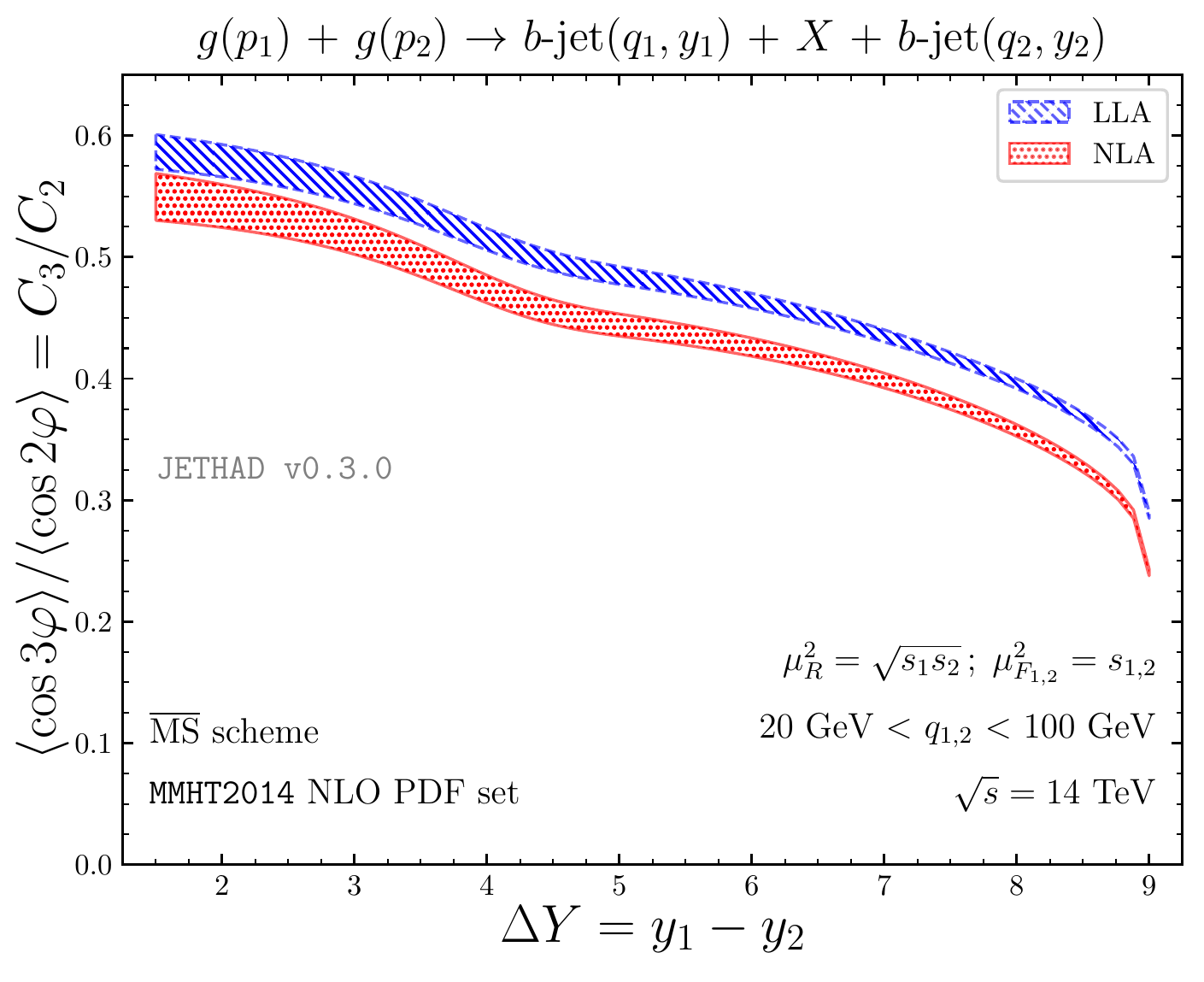}
	\caption{\emph{$\Delta Y$-dependence of $C_0$ ($b$-jet pair) for different values of $C_\mu = \mu_R^2/\sqrt{s_1s_2}\equiv \mu^2_{F_{1,2}}/s_{1,2}$ (data points have been slightly shifted along the horizontal axis for the sake of readability), with $s_{1,2} = m^2_{1,2} + \vec{q}_{1,2}^{\; 2}$ and of several ratios $R_{nm} \equiv C_{n}/C_{m}$, for 20 GeV $< q_{1,2} <$ 100 GeV and $\sqrt{s} = 14$ TeV.}}
	\label{fig:C0-Rnm}
\end{figure}

Pure LLA and NLA BFKL predictions for the $\varphi$-averaged cross section,
$C_0$, together with the leading-order $p p \rightarrow q \bar{q}$ cross
section, are presented in Table~\ref{tab:C0-scales}.
Results for $C_0$ and for several azimuthal-correlation ratios, $R_{nm}$, are
shown in Fig.~\ref{fig:C0-Rnm}.

\begin{figure}[t]
	\centering
	\includegraphics[scale=0.5,clip]{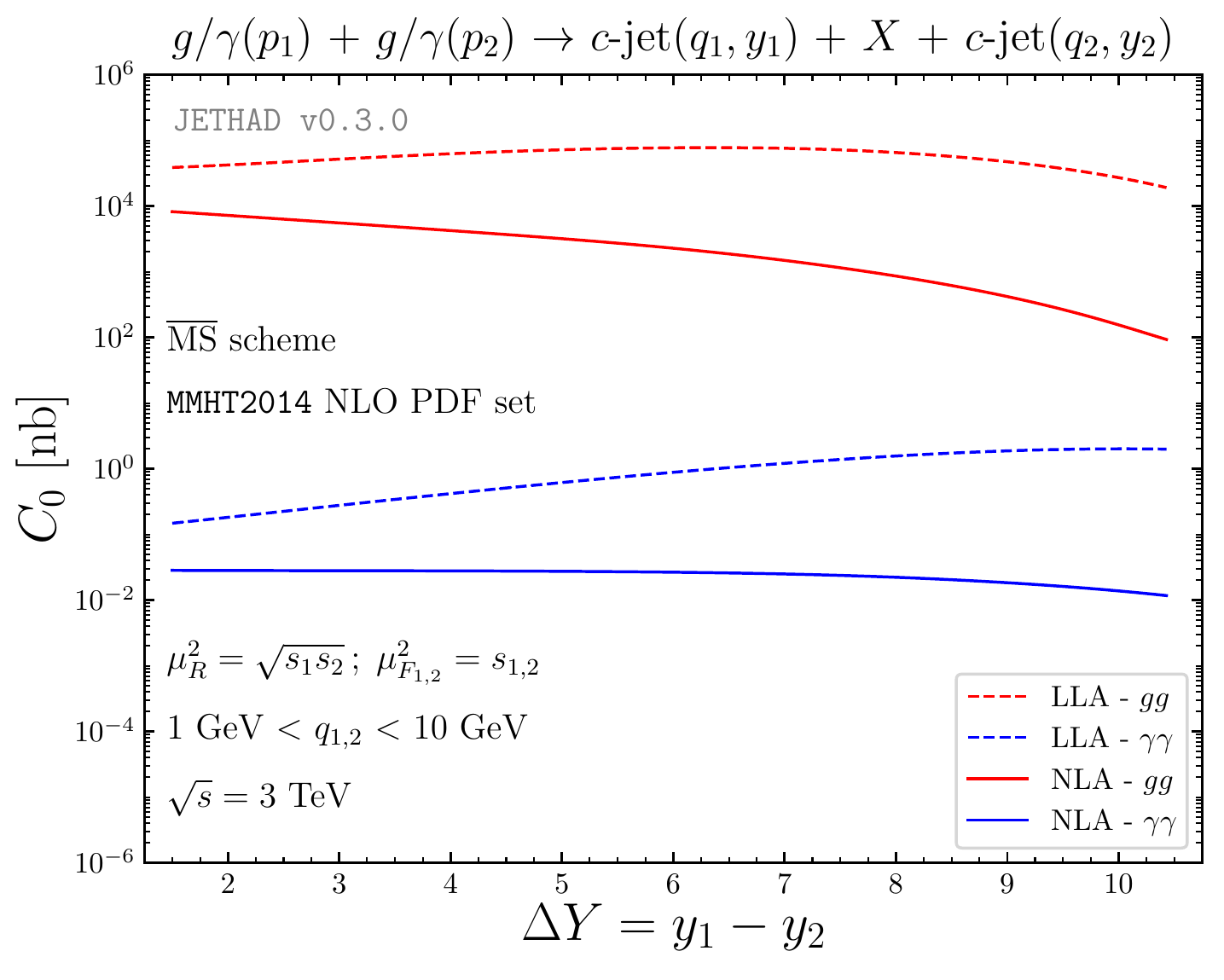}
	\caption{\emph{$\Delta Y$-dependence of $C_0$ ($c$-jet pair) for both the hadroproduction ($gg$) and the photoproduction ($\gamma \gamma$) mechanisms, for 1 GeV $< q_{1,2} <$ 10 GeV and $\sqrt{s} = 3$ TeV.}}
	\label{fig:C0_hvg}
\end{figure}

As a complementary study (see Fig.~\ref{fig:C0_hvg}), we present results for
$C_0$ in the case of charmed-jet ($c$-jet) pair emission ($m_{1,2} = m_c = 1.2$
GeV/$c^2$), comparing predictions for the hadroproduction with the ones
related to the photoproduction mechanism (see Ref.~\cite{Celiberto:2017nyx}
for details on the theoretical framework and for a recent phenomenological
analysis) in the kinematic configurations typical of
the future CLIC linear accelerator, namely 1 GeV $< q_{1,2} <$ 10 GeV, $\sqrt{s}
= 3$ TeV and $1.5 < \Delta Y < 10.5$. All calculations are done in the $\overline{\text{MS}}$ scheme.

\begin{table}[!h]
	\centering
	\caption{\emph{$\Delta Y$-dependence of the $\varphi$-averaged cross section
			$C_0$ [nb] for $\sqrt{s} = 14$ TeV.
			$C_\mu$ stands for $\mu_R^2/\sqrt{s_1 s_2} \equiv \mu_{F_{1,2}}^2/s_{1,2}$.}}
	\label{tab:C0-scales}
	\scriptsize
	\begin{tabular}{r|lllllll}
		\hline\noalign{\smallskip}
		$\Delta Y$ &
		$\tarr c \text{Box} \\ Q \bar{Q} \earr$ &
		$\tarr c \text{LLA} \\ C_\mu = 1/2 \earr$ & 
		$\tarr c \text{LLA} \\ C_\mu = 1 \earr$ &
		$\tarr c \text{LLA} \\ C_\mu = 2 \earr$ &
		$\tarr c \text{NLA} \\ C_\mu = 1/2 \earr$ &
		$\tarr c \text{NLA} \\ C_\mu = 1 \earr$ & 
		$\tarr c \text{NLA} \\ C_\mu = 2 \earr$ \\
		\noalign{\smallskip}\hline\noalign{\smallskip}
		1.5 & 33830.3 &  38.17(24) & 30.01(21) & 23.58(16) & 22.25(26) & 23.93(23) & 25.19(27) \\
		3.0 & 3368.86 &  18.118(98) & 13.191(71) & 9.838(61) & 7.245(74) & 8.205(76) & 8.172(82) \\
		4.5 & 124.333 &  6.996(33) & 4.715(23) & 3.276(16) &  2.209(20) & 2.411(17) & 2.422(19) \\       
		6.0 & 3.19206 &  1.976(10) & 1.2430(60) & 0.8044(38) & 0.4497(35) & 0.4968(35) & 0.4868(37) \\
		7.5 & 0.0610921 &  0.3317(16) & 0.19115(92) & 0.11509(57) & 0.05318(36) & 0.05785(39) & 0.05577(42) \\
		9.0 & 0.000681608 & 0.02215(10) & 0.011458(56) & 0.006340(30) & 0.002566(17) & 0.002668(16) & 0.002513(16) \\
		\noalign{\smallskip}\hline
	\end{tabular}
\end{table}
\FloatBarrier
The numerical analysis was performed using \textsc{Jethad}, a \textsc{Fortran} code recently developed, in order to support the study of inclusive semi-hard
processes. The numerical integrations are done through the interface with with the \textsc{Cern} program library~\cite{cernlib} and with the
\textsc{Cuba} integrators~\cite{Cuba:2005,ConcCuba:2015}, making extensive use
of the \textsc{Vegas}~\cite{VegasLepage:1978} and the
\textsc{WGauss}~\cite{cernlib} integrators. The numerical stability of our
predictions was crosschecked using an independent \textsc{Mathematica} code.
The gluon PDFs ($f_{g_{1,2}}$) were calculated via the \textsc{MMHT2014} NLO PDF
set~\cite{Harland-Lang:2014zoa} as implemented in the Les Houches
Accord PDF Interface (LHAPDF) 6.2.1~\cite{Buckley:2014ana}, while a two-loop
running coupling setup with $\alpha_s\left(M_Z\right)=0.11707$ with
dynamic-flavor thresholds was chosen.

The main source of uncertainty, coming from the numerical
six-dimensional integration over the variables $|\vec q_1|$, $|\vec q_2|$,
$y_1$, $\nu$, $x_1$, and $x_2$, was directly estimated by \textsc{Vegas}.
Other sources of uncertainties, related with the upper cutoff in the $\nu$-
and the $\Delta$-integration in Eq.~(\ref{Cn}) and Eq.~(\ref{I4}),
respectively, are negligible with respect to the first one. Thus, the error
estimates of our predictions are just those given by \textsc{Vegas}. In order
to quantify the uncertainty related to the renormalization scale ($\mu_R$) and
the factorization one ($\mu_{F_{1,2}}$), we simultaneously vary the square of
both of them around their ``natural'' values, $\sqrt{s_1 s_2}$ and $s_{1,2}$
respectively, in the range 1/2 to two. The parameter $C_{\mu}$ entering
Table~\ref{tab:C0-scales} gives the ratio $C_{\mu} = \mu_R^2/\sqrt{s_1 s_2}
\equiv \mu_{F_{1,2}}^2/s_{1,2}$.

\subsection{Results and discussion}
\label{HQ:analysis}
Results for the $\varphi$-averaged cross section, $C_0$, in
the $b$-jet pair production case (shown in the Table~\ref{tab:C0-scales} and in the left upper panel of Fig.~\ref{fig:C0-Rnm}) evidently manifest the usual BFKL dynamics. On one side, although the high-energy resummation predicts
a growth with energy of the partonic-subprocess cross section, its convolution
with parent-gluon PDFs (Eq.~(\ref{dsigma_pp_conv})) leads to
a falloff with $\Delta Y$ of both LLA and NLA predictions. On the other side,
next-to-leading corrections to the BFKL kernel become more and more negative
when the rapidity distance grows, thus making NLA results steadily lower than
pure LLA ones. In Table~\ref{tab:C0-scales} data describe the cross section $C_0$ is smaller than the reference ``box'' cross section for small $\Delta Y$; however, at larger rapidity differences, the BFKL mechanism with the gluonic
exchange in the $t$-channel starts to dominate. We remark that for
our two heavy-quark (or two heavy-antiquark) tagged process, the ``box''
mechanism is not a background.

The othere panels of Fig.~\ref{fig:C0-Rnm} illustrate azimuthal correlations which are always smaller than one and decrease when $\Delta Y$ grows (LLA results are always more decorrelated than NLA ones). This is an expected consequence of the larger emission of undetected partons. The cause for this narrowness, with respect to
other, recently investigated reactions, such as Mueller--Navelet jet, dihadron
or hadron-jet correlations (see, {\it e.g.},
Refs.~\cite{Caporale:2012ih,Celiberto:2017ptm,Bolognino:2018oth}), is
straightforward.
Since the two detected (anti)quarks stem from distinct vertices (each of them
together the respective antiparticle), their transverse momenta are
kinematically not constrained at all, even at leading order.

Fig.~\ref{fig:C0_hvg} presents the analysis for the $c$-jet pair production which shows that, at fixed center-of-mass energy and transverse-momentum range, predictions for $C_0$ in the hadroproduction channel
($g g$) are several orders of magnitude higher than the corresponding ones in
the photoproduction case ($\gamma \gamma$). This result is due to two competing effects. On one hand, from a ``rough'' comparison between the
($g g$) impact factor (Eq.~(\ref{eq:imp.fac2})) and the ($\gamma \gamma$) one
(Eq.~(2) of Ref.~\cite{Celiberto:2017nyx} and footnote of
Ref.~\cite{Bolognino:2019ouc}), it emerges that the two analytic
structures are quite similar, the main difference being the fact that, since
the photon cannot interact directly with the Reggeized gluon, some terms
present in the first case are missing in the second one. In both of
the two impact factors there are constants that can be factorized out in the
final form of the cross section. Since two heavy-quark impact factors enter the
expression of cross sections, one has an overall factor
\begin{equation}
	\label{k_had}
	\kappa_{(g g)} = \frac{\alpha_s^4 \, (N_c^2 - 1)}{(2 \pi N_c)^2}
\end{equation}
in the ($g g$) case and an overall factor
\begin{equation}
	\label{k_gam}
	\kappa_{(\gamma \gamma)} = \frac{\alpha_{\text{em}}^2 \, \alpha_s^2 \, e_c^4 (N_c^2 - 1)}{\pi^2}
\end{equation}
in the ($\gamma \gamma$) case, with $\alpha_{\text{em}}$ the QED coupling and $e_c$
the electric charge of the charm quark in units of the positron charge. The
ratio between the two factors is
\begin{equation}
	\label{k_had-gam}
	\kappa_{\rfrac{(g g)}{(\gamma \gamma)}} \equiv \frac{\kappa_{(g g)}}{\kappa_{(\gamma \gamma)}} \simeq 3 \div 4 \times 10^3 \;,
\end{equation}
which would explain the enhancement of the hadroproduction with respect to the
photoproduction. However, on the other hand, also the effect of the parent-particle distributions should be considered: gluon PDF~\cite{Harland-Lang:2014zoa} or equivalent-photon approximation (EPA)
photon flux (see Eq.~(8) of Ref.~\cite{Celiberto:2017nyx}). It is possible to
demostrate that the gluon PDF dominates over the photon flux in the moderate-$x$
region, while the second one prevails in the $x \to 0^+$ and $x \to 1^-$ limits.
In the realistic kinematic ranges we have considered here and in
Refs.~\cite{Celiberto:2017nyx,Bolognino:2019ouc} the relevant $x$-region
turns to be just the intermediate one, thus leading to an enhancement
of the hadroproduction mechanism with respect to the photoproduction one
(see Fig.~\ref{fig:C0_hvg}) even larger than what suggested by
the ratio $\kappa_{\rfrac{(g g)}{(\gamma \gamma)}}$.  

\subsection{Summary}
\label{HQ:summary_outlook}
The study of the inclusive hadroproduction of two heavy quarks separated by
a large rapidity interval has been proposed as a new reaction for the investigation of BFKL mechanism. We have performed an all-order resummation of the leading energy logarithms and a resummation of the next-to-leading ones entering the BFKL
Green's function. In this approximation, the cross section can be expressed
as the convolution of the partonic cross section for the collision of two
gluons producing the two heavy quarks with the respective gluon PDFs.
The cross section for this process is calculated and it is summed over the relative azimuthal angle of the two tagged quarks; we presented results for the
azimuthal angle correlations, whose behavior reveals the characteristic feature of the onset of the BFKL dynamics.
Finally, a comparison between the photoproduction and the hadroproduction
mechanism has been carried out. The process under consideration contributes to enrich the selection of semi-hard reactions that can be used as
probes of the QCD in the high-energy limit, and in particular of the BFKL
resummation mechanism, in the kinematic ranges of the LHC and of future
hadronic colliders. 

Starting from this topic it could be challenging to develop and plan the following perspectives. The first one consists in the calculation of the NLO correction to the forward heavy-quark pair impact factor, which would allow for a full NLA BFKL treatment of the process here investigated. The second one is to include into the theoretical analysis the quark fragmentation needed to match, from the theoretical side, the experimental tagging procedure of heavy-quark mesons.
Since the photoproduction channel has already been considered
(Refs.~\cite{Celiberto:2017nyx,Bolognino:2019ouc}), a process of
photo/hadro-production (when the first (anti)quark is emitted by a (quasi-)real
photon, while the second one stems from a gluon), hybrid with respect to the
previous ones, can also be examined. Moreover, it would be interesting to study semi-hard channels featuring the emission of a
single quark. In this regard, an idea is to study the single forward heavy-quark
production, convolving the corresponding impact factor with the unintegrated
gluon density (UGD) in the proton. This latter will be the key object of the next Section.
\chapter{Single forward emissions}
\label{chap3}
A better understanding of the dynamics of strong
interactions at the LHC strongly relies on getting a more and more precise
knowledge of  the structure of the proton.  In general, the  latter 
is encoded in different types of partonic distribution functions that enter
the factorization formalism for the description of the hard processes.
{\em Collinear factorization} is the most developed approach  to
calculate cross sections of inclusive reactions as a power expansion over the
hard-scale  parameter. A prominent example here is the deep inelastic
scattering (DIS) of an electron off a proton.
Its cross section, at the leading
order in the power expansion over the virtuality $Q^2$ of the exchanged photon
$\gamma^*$, is factorized as a convolution of a hard cross sections (calculable
in perturbation theory) with parton distribution functions (PDFs) of quarks
and gluons, $q_i(\zeta,\mu_F)$ and $g(\zeta,\mu_F)$, that depend on the
longitudinal momentum fraction of the proton carried by the parton, $\zeta$,
and on the factorization scale $\mu_F$, and obey DGLAP evolution
equations~\cite{DGLAP}. 
At the leading order (LO) of perturbation theory the variable $\zeta$ coincides
with the Bjorken variable $x=Q^2/(W^2+Q^2)$, where $W^2$ is the squared
center-of-mass energy of the $\gamma^* p$ system.  The collinear factorization
scheme can be also applied to the amplitudes of hard exclusive processes, where
the nonperturbative part is factorized in generalized parton
distributions~\cite{Collins:1996fb,Radyushkin:1997ki}.
At high energy, $W\gg Q\gg \Lambda_{\text{QCD}}$, the application of collinear
factorization is limited because the perturbative expansion includes in this
kinematics large logarithms of the energy that have to be resummed. Such a
resummation is incorporated in the {\em $\kappa$-factorization}~\footnote{We use the expression ``$\kappa$-factorization'' to mean what elsewhere is known also as ``$k_T$-factorization''.}. The scattering
amplitudes are basically written as a convolution of the unintegrated gluon
distribution (UGD) in the proton with the impact factor (IF) that depends on
the  considered  process.
In the DIS case the $\gamma^* \to \gamma^*$   IF is calculated fully
in perturbation theory. The UGD is a nonperturbative quantity, function of
$x$ and $\kappa$, where the latter represents the gluon momentum transverse
to the direction of the proton and is the Fourier-conjugate variable of the
transverse separation $r_d$ of the color dipole into which the virtual
photon splits. Therefore small values of $r_d$ mean large values of $\kappa$
and {\it vice versa}. The UGD, in its original definition, obeys the
BFKL~\cite{Fadin, elf2, BL} evolution equation in the $x$ variable.
Differently from collinear PDFs, the UGD is not well known and several types of 
models for it do exist, which lead to very different shapes in the
$(x,\kappa)$-plane (see, for instance, Refs.~\cite{small_x_WG,Angeles-Martinez:2015sea}).

The idea is to present our arguments that HERA data on
polarization observables in vector meson (VM) electroproduction can be used to constrain the $\kappa$-dependence of the UGD in the HERA energy range.
We will focus our attention on the polarized leptoproduction of $\rho$ (see Section~\ref{Sec_rho}) and $\phi$ mesons (see Section~\ref{Sec_phi}), {\it i.e.} the longitudinal VM production from longitudinally polarized virtual photons and the transverse VM production from transversely polarized virtual photons. Firstly, in the case of $\rho$-meson production we will concentrate on the ratio of the two dominant amplitudes~\cite{Bolognino:2018}; after that we will deal with polarized cross sections and their ratios. Secondly, in order to provide a more complete overview concerning the UGD effects on the diffractive processes variables, we will extend our investigation also to the $\phi$-meson leptoproduction~\cite{Bolognino:2019pba}. The discussion of this latter will proceed illustrating not only the effects of UGDs but also those ones due to the strange-quark mass.

The H1 and ZEUS collaborations performed a complete
analysis~\cite{Aaron:2009xp,Chekanov:2007zr} of the spin density matrix elements
describing the hard exclusive light vector meson production, which can be
expressed in terms of helicity amplitudes for this process. 
The HERA data show distinctive features for both longitudinal and transverse VM
production: the same $W$- and $t$-dependence, that are  different 
from those seen in soft exclusive reactions (like VM photoproduction).
This  supports the
idea that the same physical mechanism, involving the scattering of a small
transverse size color dipole on the proton target, is at work for both helicity
amplitudes. Contrary to DIS case, the IFs for $\gamma^*\to \text{VM}$ transitions
are not fully perturbative,  since  they include information about the VM
bound state. However, assuming the small size dipole dominance, one can
calculate the $\gamma^*\to \text{VM}$ IFs unambiguously in collinear
factorization, as a convolution of the amplitudes of perturbative subprocesses
with VM distribution amplitudes (DAs) of twist-2 and
twist-3~\cite{Anikin:2009bf}. Such approach to helicity amplitudes of VM
electroproduction was used earlier in Ref.~\cite{Anikin:2011sa}, where a rather
simple model for UGD was adopted.      

We will concentrate on the $\kappa$-factorization method. The dipole
approach is based on similar physical ideas, but formulated not in $\kappa$-but in the transverse coordinate space; this scheme is especially suitable to account for nonlinear evolution and gluon saturation effects.  Interesting developments are the results of the papers~\cite{Besse:2012ia,Besse:2013muy}, where the helicity amplitudes of VM production were considered in the dipole approach.

The discussion is organized as follows: in Section~\ref{Sec_Lepto} we describe the type of process involved; in Section~\ref{Sec_kt_fact} we introduce the $\kappa$-factorization where the UGD arises; Section~\ref{Sec_obs} is devoted to the helicity amplitude and the cross section and their formalism. We provide a sketch of a few models used for the UGD in Section~\ref{Sec_UGD}. Finally, Sections~\ref{Sec_rho} and \ref{Sec_phi} are dedicated to $\rho$ and $\phi$ electroproduction, respectively, discussing the sources of theoretical uncertainties and presenting numerical results, compared with HERA data. In particular, the study on $\phi$-meson production is focused just on the polarized cross sections and on the effects of the strange-quark mass.

\section{Leptoproduction of light vector mesons at HERA}
\label{Sec_Lepto}
A diffractive process is characterized by the exchange of vacuum quantum numbers between the colliding particles, which implies final states well separated in rapidity.

Exclusive leptoproduction of mesons is a particularly important testing ground to study diffraction processes. In the high-energy electron-proton collider HERA~\cite{Aaron:2009xp,Adloff:2000nx,Chekanov:2005cqa}, the vector meson production is described by the process
\begin{figure*}[htbp]	
	\centering
	\includegraphics[width=0.600\textwidth]{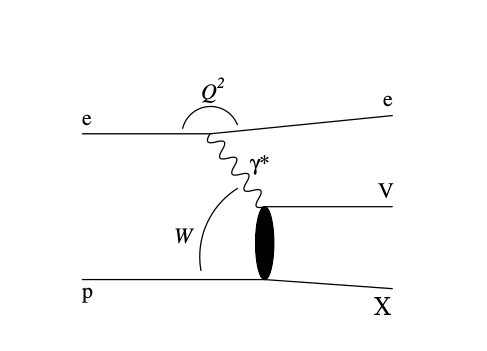}	
	\caption{\small \emph{Diffractive vector meson leptoproduction.}}
	\label{ep_collision}
\end{figure*}
\begin{equation}
e^- + p \longrightarrow e^- + V + X\,,
\end{equation}
illustrated in Fig.~\ref{ep_collision}: the highly-virtual exchanged photon, converts into a diffractively scattered vector meson $V$ (\emph{i.e.} $\rho$, $\phi$, $\omega$, J/$\Psi$, \dots) of mass $m_V$, while the incoming proton is scattered into a system $X$ of mass $m_X$, which can be a proton or a diffractively excited system.

In vector mesons production, a hard scale is provided by the photon virtuality $Q^2 \gg \Lambda_\text{QCD}^2$ (making pQCD applicable), while the reaction energy is expressed by the center-of-mass energy $W$ of the virtual photon-proton system. A necessary condition so that rapidity gap occurs is $W^2 = s \gg Q^2$, or in terms of Bjorken $x$
\begin{equation}
x = \frac{Q^2}{s} \ll 1\,,
\end{equation}
The experimental data were taken with the H1 detector in the period from 1996 to 2000 and they correspond to an integrated luminosity of 51 pb$^{-1}$. The analysis - based on about 10500 $\rho$ and 2000 $\phi$ events - have provided measurements of the production cross sections and of the spin density matrix elements, which give access also to helicity amplitudes, in terms of kinematic variables.

\begin{figure}[!h]	
	\centering
	\includegraphics[width=1.000\textwidth]{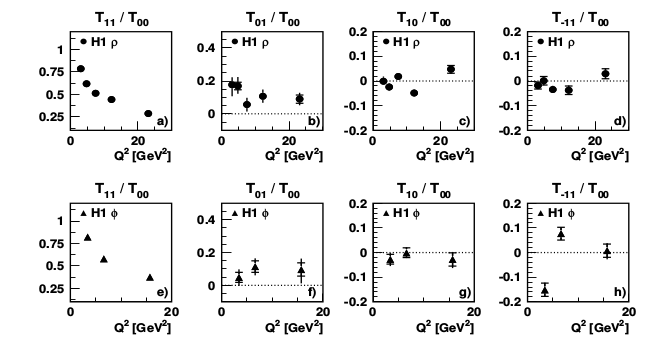}
	\caption{\small\emph{Ratios of helicity amplitudes as functions of $Q^2$ for $\rho$ (upper panel) and for $\phi$ (lower panel) production. Plots from Ref.~\cite{Aaron:2009xp}.}}
	\label{HERADATA}
\end{figure}

We will focus our attention on the $\rho$ and $\phi$ mesons production, where $\rho$ and $\phi$ are light mesons with mass $m_\rho = 776$ MeV and $m_\phi =1.020$ MeV, respectively.
The $e$--$p$ collisions at HERA, providing, through H1~\cite{Aaron:2009xp,Adloff:2000nx} and ZEUS~\cite{Chekanov:2005cqa} collaborations, useful and complete analysis, describe the hard exclusive production of the vector mesons under consideration in the process
\begin{equation}
\label{processVM}
\gamma^*(\lambda_\gamma)p\rightarrow V (\lambda_V)p\,,
\end{equation}
which can be expressed in terms of helicity amplitudes $T_{\lambda_V \lambda_\gamma}$, where $\lambda_V$ and $\lambda_\gamma$ are the helicities of the vector meson and virtual photon, respectively. They can assume values equal to $0,\pm 1$.\
A hierarchy for these observables has been obtained according to
\beq
\label{hierarchy}
T_{00} \gg T_{11} \gg T_{10} \gg T_{01} \gg T_{-11},
\eq
and it has been confirmed by the experimentalists at HERA, as illustrated in Fig.~\ref{HERADATA} for the ratios $T_{11}/T_{00}$ and $T_{01}/T_{00}$, in which the two left panels show the dominance of the amplitude $T_{00}$ respect to the amplitudes $T_{11}$ and $T_{01}$. The $Q^2$ dependence of the four amplitude ratios is here presented: the amplitude ratio $T_{11}/T_{00}$ strongly decreases with $Q^2$, while no significant $Q^2$ dependence is observed for the amplitude ratios $T_{10}/T_{00}$ and $T_{-11}/T_{00}$. Morevorer, in this context, for the first time, a $Q^2$ dependence for the amplitude ratio $T_{01}/T_{00}$ is also observed.
\FloatBarrier
The H1 and ZEUS collaborations have analysed data employing in different ranges of $Q^2$ and $W$.
In this treatment we will refer to H1 ranges for $\rho$ production: 
\begin{equation}
\label{H1range}
\begin{split}
2.5\,\text{\rm GeV$^2$} < Q^2 <60 \,\text{\rm GeV$^2$}\,,\\
35\, \text{GeV} < W < 180\,\text{GeV}\,,
\end{split}
\end{equation}
and to ZEUS ranges for $\phi$ one:
\begin{equation}
\begin{split}
	2\,\text{\rm GeV$^2$} < Q^2 <160 \,\text{\rm GeV$^2$}\,,\\
	32\, \text{GeV} < W < 180\,\text{GeV}\,.
\end{split}
\end{equation}
HERA data, giving information of the process, as $Q^2$ scaling, $t$ and $W$ dependence of the cross sections, determine relevant features for which the mechanism of the diffractive process in \eqref{processVM} is ruled by the scattering of a small transverse-size ($\sim 1/Q$) colorless quark-antiquark dipole (see Fig.~\ref{photoabs} below) on the proton target. This justifies the use of pQCD.
\section{Towards the $\kappa$-factorization}
\label{Sec_kt_fact}
\begin{figure}[htbp]	
	\centering
	\includegraphics[width=0.65\textwidth]{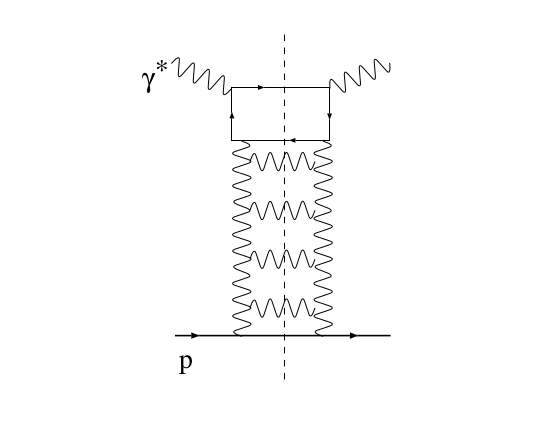}	
	\caption{\small\emph{The pQCD modeling of DIS as multiproduction of parton final states in the case of four produced gluons. For semplicity, we are showing here only one of the possible diagrams, with the two vertical gluon attached to the same fermionic line.}}
	\label{DDIS}
\end{figure}

\begin{figure}[htbp]		
	\centering
	\includegraphics[width=0.75\textwidth]{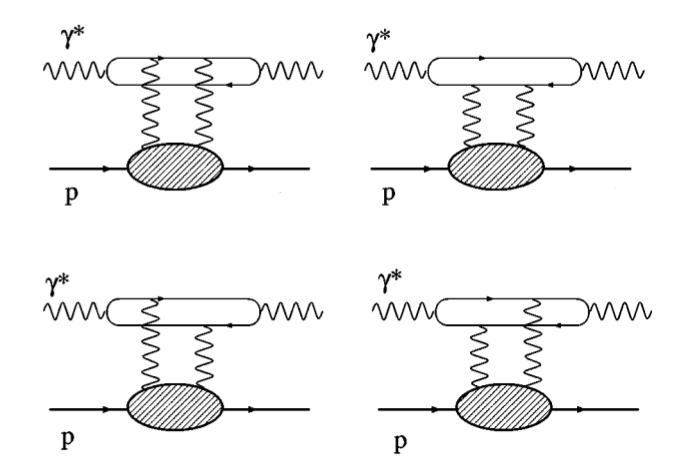}	
	\caption{\small\emph{The $\kappa$-factorization representation at small $x$. We note that the top-right diagram is the same as the one in Fig.~\ref{DDIS}.}}
	\label{photoabs}
\end{figure}

In order to get a better understanding of the $\kappa$-factorization, we can refer to a model of virtual photoabsorption in QCD. The quantity measured is the total cross section of the inclusive process $\gamma^*p \longrightarrow X$, summed over all hadronic final states $X$. Through the optical theorem, we can express the cross section as the imaginary part of the amplitude of the forward scattering $\gamma^* p \to \gamma^* p$, 
\beq
\label{optical}
\begin{split}
	\sigma_\text{tot}(\gamma^* p) & = \text{Im}\, A_\text{el}(\gamma^*p \to \gamma^*p)|_{(t=0)}\\
	& = \sum_X \int |A(\gamma^* p \to X)|^2d\tau_X\,,
\end{split}
\eq
where $A(\gamma^* p \to X)$ is the amplitude for the production of the final state $X$, whose square modulus is then integrated over the phace space and summed over all possible hadronic final states $X$.\\
In pQCD, the last line of Eq.~\eqref{optical} for sufficiently low values of $x$ can be schematized in terms of the production of a quark and antiquark pair plus any number of gluons over which there is a summation, as illustrated in Fig.~\ref{DDIS}.
The integration over the entire phase space of hadronic states $X$, in Eq.~\eqref{optical}, is replaced in the pQCD calculation by integration over the whole phase space of QCD partons,
\beq
\sum_X\int \left|M_X\right|^2 d\tau_X \Longrightarrow \sum_n \left|M_n\right|^2 \prod_i \int_0^1 \frac{dx_i}{x_i}\, d^2\,\overset{\rightarrow}{\kappa}_i\,,
\eq
where $M_X = A(\gamma^* p \to X)$, while $M_n$ is the amplitude referred to $n$ partonic states. To be specific, in our example, $n$ will include the quark-antiquark pair and $n-2$ gluons.
The integration over the transverse momenta is over the whole region
\beq
0\le \vec{\kappa}^2_i \le \frac{1}{4}W^2 = \frac{Q^2(1 - x)}{4x}\,.
\eq
For $Q^2$ large enough to guarantee $\ln Q^2 \gg \ln \frac{1}{x}$, the DGLAP resummation, which is the sum of the leading powers of $[\alpha_s \ln Q^2]^n$ generated by multigluon emission, predicts that the dominant contribution to the multiparton production cross sections comes from a restricted part of the phase space, given by
\[
1 \ge x_1 \ge x_2 \dots x_{n-1} \ge x_n \ge x \, ,\nonumber
\]
\beq
\label{k_ordered}
0 \leq \vec{\kappa}_{1}^{2} \ll \vec{\kappa}_{2}^{2}\dots \ll\vec{\kappa}_{n-1}^{2}
\ll \vec{k}^{2} \ll  Q^2\,.
\eq
At HERA, where the $Q^2$ values are not so large at very small $x$ and it may be more appropriate to consider $\ln Q^2 \ll \ln \frac{1}{x}$, however, this limitation of the transverse phase space is much restrictive: the ordered transverse momenta in~\eqref{k_ordered} does not provide the dominant contribution to the cross section. If one resums over all virtualities, a relevant part of the phase space is thrown away. The reason lies in the fact that DGLAP evolution does not include all the leading terms in this limit: it neglects those terms containing the leading power of $\log \frac{1}{x}$ but which are not accompanied by the leading power of $\ln Q^2$.\\ The resummation of terms proportional to the leading $\ln\frac{1}{x}$ (LL$\frac{1}{x}$), with full $Q^2$ dependence, is ruled the BFKL equation (see Section~\ref{Chap:BFKL}) and according to this approximation, the dominant contribute to photoabsorption comes from multigluon final states as in Fig.~\ref{DDIS}, but in a different kinematic region. In this new domain, \emph{i.e.} at finite $Q^2$ and $x \to 0$, such that $\ln \frac{1}{x} \gg \ln Q^2$, the integration is taken over the full phase space of transverse momenta of the gluons, not just the strongly-ordered part in \eqref{k_ordered}, but instead over regions such that
\beq
x_1 \gg x_2 \gg x_3 \gg \dots x_n = x,
\eq
\[
k^2_1 \simeq Q^2.
\]
In this regime, the photon production can be viewed as the result of the interaction between the $q\bar{q}$ fluctuation of the virtual photon and two hard gluons. This process is described by four diagrams as illustrated in Fig.~\ref{photoabs}, with all possible two-gluons attachments to $q\bar{q}$ pair taken into account. All of them are equally important and needed to preserve gauge invariance and color transparency. 
The upper parts of the diagrams in Fig.~\ref{photoabs}, or equivalently in Fig.~\ref{DDIS}, describe all possible contributions to the transition $\gamma^* \to \gamma^*$ and are given by the integration over the $q\bar{q}$ phace space of $|A(\gamma^* g \to q \bar{q})|^2$. This quantity defines the impact factor, $\Phi_{\gamma^* \to \gamma^*}$.\\
The remaining piece in our description, denoted by the lower shaded blobs in Fig.~\ref{photoabs} is the \emph{unintegrated gluon distribution} ${\cal F}(x,\kappa^2)$, which, by comparison with Fig.~\ref{DDIS}, involves the gluon ``ladder" exchanged in photon-proton interaction, in Fig.~\ref{DDIS} combined with the proton line~\footnote{As a matter of fact, also for the proton we should introduce an impact factor, which, however, being fully non-perturbative, can only be modeled. This is not relevant here since proton impact factor is included in the definition of unintegrated gluon distribution.}. ${\cal F}(x,\kappa^2)$ is not computable in pQCD and describes the probability distribution to find a gluon with a longitudinal momentum fraction $x$ and the transverse momentum $\kappa^2$ within the proton.\\
In the next Section, we will present the observables investigated for the two light vector meson productions and will proceed comparing different models for this non-perturbative quantity in both $\rho$ and $\phi$ leptoproductions.
\section{Observables of interest}
\label{Sec_obs}
\subsection{Helicity amplitudes}
\label{helamp}
In the exclusive leptoproduction of a vector meson, although a different particle from the incoming one, appears in the final state, the use of the $\kappa$-factorization is however legitimate. This process, indeed, can be treated as photoabsorption, similarly to $\gamma^* p \to \gamma^*p$ discussed in Sec.~\ref{Sec_kt_fact}, as long as we refer to the generalized formulation of the optical theorem.\\
In high energy limit, that is called Regge limit, in which $s = W^2 \gg Q^2 \gg \Lambda^2_\text{QCD}$ the scattering amplitude is dominated by the exchange of two gluons in the $t$-channel. In this regime, using the $\kappa$-factorization approach, or impact factor representation, the forward helicity-amplitudes have the form~\cite{Anikin:2011sa}
\beq
T_{\lambda_{V}\lambda_{\gamma}}(s;Q^2) = \frac{is}{(2\pi)^2}\int \frac{d^2\kappa}{(\kappa^2)^2} \Phi^{\gamma^{*}(\lambda_{\gamma}) \rightarrow V(\lambda_{V})}(\kappa^2,Q^2)\, {\cal F}(x,\kappa^2 )\,,
\quad x=\frac{Q^2}{s} \, ,
\label{kt-fact}
\eq
where $\Phi^{\gamma^*(\lambda_\gamma) \longrightarrow V(\lambda_V)}$ is the impact factor describing the transition $\gamma^* \to V$. It corresponds to all possible contributions of the upper part of the diagrams in Fig.~\ref{photoabs}, with the only difference that in this process, the $q\bar{q}$ fluctuations of the photon recombine into the vector meson; we find again the unintegrated gluon density $\cal{F}(x, \kappa^2)$, whose attachment to the impact factor is here represented by $\frac{1}{(\kappa^2)^2}$. Here, $\kappa$ indicates the transverse momentum of the $t$-channel exchanged gluons. In the next sections we will define specific impact factors for both $\rho$ and $\phi$ productions taking into account the helicities of the vector mesons.
\subsection{Cross section}
\label{Cross_section_th}
The general expression for the helicity amplitude in Eq.~\eqref{kt-fact} contributes to the calculation of the cross section. According to whether the polarization of helicity amplitude is the longitudinal one $T_{00}$ or the transverse one $T_{11}$, one gets differential polarized cross sections $\sigma_L$ and $\sigma_T$ respectively:

\begin{equation}
\frac{d\sigma_L}{dt} (t = 0) = \frac{\left|T_{00}(s,t = 0)\right|^2}{16\,\pi\,W^2}\,,
\end{equation}

\begin{equation}
\frac{d\sigma_T}{dt} (t = 0) = \frac{\left|T_{11}(s,t = 0)\right|^2}{16\,\pi\,W^2}\,.
\end{equation}
The $t$-dependence is ruled by non-perturbative effects of the nucleon, which can be parametrized by an exponential dependence of the differential cross section
\begin{equation}
\frac{d\sigma_{L,T}}{dt}(t) = e^{-b(Q^2)t}\,\frac{d\sigma_{L,T}}{dt}(t = 0)\,.
\end{equation}
This result leads to polarized cross sections given by
 \begin{equation}
 \sigma_L\,(\gamma^*\,p \rightarrow V\,p) = \frac{1}{16 \pi b(Q^2)}\left|\frac{T_{00}(s, t = 0)}{W^2}\right|^2\,,
 \end{equation}
 \begin{equation}
 \sigma_T\,(\gamma^*\,p \rightarrow V\,p) = \frac{1}{16 \pi b(Q^2)}\left|\frac{T_{11}(s, t = 0)}{W^2}\right|^2\,.
 \end{equation}
 The $b(Q^2)$ slope, which depends on the photon virtuality $Q^2$, will be parametrized in our analysis as follows~\cite{Nemchik:1997xb}:
 \begin{equation}
 \label{slope_B}
 b(Q^2) = \beta_0 - \beta_1\,\log\left[\frac{Q^2+m_V^2}{m^2_{J/\psi}}\right]+\frac{\beta_2}{Q^2+m_V^2}\,,
 \end{equation}
 where the constants $\beta_0$,  $\beta_1$ and $\beta_2$ assume different values according to the vector meson involved in the process. In Sections~\ref{Sec_rho} and~\ref{Sec_phi} we will explicit the choice of these parameters.
 Our analysis will provide results for each polarized cross section and for their ratio.
 
\section{UGD models}
\label{Sec_UGD}

We have considered a selection of several models of UGD, without
pretension to exhaustive coverage, but with the aim of comparing
(sometimes radically) different approaches. We refer the reader to
the original papers for details on the derivation of each model and limit
ourselves to presenting here just the functional form ${\cal F}(x,\kappa^2)$
of the UGD as we implemented it in the numerical analysis.

\subsection{An $x$-independent model (ABIPSW)}

The simplest UGD model is $x$-independent and merely coincides with
the proton impact factor~\cite{Anikin:2011sa}:
\begin{equation}
\label{ABIPSW}
{\cal F}(x,\kappa^2)= \frac{A}{(2\pi)^2\,M^2}
\left[\frac{\kappa^2}{M^2+\kappa^2}\right]\,,
\end{equation}
where $M$ corresponds to the non-perturbative hadronic scale, fixed as~$M = 1$~GeV. The constant $A$ is negligible when we consider the ratio $T_{11}/T_{00}$ for the $\rho$-meson leptoproduction, but become essential to calculate the cross sections. Therefore, in Section~\ref{Theo_rho} a method to fix the parameter $A$ will be presented.

\subsection{Gluon momentum derivative}

This UGD is given by
\begin{equation}
\label{xgluon}
{\cal F}(x, \kappa^2) = \frac{dxg(x, \kappa^2)}{d\ln \kappa^2}
\end{equation}
and encompasses the collinear gluon density $g(x, \mu_F^2)$, taken at
$\mu_F^2=\kappa^2$. It is based on the obvious requirement that, when
integrated over $\kappa^2$ up to some factorization scale, the UGD must
give the collinear gluon density. We have employed the CT14
parametrization~\cite{Dulat:2015mca}, using the appropriate cutoff
$\kappa_{\text{min}} = 0.3$~GeV (see Section~\ref{tools} for further details).

\subsection{Ivanov--Nikolaev' (IN) UGD: a soft-hard model}

The UGD proposed in Ref.~\cite{Ivanov:2000cm} is developed with the purpose
of probing different regions of the transverse momentum. In the large-$\kappa$
region, DGLAP parametrizations for $g(x, \kappa^2)$ are employed. Moreover,
for the extrapolation of the hard gluon densities to small $\kappa^2$, an
Ansatz is made~\cite{Nikolaev:1994cd}, which describes the color gauge
invariance constraints on the radiation of soft gluons by color singlet
targets. The gluon density at small $\kappa^2$ is supplemented by a
non-perturbative soft component, according to the color-dipole
phenomenology.

This model of UGD has the following form:
\begin{equation}
{\cal F}(x,\kappa^2)= {\cal F}^{(B)}_\text{soft}(x,\kappa^2) 
{\kappa_{s}^2 \over 
	\kappa^2 +\kappa_{s}^2} + {\cal F}_\text{hard}(x,\kappa^2) 
{\kappa^2 \over 
	\kappa^2 +\kappa_{h}^2}\,,
\label{eq:4.7}
\end{equation}
where $\kappa_{s}^2 = 3$ GeV$^2$ and $\kappa_{h}^2 = [1 + 0.047\log^2(1/x)]^{1/2}$.

The soft term reads
\begin{equation}
\label{softterm}
{\cal F}^{(B)}_\text{soft}(x,\kappa^2) = a_\text{soft}
C_{F} N_{c} {\alpha_{s}(\kappa^2) \over \pi} \left( {\kappa^2 \over 
	\kappa^2 +\mu_\text{soft}^{2}}\right)^2 V_{\text N}(\kappa)\,,
\end{equation}
where $C_{F} = \dfrac{N_{c}^2 -1}{2N_{c}}$ and $\mu_\text{soft} = 0.1$ GeV. The
parameter $a_\text{soft} = 2$ gives a measure of how important is the soft part
compared to the hard one. On the other hand, the hard component reads
\begin{equation}
\label{hardterm}
{\cal F}_\text{hard}(x,\kappa^2)= 
{\cal F}^{(B)}_{\text{pt}}(\kappa^2){{\cal F}_{\text{pt}}(x,Q_{c}^{2})
	\over {\cal F}_{\text{pt}}^{(B)}(Q_{c}^{2})}
\theta(Q_{c}^{2}-\kappa^{2}) +{\cal F}_{\text{pt}}(x,\kappa^2)
\theta(\kappa^{2}-Q_{c}^{2})\,,
\end{equation}	
where ${\cal F}_{\text{pt}}(x, \kappa^2)$ is related to the standard gluon parton
distribution as in Eq.~\eqref{xgluon} and $Q_{c}^2 = 3.26$ GeV$^2$ 
(see Section~\ref{tools} for further details).
We refer to Ref.~\cite{Ivanov:2000cm} for the expressions of the vertex
function $V_{\text N}(\kappa)$ and of ${\cal F}^{(B)}_{\text{pt}}(\kappa^2)$.
Another relevant feature of this model is given by the choice of the coupling
constant. In this regard, the infrared freezing of strong coupling at leading
order (LO) is imposed by fixing $\Lambda_\text{QCD} = 200$ MeV:
\begin{equation}
\label{frozen}
\alpha_s(\mu^2) = \text{min} \left\{0.82, \, \frac{4 \pi}{\beta_0
	\log \left(\frac{\mu^2}{\Lambda^2_\text{QCD}}\right)}\right\}.
\end{equation}

We stress that this model was successfully tested on the {\em unpolarized}
electroproduction of VMs at HERA.

\subsection{Hentschinski--Sabio Vera--Salas' (HSS) model}

This model, originally used in the study of DIS structure
functions~\cite{Hentschinski:2012kr}, takes the form of a convolution between
the BFKL gluon Green's function and a LO proton impact factor. It has been
employed in the description of single-bottom quark production at LHC
in Ref.~\cite{Chachamis:2015ona} and to investigate the photoproduction of
$J/\Psi$ and $\Upsilon$ in Refs.~\cite{Bautista:2016xnp,Garcia:2019tne,Hentschinski:2020yfm}. We implemented
the formula given in Ref.~\cite{Chachamis:2015ona} (up to a $\kappa^2$ overall
factor needed to match our definition), which reads
\begin{equation}
\label{HentsUGD}
{\cal F}(x, \kappa^2, M_h) = \int_{-\infty}^{\infty}
\frac{d\nu}{2\pi^2}\ {\cal C} \  \frac{\Gamma(\delta - i\nu -\frac{1}{2})}
{\Gamma(\delta)}\ \left(\frac{1}{x}\right)^{\chi\left(\frac{1}{2}+i\nu\right)}
\left(\frac{\kappa^2}{Q^2_0}\right)^{\frac{1}{2}+i\nu}
\end{equation}
\[
\times \left\{ 1 +\frac{\bar{\alpha}^2_s \beta_0 \chi_0\left(\frac{1}{2}
	+i\nu\right)}{8 N_c}\log\left(\frac{1}{x}\right)
\left[-\psi\left(\delta-\frac{1}{2} - i\nu\right)
-\log\frac{\kappa^2}{M_h^2}\right]\right\}\,,
\]
where $\beta_0=\frac{11 N_c-2 N_f}{3}$, with $N_f$ the number of
active quarks (put equal to four in the following),
$\bar{\alpha}_s = \dfrac{\alpha_s\left(\mu^2\right) N_c}{\pi}$,
with $\mu^2 = Q_0 M_h$, and $\chi_0(\frac{1}{2} + i\nu)\equiv \chi_0(\gamma)
= 2\psi(1) - \psi(\gamma) - \psi(1-\gamma)$ is  (up to the factor $\bar\alpha_s$) the LO eigenvalue of the BFKL
kernel, with $\psi(\gamma)$ the logarithmic derivative of Euler Gamma
function. Here, $M_h$ plays the role of the hard scale which can be identified
with the photon virtuality, $\sqrt{Q^2}$.
In Eq.~\eqref{HentsUGD}, $\chi(\gamma)$ (with $\gamma = \frac{1}{2} + i\nu$)
is the NLO eigenvalue of the BFKL kernel, collinearly improved and BLM
optimized; it reads
\begin{equation}
\chi(\gamma) = \bar{\alpha}_s\chi_0(\gamma)+\bar{\alpha}^2_s\chi_1(\gamma)
-\frac{1}{2}\bar{\alpha}^2_s\chi^\prime_0(\gamma)\,\chi_0(\gamma)
+ \chi_{RG}(\bar{\alpha}_s, \gamma)\,,
\end{equation}
with $\chi_1(\gamma)$ and $\chi_{RG}(\bar{\alpha}_s, \gamma)$ given in
Section~2 of Ref.~\cite{Chachamis:2015ona}.

This UGD model is characterized by a peculiar parametrization for the proton
impact factor, whose expression is
\begin{equation}
\Phi_p(q, Q^2_0) = \frac{{\cal C}}{2\pi \Gamma(\delta)}
\left(\frac{q^2}{Q^2_0}\right)^\delta e^{-\frac{q^2}{Q^2_0}}\,,
\end{equation}
which depends on three parameters $Q_0$, $\delta$ and ${\cal C}$ which
were fitted to the combined HERA data for the $F_2(x)$ proton structure
function. We adopted here the so called
{\em kinematically improved} values (see Section~\ref{tools} for further
details) and given by
\begin{equation}
\label{ki}
Q_0 = 0.28\,\text{GeV}, \qquad \delta = 6.5, \qquad {\cal C} = 2.35 \;.
\end{equation}

\subsection{Golec-Biernat--W{\"u}sthoff' (GBW) UGD}

This UGD parametrization derives from the effective dipole cross section
$\hat{\sigma}(x,r)$ for the scattering of a $q\bar{q}$ pair off a
nucleon~\cite{GolecBiernat:1998js},
\begin{equation}
\hat{\sigma}(x, r^2) = \sigma_0 \left\{1-\exp\left(-\frac{r^2}{4R^2_0(x)}
\right)\right\}\,,
\end{equation}
through a reverse Fourier transform of the expression 
\begin{equation}
\sigma_0 \left\{1-\exp\left(-\frac{r^2}{4R^2_0(x)}
\right)\right\}=\int \frac{d^2\kappa}{\kappa^4} {\cal F}(x,\kappa^2)
\left(1-\exp(i \vec{\kappa}\cdot\vec{r})\right)\left(1-\exp(-i \vec{\kappa}
\cdot\vec{r})\right)\,,
\end{equation}
\begin{equation}
{\cal F}(x,\kappa^2)= \kappa^4 \sigma_0 \frac{R^2_0(x)}{2\pi}
e^{-\kappa^2 R^2_0(x)}\,,
\end{equation}
with 
\begin{equation}
R^2_0(x) = \frac{1}{{\text{GeV}}^2} \left(\frac{x}{x_0}\right)^{\lambda_p}
\end{equation}
and  the following values
\begin{equation}
\sigma_0 = 23.03\,\text{mb}, \qquad \lambda_p = 0.288, \qquad x_0 = 3.04 \cdot 10^{-4}\,.
\end{equation}
The normalization $\sigma_0$ and the parameters $x_0$ and $\lambda_p > 0$ of
$R^2_0(x)$ have been determined by a global fit to $F_2(x)$ in the
region $x < 0.01$.

\subsection{Watt--Martin--Ryskin' (WMR) model}

The UGD introduced in Ref.~\cite{Watt:2003mx} reads
\[
{\cal F}(x,\kappa^2,\mu^2) = T_g(\kappa^2,\mu^2)\,\frac{\alpha_s(\kappa^2)}
{2\pi}\,\int_x^1\!dz\;\left[\sum_q P_{gq}(z)\,\frac{x}{z}q\left(\frac{x}{z},
\kappa^2\right) + \right.\nonumber
\]
\begin{equation}
\label{WMR_UGD}
\left. \hspace{6.5cm} P_{gg}(z)\,\frac{x}{z}g\left(\frac{x}{z},\kappa^2\right)\,\Theta\left(\frac{\mu}{\mu+\kappa}-z\right)\,\right]\,,
\end{equation}
where the term
\begin{equation}
\label{WMR_Tg}
T_g(\kappa^2,\mu^2) \!= \exp\!\left(-\int_{\kappa^2}^{\mu^2}\!d\kappa_t^2\,
\frac{\alpha_s(\kappa_t^2)}{2\pi}\,\!\left( \int_{z^\prime_{{\text{min}}}}^{z^\prime_{{\text{max}}}}
\!dz^\prime\;z^\prime \,P_{gg}(z^\prime ) + N_f\!\!\int_0^1\!dz^\prime\,P_{qg}(z^\prime)
\right)\right)\,\!,
\end{equation}
gives the probability of evolving from the scale $\kappa$ to the
scale $\mu$ without parton emission. Here $z^\prime_{\mathrm{max}}\equiv
1-z^\prime_{\mathrm{min}}=\mu/(\mu+\kappa_t)$; $N_f$ is the number of active quarks.
This UGD model depends on an extra-scale $\mu$, which we fixed at $Q$.
The splitting functions $P_{qg}(z)$ and $P_{gg}(z)$ are given by
\[
P_{qg}(z) = T_R\,[z^2 + (1-z)^2]\;,
\]
\[
P_{gg}(z) = 2\,C_A \left[\dfrac{1}{(1-z)_+} + \dfrac{1}{z}- 2 +z(1-z)\right]
+ \left(\frac{11}{6}C_A - \frac{N_f}{3}\right) \delta(1 -z)\;,
\]
with the plus-prescription defined as
\begin{equation}
\label{pluspre}
\int_{a}^{1} dz \frac{F(z)}{(1-z)_+} = \int_{a}^{1} dz \frac{F(z) - F(1)}{(1-z)}
- \int_{0}^{a} dz \frac{F(1)}{(1 -z)}\,.
\end{equation}

\section{Polarized $\rho$-meson leptoproduction}
\label{Sec_rho}
The H1 and ZEUS collaborations have provided extended analyses
of the helicity structure in the hard exclusive production of the $\rho$ meson
in $ep$ collisions through the subprocess
\begin{equation}
\label{process_rho}
\gamma^*(\lambda_\gamma)p\rightarrow \rho (\lambda_\rho)p\,.
\end{equation}
Here $\lambda_\rho$ and $\lambda_\gamma$ represent the meson and photon
helicities, respectively, and can take the values 0 (longitudinal polarization)
and $\pm 1$ (transverse polarizations). The helicity amplitudes 
$T_{\lambda_\rho \lambda_\gamma}$ extracted at HERA~\cite{Aaron:2009xp} exhibit the
hierarchy shown in~\eqref{hierarchy}, that follows from the dominance of a  small-size 
dipole scattering mechanism, as discussed first in Ref.~\cite{Ivanov:1998gk}. As we previously mentioned, the H1 and ZEUS collaborations have analyzed data in different ranges of $Q^2$
and $W$. In what follows we will refer only to the H1 ranges, specified in~\eqref{H1range} and will present the full scenario to study the $\rho$-meson leptoproduction as a channel to test the UGDs~\cite{Bolognino:2018mlw,Bolognino:2019bko,Celiberto:2019slj}.
\subsection{Theoretical framework}
\label{Theo_rho}
In the high-energy regime, $s\equiv W^2\gg Q^2\gg\Lambda_{\text{QCD}}^2$, which
implies small $x=Q^2/W^2$, the forward helicity amplitude for the
$\rho$-meson electroproduction can be written, in $\kappa$-factorization, as
the convolution of the $\gamma^*\rightarrow \rho$ IF,
$\Phi^{\gamma^*(\lambda_\gamma)\rightarrow\rho(\lambda_\rho)}(\kappa^2,Q^2)$,
with the UGD, ${\cal F}(x,\kappa^2)$. Its expression reads
\begin{equation}
\label{amplitude}
T_{\lambda_\rho\lambda_\gamma}(s,Q^2) = \frac{is}{(2\pi)^2}\int \dfrac{d^2\kappa}
{(\kappa^2)^2}\Phi^{\gamma^*(\lambda_\gamma)\rightarrow\rho(\lambda_\rho)}(\kappa^2,Q^2)
{\cal F}(x,\kappa^2),\quad \text{$x=\frac{Q^2}{s}$}\,.
\end{equation}

Defining $\alpha = \frac{\kappa^2}{Q^2}$ and $B =~2\pi \alpha_s
\frac{e}{\sqrt{2}}f_\rho$, the expression for the IFs takes the
following forms (see Ref.~\cite{Anikin:2009bf} for the derivation):

\begin{itemize}
	
	\item longitudinal case
	\begin{equation}
	\label{Phi_LL}
	\Phi_{\gamma_L\rightarrow\rho_L}(\kappa,Q;\mu^2) = 2 B\frac{\sqrt{N_c^2-1}}{Q\,N_c}
	\int^{1}_{0}dy\, \varphi_1(y;\mu^2)\left(\frac{\alpha}{\alpha + y\bar{y}}\right)
	\,,
	\end{equation}
	where $N_c$ denotes the number of colors and
	$\varphi_1(y;\mu^2)$ is the twist-2 DA which, up to
	the second order in the expansion in Gegenbauer polynomials,
	reads~\cite{Ball:1998sk}
	\begin{equation}
	\label{phi}
	\varphi_1(y; \mu^2) = 6y\bar{y}\left(1+a_2(\mu^2)\frac{3}{2}
	\left(5 (y-\bar{y})^2-1\right)\right)\,;
	\end{equation}
	
	\item transverse case
	\[
	\Phi_{\gamma_T\rightarrow\rho_T}(\alpha,Q;\mu^2)=
	\dfrac{(\epsilon_{\gamma}\cdot\epsilon^{*}_{\rho}) \, 2 B m_{\rho}
		\sqrt{N_c^2-1}}{2 N_c Q^2}
	\]
	\begin{equation}
	\times\left\{-\int^{1}_{0} dy \frac{\alpha (\alpha +2 y \bar{y})}{y\bar{y}
		(\alpha+y\bar{y})^2}\right. \left[(y-\bar{y})\varphi_1^T(y;\mu^2)
	+ \varphi_A^T(y;\mu^2)\right]
	\label{Phi_TT}
	\end{equation}
	\[
	+\int^{1}_{0}dy_2\int^{y_2}_{0}dy_1 \frac{y_1\bar{y}_1\alpha}
	{\alpha+y_1\bar{y}_1}\left[\frac{2-N_c/C_F}{\alpha(y_1+\bar{y}_2)
		+y_1\bar{y}_2}
	-\frac{N_c}{C_F}\frac{1}{y_2 \alpha+y_1(y_2-y_1)}\right]
	\]
	\[
	\left.\times M(y_1,y_2;\mu^2)-\!\int^{1}_{0}dy_2\!\int^{y_2}_{0}dy_1\! \left[\frac{2+N_c/C_F}{\bar{y}_1}
	+\frac{y_1}{\alpha+y_1\bar{y}_1}\right.\right.
	\]
	\[
	\!\left.\left.\times\left(\frac{(2-N_c/C_F)
		y_1\alpha}{\alpha(y_1+\bar{y}_2)+y_1\bar{y}_2}-2\right)-\frac{N_c}{C_F}\frac{(y_2-y_1)\bar{y}_2}{\bar{y}_1}\right.\right.
	\]
	\[
	\left.\left.\times\frac{1}{\alpha\bar{y}_1+(y_2-y_1)\bar{y}_2}\right]\!S(y_1,y_2;\mu^2)\right\}\,,
	\]
	where $C_F=\frac{N_c^2-1}{2N_c}$, while the functions $M(y_1,y_2;\mu^2)$
	and $S(y_1,y_2;\mu^2)$ are defined in Eqs.~(12)-(13) of
	Ref.~\cite{Anikin:2011sa}
	and are combinations of the twist-3 DAs $B(y_1,y_2;\mu^2)$ and
	$D(y_1,y_2;\mu^2)$ (see Ref.~\cite{Ball:1998sk}), given by
	\begin{align}
	\label{BD}
	B(y_1,y_2;\mu^2) & =-5040 y_1 \bar{y}_2 (y_1-\bar{y}_2) (y_2-y_1) \notag\\
	D(y_1,y_2;\mu^2) & =-360 y_1\bar{y}_2(y_2-y_1)
	\left(1+\frac{\omega^{A}_{\{1,0\}}(\mu^2)}{2}\left(7\left(y_2-y_1\right)-3\right)
	\right)\,.
	\end{align}
	
\end{itemize}

In Eqs.~\eqref{phi} and~\eqref{BD} the functional dependence of
$a_2$, $\omega^{A}_{\{1,0\}}$, $\zeta^{A}_{3\rho}$, and $\zeta^{V}_{3\rho}$ on the
factorization scale $\mu^2$ can be determined from the corresponding known
evolution equations~\cite{Ball:1998sk}, using some suitable initial condition
at a scale $\mu_0$. 

Note that the $\kappa$-dependence of the IFs is different in the cases of
longitudinal and transverse polarizations and this poses a strong constraint
on the $\kappa$-dependence of the UGD in the HERA energy range. The main point will be to demonstrate, considering different models for UGD, that the uncertainties of the theoretical description do not prevent us from some, at least qualitative, conclusions about the $\kappa$-shape of the UGD. 

In Eqs.~\eqref{phi} and~\eqref{BD} the functional dependence of
$a_2$, $\omega^{A}_{\{1,0\}}$, $\zeta^{A}_{3\rho}$, and $\zeta^{V}_{3\rho}$ on the
factorization scale $\mu^2$ can be determined from the corresponding known
evolution equations~\cite{Ball:1998sk}, using some suitable initial condition
at a scale $\mu_0$. 

The DAs $\varphi^T_1(y;\mu^2)$ and $\varphi^T_A(y;\mu^2)$ in Eq.~\eqref{Phi_TT}
encompass both genuine twist-3 and Wandzura-Wilczek~(WW)
contributions~\cite{Anikin:2011sa,Ball:1998sk}. The former are related to
$B(y_1,y_2;\mu^2)$ and $D(y_1,y_2;\mu^2)$; the latter are those obtained
in the approximation in which $B(y_1,y_2;\mu^2)=D(y_1,y_2;\mu^2)=0$, and in
this case read~\footnote{For asymptotic form of the twist-2 DA,
	$\varphi_1(y)=\varphi_1^{\text{as}}(y)=6y\bar y$, these equations give
	$\varphi^{T\;WW,\;\text{as}}_A(y)=-3/2y\bar y$ and $\varphi^{T\;WW,\;\text{as}}_1(y)=
	-3/2 y\bar y (2y -1)$.}
\[
\label{WWT}
\varphi^{T\;WW}_A(y;\mu^2)= \frac{1}{2}\left[-\bar y
\int_0^{y}\,dv \frac{\varphi_1(v;\mu^2)}{\bar v} -
y \int_{y}^1\,dv \frac{\varphi_1(v;\mu^2)}{v}   \right]\;,
\]
\begin{equation}
\varphi^{T\;WW}_1(y;\mu^2)= \frac{1}{2}\left[
-\bar y \int_0^{y}\,dv \frac{\varphi_1(v;\mu^2)}{\bar v} +
y \int_{y}^1\,dv \frac{\varphi_1(v;\mu^2)}{v}   \right]\;.
\end{equation}

The other interesting point will be the extension of information, collected by the helicity-amplitude analysis, reachable from the calculation of the cross section. As a matter of fact, the expressions for the polarized cross sections $\sigma_L$ and $\sigma_T$, calculated using the Eqs.~\eqref{Phi_LL}, \eqref{phi} in Eq.~\eqref{amplitude} and the Eqs.~\eqref{Phi_TT}, \eqref{BD} in Eq.~\eqref{amplitude} respectively, are
\begin{equation}
\label{sigL_rho}
\sigma_L\,(\gamma^*\,p \rightarrow \rho\,p) = \frac{1}{16 \pi b(Q^2)}\left|\frac{T_{00}(s, t = 0)}{W^2}\right|^2\,,
\end{equation}
\begin{equation}
\label{sigT_rho}
\sigma_T\,(\gamma^*\,p \rightarrow \rho\,p) = \frac{1}{16 \pi b(Q^2)}\left|\frac{T_{11}(s, t = 0)}{W^2}\right|^2\,,
\end{equation}
where $b(Q^2)$ in Eqs.~\eqref{sigL_rho} and~\eqref{sigT_rho}, as anticipated in Section~\ref{Cross_section_th}, is the slope described by the parametrization~\cite{Nemchik:1997xb}:
\begin{equation}
\label{slope_B_rho}
b(Q^2) = \beta_0 - \beta_1\,\log\left[\frac{Q^2+m_\rho^2}{m^2_{J/\psi}}\right]+\frac{\beta_2}{Q^2+m_\rho^2}\,,
\end{equation}
fixing the values of the constants as follows: $\beta_0 = 6.5$ GeV$^{-2}$, $\beta_1 = 1.2$  GeV$^{-2}$ and $\beta_2 = 1.6$.

The advantage of considering polarized cross sections relies on the possibility to constrain, not only the shape and the behavior of results, but specially the \emph{normalization}, now essential and which is clearly irrelevant for the evaluation of the helicity-amplitude ratio $T_{11}/T_{00}$. Although all UGDs described in Section~\ref{Sec_UGD} present fixed values for their parameters, the ABIPSW model, one of the first UGD parametrization adopted in phenomenological analysis~\cite{Forshaw}, is missing of one of its two characteristic parameters. As regards the helicity-amplitude ratio $T_{11}/T_{00}$, results for the ABIPSW UGD model show a fair agreement with data~\cite{Anikin:2011sa}, this suggesting that the shape, controlled by the $M$ parameter (see Eq.~\eqref{ABIPSW}), has been already guessed. Therefore, the best value of the normalization parameter, $A$, can be obtained via a simple global fit on data of both polarized cross sections, $\sigma_L$ and $\sigma_T$. Being $A$, \emph{de facto}, a multiplicative factor in Eq.~\eqref{ABIPSW}, its fit can be done analytically.

Let us consider the experimental HERA data set~\cite{Aaron:2009xp} at $W = 75$ GeV. Our dataset is defined as $\{d_{\lambda, i}\} = \{\sigma_{\lambda, i}\}_{i \in\{Q^2\}}^{\lambda = L,T}$ with their errors $\{e_{\lambda, i}\}$.\\
It is useful to express the ABIPSW UGD model in Eq.~\eqref{ABIPSW} as
\begin{equation}
\cal F(\kappa^2, A, M_h) \equiv \cal F(\kappa^2, A = 1, M_h)\, A
\end{equation}
and recognizing that the polarized cross section $\sigma_\lambda \propto |T_\lambda|^2$ can be written as a function of the normalization parameter $A$, we can get:
\begin{equation}
\sigma_\lambda(A) = |A|^2 S_\lambda(A = 1) \equiv A^2 \sigma_\lambda(1)\,.
\end{equation}
Now the normalization constant, $A$, can be obtained by calculating the $\chi^2$:
\begin{equation}
\label{chi2}
	\chi^2 = \sum_{\lambda,i} \frac{(d_{\lambda,i} - t_{\lambda,i}(A))^2}{e_{\lambda,i}^2}\,.
\end{equation}
Here $\{t_{\lambda,i}\}$ is the theory data set, such that:
\begin{equation}
\label{ti_A}
	t_{\lambda,i}(A) = \sigma_\lambda(A)|_{Q^2 = i} = A^2 \sigma_\lambda(1)|_{Q^2 = i} \equiv A^2 \tilde{t}_{\lambda,i}\,,
\end{equation}
where the $\{\tilde{t}_{{\lambda,i}}\}$ set corresponds to the theory data calculated for $A = 1$.\\
Hence, replacing the Eq.~\eqref{ti_A} in Eq.~\eqref{chi2}, we get
\begin{equation}
\tilde{\chi}^2 = \sum_{\lambda,i} \frac{(d_{\lambda,i} - A^2\tilde{t}_{{\lambda,i}})^2}{e_{\lambda,i}^2} = \sum_{\lambda,i} \frac{d_{\lambda,i}^2 - 2\,A^2d_{\lambda,i}\tilde{t}_{{\lambda,i}} + A^4\tilde{t}_{{\lambda,i}}^2}{e_{\lambda,i}^2}\,,
\end{equation}
so that
\begin{equation}
\frac{d \tilde{\chi}^2(A^2)}{d A^2} = 0
\end{equation}
provides the relation for the normalization constant
\begin{equation}
A = \sqrt{\frac{\sum_{\lambda,i} 2\,d_{\lambda,i}\tilde{t}_{{\lambda,i}} }{2\,\sum_{\lambda,i}\tilde{t}_{{\lambda,i}}^2}}\,.
\end{equation}
\subsection{Theoretical uncertainties and approximations}
\label{uncertainties}

There are four sources of uncertainty and/or approximation in our analysis,
as based on the above expressions for the helicity amplitudes.

(i) The $\gamma^*\to V$ IF is a function of $Q^2$ and $\kappa$ which is not
fully perturbative and includes also physics of large distances. Here we use
collinear factorization to express the IF as a convolution -- integration over
longitudinal momentum fraction --  of the nonperturbative twist-2 and -3
DAs and a perturbative hard part.

In the region of large $\kappa$, $\kappa\sim Q$, which corresponds to
the range of small dipole sizes, the IFs for the production of both the
longitudinally and transversely polarized meson are well described in our
collinear factorization scheme. The neglected contributions are relatively
suppressed as powers of $\Lambda_{\text{QCD}}/Q$ and are therefore neglected.

The region of small $\kappa$, $\kappa\ll Q$, is also present in our
$\kappa$-factorization formulas and corresponds to the range of larger dipole
sizes $r_d$. Can we calculate also in this case our IFs as convolution of
the perturbative hard part with the meson DAs?

The situation here is different in the cases of longitudinal and transverse
polarizations. In the longitudinal case, we have, in fact, small $r_d$
dominance in the region of all $\kappa$. Note that,  as  calculated in our
scheme, the longitudinal IF divided by $\kappa^2$ is finite for $\kappa \to 0$, and, therefore, the  neglected contributions for our longitudinal IF are
suppressed by powers of $\Lambda_{\text{QCD}}/Q$ in the region of all $\kappa$.
In relation with that, we note here that the longitudinal VM electroproduction
can be described not only in $\kappa$-, but also fully in QCD collinear
factorization (in terms of generalized parton distributions).

For the transverse polarization the situation is different, since
the transverse IF divided by $\kappa^2$ behaves like $\log(\kappa^2/Q^2)$,
which means that the collinear limit, $\kappa\to 0$, is not safe and one
cannot describe the transverse VM electroproduction fully in QCD collinear
factorization. One can easily trace that this behavior 
$\sim \log(\kappa^2/Q^2)$  for  $\kappa \to 0$ of the transverse IF appears
due to the  integration over the  longitudinal fraction $z$ and its
logarithmic divergence near the end points $z\to 0, 1$.
The light-cone wave-function of the virtual photon, that controls the hard part
of the IF, has a scale $Q^2 r_d^2 z (1-z)$. This means that, at $\kappa \to 0$,
when the endpoint region of the $z$-integration is important, large values of
$Q$ do not mean automatically that the small-$r_d$ region is dominant, but
instead both large and small $r_d$ contribute. Therefore we do not control
the accuracy of our IF calculation for $\kappa \to 0$ in the transverse
polarization case.

However, in $\kappa$-factorization the small-$\kappa$ region is only a corner
of the $\kappa$ integration domain and the importance of this corner is a
matter of investigation. On the experimental side we do not see indications
that large $r_d$'s, and therefore small $\kappa$'s, are dominant; indeed,
HERA data show similar $t$- and $W$-dependence for the both longitudinal and
transverse helicity amplitudes. In our phenomenological analysis we will check
the importance of the region of small $\kappa$'s by studying the dependence of
our predictions on the $\kappa$ lower cut value.

(ii) Another source of uncertainty comes from the adopted form of the 
light-cone DAs.

We considered, for the sake of simplicity, the so called {\em asymptotic}
choice for the twist-2 DA given in Eq.~\eqref{phi}, corresponding
to fixing $a_2(\mu^2)=0$. The impact of this approximation was estimated by
letting $a_2(\mu_0^2)$ take a non-zero value as large as 0.6 at
$\mu_0^2=1$~GeV$^2$ in the analysis with one specific model for UGD.

We used typically twist-3 DAs in the Wandzura-Wilczek approximation, but
considered in one case the effect of the inclusion of the genuine twist-3
contribution to check the validity of this approximation.

(iii) We calculate the {\em forward} amplitudes for both longitudinal and
transverse case. The experimental analysis showed that the $t$-dependence is
similar for the two helicity amplitudes, the measured values of the slope
parameter have, within errors, the same values for the both polarizations cases.
Therefore in $T_{11}/T_{00}$ ratio  considered here  the $t$-dependence
drops. 

(iv) The expression in Eq.~\eqref{amplitude} represents, as a matter of fact,
the {\em imaginary part} of the amplitude and not the full amplitude. The real
parts of the amplitudes at high energy are smaller: they are suppressed in
comparison to the imaginary parts by the factor $\sim 1/\log s$, and they are
related to the latter by dispersion relations. Here again we appeal to the
results of the experimental analysis, that showed similar $W$-dependence for
both helicity amplitudes, that means the effective cancellation of the
contribution  from the real parts of the amplitudes in the ratio
$T_{11}/T_{00}$.

\subsection{Numerical analysis}
\label{analysis}
In this Section we present our results for the helicity-amplitude ratio
$T_{11}/T_{00}$, polarized cross sections $\sigma_L$ and $\sigma_T$ and their ratio $\sigma_L/\sigma_T$, as obtained with all the UGD models presented above, and compare them with HERA data.

\begin{figure}[tb]
	\centering
	
	\includegraphics[scale=0.50,clip]{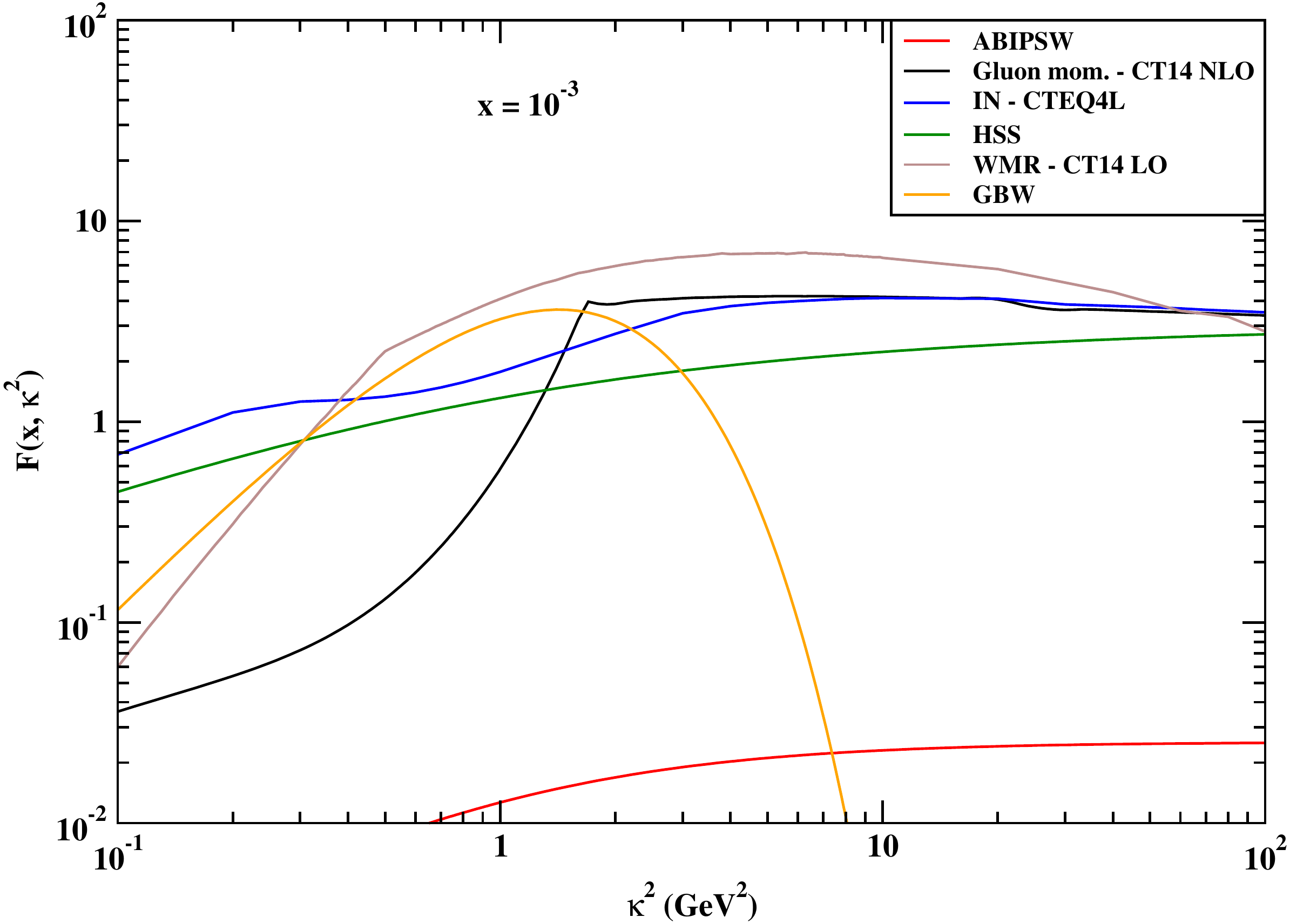}
	
	\includegraphics[scale=0.50,clip]{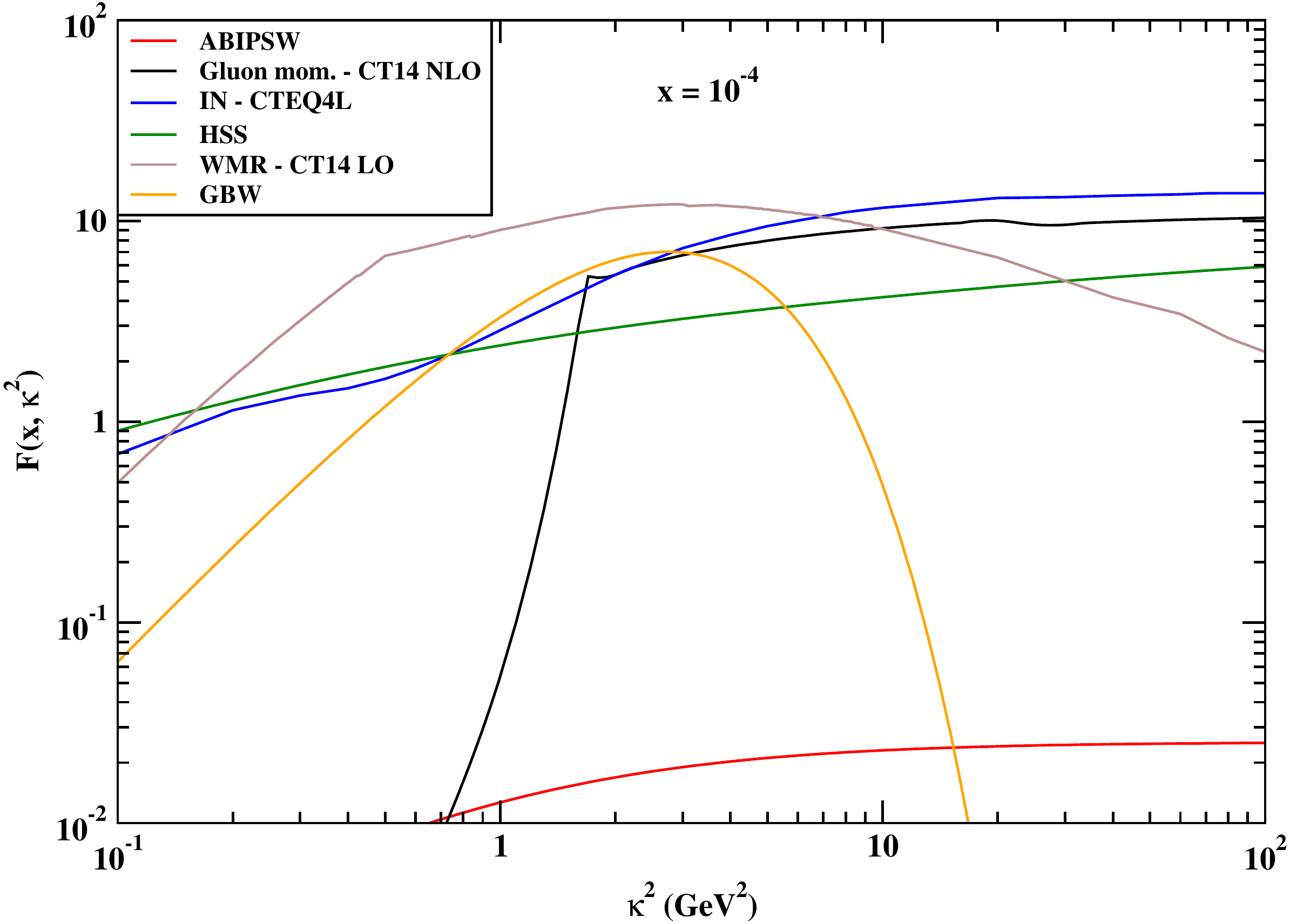}
	
	\caption{\emph{$\kappa^2$-dependence of all UGD models for $x = 10^{-3}$ and $10^{-4}$.}}
	\label{fig:UGDs_vs_k2}
\end{figure}

We preliminarily present a plot, Fig.~\ref{fig:UGDs_vs_k2}, with the
$\kappa^2$-dependence of all the considered UGD models, for two different
values of $x$. The plot clearly exhibits the marked difference in the
$\kappa^2$-shape of the six UGDs.

\begin{figure}[tb]
	\centering
	
	\includegraphics[scale=0.50,clip]{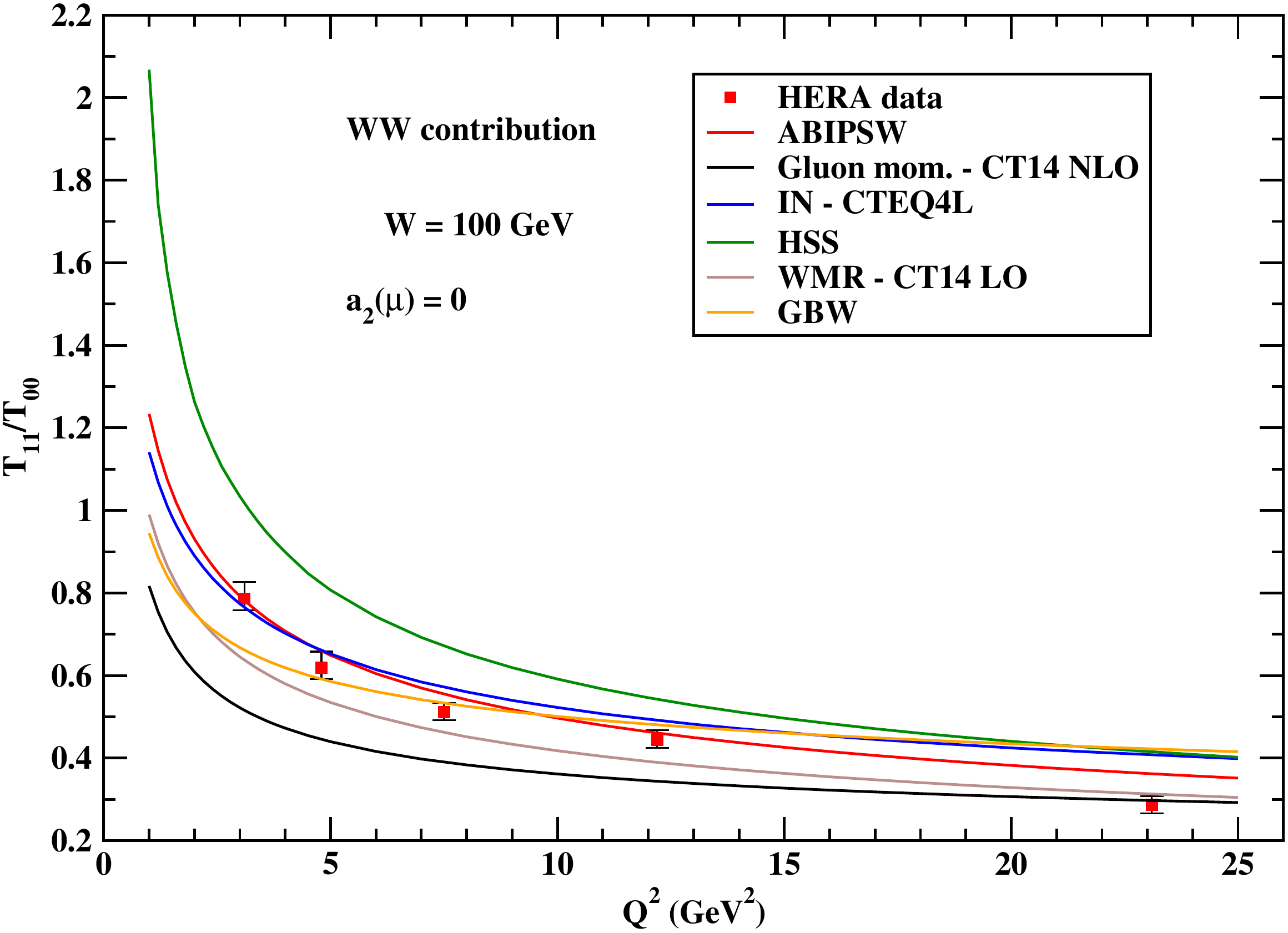}
	
	\caption{\emph{$Q^2$-dependence of the helicity-amplitude ratio $T_{11}/T_{00}$ for
			all the considered UGD models at $W = 100$ GeV. In the twist-2 DA we have
			put $a_2(\mu_0 = 1 \mbox{ GeV}) = 0$ and the $T_{11}$ amplitude has
			been calculated in the WW approximation.}}
	\label{fig:ratio_all}
\end{figure}

\begin{figure}[tb]
	\centering
	
	\includegraphics[scale=0.50,clip]{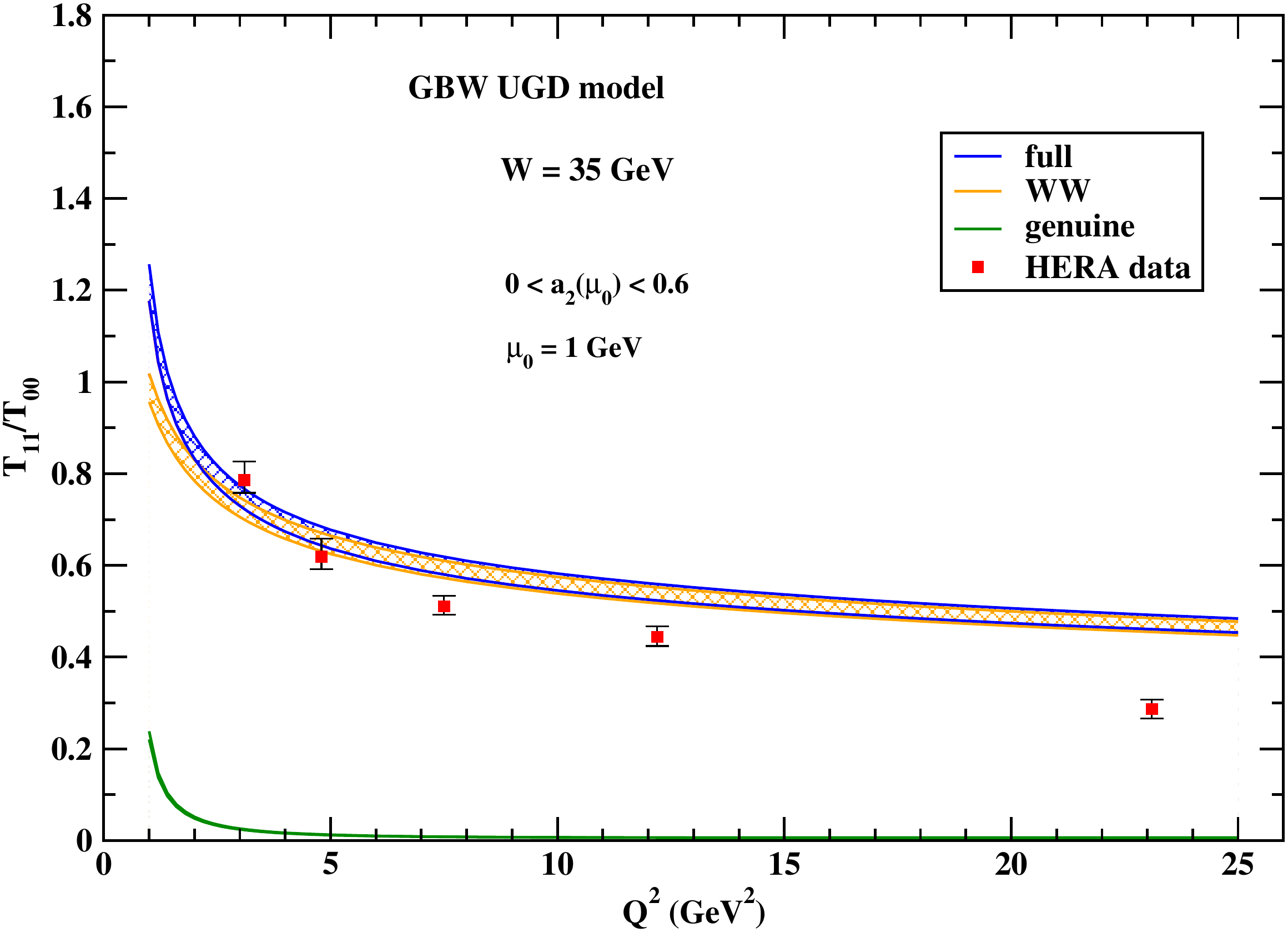}
	
	\includegraphics[scale=0.50,clip]{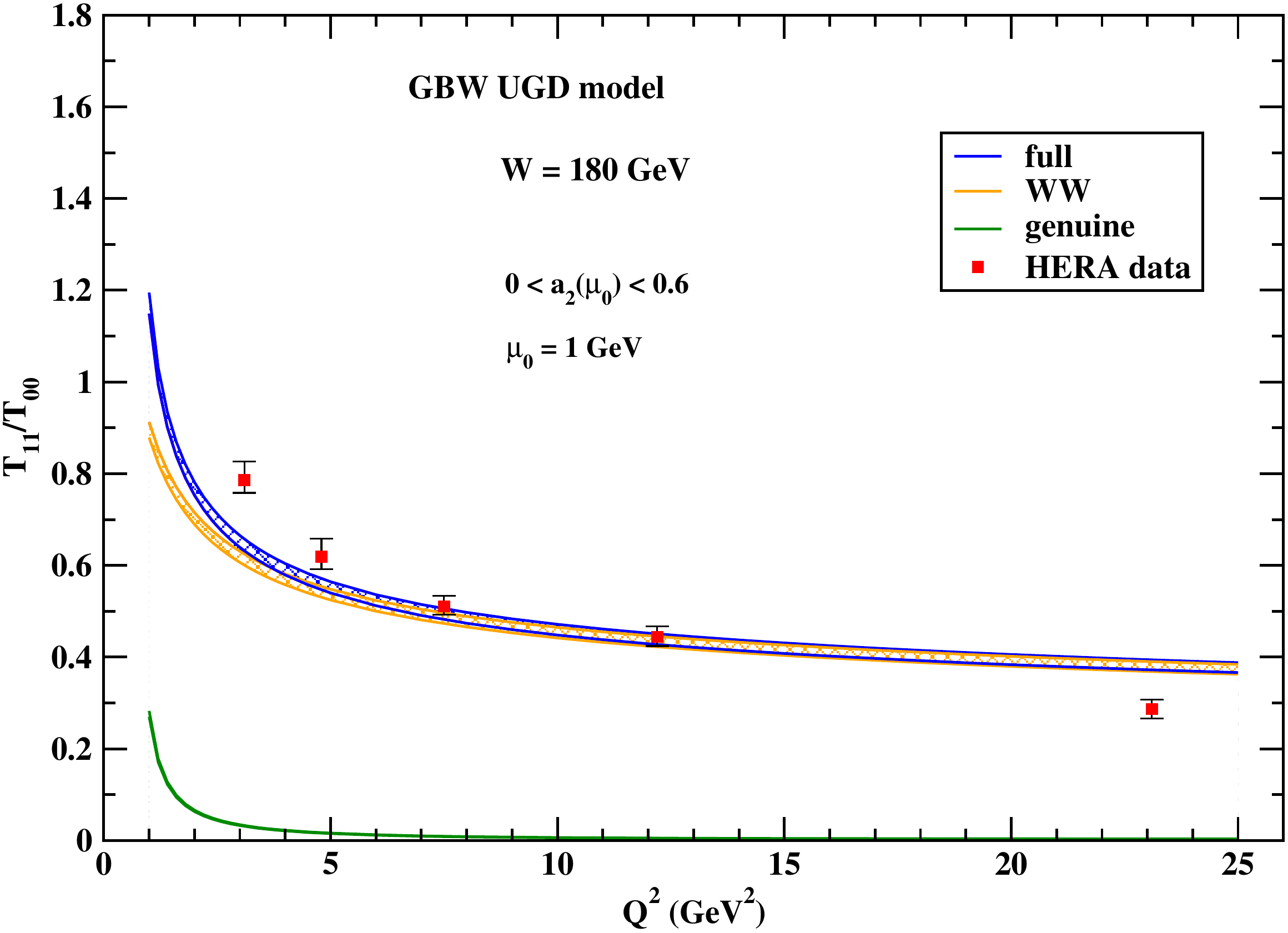}
	
	\caption{\emph{$Q^2$-dependence of the helicity-amplitude ratio $T_{11}/T_{00}$ for
			the GBW UGD model at $W = 35$ (top) and 180~GeV (bottom). The full, WW and
			genuine contributions are shown. The shaded bands give the effect of
			varying $a_2(\mu_0 = 1 \mbox{ GeV})$ between 0. and 0.6.}}
	\label{fig:ratio_GBW_evolved}
\end{figure}

In Fig.~\ref{fig:ratio_all} we compare the $Q^2$-dependence of $T_{11}/T_{00}$
for all six models at $W = 100$~GeV, together with the experimental result.
We used here the asymptotic twist-2 DA ($a_2(\mu^2)=0$) and the WW approximation
for twist-3 contributions. Theoretical results are spread over a large interval,
thus supporting our claim that the observable $T_{11}/T_{00}$ is potentially
able to strongly constrain the $\kappa$-dependence of the UGD. None of the
models is able to reproduce data over the entire $Q^2$ range; the
$x$-independent ABIPSW model, the IN model~\footnote{As a result of a correction in the numerical implementation, we note a significantly improved agreement between IN model and the experimental data with respect to the Fig.~(2) in the Ref.~\cite{Bolognino:2018}.}, and the GBW model seem to better catch the
intermediate-$Q^2$ behavior of data.\\
For the sake of semplicity and recognizing in the GBW model intriguing features and parameters worthy to be probed further, we propose an exhaustive phenomenological analysis of this UGD parametrization.\\
To gauge the impact of the approximation made in the DAs, we calculated
the $T_{11}/T_{00}$ ratio with the GBW model, at $W = 35$ and 180~GeV,
by varying $a_2(\mu_0=1 \ {\text{GeV}})$ in the range 0. to 0.6 and properly
taking into account its evolution. Moreover, for the same UGD model, we
relaxed the WW approximation in $T_{11}$ and considered also the genuine
twist-3 contribution. All that is summarized in
Fig.~\ref{fig:ratio_GBW_evolved}, which indicates that the approximations
made are quite reliable.

\begin{figure}[tb]
	\centering
	
	\includegraphics[scale=0.50,clip]{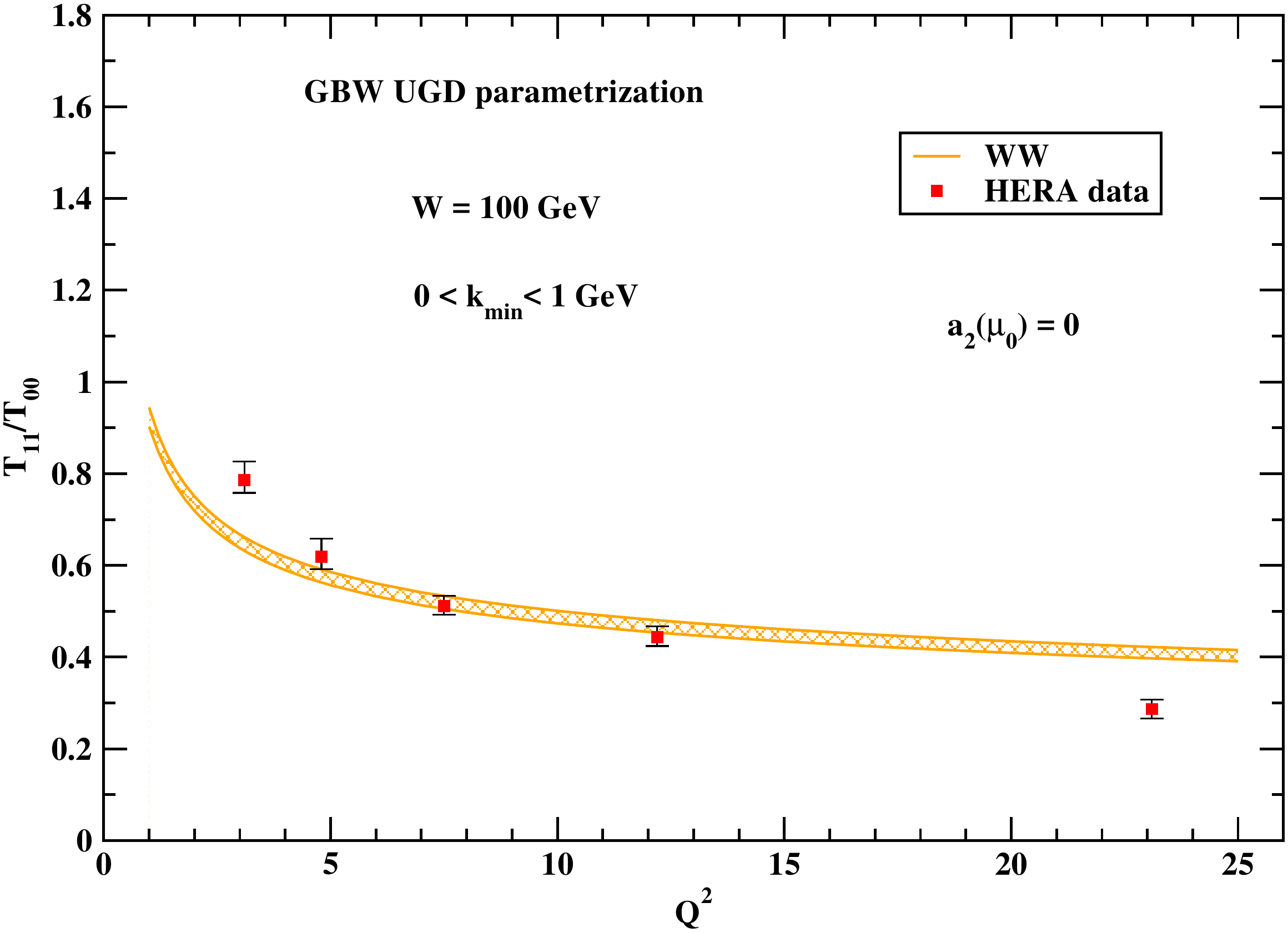}
	
	\caption{\emph{$Q^2$-dependence of the helicity-amplitude ratio $T_{11}/T_{00}$ for
			the GBW UGD model at $W = 100$~GeV. The band is the effect of a lower
			cutoff in the $\kappa$-integration, ranging from 0. to 1~GeV. In the
			twist-2 DA we have put $a_2(\mu_0 = 1 \mbox{ GeV}) = 0$ and the $T_{11}$
			amplitude has been calculated in the WW approximation.}}
	\label{fig:ratio_GBW_kmin}
\end{figure}

\begin{figure}
\centering

\includegraphics[scale=0.80,clip]{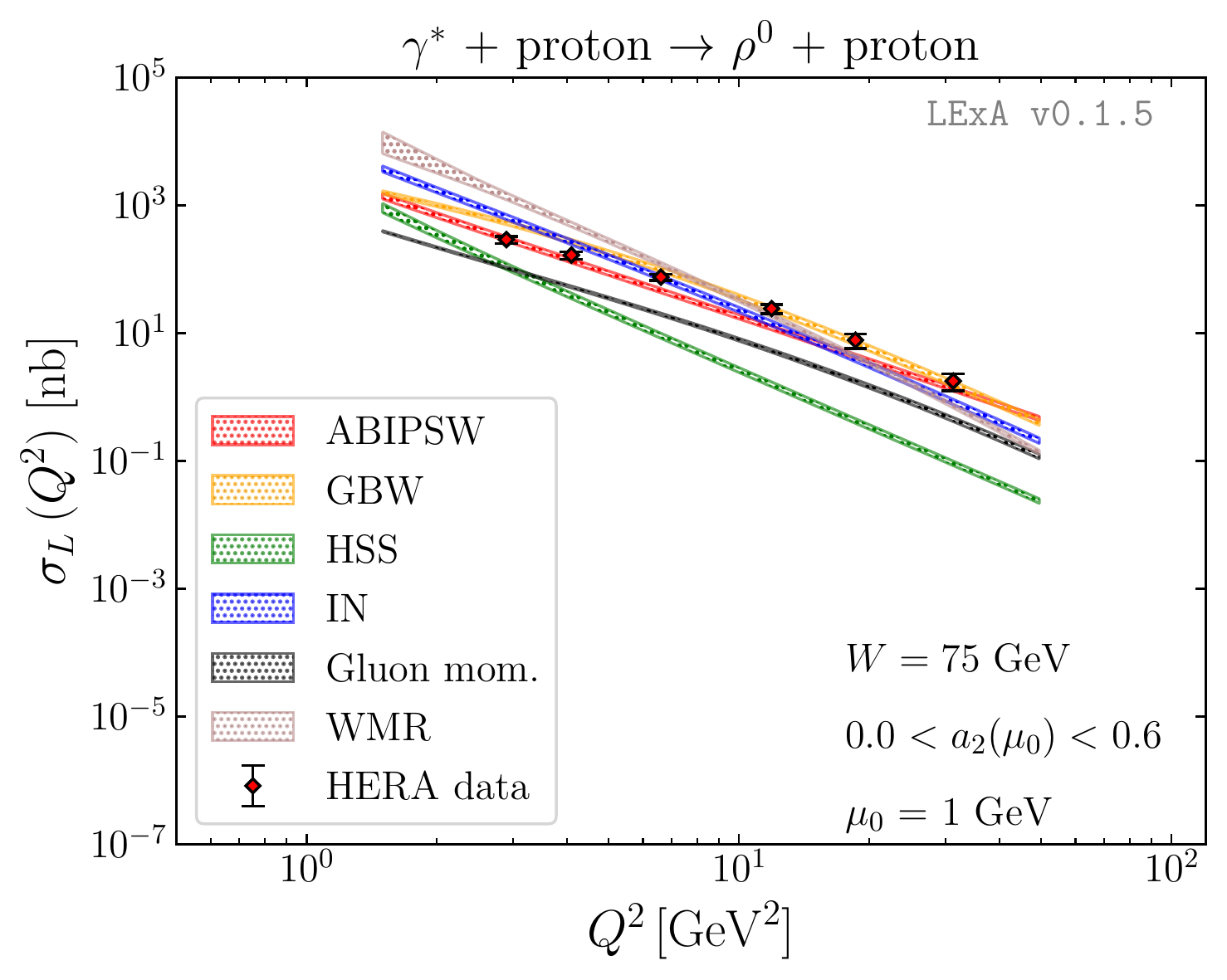}
\includegraphics[scale=0.80,clip]{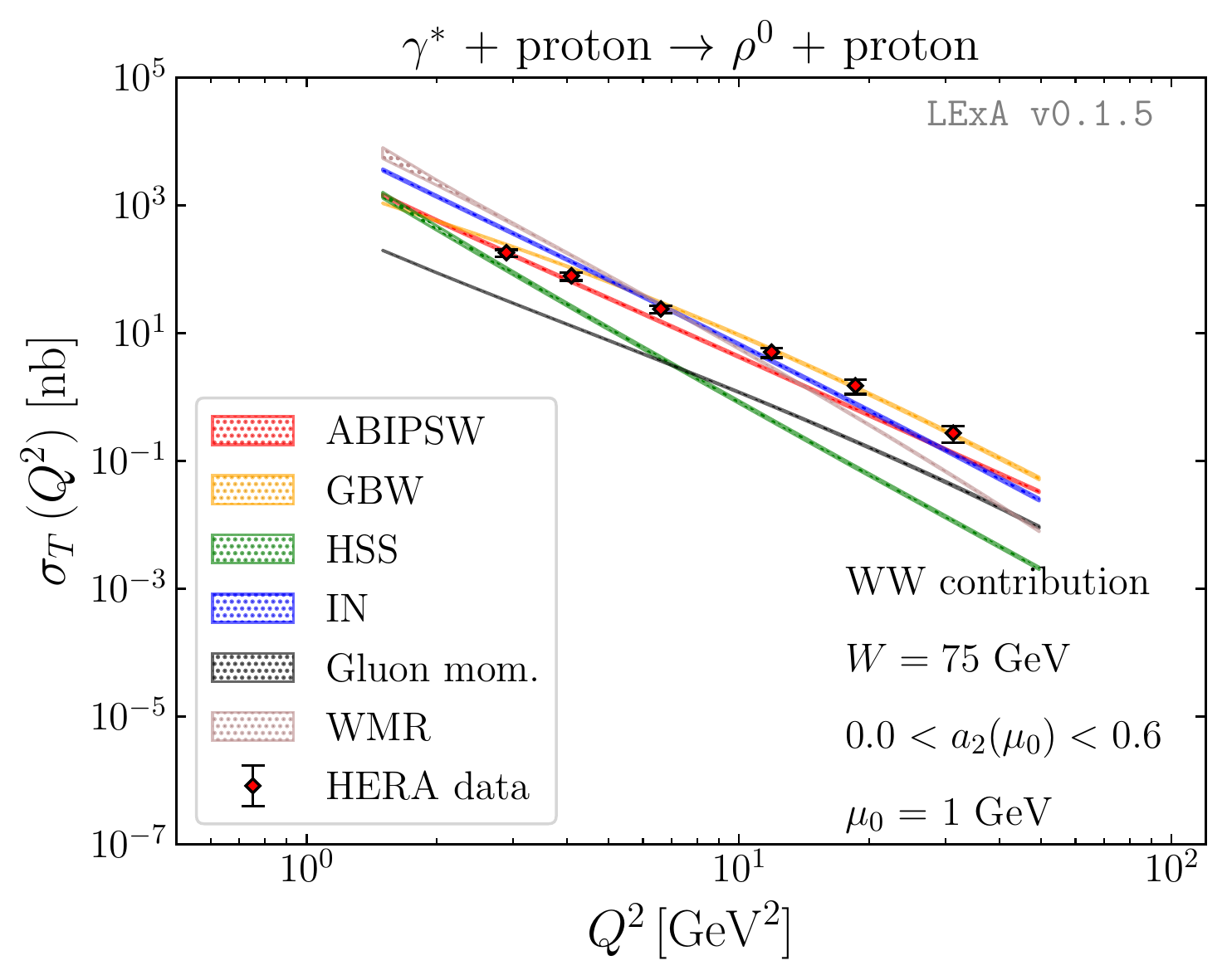}
\caption{\emph{$Q^2$-dependence of the polarized cross sections $\sigma_L$ (top panel) and $\sigma_T$ (bottom panel) for all the considered UGD models at $W = 75$ GeV. The bands give the effect of varying $a_2(\mu_0 = 1\,GeV)$ between $0.$ and $0.6$~\cite{Bolognino_new}.}}
\label{fig:sigmaLT_all}
\end{figure}

\begin{figure}
	\centering
	
	\includegraphics[scale=0.80,clip]{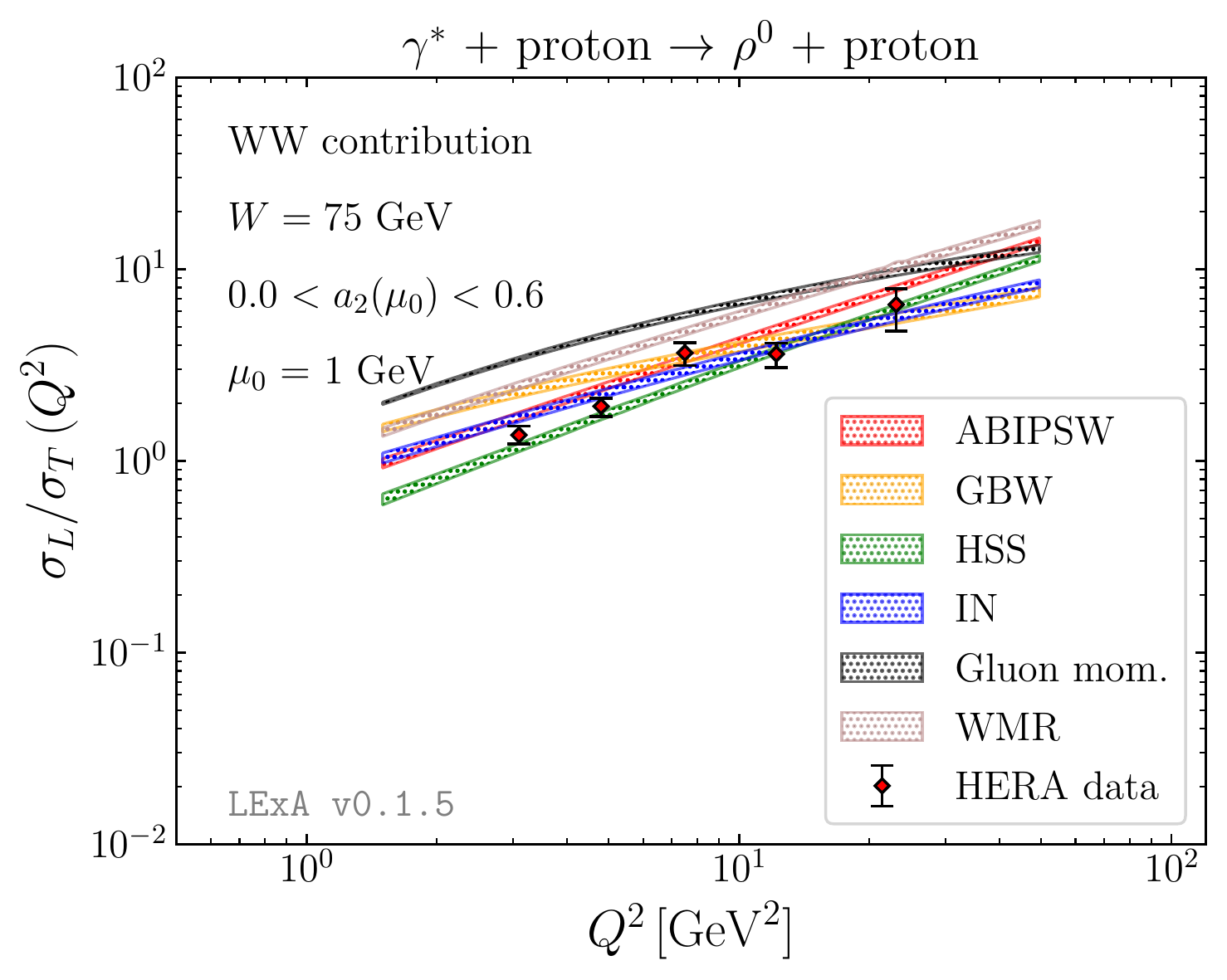}
	\caption{\emph{$Q^2$-dependence of the polarized cross section ratio $\sigma_L/\sigma_T$ for all the considered UGD models at $W = 75$ GeV. Here $\sigma_T$ is calculated in the $WW$ approximation~\cite{Bolognino_new}.}}
	\label{fig:sigmaR_all}
\end{figure}

\begin{figure}
	\centering
	
	\includegraphics[scale=0.80,clip]{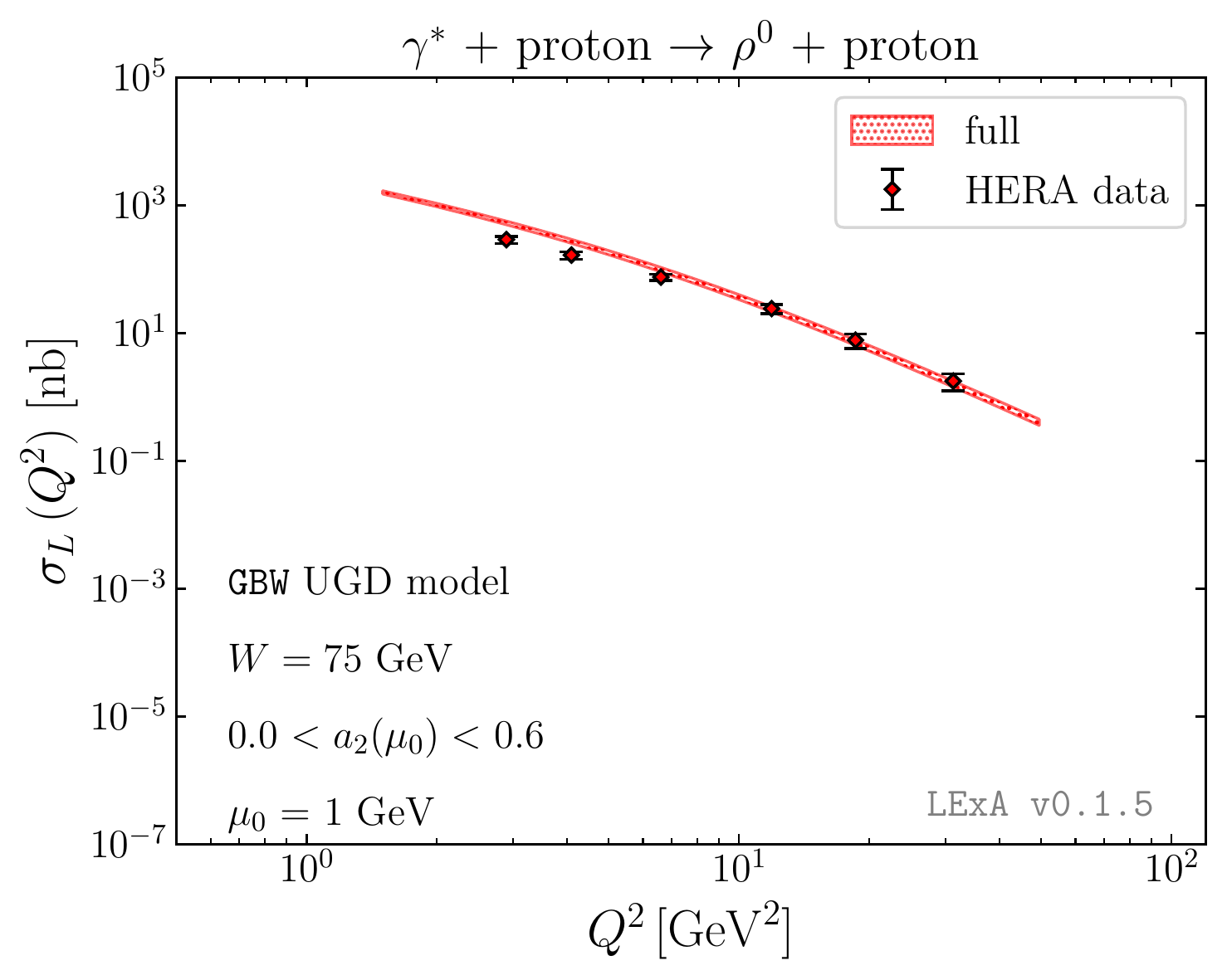}
	\includegraphics[scale=0.80,clip]{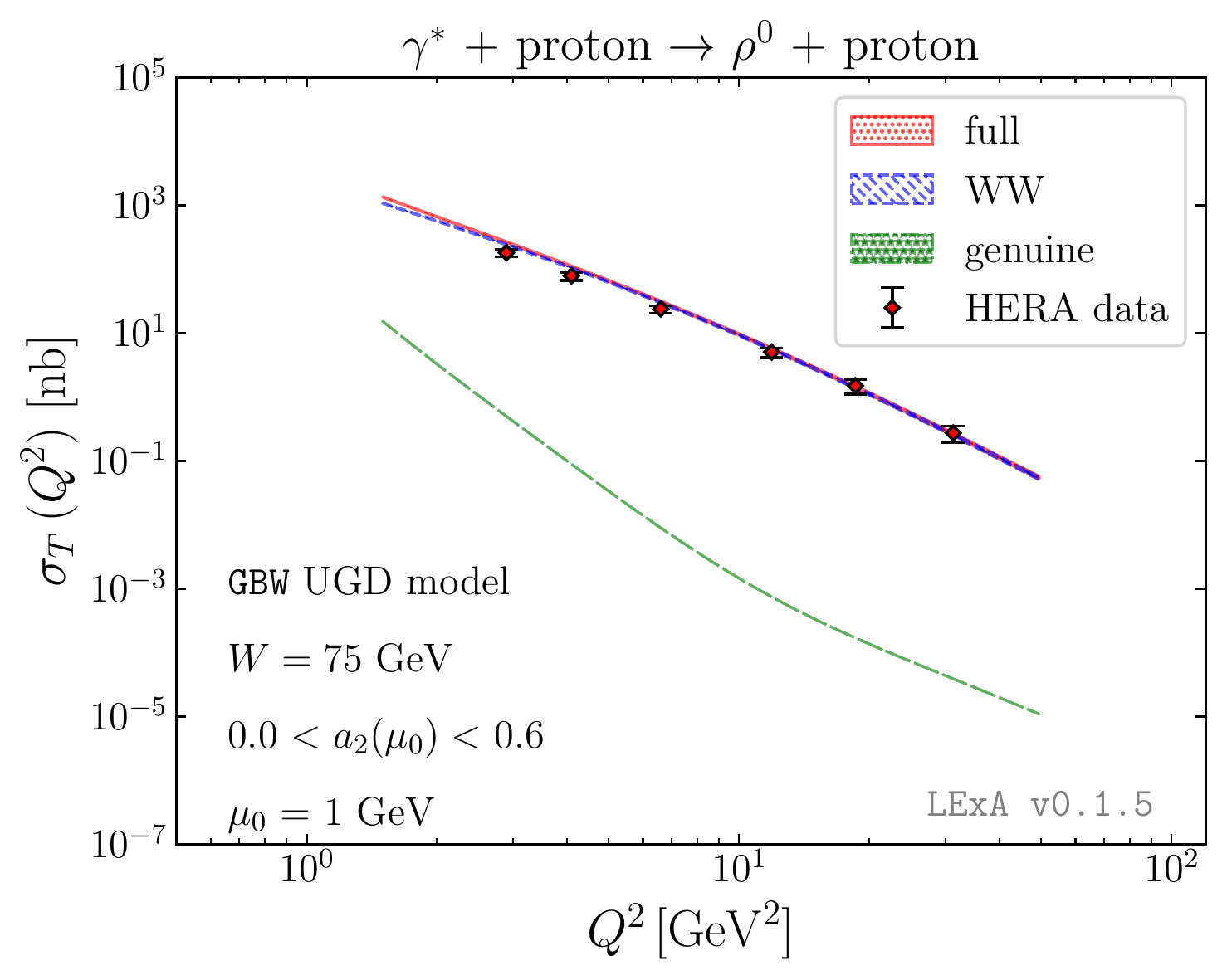}
	\caption{\emph{$Q^2$-dependence of the polarized cross sections $\sigma_L$ (top panel) and $\sigma_T$ (bottom panel) for the GBW UGD model at $W = 75$ GeV. The full, WW and genuine contributions are shown for $\sigma_T$~\cite{Bolognino_new}.}}
	\label{fig:sigmaLT_GBW}
\end{figure}

\begin{figure}
	\centering
	
	\includegraphics[scale=0.80,clip]{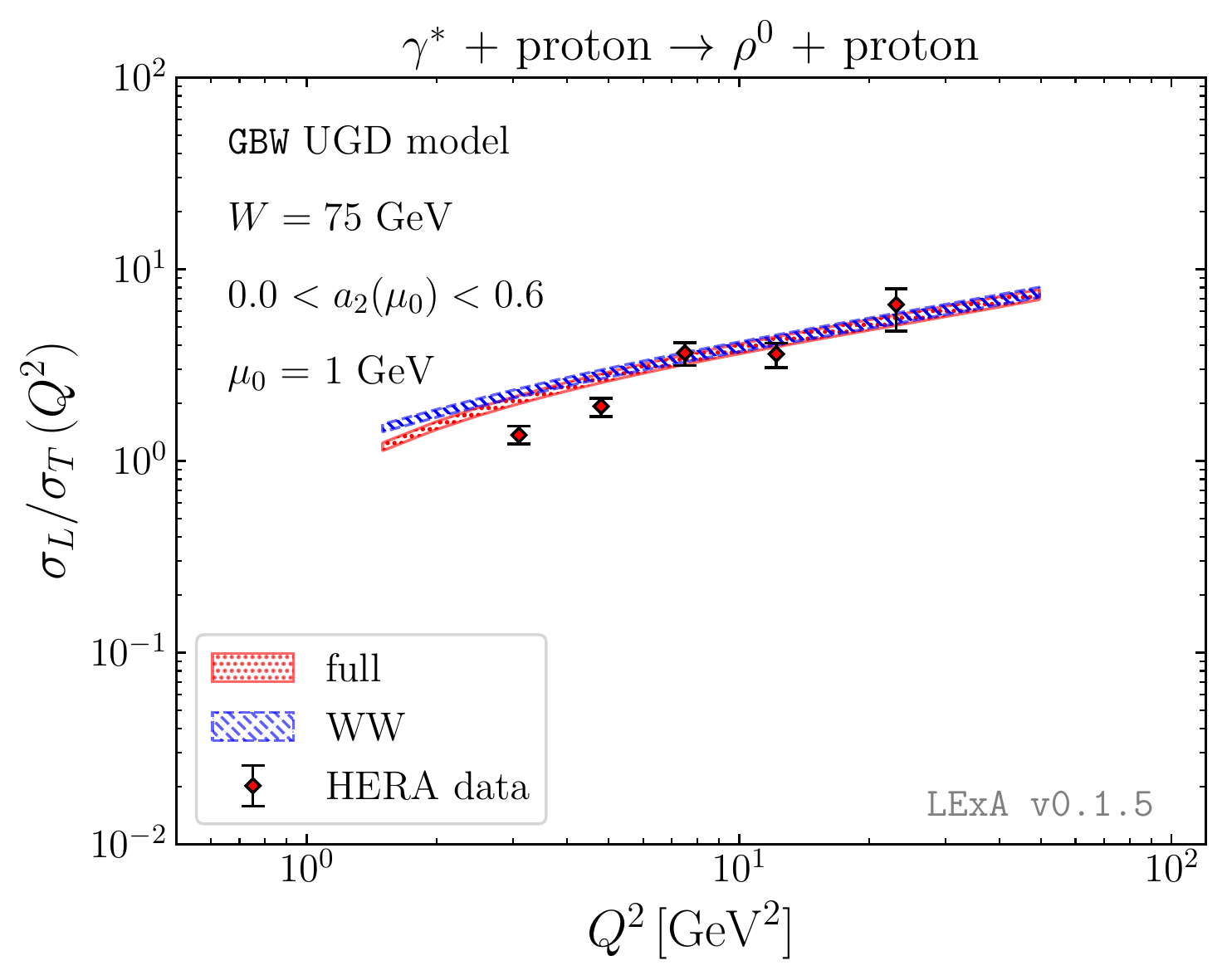}
	\caption{\emph{$Q^2$-dependence of the polarized cross section ratio $\sigma_L/\sigma_T$ for the GBW UGD model at $W = 75$ GeV. The bands give the effect of varying $a_2(\mu_0 = 1\,GeV)$ between $0.$ and $0.6$. The full and WW contributions are shown~\cite{Bolognino_new}.}}
	\label{fig:sigmaR_GBW}
\end{figure}

The stability of $T_{11}/T_{00}$ under the lower cut-off for $\kappa$, in the
range 0 $< \kappa_{\text{min}} < 1$~GeV, has been investigated. This is a
fundamental test since, if passed, it underpins the main underlying
assumption of this work, namely that {\em both} the helicity amplitudes
considered here are dominated by the large $\kappa$ region. In
Fig.~\ref{fig:ratio_GBW_kmin} we show the result of this test for the
GBW model at $W = 100$~GeV; similar plots can be obtained with the other
UGD models, with the only exception of the IN model.
There is a clear indication that the small-$\kappa$ region
gives only a marginal contribution.\\
The behavior of the polarized cross sections $\sigma_L$ and $\sigma_T$ and the ratio $\sigma_L/\sigma_T$~\cite{Bolognino_new} in terms of $Q^2$ for all UGDs at $W = 75$ GeV is probed in Figs.~\ref{fig:sigmaLT_all} and~\ref{fig:sigmaR_all}, taking into account the variation of the Gegenbauer coefficient $a_2(\mu_0)$ in the same range used for the helicity-amplitude ratio analysis. The normalization parameter $A$ that completes the definition of the ABIPSW UGD, calculated according to Section~\ref{Theo_rho}, has the value $A = 148.137$~\footnote{The uncertainty of the cross section due to the normalization parameter $A$ is $0.1\%$.}. The comparison exhibits a partial agreement with the experimental data, where once again none of the proposed models is able to describe the whole $Q^2$ region, however we can specify which UGD model is more suitable for the description of the polarized cross sections and which, indeed, for their ratio in the $Q^2$ intermediate range. On one side, observing Fig.~\ref{fig:sigmaLT_GBW}, the GBW model, considered in its \emph{standard} definition (\emph{i.e.} without the evolution of saturation scale~\footnote{The discussion about saturation effects in this exclusive process is treated in details in Ref.~\cite{Besse:2013muy}.}), appears the UGD that allows us to match data for the single polarized cross sections, $\sigma_L$ and $\sigma_T$, in the most accurate way. Here the genuine twist-3 contribution is considered. On the other side, although the predictions for the cross section ratio $\sigma_L/\sigma_T$ in Fig.~\ref{fig:sigmaR_GBW} with the GBW model is quite resonable if we regard increasing values of $Q^2 \sim 10$ GeV$^2$, the IN UGD model in Fig.~\ref{fig:sigmaR_all} is able to slightly better catch also the low $Q^2$ region of data.

\FloatBarrier

\FloatBarrier

\subsection{Tools and systematics}
\label{tools}

All numerical calculations we done in \textsc{Fortran}, making use of specific
CERNLIB routines~\cite{cernlib} to perform numerical integrations and the
computation of (poly-)gamma functions. In order to deal with all the considered UGD models, we found advantageous to create a modular library which allowed us to bind and call all sets via a unique and simple interface, LeXa (Leptonic Exclusive Amplitudes), which is in turn an interface of {\tt JETHAD}~\cite{Celiberto:2020wpk} and serving at the same time as a working environment for the creation of new, user-customized UGD parametrizations. 

The uncertainty coming from the numerical 2-dimensional over $\kappa$ and $y$
in Eqs.~\eqref{amplitude}, \eqref{Phi_LL} and \eqref{Phi_TT} was directly
estimated by the {\tt Dadmul} integrator~\cite{cernlib} and it was constantly
kept below 0.5\%. In the case of the HSS and WMR models, one should take into
account of an extra-source of systematic uncertainties coming from the
integration on $\nu$ (Eq.~(\ref{HentsUGD})) and on $z$ and $\kappa_t^2$
(Eqs.~(\ref{WMR_UGD}) and~(\ref{WMR_Tg})), respectively. Even in this case,
we managed to keep the numerical error very small.

Furthermore, it is worth to note that for all the UGD models involving the use
of standard PDF parametrizations, it was needed to put a lower cut-off in
$\kappa$, in order to respect the kinematical regime where each set has been
extracted. We gauged the effect of using different PDF parametrizations by making
tests with the most popular sets extracted from global fits, namely
MMHT14~\cite{Harland-Lang:2014zoa}, CT14~\cite{Dulat:2015mca} and
NNPDF3.0~\cite{Ball:2014uwa}, as provided by the LHAPDF Interface
6.2.1~\cite{Buckley:2014ana}, after imposing a provisional cut-off of
$\kappa_{\text{min}}^{\text{(test)}} = 1$ GeV. We checked that the discrepancy among the
various cases is small or negligible. Then, we did the final calculations by
using the CT14 parametrization, which allowed us to integrate over $\kappa$ down
to $\kappa_{\text{min}} = 0.3$ GeV. Results with the gluon momentum derivative and
the WMR model were obtained by imposing such a cut-off, while we adopted a
dynamic strategy as for the IN one: we cut the contribution coming from the
IN hard component (Eq.~(\ref{hardterm})), at $\kappa_{\text{min}}$, while no
cut-off was imposed for the soft component  (Eq.~(\ref{softterm})). With
respect to this model, we made further tests by considering the effect of
using different DGLAP inputs (see Table I of Ref.~\cite{Ivanov:2000cm}) for
the parameters $Q_{c}$, $\kappa_{h}$ and $\mu_{\text{soft}}$ entering
Eqs.~(\ref{softterm}) and~(\ref{hardterm}). No significant discrepancy among
them was found, so we gave our results for the IN model by using the so-called
CTEQ4L DGLAP input.

Following the definition of IN and WMR models, the PDF set was taken at LO,
while the NLO one was employed in the gluon momentum derivative parametrization.
Moreover, as for the HSS UGD parametrization, we checked that the discrepancy
between the so-called {\em improved} setup given in Eq.~(\ref{ki}) and the
standard one (see, {\it e.g.} Ref.~\cite{Chachamis:2015ona}) is negligible when
considering the helicity-amplitude ratio $T_{11}/T_{00}$.
\subsection{Discussion}
\label{disc_rho}
We have proposed the study of observables that have been well measured in the
experiments at HERA (and could be studied in possible future electron-proton
colliders) -- the dominant helicity-amplitudes ratio, the polarized cross sections and their ratio for the electroproduction
of vector mesons -- as a nontrivial testfield to discriminate the models for
the unintegrated gluon distribution in the proton.

The main motivation of our study are the features, observed at HERA,  of
polarization observables for exclusive vector meson electroproduction. In the
cases of both longitudinal and transverse polarizations, the measured cross
sections demonstrate similar dependencies on kinematic variables: specific
$Q^2$ scaling, $t$- and $W$-dependencies that are distinct from the ones seen
in soft diffractive exclusive processes.  This  indicates that the dominant
physical mechanism in both cases is the scattering of a small transverse-size,
$\sim 1/Q$, dipole on a proton. 

On the theoretical side we have a description in $\kappa$-factorization,
where the nonperturbative physics is encoded in the unintegrated gluon
distribution, ${\cal F}(x,\kappa^2)$, and in the vector meson twist-2 and twist-3
DAs (which includes both WW and genuine twist-3 contributions), that
parameterize the probability amplitudes for the transition of 2- and 3- parton
small-transverse-size colorless states to the vector meson.   

In our analysis we  have  considered six models for ${\cal F}(x,\kappa^2)$, which
exhibit rather different shape of $\kappa$-dependence in the  region,
$\kappa^2\sim$ few $\text{GeV}^2$, relevant for the kinematic of the $\rho$-meson
electroproduction at HERA, as shown in Fig.~\ref{fig:UGDs_vs_k2}. 

In our numerical study we  have  found rather weak sensitivity of our predictions for
the helicity-amplitude ratio to the physics encoded in the meson DAs (though
values of longitudinal and transverse amplitudes separately depend strongly
on the model for DAs). As an example, in Fig.~\ref{fig:ratio_GBW_evolved} we
have presented  results for the GBW model of ${\cal F}(x,\kappa^2)$. Here the dominance
of the WW contribution over the genuine twist-3 one is clearly seen. Besides,
we  have  found rather moderate dependence of our observable on the shape of twist-2
DA. Indeed, varying the value of $a_2$ in a wide range         
in comparison to the value $a_2(\mu_0)=0.18 \pm 0.10$ obtained from the QCD sum rules~\cite{Ball:1998sk}, and the one calculated recently on the lattice in Ref.~\cite{Braun:2016wnx}, $a_2(\mu=2 \mbox{ GeV})=0.132 \pm 0.027$, we have found small variation of the amplitude ratio, on the level of the experimental errors.  

Another important issue is the small-size color dipole dominance that allows us
to use results for the $\gamma^*\to \rho$ IF calculated unambiguously in terms
of the meson DAs. To clarify this question we  have introduced a cut-off in the
$\kappa$-integration and studied the stability of our predictions on the
excluded region of small gluon transverse momenta. In
Fig.~\ref{fig:ratio_GBW_kmin}, considering again GBW model as an example, we
have shown  that the sensitivity of our predictions to the region of small $\kappa$ is
indeed not strong, the variation of our results is lower than or comparable to the
data errors.     

In this way we  have seen  that the dominance of the small-size dipole production
mechanism is supported both by the qualitative features of the data and by the
theoretical calculations in  $\kappa$-factorization.     
This gives evidence to our main statement that, having precise HERA data on the
helicity-amplitude ratio, one can obtain important information about the
$\kappa$-shape of the UGD. To demonstrate this in Fig.~\ref{fig:ratio_all} we
have confronted  HERA data with the predictions calculated with six different UGD
models. We  have seen  that none of the models is able to reproduce data over the
entire $Q^2$ range and that HERA data on the transverse to longitudinal
amplitudes ratio are really precise enough to discriminate predictions of
different UGD models. Moreover, the investigation of polarized cross sections in Fig.~\ref{fig:sigmaLT_GBW} has provided with a more complete anaysis, confirming that the GBW model is a proper UGD to describe these observables. The theoretical prediction obtained for the IN model and illustrated in Fig.~\ref{fig:sigmaR_all}, instead, can be seen as a suitable UGD parametrization to test the polarized cross-section ratio $\sigma_L/\sigma_T$.

Our work is closely related to the study of Ref.~\cite{Besse:2013muy}, where the
same process was investigated in much detail in the dipole approach. In this
case the process helicity amplitudes are factorized in terms of the dipole
cross section $\hat \sigma (x,r)$. The  $\kappa$-factorization and the dipole
approach are mathematically related through a Fourier transformation, but the latter approach represents the most natural language to discuss saturation
effects, due to a distinct picture of saturation for the $\hat \sigma (x,r)$
dependence on $r$ for the dipole sizes that exceed the reverse saturation scale,
$r\geq 1/Q_S(x)$. Besides, nonlinear evolution equations that determine the $x$-dependence of $\hat \sigma (x,r)$ and include saturation effects are
formulated in the transverse coordinate space.  

In Ref.~\cite{Besse:2013muy} several models for $\hat \sigma (x,r)$
(including the GBW model adopted by us here) that include saturation
effects, and whose parameters were fitted to describe inclusive DIS data, were
considered. They were used to make predictions for vector meson exclusive
production at HERA kinematics. It was found in Ref.~\cite{Besse:2013muy} that the
predictions of GBW and of other more elaborated dipole models are close to
each other and give rather good, but not excellent, description of HERA data at
virtualities bigger than $Q^2\approx 5\,\text{GeV}^2$. 

Another interesting issue studied in Ref.~\cite{Besse:2013muy} is the radial
distribution of the dipoles, that contributes to the longitudinal and the
transverse helicity amplitudes for $\rho$-production. It was shown that for
large $Q^2$ both helicity amplitudes are dominated by the contributions of
small size dipoles, which is another source of evidence in favour of
the small-size dipole mechanism for the hard vector meson electroproduction
at HERA.  Besides, as it is shown in Ref.~\cite{Besse:2013muy}, in the case of
large $Q^2$, see the right panels of Fig.~17 in Ref.~\cite{Besse:2013muy} for $Q^2=10\,\text{GeV}^2$, the  relevant values of $r$ are considerably lower than
those where the dipole cross section $\hat \sigma (x,r)$ starts to saturate.
This is perhaps not surprising, since the estimated value of the saturation
scale at HERA energies is not big, $Q^2_S\sim 1\,\text{GeV}^2$. Therefore one can
anticipate that saturation effects for hard vector meson electroproduction at
HERA do not play a crucial role. The region of large values of $r$, where
the dipole cross section saturates, represents only a corner of $r$-integration
region for both helicity amplitudes in the dipole approach.~\footnote{The
	situation is different in the region of smaller $Q^2$, where the saturation
	region constitutes an essential part of $r$-integration range, see the left
	panels of Fig.~17 in Ref.~\cite{Besse:2013muy}. But in that case one cannot rely
	on the twist expansion in the calculation of the $\gamma^*\to \rho$
	transition, which is expressed in terms of the lowest twist-2 and twist-3
	DAs only. It would be very interesting to consider the same process at
	smaller values of $x$, where the saturation scale is bigger and saturation
	effects are expected to be more pronounced, but this would require
	experiments at larger energies.} In the language of $\kappa$-factorization,
that we use in this work, the saturation region is related to the
$\kappa$-integration region of small $\kappa$. Our calculations for the GBW
model with $\kappa$ cutoff, see Fig.~\ref{fig:ratio_GBW_kmin}, show that,
indeed, the helicity-amplitude ratio for hard meson electroproduction at HERA is not very sensitive to this saturation region. Therefore we believe that
HERA data allow to obtain nontrivial information on the UGD shape (or
equivalently, about  the $r$ shape of the dipole cross section) in the
kinematic range where the {\it linear} evolution regime is still dominant.

Further tests of models
for the unintegrated gluon distribution, as well as possible new model proposals, take into due account our suggestion to utilize the important information encoded in the HERA data on the helicity structure and on cross sections in the light vector meson electroproduction. Therefore in the next Section we propose a phenomenological study on the UGD via the $\phi$-meson leptoproduction channel. This time we focus on the comparison with the light-cone wave-function formalism (Ref.~\cite{Cisek:2010jk}), using just two of the UGD models listed in Section~\ref{Sec_UGD} and switching our attention on the effects of the strange-quark mass, characterizing the $\phi$ meson.

\section{Polarized $\phi$-meson leptoproduction}
\label{Sec_phi}

The diffractive electroproduction of vector mesons, $\gamma^*\,p \to V p$,
has attracted much attention at the HERA collider (for a review
see e.g. Ref.~\cite{Ivanov:2004ax}) and is to be expected an important subject in future experiments e.g. at an electron-ion collider (EIC)~\cite{Accardi:2012qut}.
We are interested in the limit of large $\gamma^*\,p$ center-of-mass energy $W$, $s\equiv W^2\gg Q^2\gg\Lambda_{\text{QCD}}^2$, 
which implies small gluon longitudinal fraction $x=(Q^2+m_V^2)/(W^2+ Q^2 - m_p^2) \sim Q^2/W^2$.
In this kinematics, the photon virtuality $Q^2$ gives a handle on the dominant size of color dipoles in the $\gamma^* \to V$ transition and thus allows to study a transition from the hard, perturbative (small dipole), to the soft,
nonperturbative (large dipole), regimes of scattering.
In momentum space, the color dipole approach has its correspondence
in the $\kappa$-factorization, where the main ingredient is the UGD.
At large  photon virtualities the diffractive cross section is a
sensitive probe of the proton UGD.

The $\kappa$-factorization formalism reviewed in Ref.~\cite{Ivanov:2004ax}, 
includes besides the transverse momentum of gluons also the 
transverse momentum of quark and antiquark in the vector meson as
encoded in the light-cone wave-function of the meson.
This approach was used with some success in Ref.~\cite{Cisek:2010jk}. At very large $Q^2$ one may expect, that the 
relative transverse motion of (anti-)quarks in the bound state becomes negligible, and can be integrated out. Then, regarding the vector meson
only a dependence on the longitudinal momentum fraction of quarks
encoded in DAs is left. 

A quite general factorization formalism of vector meson production
in deep inelastic scattering was formulated in Refs.~\cite{Anikin:2009bf,Anikin:2011sa} and has been recently applied
in Ref.~\cite{Bolognino:2018} to the diffractive deep inelastic production
of $\rho$ mesons. While the production of longitudinal vector mesons
involves the leading twist DA, similar to the one introduced for other
hadronic processes in Refs.~\cite{Radyushkin:1977gp,Lepage:1979zb,Lepage:1980fj,Chernyak:1983ej},
for transverse vector mesons higher twists are involved and corresponding DAs studied in Ref.~\cite{Ball:1998sk} are needed.
The recent analysis of helicity amplitudes for $\rho$-meson production~\cite{Bolognino:2018} showed that this approach may be useful in testing UGDs. Therefore, in order to test the formalism further, we wish to focus on and investigate the exclusive photoproduction of $\phi$ meson~\cite{Bolognino:2019pba}:
\begin{center}
	$\gamma^* \, p \rightarrow \phi \, p$ \,.
\end{center}
Corresponding experimental data were obtained by 
the H1~\cite{Aaron:2009xp, Adloff:2000nx} and ZEUS\cite{Chekanov:2005cqa} collaborations at HERA.
We will show how the $\kappa$-factorization approach 
of~\cite{Ivanov:2000cm} matches to the higher-twist DA expansion,
at least in the Wandzura-Wilczek (WW) approximation, where no explicit 
$q \bar q g$ contributions are included.
Future applications at EIC may require the inclusion of
next-to-leading-order contributions. Here the approach based on DAs
has the advantage that for the case of collinear partons in the final state, the necessary techniques for the calculation of NLO impact factors are well advanced~\cite{Ivanov:2005gn,Ivanov:2004pp}.

This Section is organized as follows: the first part is devoted to the summary of the theoretical
framework of calculating the helicity amplitudes within 
the  $\kappa$-factorization approach.
The second part shows the cross sections of the process and 
the effects due to UGDs and/or due to the strange-quark mass.
Here we take advantage of the fact that we can rather straightforwardly
derive the massive impact factor in the WW approximation from the 
light-cone wave function approach.
As a byproduct we show how higher twist DAs can be obtained from the
light-cone wave-functions which may be interesting for the application
of various light-cone models also to other vector mesons. 

A comparison with the H1 and ZEUS measurements will be presented and then we will give our conclusions.

\subsection{Theoretical framework}
\label{theo_phi}
In the high-energy regime, $s\equiv W^2\gg Q^2\gg\Lambda_{\text{QCD}}^2$, which implies small $x=(Q^2+m_V^2)/(W^2+ Q^2 - m_p^2) \sim Q^2/W^2$, the forward helicity amplitude $T_{\lambda_V\lambda_\gamma}$ can be expressed, in the $\kappa$-factorization, as
the convolution of the $\gamma^*\rightarrow V$ impact factor (IF),
$\Phi^{\gamma^* \to V}_{\lambda_V,\lambda_\gamma}(\kappa^2,Q^2)$, with the UGD, ${\cal F}(x,\kappa^2)$.

Our normalization of the impact factor is chosen, such that the forward amplitude 
for the $\gamma^*\,p \to V\,p $ process reads
\begin{equation}
\label{amplitude_phi}
\Im m T_{\lambda_V\lambda_\gamma}(s,Q^2) = s \int \dfrac{d^2\kappa}
{(\kappa^2)^2}\Phi^{\gamma^* \to V}_{\lambda_V,\lambda_\gamma}(\kappa^2,Q^2)
{\cal F}(x,\kappa^2)\,.
\end{equation}
Here, the UGD is related to the collinear 
gluon parton distribution as
\begin{eqnarray}
xg(x,\mu^2) = \int^{\mu^2} {d \kappa^2 \over \kappa^2} {\cal F}(x,\kappa^2) \, .
\end{eqnarray}

We now turn to the two different approaches to the impact factors 
which we want to compare in this work.

\subsection{Distribution amplitude expansion}
\label{DA_exp}
We start with the scheme based on the collinear factorization of the
meson structure which was worked out in Refs.~\cite{Anikin:2009bf,Anikin:2011sa} and was used recently for $\rho$-meson electroproduction in Ref.~\cite{Bolognino:2018}, whose we gave a detailed discussion in Section~\ref{Sec_rho}. 
This approach starts from the observation, that at large $Q^2$ 
the transverse internal motion of partons in the meson can be neglected.

The longitudinal impact factor is expressed in terms of the 
standard twist-2 DA.
In the normalization adopted by us, the IF for the $L \to L$ transition 
reads
\begin{equation}
\label{Phi_LLphi}
\Phi^{\gamma^* \to V}_{0,0}(\kappa^2,Q^2) = {4 \pi \alpha_S e_q \sqrt{4 \pi \alpha_{\text{em}}} f_V \over N_c Q}
\int^{1}_{0}dy\, \varphi_1(y;\mu^2)\left(\frac{\alpha}{\alpha + y\bar{y}}\right)
\,,
\end{equation}
where $\alpha = \kappa^2/Q^2$, $y$ is the fraction of the mesons's lightcone-plus momentum 
carried by the quark, $\bar y = 1-y$, and $\varphi_1(y;\mu^2)$ is the twist-2 DA. It is normalized as
\begin{eqnarray}
\int_0^1 dy \, \varphi_1(y;\mu^2) = 1
\end{eqnarray}
and we recall its asymptotic form 
\begin{equation}
\label{phi_phi}
\varphi_1(y; \mu^2) \xrightarrow{\mu^2\rightarrow \infty} \varphi_1^{as}(y) = 6y\bar{y} \, .
\end{equation}
The expression for the transverse case is
\[
\hspace{-6.8cm}\Phi^{\gamma^* \to V}_{+,+}(\kappa^2,Q^2) = {2 \pi \alpha_S e_q \sqrt{4 \pi \alpha_{\text{em}}} f_V m_V\over N_c Q^2}
\]
\[
\hspace{-0.6cm}\times \left\{ \int^{1}_{0} dy \frac{\alpha (\alpha +2 y \bar{y})}{y\bar{y}
	(\alpha+y\bar{y})^2}\right. \left[(y-\bar{y})\varphi_1^T(y;\mu^2)
+ \varphi_A^T(y;\mu^2)\right]
\]
\[
\hspace{-5.7cm}-\int^{1}_{0}dy_2\int^{y_2}_{0}dy_1 \frac{y_1\bar{y}_1\alpha}
{\alpha+y_1\bar{y}_1}
\]
\[
\hspace{-0.08cm}\times \left[\frac{2-N_c/C_F}{\alpha(y_1+\bar{y}_2)
	+y_1\bar{y}_2}
-\frac{N_c/C_F}{y_2 \alpha+y_1(y_2-y_1)}\right]M(y_1,y_2;\mu^2)
\]
\[
\hspace{1.8cm}+ \int^{1}_{0}dy_2\int^{y_2}_{0}dy_1 \Big[ {2+N_c/C_F \over \bar{y}_1}
+ {y_1 \over \alpha+y_1\bar{y}_1} 
\left({(2-N_c/C_F)y_1\alpha \over
	\alpha(y_1+\bar{y}_2) + y_1\bar{y}_2}-2\right)
\]
\begin{equation}
\hspace{-1.2cm}-\frac{N_c}{C_F}\frac{(y_2-y_1)\bar{y}_2}{\bar{y}_1}\frac{1}{\alpha\bar{y}_1+(y_2-y_1)\bar{y}_2}\Big]\, S(y_1,y_2;\mu^2) \Big\}\,,
\label{Phi_TTphi}
\end{equation}
where:
\begin{equation}
C_F=\frac{N_c^2-1}{2N_c}\,,
\end{equation}
\begin{equation}
B(y_1,y_2;\mu^2)  =-5040 y_1 \bar{y}_2 (y_1-\bar{y}_2) (y_2-y_1)\,,
\end{equation}
\begin{equation}
D(y_1,y_2;\mu^2)  =-360 y_1\bar{y}_2(y_2-y_1)
\left(1+\frac{\omega^{A}_{\{1,0\}}(\mu^2)}{2}\left(7\left(y_2-y_1\right)-3\right)
\right)\,,
\end{equation} 
and where the three-body DAs read:
\begin{equation}
M(y_1,y_2;\mu^2) = \zeta^{V}_{3V}(\mu^2) B(y_1,y_2;\mu^2) - \zeta^{A}_{3V}(\mu^2) D(y_1,y_2;\mu^2)\,,
\end{equation}
\begin{equation}
S(y_1,y_2;\mu^2) = \zeta^{V}_{3V}(\mu^2) B(y_1,y_2;\mu^2) + \zeta^{A}_{3V}(\mu^2) D(y_1,y_2;\mu^2)\,
\end{equation}
with the dimensionless coupling constants $\zeta^{V}_{3V}(\mu^2)$ and $\zeta^{A}_{3V}(\mu^2)$ defined as
\begin{equation}
\label{zeta}
\zeta^{V}_{3V}(\mu^2) = \frac{f^{V}_{3V}(\mu^2)}{f_V}\,, \qquad 	\zeta^{A}_{3V}(\mu^2) = \frac{f^{A}_{3V}(\mu^2)}{f_V}\,.
\end{equation}
The dependence on the factorization scale $\mu^2$ can be determined from evolution equations~\cite{Ball:1998sk} (see also Appendix B in Ref.~\cite{Anikin:2011sa}), with the initial condition at a renormalization scale $\mu_0 = 1$ GeV.\\
The DAs $\varphi^T_1(y;\mu^2)$ and $\varphi^T_A(y;\mu^2)$ in Eq.~\eqref{Phi_TTphi}
encompass both genuine twist-3 and Wandzura-Wilczek~(WW)
contributions~\footnote{Genuine terms are related to
	$B(y_1,y_2;\mu^2)$ and $D(y_1,y_2;\mu^2)$; WW contributions, instead, are those obtained
	in the approximation in which $B(y_1,y_2;\mu^2)=D(y_1,y_2;\mu^2)=0$. For their expressions in this last case see Eq.~\eqref{BD} and Ref.~\cite{Anikin:2011sa}.}~\cite{Ball:1998sk}. 
\subsection{Light-cone wave function (LCWF) approach}
\label{LCWF}
In the light-cone $\kappa$-factorization approach, the calculation proceeds in a 
slightly different way. Here one calculates the amplitude for the 
$ \gamma^*\,p \to q \bar q p$ diffractive process and projects the 
final state $q \bar q$ pair onto the vector meson state.
We treat the $\phi$-meson as a pure $s \bar s$ state. The meson of momentum $P= (P_+, m_\phi^2 /(2 P_+), 0)$ is described by the $s \bar s$ light-cone wave-function as 
\begin{eqnarray}
\ket{\phi, P_+, \lambda_V} = \int {dy d^2 k \over y \bar y} \, \Psi^{(\lambda_V)}_{\lambda \bar \lambda}(y,k) \, \ket{ s(yP_+,k,\lambda)  \bar s(\bar yP_+,-k,\bar \lambda)} + \dots
\end{eqnarray} 

The amplitude for diffractive vector meson production then takes the form
\begin{eqnarray}
\Im m   T_{\lambda_V,\lambda_\gamma} (s,Q^2) 
= s \, \int {dy d^2 k \over y \bar y 16 \pi^3} \sum_{\lambda \bar \lambda} {\cal M}^{(\lambda_\gamma)}_{\lambda \bar \lambda}(\gamma^*p \to s \bar s p) \Psi^{(\lambda_V)*}_{\lambda \bar \lambda} (y,k) \, .
\end{eqnarray}
The explicit expressions for the diffractive amplitudes can be found in Ref.~\cite{Ivanov:2004ax}. Here we are interested only in the forward scattering limit of vanishing transverse momentum transfer, where only the helicity conserving amplitudes with $\lambda_V = \lambda_\gamma$ contribute.

We can easily read off the following expressions for the impact factors of interest.
The longitudinal transition IF reads 
\begin{eqnarray}
\Phi^{\gamma^* \to \phi}_{0,0}(\bkappa^2,Q^2) 
&=& \sqrt{4 \pi \alpha_{\text{em}}} e_q \, 8 \pi  \alpha_S(\mu^2) Q \int {dy d^2 k \over \sqrt{y \bar y} 16 \pi^3} I_0(k,\kappa) 
\nonumber \\
&\times& 
y \bar y \Big\{ \Psi^{(0)*}_{+-}(y,k) +  \Psi^{(0)*}_{-+}(y,k) \Big \}\,.
\end{eqnarray}
For the transverse transition IF we obtain 
\begin{eqnarray}
\Phi^{\gamma^* \to \phi}_{\pm,\pm} (\kappa^2,Q^2) &=&  
\sqrt{4 \pi \alpha_{\text{em}}} e_q
4 \pi \alpha_S(\mu^2)  \int {dy d^2 k \over \sqrt{y \bar y} 16 \pi^3}  \nonumber \\
&\times& 
\Big[ (e(\pm) \cdot I_1 (k,\kappa))
\Big\{ (y - \bar y) \Big( \Psi^{(\pm)*}_{+-}(y,k) +  \Psi^{(\pm)*}_{-+}(y,k) \Big) 
\nonumber \\
&& + \Psi^{(\pm)*}_{+-}(y,k) -  \Psi^{(\pm)*}_{-+}(y,k)\Big\} 
+ \sqrt{2} m_q I_0(k,\kappa)  \Psi^{(\pm)*}_{++}(y,k) \Big]
\, .
\nonumber \\
\end{eqnarray}
Here 
\begin{eqnarray}
I_0(k,\kappa) = {1 \over k^2 + \varepsilon^2} - {1 \over (k+ \kappa)^2 + \varepsilon^2} , \quad
I_1 (k,\kappa) = {k \over k^2 + \varepsilon^2} - {k + \kappa \over (k+ \kappa)^2 + \varepsilon^2}, 
\end{eqnarray}
and $\varepsilon^2 = m_q^2 + y \bar y Q^2$.
We now want to compare these results with the twist expansion
approach presented in the previous chapters.
To this end, we should expand the impact factors around the 
limit of collinear kinematics for the $q \bar q$-pair.
While an analogous expansion around the small-$\kappa$ limit,
has been discussed in great detail, the analogous comparison to
leading and higher twist DAs is up to now 
missing. 

Expanding in $k^2/(\kappa^2 + \varepsilon^2) \ll 1$, we obtain
\begin{eqnarray}
I_0(k,\kappa) \approx {1 \over \varepsilon^2 } 
- {1 \over \kappa^2 + \varepsilon^2} = 
{\kappa^2 \over \varepsilon^2 (\kappa^2 + \varepsilon^2)} \, ,
\end{eqnarray}
and 
\begin{eqnarray}
I_1 (k,\kappa) \approx k {\kappa^2 \over \varepsilon^2 (\kappa^2 + \varepsilon^2)}  + {2 (k \cdot \kappa) \kappa \over (\kappa^2 + \varepsilon^2)^2} \to {\kappa^2 (\kappa^2 + 2 \varepsilon^2) \over \varepsilon^2 (\kappa^2 + \varepsilon^2)^2} k \, ,
\end{eqnarray}
where we performed the azimuthal average in the last step.
Inserting the expanded $I_0$ into the LL IF, we find
\[
\Phi^{\gamma^* \to \phi}_{0,0}(\kappa^2,Q^2) 
= \sqrt{4 \pi \alpha_{\text{em}}} e_q \, 8 \pi  \alpha_S(\mu^2) Q
\int_0^1 dy \, y \bar y {\kappa^2 \over \varepsilon^2 (\kappa^2 + \varepsilon^2)} \,
\]
\begin{equation}
\times {1 \over \sqrt{y \bar y}} \int{d^2 k \over 16 \pi^3}
\Big\{ \Psi^{(0)*}_{+-}(y,k) +  \Psi^{(0)*}_{-+}(y,k) \Big \} \theta(\mu^2 - k^2) \,
\label{IF_LL_expI0}
 \end{equation}
\[
= \sqrt{4 \pi \alpha_{\text{em}}} e_q { 4 \pi \alpha_S(\mu^2) f_V  \over N_c Q }\int_0^1 dy { y \bar y \over (y \bar y + \tau)} {\alpha \over (\alpha + y \bar y + \tau)} \, \varphi_1(y,\mu^2) \, . 
\]

Here we introduced the variables $\alpha = \kappa^2/Q^2$ and
$\tau = m_q^2/Q^2$. We see that we have obtained a generalization
to finite quark mass of the impact factor of Eq.~\eqref{Phi_LLphi}.
The helicity combination of the LCWF which appears under the $k$ integral gives rise to the leading twist DA 
of the longitudinally polarized vector meson, defined following the rules of Ref.~\cite{Lepage:1980fj} as
\begin{eqnarray}
f_V \varphi_1(y,\mu_0^2) = {2 N_c \over \sqrt{y \bar y}} \, \int {d^2k \over 16 \pi^3} \theta(\mu_0^2 - k^2)
\Big\{ \Psi^{(0)*}_{+-}(y,k) +  \Psi^{(0)*}_{-+}(y,k) \Big \}\,. 
\end{eqnarray} 
The scale $\mu^2$ in Eq.~\eqref{IF_LL_expI0} must be chosen such that the small-$k$ expansion is valid, i.e. $\mu^2 \sim (Q^2 + m_\phi^2)/4$.

We can now follow a similar strategy for the transverse IF.
To that end we introduce the following representations of the
higher twist DA's: 
\begin{eqnarray}
f_V \varphi_1^T(y,\mu_0^2) &=& {2 N_c \over \sqrt{y \bar y}  }
\int {d^2 k \over 16 \pi^3} \theta(\mu_0^2 - k^2)
(e(\pm) \cdot k)\Big\{ \Psi^{(\pm)*}_{+-}(y,k) +  \Psi^{(\pm)*}_{-+}(y,k) \Big \}\,, \nonumber \\
f_V \varphi_A^T(y,\mu_0^2) &=& {2 N_c \over \sqrt{y \bar y}  }
\int {d^2 k \over 16 \pi^3} \theta(\mu_0^2 - k^2)
(e(\pm) \cdot k)\Big\{ \Psi^{(\pm)*}_{+-}(y,k) -  \Psi^{(\pm)*}_{-+}(y,k) \Big \}\,, \nonumber \\
f_V \varphi_m(y,\mu_0^2) &=& 	{2 N_c \over \sqrt{y \bar y}  }
\int {d^2k \over 16 \pi^3} \theta(\mu_0^2 - k^2) \sqrt{2} m_q \Psi^{(\pm)*}_{++}(z,k)\,.
\end{eqnarray}
We notice, that
\begin{eqnarray}
\int_0^1 dy \, \varphi(y,\mu_0^2) = 1, \qquad \int_0^1 dy \, \varphi_1^T(y,\mu_0^2) = 0. 
\end{eqnarray}
The transverse IF that we derive is again a massive generalization of Eq.~\eqref{Phi_TTphi} and reads
\begin{eqnarray}
\Phi^{\gamma \to \phi}_{\pm,\pm}(\bkappa^2,Q^2) &=& \sqrt{4 \pi \alpha_{\text{em}}} e_q { 2 \pi \alpha_S(\mu^2) f_V  \over N_c Q^2 }
\int_0^1 {dy \over y \bar y + \tau} \Big\{ {\alpha (\alpha + 2 y \bar y + 2 \tau) \over (\alpha + y \bar y +\tau)^2} \nonumber \\
&\times& \hspace{-0.2cm}
\Big( (y - \bar y) \varphi_1^T(y,\mu^2) + \varphi_1^A(y,\mu^2) 
\Big) 
+ {\alpha \over \alpha + y \bar y + \tau} \varphi_m(y,\mu^2)
\Big \}.
\nonumber \\
\end{eqnarray}
We realize that up to the DA $\varphi_m$, which vanishes in the massless limit, the structure of the IF is exactly the same
as for the one of Eq.~\eqref{Phi_TTphi} neglecting the so-called genuine three particle distributions. The latter obviously
can appear only at the level of the $q \bar q g$-Fock state.

We now wish to give some explicit expressions for the DA's in
question. To this end, we use the $V \to q \bar q$ vertex
from Ref.~\cite{Ivanov:2004ax}, where the $\phi$-meson is treated as a pure $S$-wave bound state of strange quark and antiquark. 
For the relevant combinations of light-cone wave-functions we obtain in the case of the longitudinally polarized vector meson:
\begin{eqnarray}
\Psi^{(0)*}_{+-}(y,k) +  \Psi^{(0)*}_{-+}(y,k) = - 4M \sqrt{y \bar y} \Big\{ 1 + {(y- \bar y)^2 \over 4 y \bar y} {2 m_q \over M + 2 m_q}   
\Big\} \psi(y,k) \, .
\end{eqnarray}
The radial wave function $\psi(z,k)$ is normalized as
\begin{eqnarray}
N_c \int {dz d^2 k \over y \bar y 16 \pi^3} 2 M^2 \, |\psi(y,k)|^2 = 1\, .
\end{eqnarray}
Above $M^2 = (k^2 + m_q^2)/(y \bar y)$ is the invariant mass of the $s \bar s$-system.
We can now express the leading twist DA through the radial wave function as
\begin{eqnarray}
f_V \varphi_1(y,\mu_0^2) = {N_c \over 2 \pi^2}  \int_0^{\mu_0^2}  dk^2 
M \Big\{ 1 + {(y - \bar y)^2 \over 4 y \bar y} {2 m_q \over M + 2 m_q}   
\Big\} \psi(y,k) \, .
\end{eqnarray}
Now, for the higher twist DA's of the transversely polarized vector meson, we obtain
\begin{eqnarray}
f_V \varphi_1^T(y,\mu_0^2) &=& (y - \bar y){N_c \over 8 \pi^2}  \int_0^{\mu_0^2}  dk^2 \, k^2 
{M \over M + 2 m_q} {\psi(y,k) \over y \bar y} \,, \nonumber \\
f_V \varphi_A^T(y,\mu_0^2) &=& {N_c \over 4 \pi^2}  \int_0^{\mu_0^2}  dk^2 \, k^2  {\psi(y,k) \over y \bar y} \,, \nonumber \\ 
f_V \varphi_m(y,\mu_0^2) &=& m_q^2 {N_c \over 4 \pi^2}  \int_0^{\mu_0^2}  dk^2 \, 
\Big\{ 1 + {k^2 \over m_q (M+2 m_q)} \Big \} {\psi(y,k) \over y \bar y} \,. \nonumber \\ 
\end{eqnarray}

\subsection{Useful parameters, cross section and $b$-slope}
\label{par_cs_b}
Typical constants for $\phi$ meson, entering DAs and IFs, used in numerical computations, are provided in the following tables:
\begin{table}[htb]
	$$
	\renewcommand{\arraystretch}{1.4}
	\addtolength{\arraycolsep}{3pt}
	\begin{array}{|c|c|}
	\hline
	V & \phi\\ \hline
	f_V [$\text{GeV}$] & 0.254\\
	\zeta_{3V}^A & 0.032\\
	\zeta_{3V}^V  & 0.013\\
	\omega_{1,0}^A & -2.1 \\
	\omega_{1,0}^V & 28/3 \\
	\hline
	\end{array}
	$$
	\caption[]{\emph{Experimental value of coupling to the vector
			current \protect{\cite{PDG}}\label{tab:B1} (first row); couplings entering the vector meson DAs at the scale $\mu_0 = 1$ GeV.}}
	\renewcommand{\arraystretch}{1.4}
	\addtolength{\arraycolsep}{3pt}
\end{table}
\FloatBarrier
\begin{table}[htb]
	$$
	\renewcommand{\arraystretch}{1.4}
	\addtolength{\arraycolsep}{3pt}
	\begin{array}{|c|c|}
	\hline
	V & \phi\\ \hline
	m_Vf^{A}_{3V}[$\text{GeV$^2$}$] & 3.37\cdot10^{-3} \\
	m_Vf^{V}_{3V}[$\text{GeV$^2$}$] & 5.26\cdot10^{-3} \\
	\hline
	\end{array}
	$$
	\caption[]{\emph{Decay constants obtained from Eq.~\eqref{zeta}.}}
	\renewcommand{\arraystretch}{1.4}
	\addtolength{\arraycolsep}{3pt}
\end{table}	\renewcommand{\arraystretch}{1.4}
\FloatBarrier
The imaginary part of the amplitude in Eq.~\eqref{amplitude} which enters the expression of the cross section for transverse and longitudinal polarization, can be written as
\begin{equation}
\sigma_L\,(\gamma^*\,p \rightarrow \phi\,p) = \frac{1}{16 \pi b(Q^2)}\left|\frac{T_{00}(s,Q^2)}{W^2}\right|^2\,,
\end{equation}
\begin{equation}
\sigma_T\,(\gamma^*\,p \rightarrow \phi\,p) = \frac{1}{16 \pi b(Q^2)}\left|\frac{T_{11}(s,Q^2)}{W^2}\right|^2\,,
\end{equation}
where $b(Q^2)$ is the slope parameter which depends on the virtuality of the photon, parametrized following the Eq.~\eqref{slope_B} presented in the Ref.~\cite{Nemchik:1997xb} and introduced in the Section~\ref{Cross_section_th}:
\begin{equation}
\label{slope_B_phi}
b(Q^2) = \beta_0 - \beta_1\,\log\left[\frac{Q^2+m_\phi^2}{m^2_{J/\psi}}\right]+\frac{\beta_2}{Q^2+m_\phi^2}\,,
\end{equation}
with $\beta_0 = 7.0$ GeV$^{-2}$, $\beta_1 = 1.1$  GeV$^{-2}$ and $\beta_2 = 1.1$.
\begin{figure}[h!]
	\centering	
	\includegraphics[scale=0.50,clip]{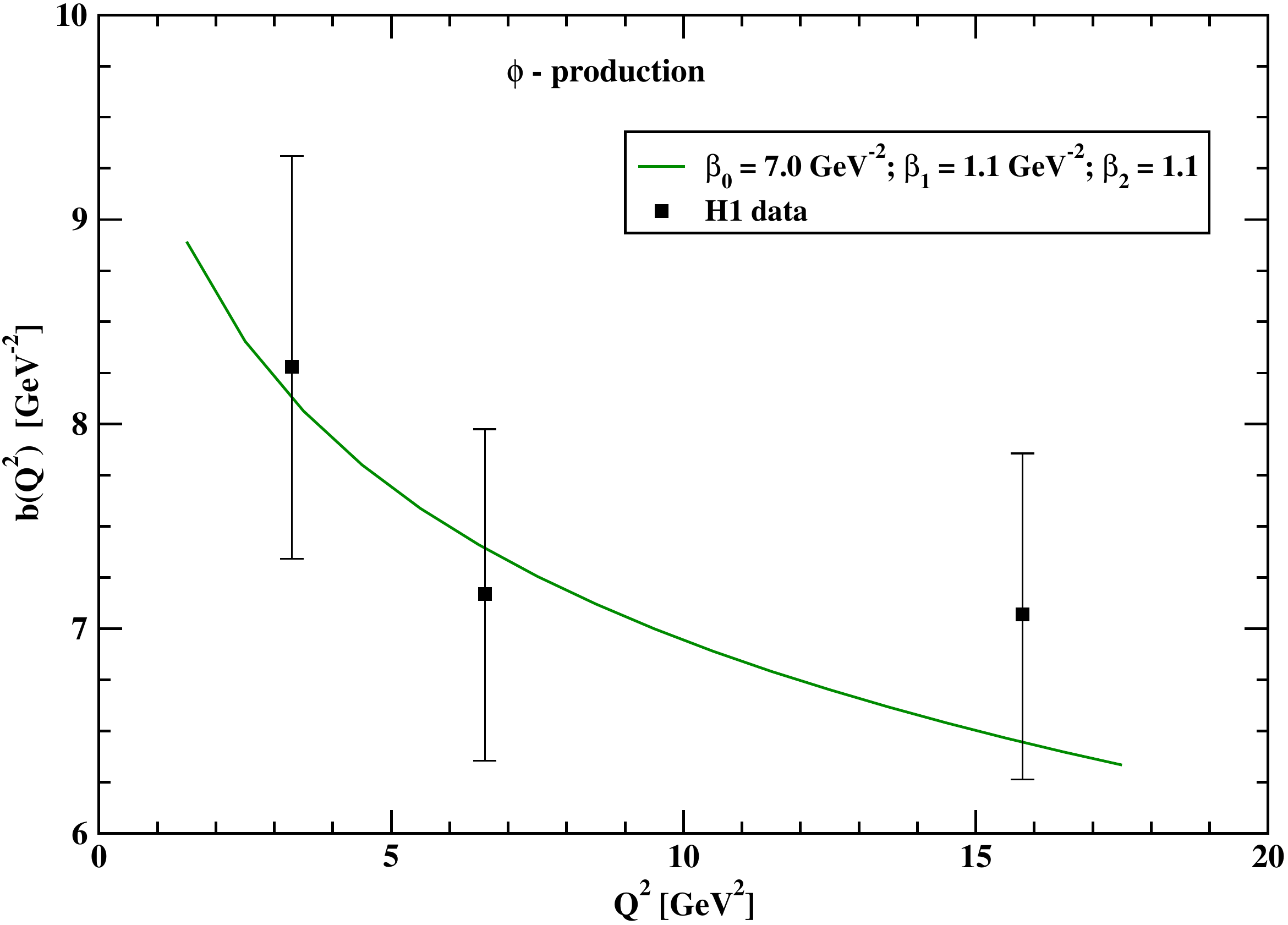}
	\caption{\emph{$Q^2$-dependence of the b-slope for $\phi$-meson production in the $\gamma^*\,p \to \phi\,p$ reaction. Due to the high uncertainty of the experimental data, we keep the standard choice of $\beta_0$, $\beta_1$ and $\beta_2$ parameters from Ref.~\cite{Nemchik:1997xb} for all our results.}}
	\label{fig:bslope}
\end{figure} 
\FloatBarrier
The full cross section is a sum of longitudinal and transverse
components, and it reads
\begin{equation}
\sigma_{tot} (\gamma^*\,p \rightarrow V\,p) = \sigma_T + \epsilon\,\sigma_L\,,
\end{equation}
where $\epsilon \approx 1$ due to HERA kinematics.

\subsection{Numerical analysis}
\label{analysis_phi}

\begin{figure}[h!]
	\centering
	\includegraphics[scale=0.50,clip]{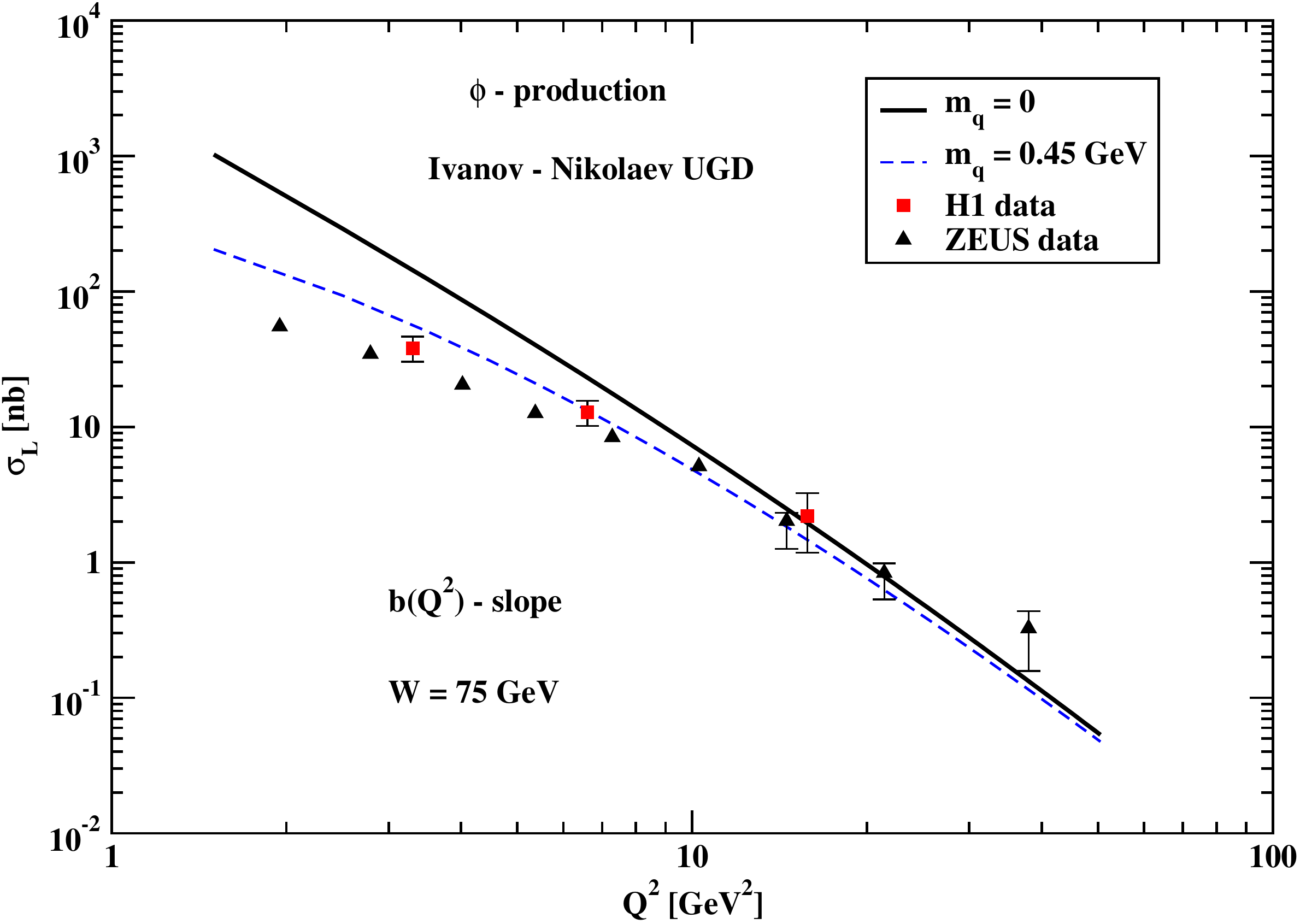}
	\includegraphics[scale=0.50,clip]{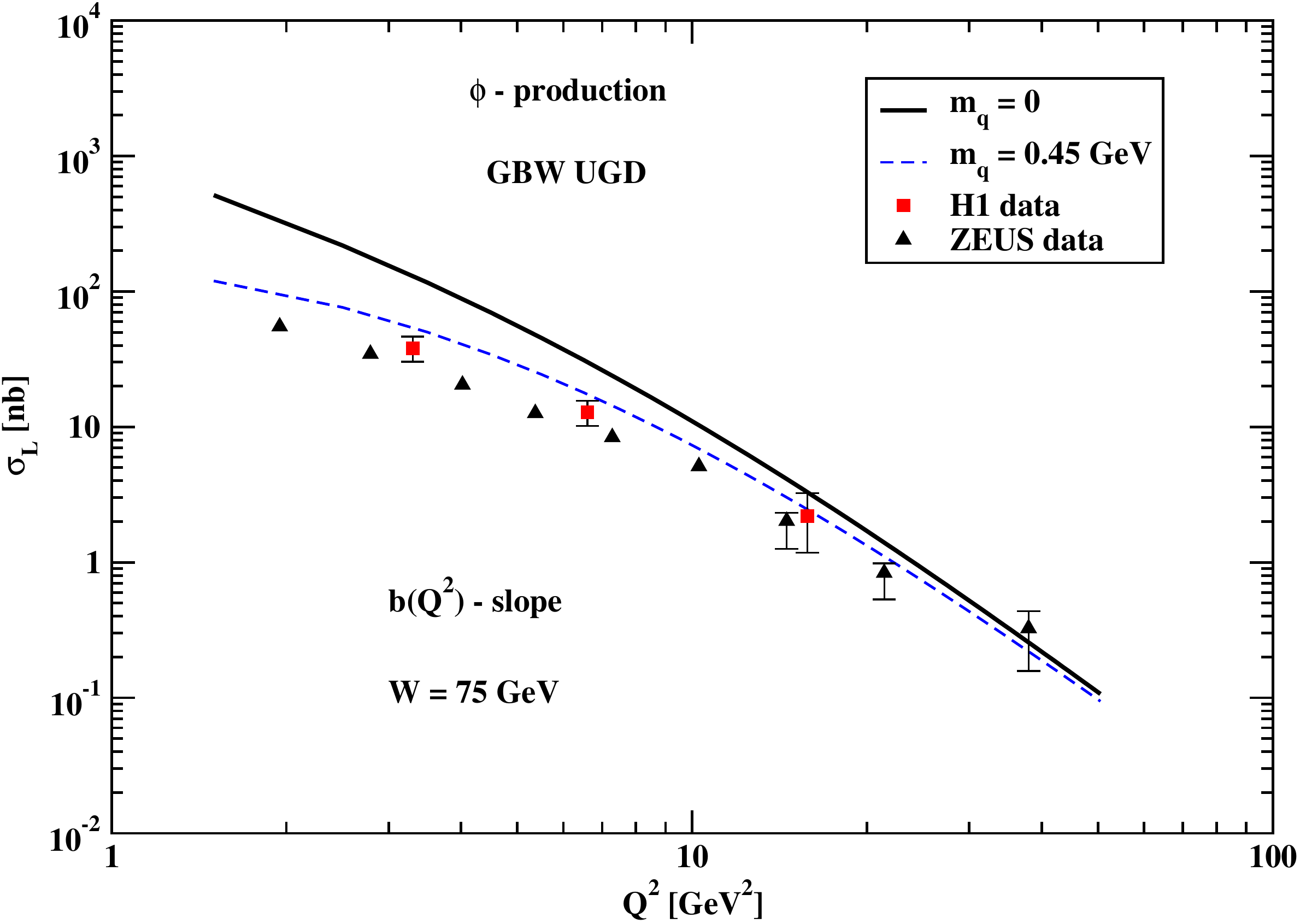}
	\caption{\emph{$Q^2$-dependence of longitudinal cross section $\sigma_L$ of $\phi$-meson production, at $W = 75$~GeV, in comparison with experimental data of the H1~\cite{Aaron:2009xp} and ZEUS~\cite{Chekanov:2005cqa} collaborations. The result is obtained within the $\kappa$-factorization using the Ivanov-Nikolaev UGD model (top panel) and the GBW one (bottom panel). The solid lines are for the case when the strange-quark mass is neglected and the dashed lines for the strange-quark mass fixed at $m_q = 0.45$ GeV. In both cases, the asymptotic DA is used.}}
	\label{fig:sigmaL_IN_DAs_mq}
\end{figure}

In what follows we present theoretical predictions adopting two different UGD models between those ones illustrated in Section \ref{Sec_UGD}:
\begin{itemize}
	\item the Ivanov-Nikolaev parametrization, endowed with soft and hard components to probe both large and small transverse momentum region (see Ref.~\cite{Ivanov:2000cm} for further details);
	\item the model provided by Golec-Biernat and W\"usthoff (GBW), which derives from the dipole cross section for the scattering of a $q\bar{q}$ pair off a nucleon~\cite{GolecBiernat:1998js}.
\end{itemize}
We start from calculating longitudinal cross section using 
the formalism described in the previous section.
In Fig.~\ref{fig:sigmaL_IN_DAs_mq} we show results of our calculation
for the Ivanov-Nikolaev (left panel) and GBW (right panel) UGDs.
In order to get these predictions, the asymptotic DA has been used. This calculation has been obtained for $W$ = 75~GeV.
We observe that the cross sections obtained for massless strange quarks
(black solid line) overestimate the experimental cross section below $Q^2 <$ 10 GeV$^2$.
\begin{figure}[htb]
	\centering
	\includegraphics[scale=0.50,clip]{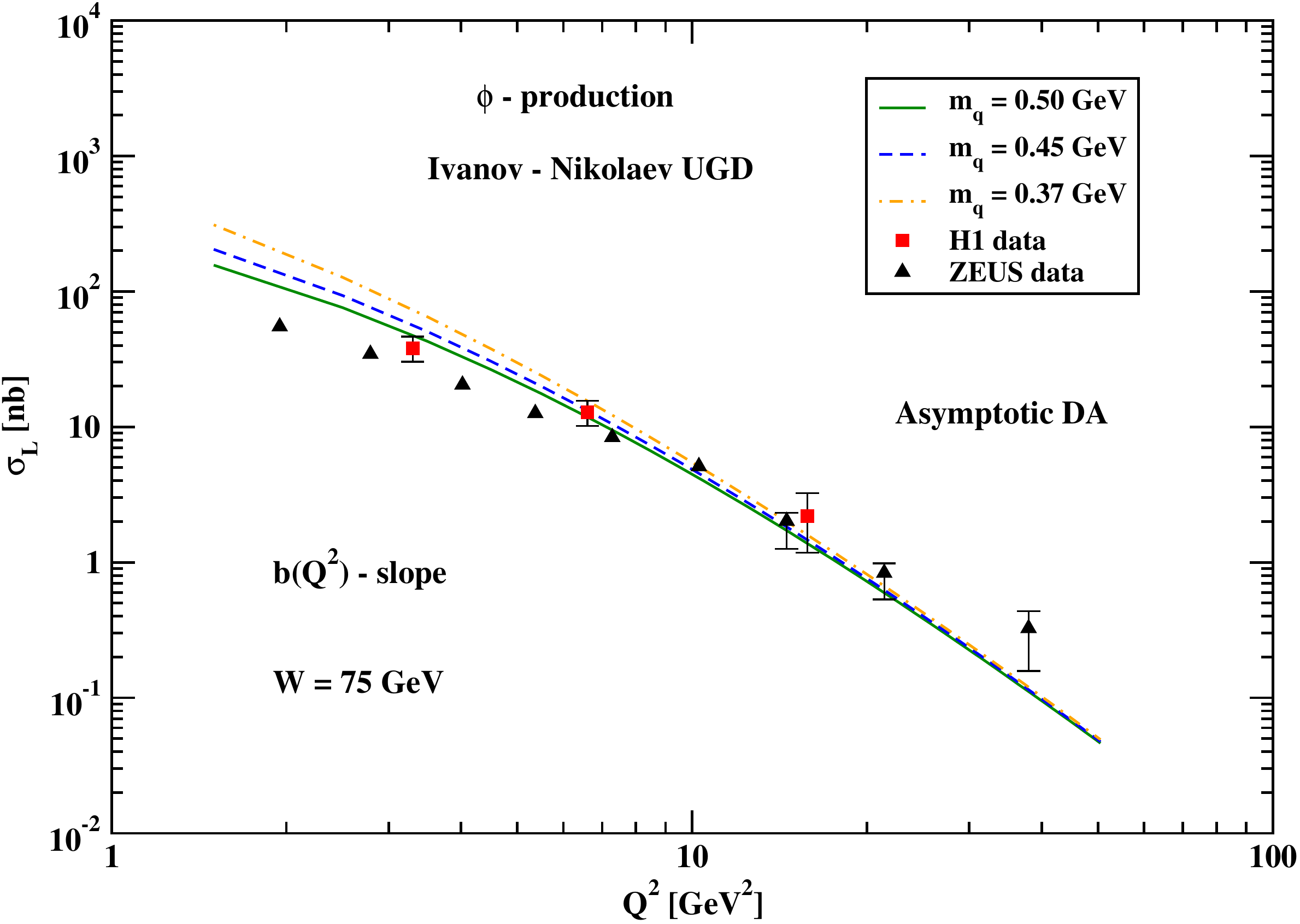}
	\includegraphics[scale=0.50,clip]{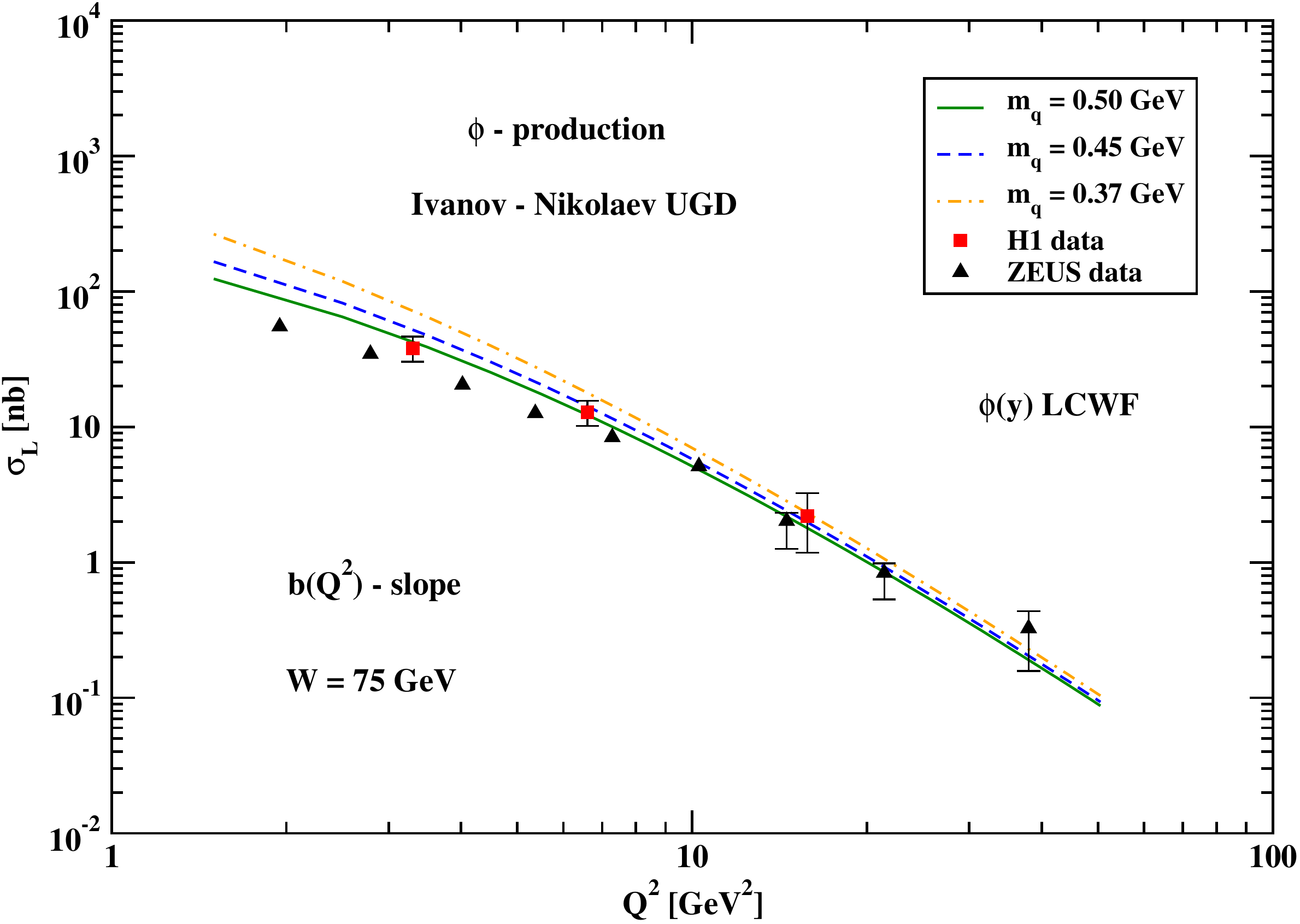}
	\caption{\emph{$\sigma_L$ for the asymptotic DA (top panel) and for the LCWF DA (bottom panel). Results for three different strange-quark-mass values are shown. Predictions are given using the Ivanov-Nikolaev UGD model.}}
	\label{fig:systematics_sigL_asy_da}
\end{figure}
In both cases we present also our results when using quarks/antiquarks
with effective masses (as described in the previous section).
Then a good description of the experimental data is obtained for both the Ivanov-Nikolaev and GBW UGDs.
\begin{figure}[htb]
	\centering
	\includegraphics[scale=0.50,clip]{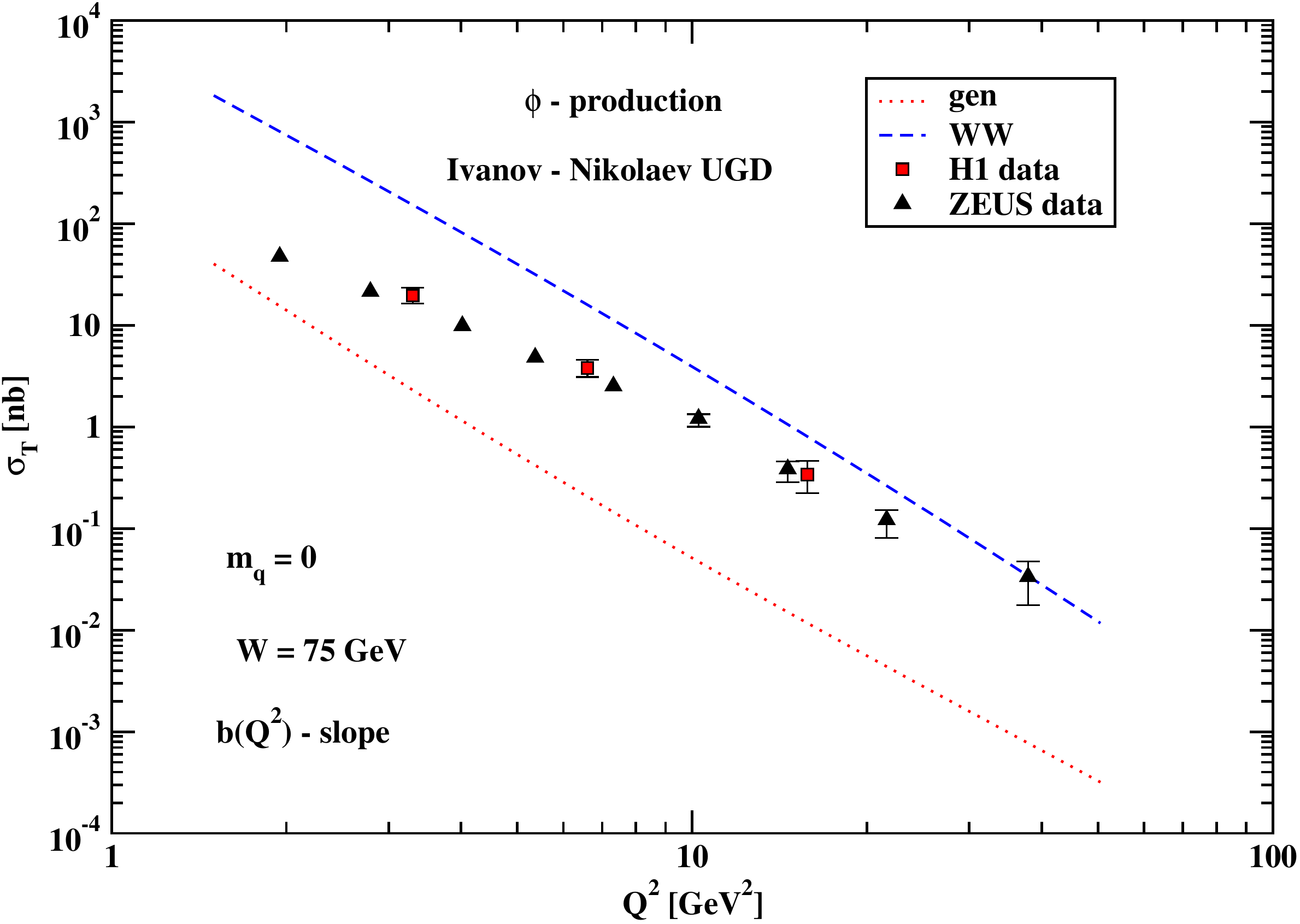}
	\caption{\emph{$Q^2$-dependence of the transverse cross section $\sigma_T$ neglecting the quark mass $m_q$. The WW and the genuine three-parton contributions are shown separately.}}
	\label{fig:WW_GEN_sigT}
\end{figure}
How much the cross section depends on the quark mass is discussed in 
Fig.~\ref{fig:systematics_sigL_asy_da}
for asymptotic (left panel) and LCWF (right panel) DAs, respectively.
The best description of the data is obtained with $m_q$ = 0.5 GeV.
A similar result was found in Ref.~\cite{Cisek:2010jk} within the $\kappa$-factorization
approach with a Gaussian $s\bar{s}$ light-cone wave-function for the $\phi$ meson.\\
Now we pass to the transverse cross section as a function of photon virtuality.
\begin{figure}[htb]
	\centering	
	\includegraphics[scale=0.50,clip]{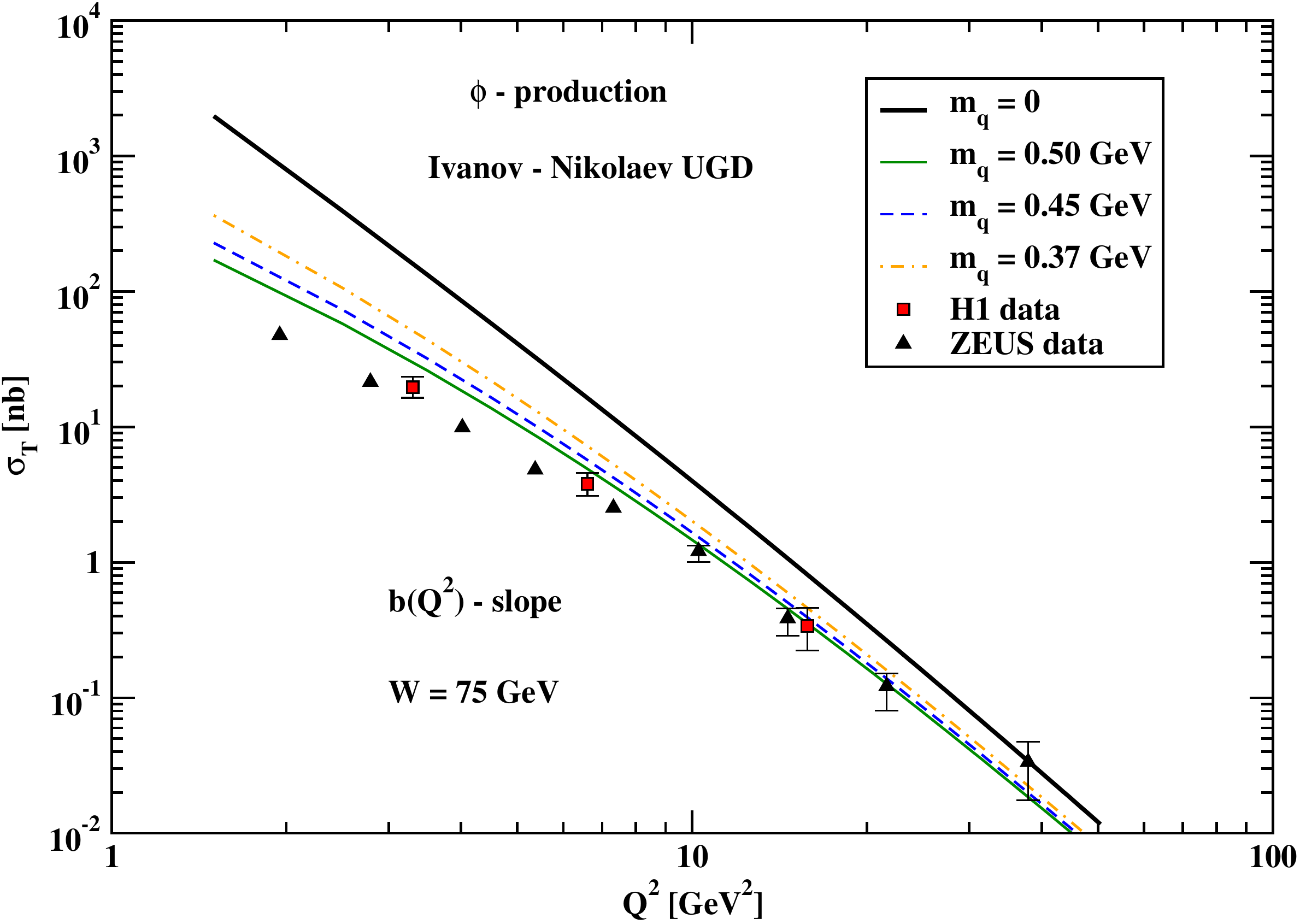}
	\includegraphics[scale=0.50,clip]{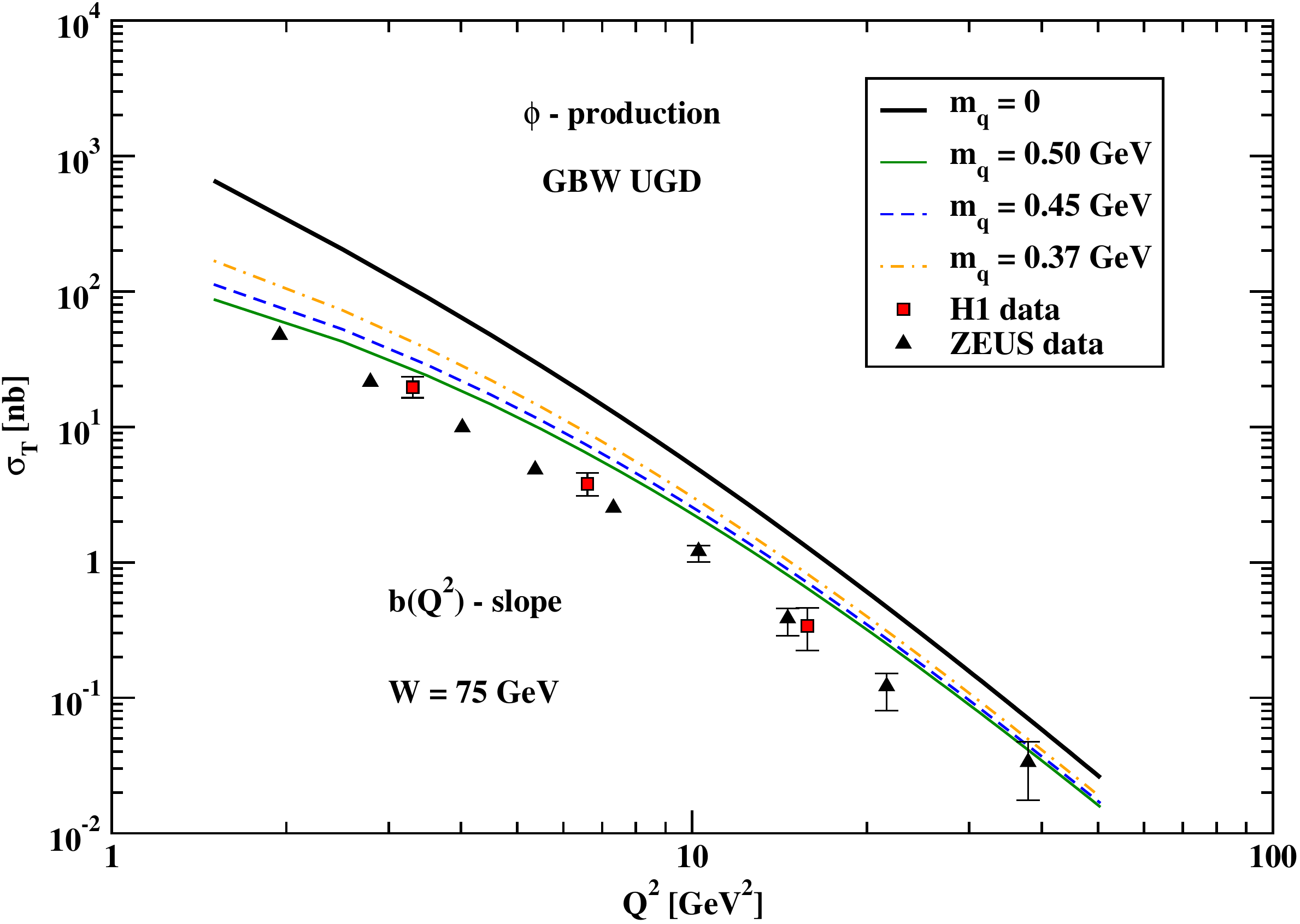}
	\caption{\emph{$Q^2$-dependence of the transverse cross section $\sigma_T$ for the $\phi$-meson production in the $\gamma^*\,p \to \phi\,p$ reaction, at $W = 75$~GeV, in comparison with the H1~\cite{Aaron:2009xp} and ZEUS~\cite{Chekanov:2005cqa} experimental data. The result is obtained within the $\kappa$-factorization using the Ivanov-Nikolaev UGD model (top panel) and the GBW one (bottom panel). The thick solid line is for the case when the strange-quark mass is neglected. The thin lines show the result for the three different values of the strange-quark mass. In both cases, the curves were obtained with the asymptotic choice of the DA.}}
	\label{fig:sigT_GBW_IN}
\end{figure}
In Fig.~\ref{fig:WW_GEN_sigT} we show the cross section separately for the 
Wandzura-Wilczek and genuine three-parton contributions. In this calculation massless quarks
were used. We observe that the transverse cross section for the genuine
three-parton contribution is rather small. However, the WW contribution for massless quarks, 
similarly as for the longitudinal cross section, overpredicts the H1 and ZEUS
data. Can this be explained as due to the mass effect discussed in the
previous section?\\
In Fig.~\ref{fig:sigT_GBW_IN} we show how the WW contribution changes when
including the mass effect discussed in the previous section.
Inclusion of the mass effect improves the description of the H1 and ZEUS experimental
transverse cross section. The description is, however, not perfect. In Fig.~\ref{fig:sigT_GBW_IN} we present a similar result for both UGD models.
Unlike for the the longitudinal cross section, here the GBW overpredicts
the experimental data in the whole range of virtuality, while the Ivanov-Nikolaev model only at smaller $Q^2$ values. A reasonable result is obtained when including mass effect.\\
We wish to show also results for the $\sigma_L/\sigma_T$ ratio (see Fig.~\ref{fig:ratio_comparison_UGD}) as a function of photon
virtuality $Q^2$ for the Ivanov-Nikolaev and GBW UGDs.
In this calculation the quark mass was fixed for $m_q$ = 0.45 GeV.
The Ivanov-Nikolaev UGD describes much better the H1 and ZEUS data.

 \begin{figure}[htb]
 	\centering	
 	\includegraphics[scale=0.50,clip]{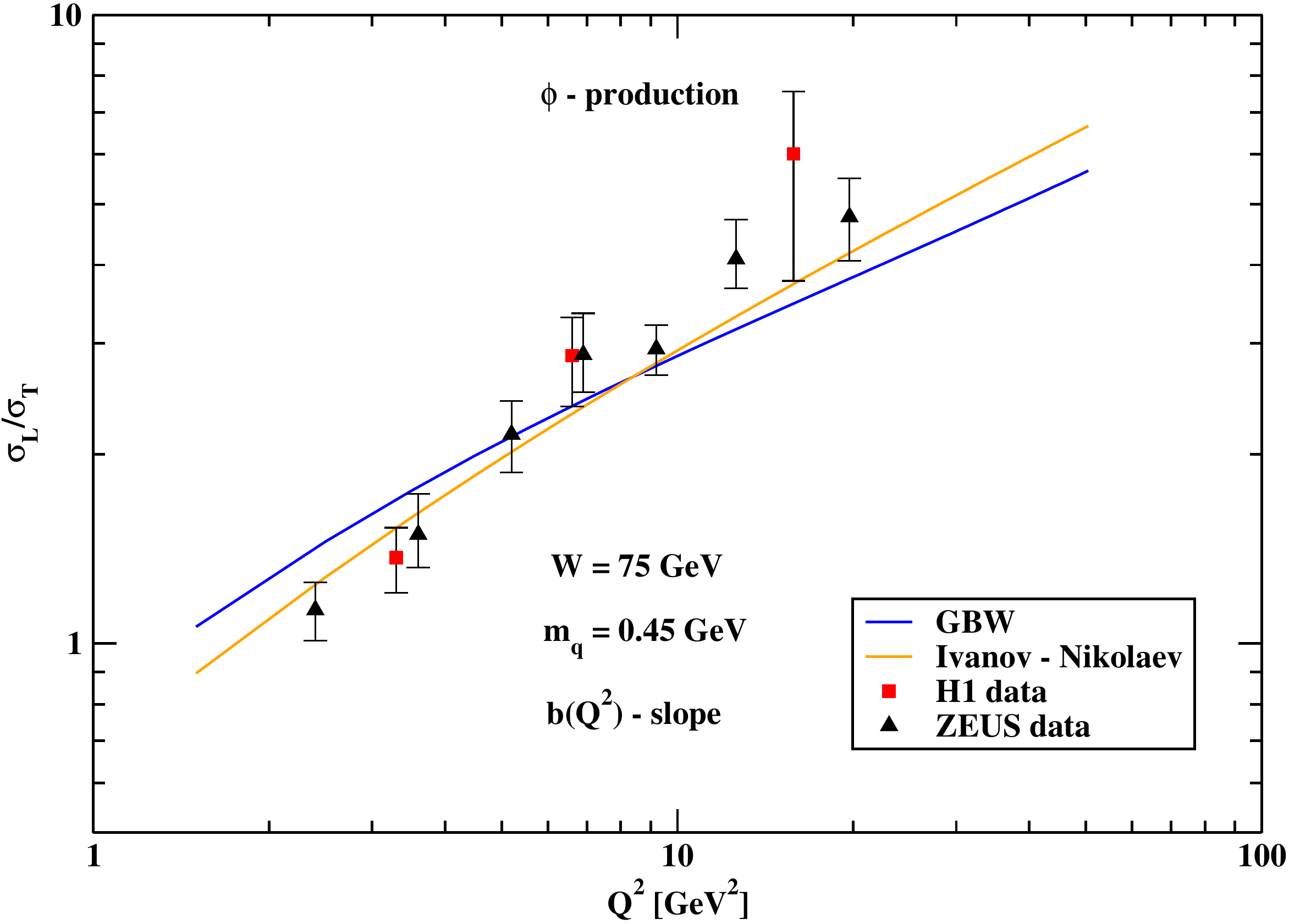}
 	\caption{\emph{$Q^2$-dependence of cross section ratio $\sigma_L/\sigma_T$ for the $\phi$-meson production in the $\gamma^*\,p \to \phi\,p$ reaction, at $W = 75$~GeV, in comparison with experimental data of the H1~\cite{Aaron:2009xp} and ZEUS~\cite{Chekanov:2005cqa} collaboration. The prediction is performed in the $\kappa$-factorization approach, using both UGD models: the Ivanov-Nikolaev and the GBW one. The strange-quark mass is fixed here at $m_q = 0.45$ GeV.}}
 	\label{fig:ratio_comparison_UGD}
 \end{figure} 
 \begin{figure}[htb]
 	\centering
 	\includegraphics[scale=0.50,clip]{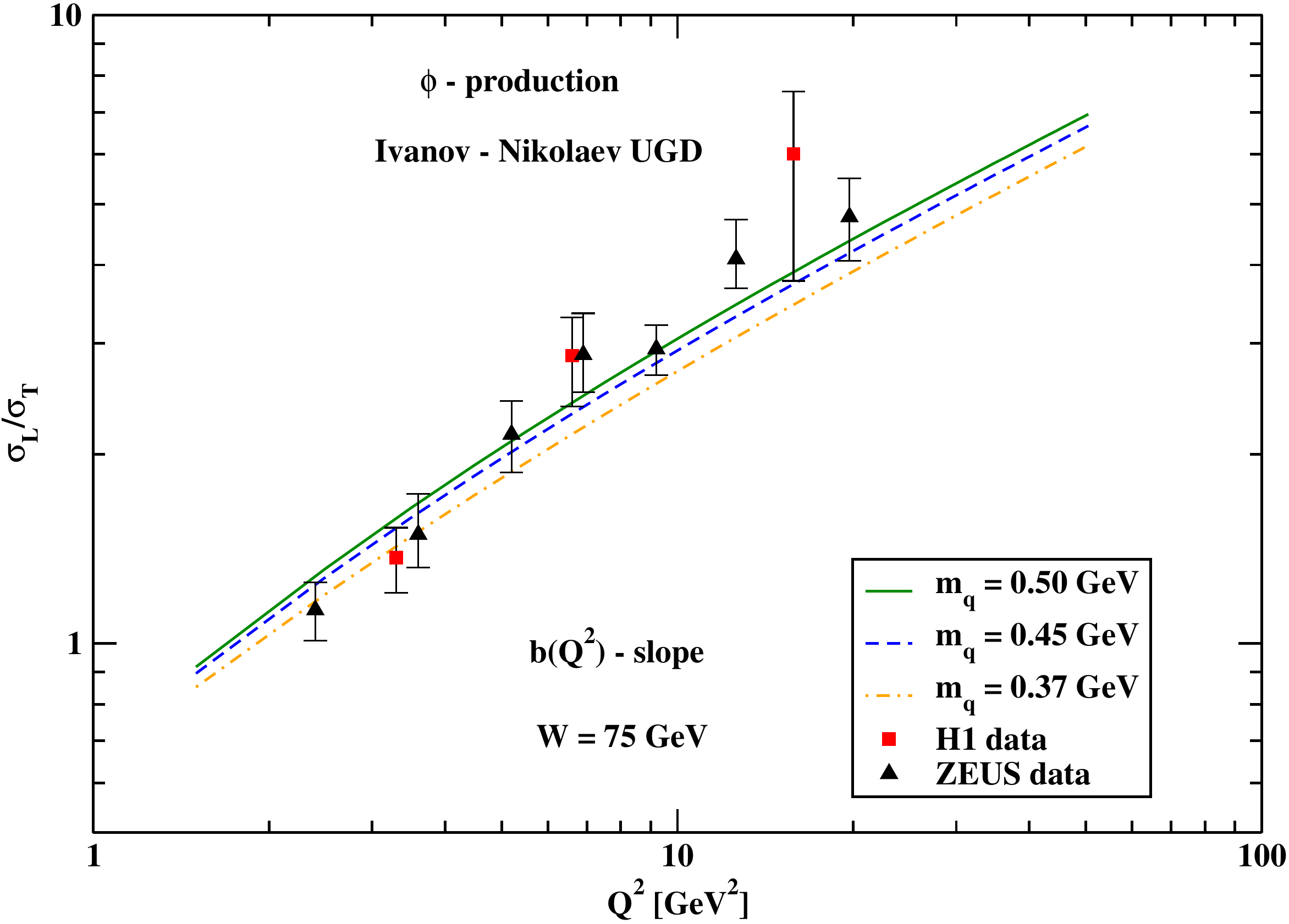}
 	\includegraphics[scale=0.50,clip]{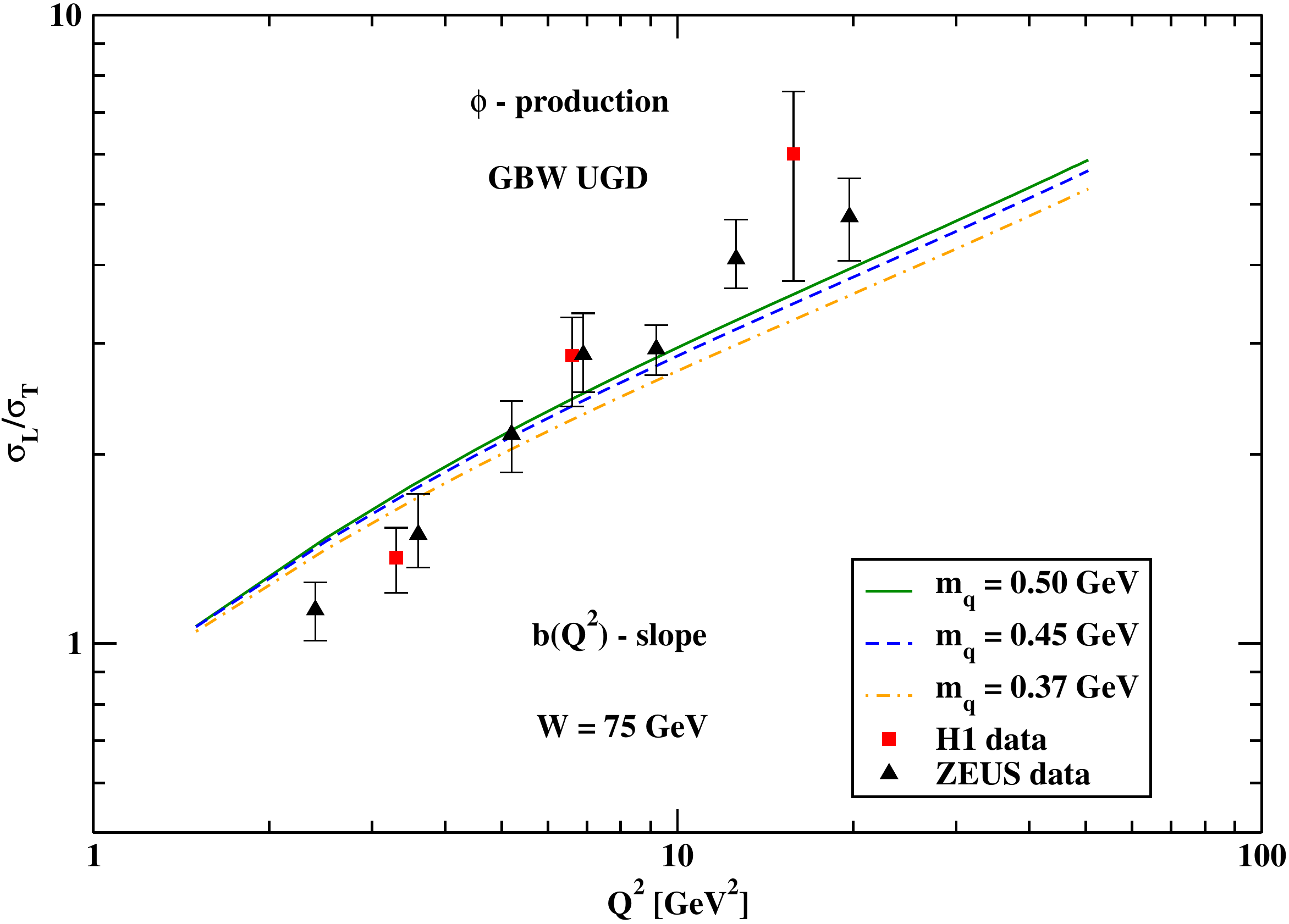}
 	\caption{$Q^2$-dependence of the cross section ratio $\sigma_L/\sigma_T$ for three different quark-mass values $m_q$ using the Ivanov-Nikolaev (top panel) and the GBW (bottom panel) UGDs.}
 	\label{fig:ratio_IN_comparison_masses}
 \end{figure} 
 
 Fig.~\ref{fig:ratio_IN_comparison_masses} shows the dependence of the ratio on the effective quark mass parameter.
 The ratio is much less sensitive to the quark mass than the polarized cross sections
 $\sigma_L$ and/or $\sigma_T$ separately. So the extraction of the mass
 parameter from the normalized cross section is preferred.
 
\begin{figure}[htb]
 	\centering
 	\includegraphics[scale=0.50,clip]{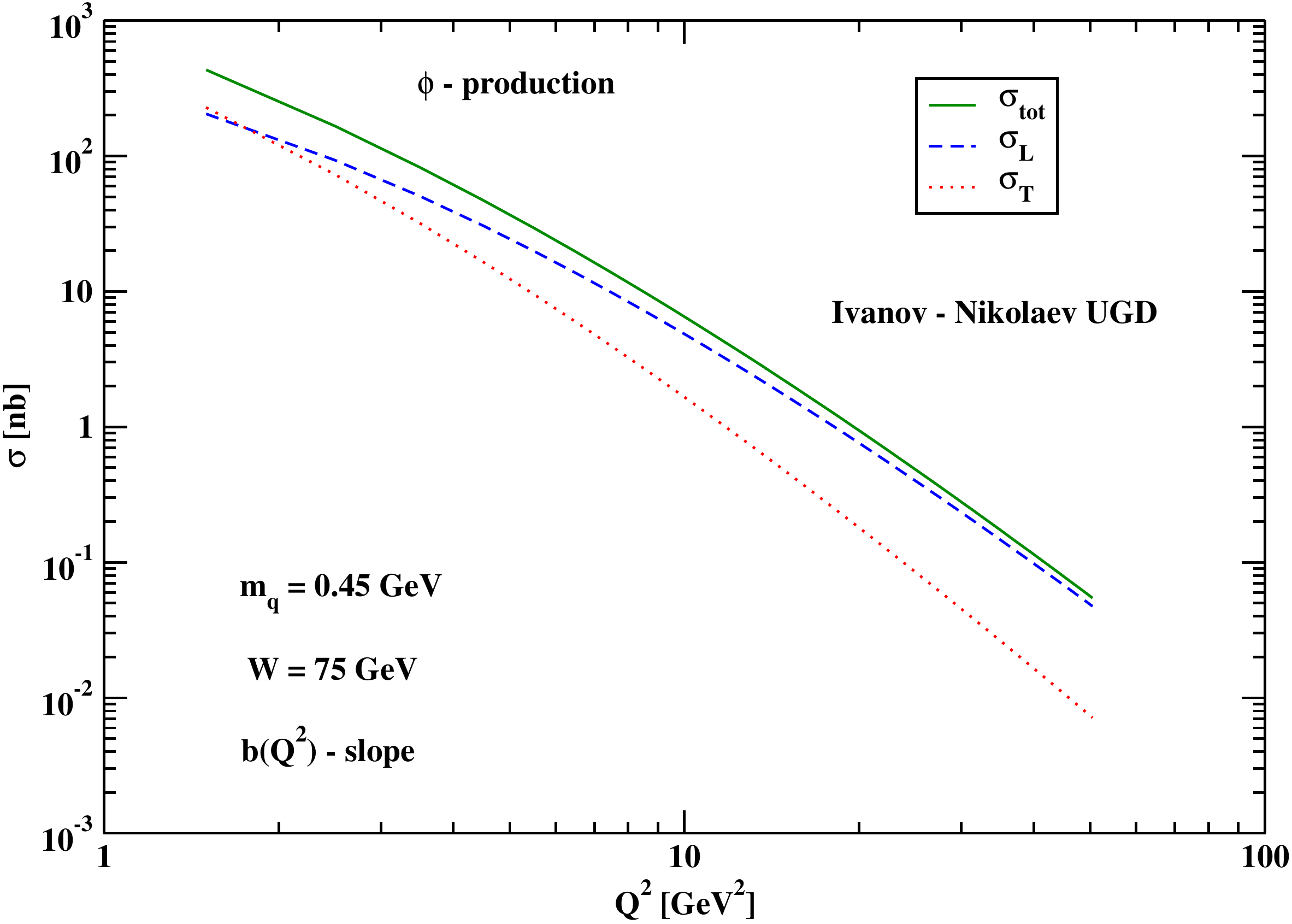}
 	\includegraphics[scale=0.50,clip]{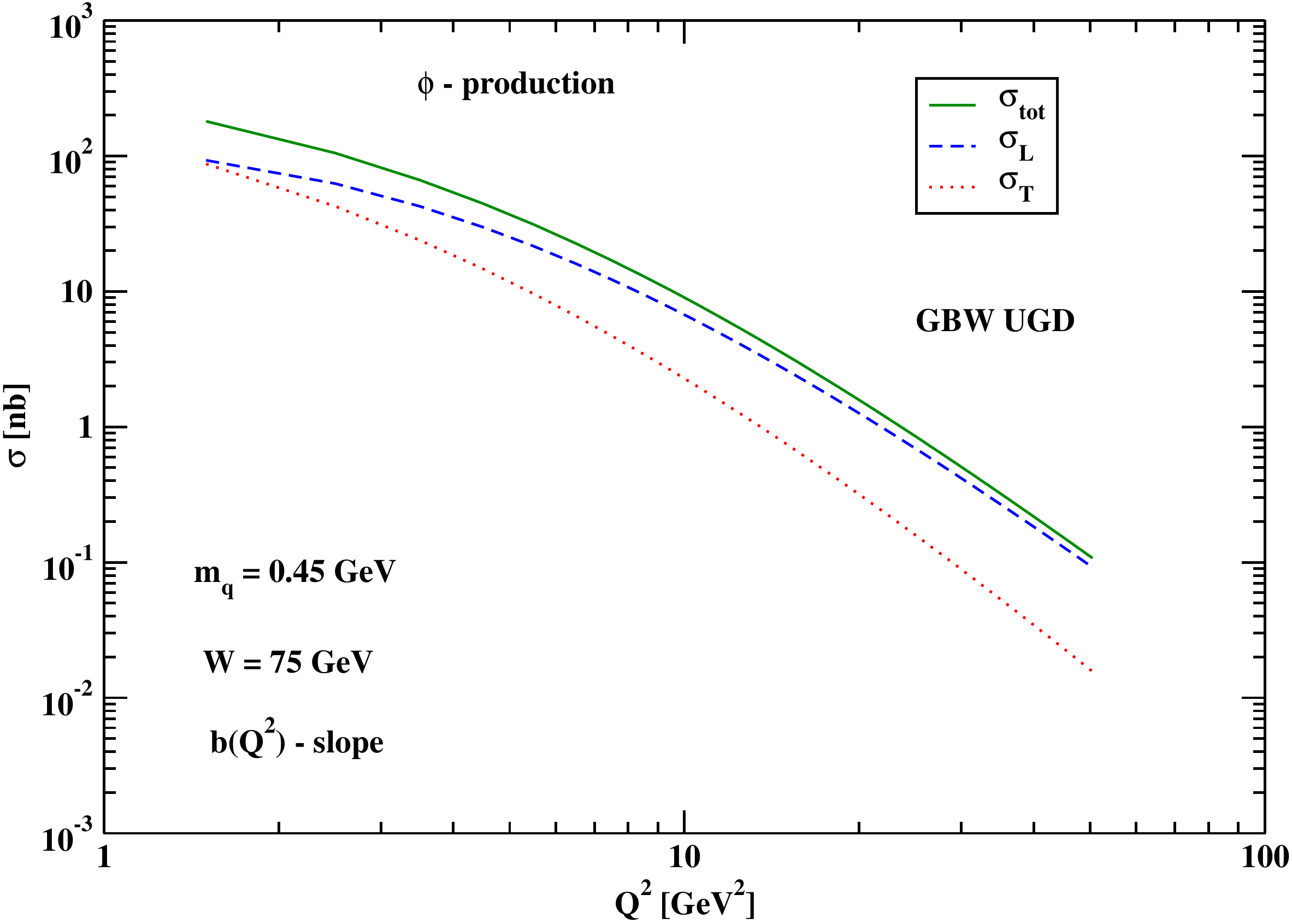}
 	\caption{\emph{Longitudinal, transverse and total cross sections as functions of $Q^2$ using the Ivanov-Nikolaev (top panel) and the GBW (bottom panel) UGDs.}}
 	\label{fig:all_sigma_GBW_IN}
 \end{figure}

\begin{figure}[htb]
 	\centering
 	\includegraphics[scale=0.50,clip]{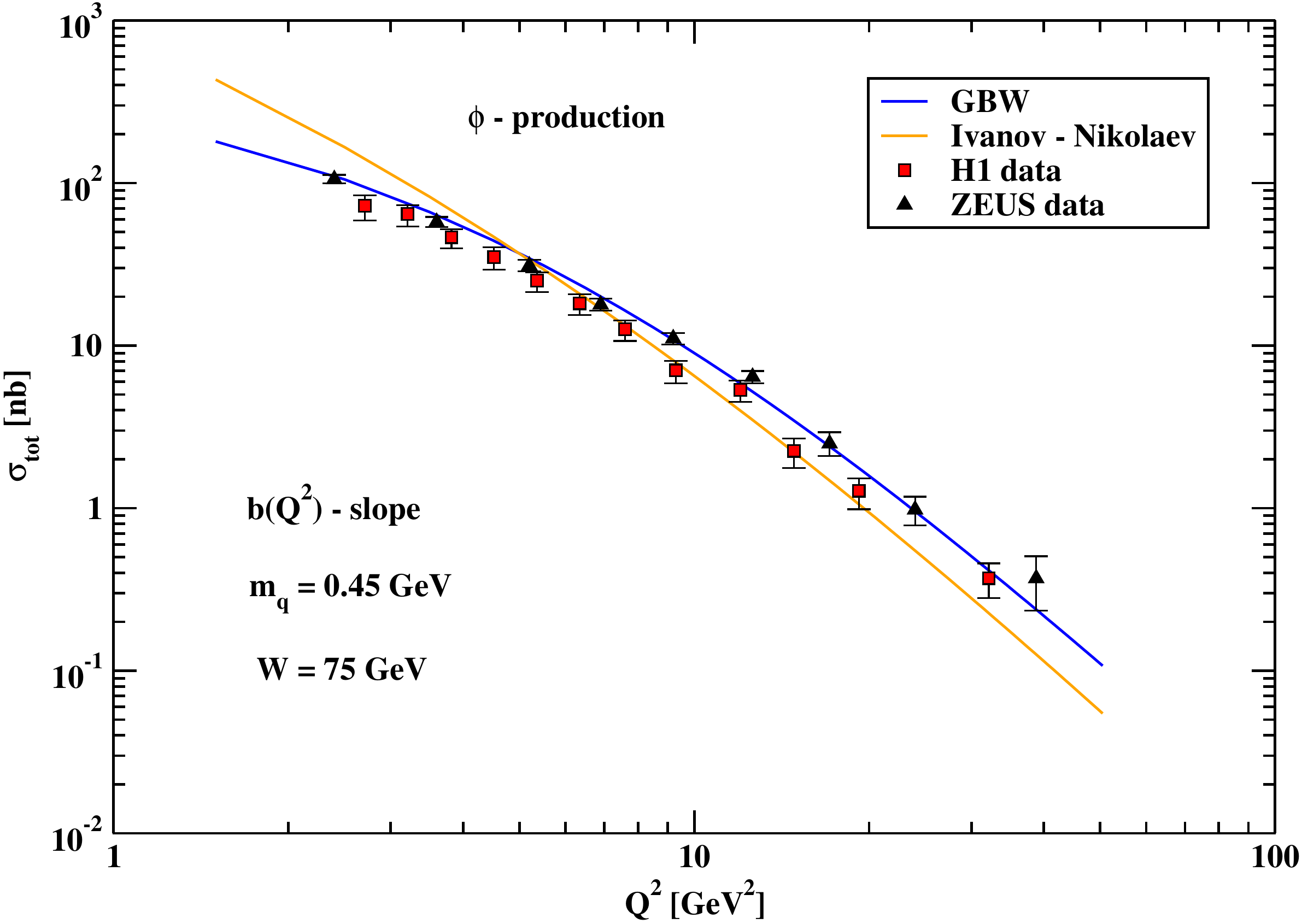}
 	\caption{\emph{$Q^2$-dependence of total cross section $\sigma_{\text{tot}}$ at $W = 75$ GeV for both UGD models in comparison with the H1~\cite{Aaron:2009xp} and ZEUS~\cite{Chekanov:2005cqa} experimental data.}}
 	\label{fig:ZEUS_comparison}
 \end{figure}

 Now we shall show the total cross section $\sigma_{\text{tot}}$ as a function of virtuality.
 In Fig.~\ref{fig:all_sigma_GBW_IN} we show both longitudinal and transverse components as well as
 their sum. The transverse cross section is somewhat steeper (falls
 faster with virtuality) than the longitudinal one.
 The comparison with the HERA data is presented in Fig.~\ref{fig:ZEUS_comparison}.
 The GBW UGD better describes the experimental data at small photon virtualities. There seems to be a small inconsistency of the
 H1 and ZEUS data at larger virtualities.
 
 \subsubsection{Skewness corrections}
 \label{skew}
\begin{figure}[htb]
	\centering
	\includegraphics[scale=0.50,clip]{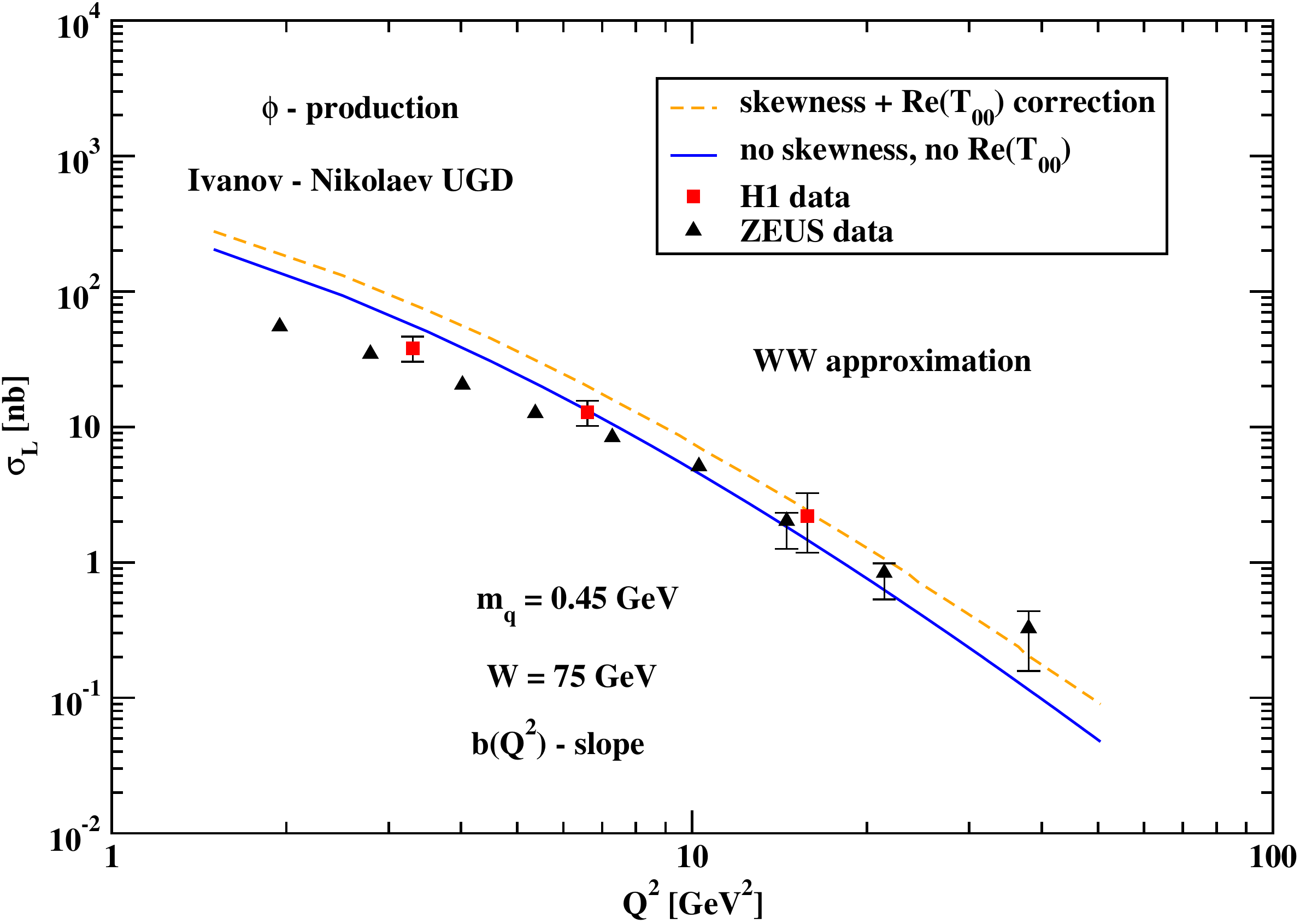}
	\includegraphics[scale=0.50,clip]{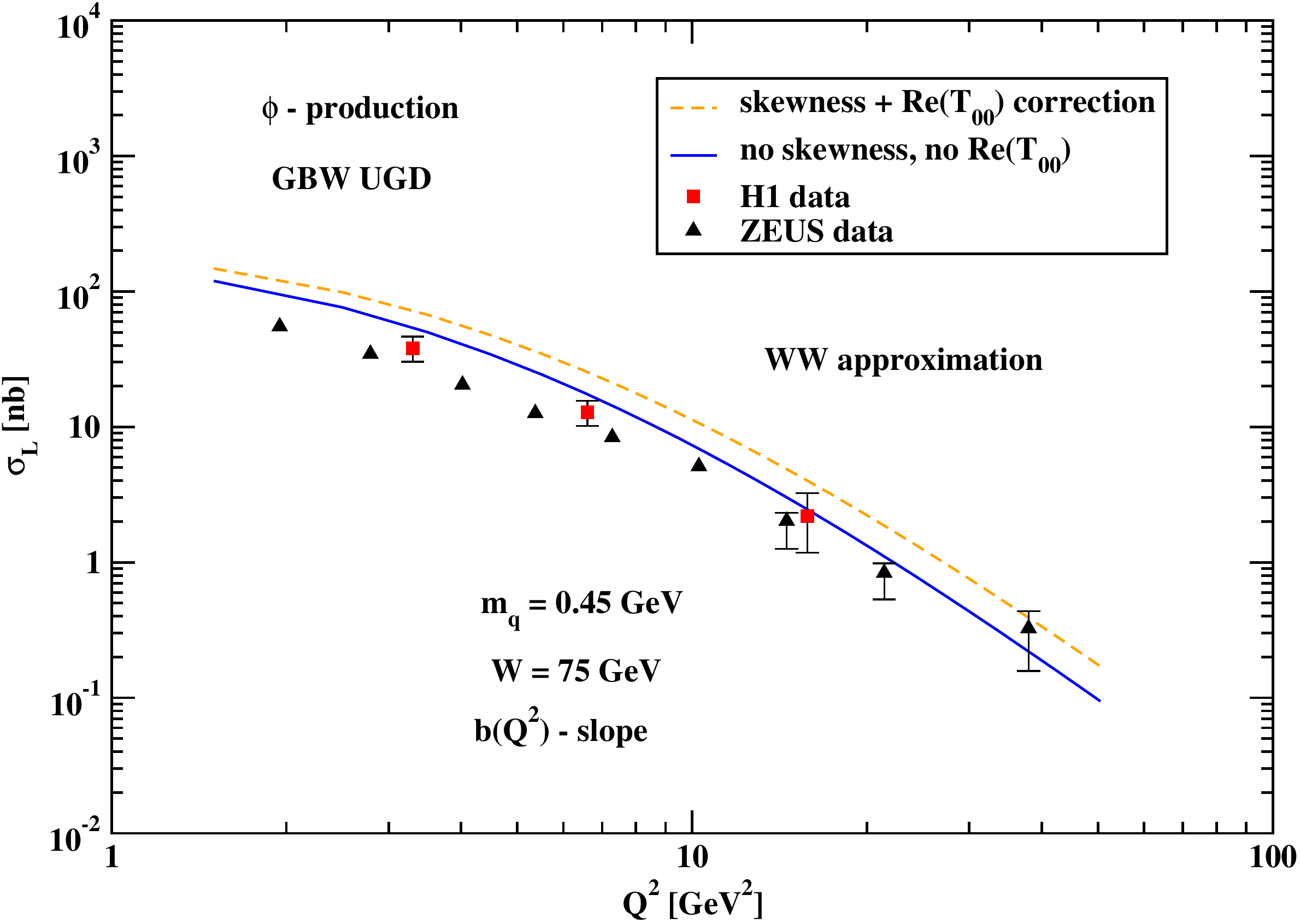}
	\caption{\emph{An estimation of the skewness effects and the real part of the amplitude for the longitudinal cross section $\sigma_L$ calculated in the WW approximation using the Ivanov-Nikolaev (top panel) and GBW (bottom panel) UGDs.}}
	\label{fig:sigL_skewness}	
\end{figure}

\begin{figure}[htb]
	\centering
	\includegraphics[scale=0.50,clip]{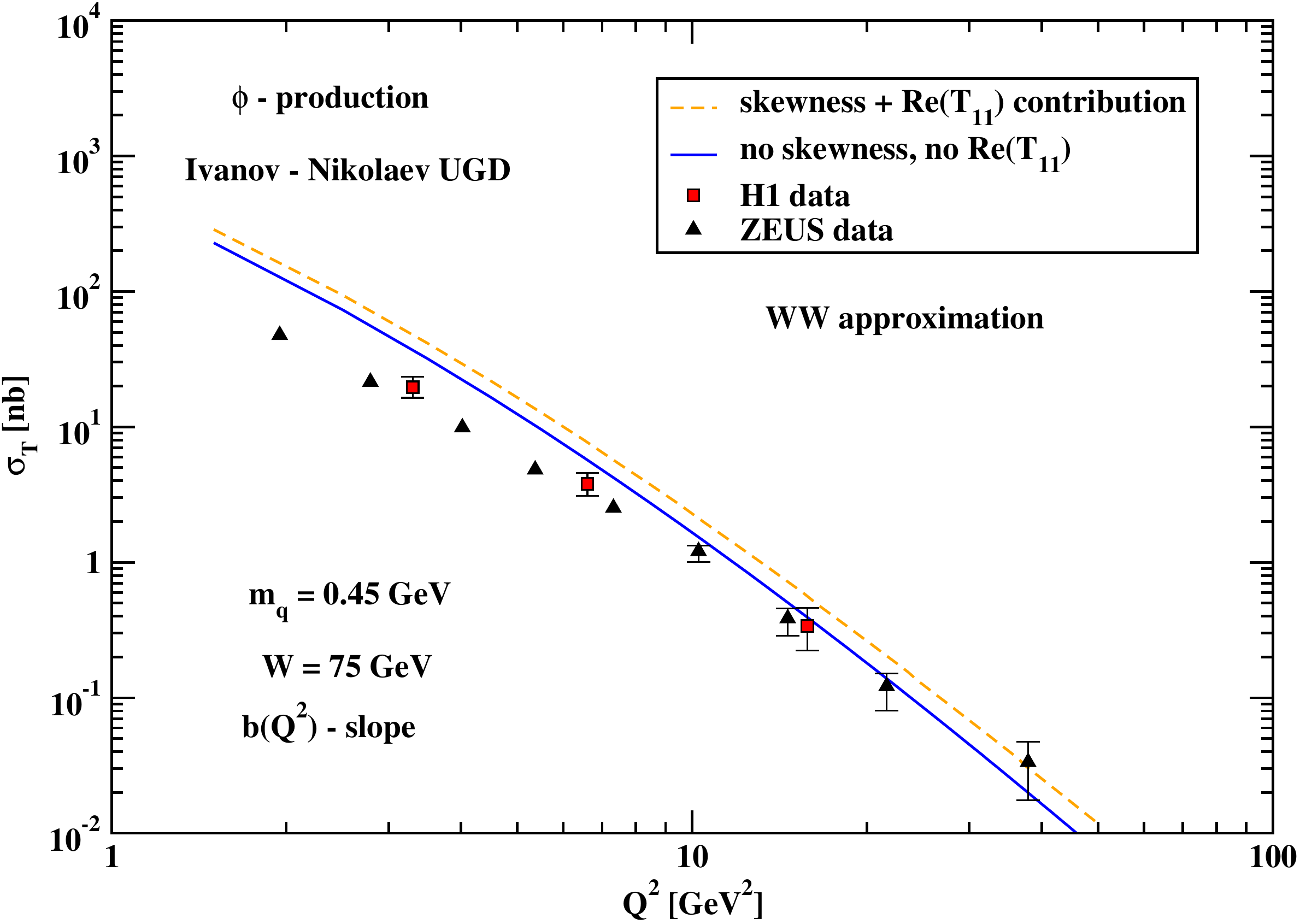}
	\includegraphics[scale=0.50,clip]{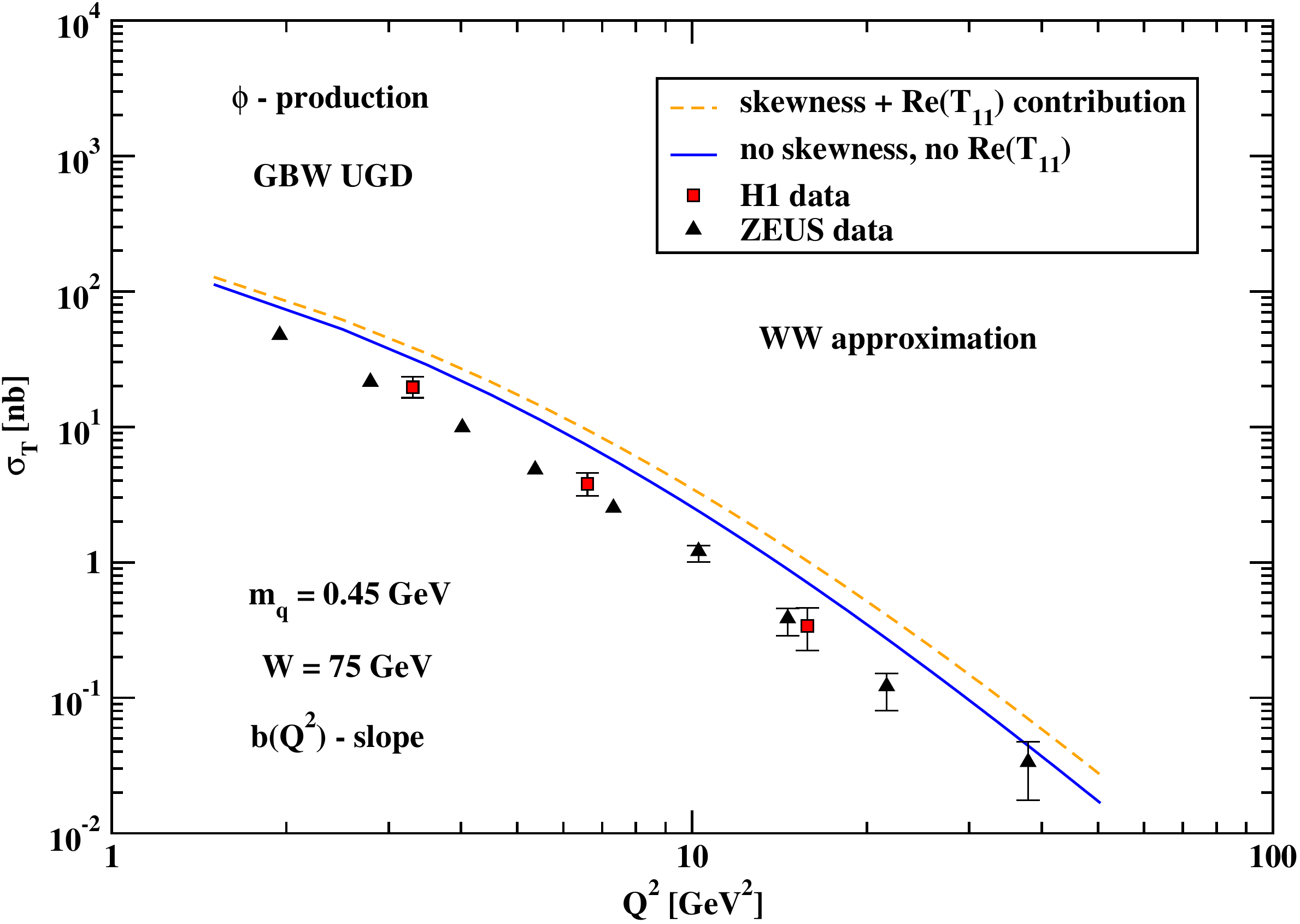}
	\caption{\emph{An estimation of the skewness effects and the real part of the amplitude for the transverse cross section $\sigma_T$  calculated in the WW approximation using the Ivanov-Nikolaev (top panel) and GBW (bottom panel) UGDs.}}
	\label{fig:sigT_skewness}	
\end{figure}

 So far we did not consider skewness effects and the real part of the $\gamma^*\,p \to \phi\,p$ amplitude.
 Both these corrections can be calculated from the energy dependence of
 the forward amplitude. Defining
 \begin{eqnarray}
 \Delta_\PP = {\partial \log 
 	\Big(	\Im m T_{\lambda_V\lambda_\gamma}(s,Q^2)/s \Big)
 	\over \partial \log(1/x)} \, , 
 \end{eqnarray}
 we can calculate the real part from 
 \begin{eqnarray}
 \rho = {\Re e  T_{\lambda_V\lambda_\gamma}(s,Q^2) \over \Im m  T_{\lambda_V\lambda_\gamma}(s,Q^2)} = \tan\Big( {\pi \Delta_\PP \over 2} \Big) \, .
 \end{eqnarray}
 The skewness correction is obtained from multiplying the forward amplitude by the factor~\cite{Shuvaev:1999ce}:
 \begin{eqnarray}
 R_{\text{skewed}} = { 2^{2 \Delta_\PP + 3} \over \sqrt{\pi} } \cdot {\Gamma(\Delta_\PP + 5/2) \over \Gamma(\Delta_\PP+ 4)} \, .
 \end{eqnarray}
 Now we wish to show our estimates of these corrections.
 We show results for longitudinal (Fig.~\ref{fig:sigL_skewness}) and transverse (Fig.~\ref{fig:sigT_skewness}) components separately. The effect is not too big but cannot
 be neglected. The effect of the skewness is much larger than the effect of the inclusion of the real part.
 
\begin{figure}[htb]
	\centering
	\includegraphics[scale=0.50,clip]{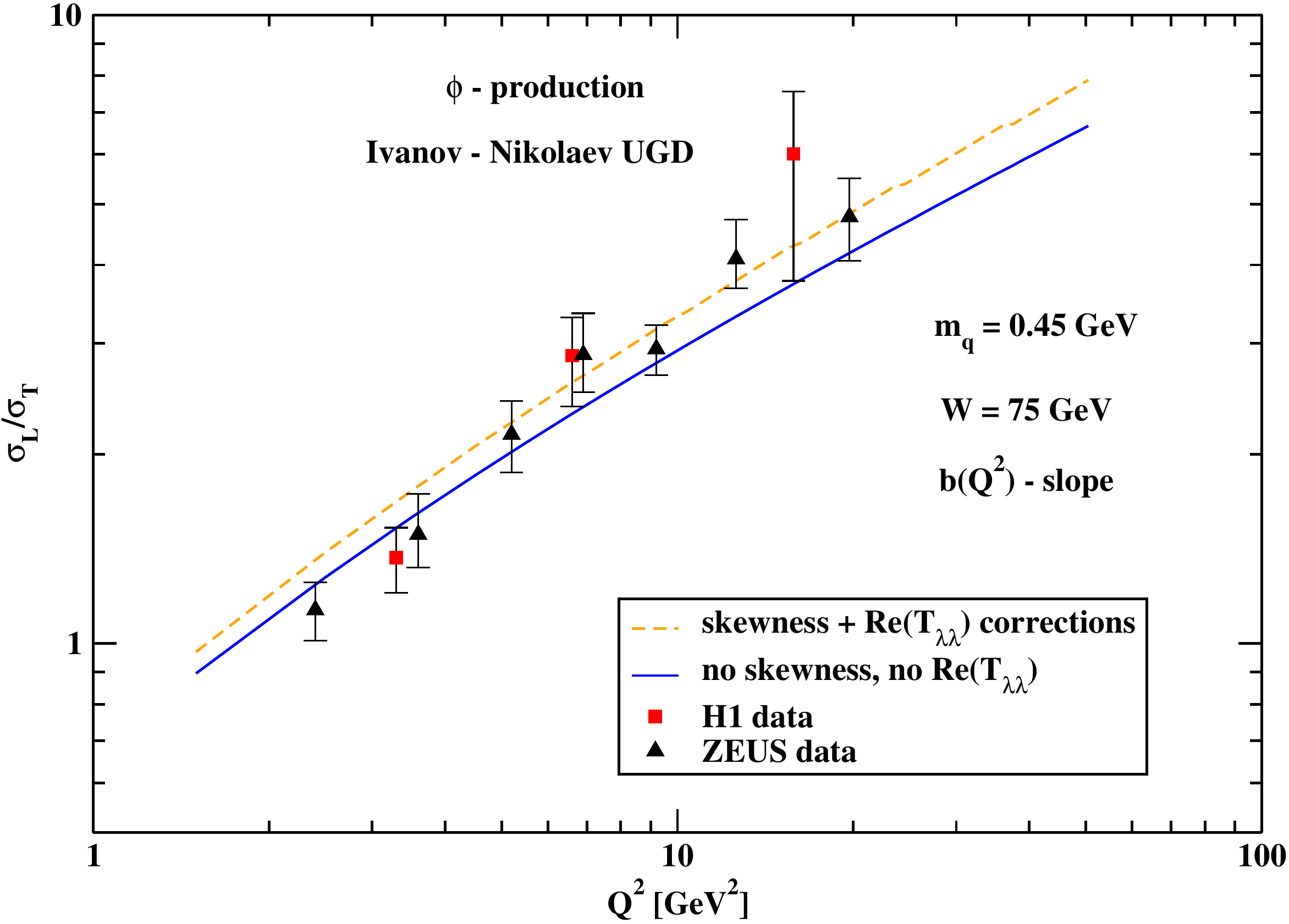}
	\caption{\emph{An estimation of the skewness effects and the real part of the amplitude are shown for the cross sections ratio $\sigma_L/\sigma_T$ through the Ivanov-Nikolaev UGD.}}
	\label{fig:ratio_comparison_skewness}	
\end{figure}

 We observe that the effect of the skewness does not cancel in the $\sigma_L/\sigma_T$ ratio as can
 be seen in Fig.~\ref{fig:ratio_comparison_skewness}.
 
 \FloatBarrier

\FloatBarrier
\subsection{Discussion}
\label{discussion_phi}
We have used a recently formulated hybrid 
formalism for the production of $\phi$ meson in the 
$\gamma^*\,p \to \phi\,p$ reaction using unintegrated gluon distributions and meson distribution amplitudes.
In this formalism the $\gamma^*  \to \phi$ impact factor
is calculated in collinear-factorization using (collinear)
distribution amplitudes. So far this formalism was used
only for massless quarks/antiquarks (e.g. for $\rho$-meson production).
Both twist-2 and twist-3 contributions are included.
The impact factor for the $p \to p$ transition are expressed in terms
of UGDs. Two different UGD models have been used.

We have shown that for massless quarks the genuine three-parton contribution is more
than order of magnitude smaller than the WW one.
Therefore in this paper we have concentrated on the WW component. 

We have observed too quick rise of the cross section
when going to smaller photon virtualities compared to the experimental
data measured by the H1 and ZEUS collaborations at HERA.
This was attributed to the massless quarks/antiquarks.
We have proposed how to include effective quark masses into the formalism.
Corresponding distribution amplitudes were calculated
and have been used in the present approach.
With effective quark mass $m_q \sim$ 0.5 GeV a good description
of the H1 and ZEUS data has been achieved
for the Ivanov-Nikolaev and GBW UGDs down to $Q^2 \sim$ 4~GeV$^2$.
This value of the strange quark mass is similar as the one found 
in Ref.~\cite{Cisek:2010jk},
where the $\kappa$-factorization formalism with $s \bar s$ light-cone wave-function of the $\phi$ meson was used for real photoproduction. 

We have estimated also the skewness effect which turned out to be not
too big but not negligible. We have shown some residual effect of
the skewness for the ratio of longitudinal-to-transverse cross sections.

\chapter*{Conclusions and Outlook} 
\rhead[\fancyplain{}{\bfseries
	CONCLUSIONS AND OUTLOOK}]{\fancyplain{}{\bfseries\thepage}}
\lhead[\fancyplain{}{\bfseries\thepage}]{\fancyplain{}{\bfseries
		CONCLUSIONS AND OUTLOOK}}
\addcontentsline{toc}{chapter}{Conclusions and Outlook} 
We proposed a double-way strategy to investigate the QCD in the high-energy regime through two different semi-hard reaction categories, which can offer significant contributions to the high-energy phenomenology. The former is represented by the inclusive forward/backward processes, with at least an undetected system, in order to further probe the BFKL formalism. As we know, among the most analyzed reactions through the BFKL approach, the Mueller--Navelet jets process emerges. For this reason, in Chapter~\ref{Chap:Incl_emiss}, first, we reminded all useful ingredients related to the inclusive production of two jets well separated in rapidity, and then we presented less inclusive reactions. In particular, Sec.~\ref{jethad} is devoted to the study of the inclusive hadron-jet process in hadroproduction~\cite{Bolognino:2018oth}, characterized by a charged light hadron and a jet with high transverse momenta, separated by a large rapidity gap. We gave predictions for cross section averaged over the azimuthal angle between the identified jet and hadron and for the ratios of azimuthal coefficients, considering both acceptances of CMS and CASTOR detectors. Results are calculated in the NLA accuracy, at $7$ and $13$ TeV, in the LHC domain. Due to the natural features of the objects produced in this channel, asymmetric cuts were naturally considered, thus enhancing the BFKL effects. In Sec.~\ref{heavy}, we presented the inclusive production of forward heavy-quark pair, widely separated in rapidity, in hadroproduction~\cite{Bolognino:2019yls}. Results describing the cross section and the azimuthal coefficients, in the same kinematical regime of the reaction in which a heavy-quark pair is produced in photoproduction, have shown higher cross section values. The other semi-hard channel of interest is represented by single forward emissions in lepton-proton scatterings. In this regard, we illustrated how the exclusive production of light vector mesons, $\rho$~\cite{Bolognino:2018} and $\phi$~\cite{Bolognino:2019pba}, revealed to be a challenging testing ground for the analyses of the UGD models, in the framework of the high-energy factorization. Predictions for the helicity-amplitude ratio and for polarized cross sections were compared with HERA experimental data, providing information related to the $\kappa$-shape of the UGD. Moreover, the observables employed in these studies allowed us to discriminate which models among those ones presented in this treatment better describe the HERA data. With the desire to enrich the study of UGD models, we proposed also a systematic analysis of the $\phi$-meson leptoproduction~\cite{Bolognino:2019pba}, presenting theoretical predictions for the polarized cross sections and their ratio. The study was focused on complementary tests, such as quark-mass effects and the comparison between DAs and LCWF~\cite{Cisek:2010jk} approaches.\\
The study of high-energy phenomenology, with the aim to extend our knowledge regarding BFKL dynamics and strong interactions, can be enriched by a selection of new semi-hard reactions. For this reason, we propose here some future developments. The field related to two-body emissions can be extended to the investigation of theoretical production mechanism~\cite{Brambilla:2010cs,Bodwin:2013nua,Andronic:2015wma} for the quark-bound states, taking inspiration from the most popular models proposed so far~\cite{Fritzsch:1977ay,Halzen:1977rs,Bodwin:1994jh}.
It is worth to highlight that processes as the inclusive hadroproduction of a Higgs boson and a jet well separated in rapidity~\cite{Celiberto:2020tmb} (further investigations are proposed in Ref.~\cite{Hentschinski:2020tbi} for forward Higgs production at NLO accuracy within high-energy factorization), are able to stabilize the BFKL series: large transverse mass of the Higgs boson suppress the higher order corrections, thus affording us a peerless opportunity to study energy scales around their natural values. In this way we have the chance to perform precision calculations via BFKL resummation. However, studying the Higgs transverse-momentum distribution, different kinematic regions are explored, allowing the applicability of specific resummation approaches. This hints the possibility to build a joined approach where different resummations are involved and play their role. For instance, we remind the Altarelli--Ball--Forte (ABF) formalism~\cite{Ball:1995vc,Ball:1997vf,Altarelli:2001ji,Altarelli:2003hk,Altarelli:2005ni,Altarelli:2008aj,White:2006yh}, where DGLAP and BFKL inputs are combined to enhance the perturbative accuracy of the resummed series. Another possibility is represented by the Catani--Ciafaloni--Fiorani--Marchesini (CCFM) scheme~\cite{Ciafaloni:1987ur,Catani:1989sg,Catani:1989yc,Marchesini:1994wr}, which embodies the DGLAP equation at intermediate-$x$ and the BFKL one at small-$x$, giving a global evolution framework for UGDs. A generalized form of CCFM formalism to consider non-linear effects in the gluon evolution has been proposed in Ref.~\cite{Kutak:2011fu} (CCFM--K equation) and recently used for phenomenological studies on the Drell--Yan channel~\cite{Golec-Biernat:2019scr}.\\
Numerous advantages can be obtained also from single forward emission channels. On the one side, further phenomenological studies can be proposed continuing to test UGD models, as well as considering other processes as testfield of this parton density, such as heavy-quark or heavy-meson productions. On the other side, an intriguing way to develop this field would be planning the UGD extraction from global fit, with the possibility to select both inclusive and exclusive channels, paying attention to transverse-momentum dependent (TMD) inputs in the BFKL evolution framework. This research line could open further analyses on the theoretical connection between the two parton densities, UGDs and TMDs.

\clearpage{\pagestyle{empty}\cleardoublepage}

\renewcommand{\chaptermark}[1]{\markright{\thechapter \ #1}{}}
\lhead[\fancyplain{}{\bfseries\thepage}]{\fancyplain{}{\bfseries\rightmark}}
\appendix                               



\end{document}